\PassOptionsToPackage{pdfpagelabels=false}{hyperref} 
\documentclass[fleqn,usenatbib]{mnras}

\usepackage{newtxtext}
\usepackage[varg]{newtxmath}

\usepackage[T1]{fontenc}
\usepackage{ae,aecompl}
\usepackage[table]{xcolor}
\usepackage{multirow}
\usepackage{lscape} 

\usepackage{graphicx}	
\usepackage{color}
\usepackage{journals}

\newcommand{\Msun}{${\rm M_\odot}$}
\newcommand{\msun}{{\rm M_\odot}}

\newcommand{\izw}{I$\,$Zw$\,$18}

\definecolor{darkgreen}{rgb}{0.0,0.5,0.0}
\definecolor{darkred}{rgb}{0.7,0.0,0.0}
\definecolor{brown}{rgb}{0.95,.35,0.}
\definecolor{grey}{rgb}{0.4,0.5,0.55}
\definecolor{royalblue}{rgb}{0,0.5,1}
\definecolor{violet}{rgb}{0.5,0,0.7}
\definecolor{lightgrey}{rgb}{0.85,0.9,0.95}

\newcolumntype{a}{>{\columncolor{grey}}c}
\definecolor{cgcol}{rgb}{0.0,0.0,0.627}
\definecolor{mnwcol}{rgb}{0.0,0.565,0.0}
\definecolor{bwccol}{rgb}{0.561,0.0,0.561}
\definecolor{mgpcol}{rgb}{0.942,0.,0.}
\definecolor{guocol}{rgb}{0,0.56,0.56}

\title[the properties of very young galaxies in the local Universe]{The properties and environment of very young galaxies in the local Universe} 

\author[M. Trevisan et al.]
{M. Trevisan$^{1,2}$\thanks{E-mail:marina.trevisan@ufrgs.br},
G. A. Mamon$^2$,
T. X. Thuan$^{3,2}$,
F. Ferrari$^{4}$,
L.~S.~Pilyugin$^5$,
A.~Ranjan$^{6,2}$
 \\ 
$^1$Universidade Federal do Rio Grande do Sul -- Departamento de Astronomia -- 91501-970, Porto Alegre-RS, Brazil\\
$^2$Institut d'Astrophysique de Paris (UMR 7095: CNRS \& Sorbonne Universit\'e), 98 bis Bd Arago, F-75014 Paris, France\\ 
$^3$Astronomy Department, University of Virginia, P.O. Box 400325, Charlottesville, VA 22904-4325\\ 
$^4$Instituto de Matem\'atica Estat\'\i stica e F\'\i sica, Universidade Federal do Rio Grande, 96201-900, Rio Grande, RS, Brazil \\
$^5$Main Astronomical Observatory of National Academy of Sciences of Ukraine, 27 Zabolotnogo str., 03680 Kiev, Ukraine\\
$^6$Korea Astronomy and Space Science Institute, 776 Daedeok-daero, Yuseong-gu, Daejeon 34055, Republic of Korea \\
}

\date{Accepted yyyy month dd. Received yyyy month dd; in original form yyyy month dd} 

\pubyear{2018}

\begin{document}

\label{firstpage}
\pagerange{\pageref{firstpage}--\pageref{lastpage}} 

\maketitle

\begin{abstract}
In the local Universe, there is a handful of dwarf compact star-forming galaxies with extremely low oxygen abundances. It has been proposed that they are young, having formed a large fraction of their stellar mass during their last few hundred Myr. However, little is known about the fraction of young stellar populations in more massive galaxies. In a previous article, we analyzed  280\,000 SDSS spectra to identify a surprisingly large sample of more massive Very Young Galaxies (VYGs), defined to have formed at least $50\%$ of their stellar mass within the last 1 Gyr. Here, we investigate in detail the properties of a subsample of 207 galaxies that are VYGs according to all three of our spectral models. We compare their properties with those of control sample galaxies (CSGs). We find that VYGs tend to have higher surface brightness and to be more compact, dusty, asymmetric and clumpy than CSGs. Analysis of a subsample with H{\sc i} detections reveals that VYGs are more gas-rich than CSGs. VYGs tend to reside more in the inner parts of low-mass groups and are twice as likely to be interacting with a neighbour galaxy than CSGs. On the other hand, VYGs and CSGs have similar gas metallicities and large scale environments (relative to filaments and voids).  These results suggest that gas-rich interactions and mergers are the main mechanisms responsible for the recent triggering of star formation in low-redshift VYGs, except for the lowest mass VYGs, where the starbursts may arise from a mixture of mergers and gas infall.
\end{abstract}


\section{Introduction}
\label{sec:introduction}
In the past decades, very deep photometric and spectroscopic surveys have
enabled astronomers to trace the cosmic star-formation history (CSFH) from these early epochs to the present \citep[][and references therein]{Madau.Dickinson:2014}. These
observations show that the cosmic star formation rate density peaked
approximately $3.5\,$Gyr after the Big Bang ($z \sim 1.9$) and declined
exponentially afterwards, with less than $\sim 2\%$ of the stars in the
Universe being formed within the last $1\,$Gyr. However, the CSFH is averaged
over large volumes and might not represent the star formation histories (SFHs) of individual galaxies. On the contrary, the SFHs of galaxies are observed to vary noticeably with galaxy properties such as mass \citep{Balogh.etal:2009,McGee.etal:2011,Trevisan.etal:2012,Woo+13}, morphology \citep*[e.g.][]{Ferrari.etal:2015} and with the environment where the galaxies reside \citep{Weinmann+06,Peng+10,vonderLinden+10}. Among the large variety of assembly histories, the systems that formed more than half of their stellar mass in the last Gyr, which we call hereafter \emph{very young galaxies} (VYGs), are of particular interest because they offer us a very close view of galaxy formation, allowing us to identify the physical mechanisms that govern the recent growth of stellar mass via gas accretion or gas-rich mergers. 

In the local Universe, there are a few low-mass  star-forming galaxies that have extremely low oxygen abundances, an indication that they could have formed most of their stars only recently ($<\,1\,$Gyr). A well-studied example is the galaxy \izw, with $12+\log ({\rm O/H})\sim 7.2$ \citep{Skillman.Kennicutt:1993, Izotov.Thuan:1998}, i.e., $\sim 3\%$ of solar metallicity, adopting $12+\log ({\rm O/H})= 8.7$ for the Sun
(Asplund et al. 2009). Using the Hubble Space Telescope (HST) to 
resolve the stellar content
of \izw\ and construct colour-magnitude diagrams, \citet{Izotov.Thuan:2004}
estimated that most of its stellar mass was formed within the last
$500\,$Myr. However, deeper HST images revealed the presence of an older
stellar component (ages $\gtrsim 1-2\,$Gyr, \citealp{Aloisi.etal:2007,
ContrerasRamos.etal:2011}). Nevertheless, the fraction of the total stellar
mass corresponding to this old population remains undetermined, and if it is
less than $50\%$, galaxies such as \izw\ could still be classified as
VYGs. Other examples of star-forming compact dwarf galaxies that are
extremely metal-poor and that have had a considerable fraction of their
stellar mass formed during their most recent burst of star formation (SF), a few
million years ago, are J0811+4730, the most metal-deficient star-forming
galaxy known \citep{Izotov.etal:2018}
and J1234+3901 \citep*{Izotov+19}.

Recent works have suggested that not only dwarf galaxies like \izw, with
stellar masses $M_{\star}\sim10^7-10^8\, {\rm M}_{\odot}$, but more massive
systems can also be very young. 
Using low resolution ($R\,\sim\,30$) optical
spectral energy distributions (SEDs) combined with $ugrizJK$ broad-band
photometry, \citet*{Dressler.etal:2018} derived the SFHs of galaxies at $0.45
< z < 0.75$ and identified a population of objects with $M_{\star} >
10^{10}\,$\Msun\ that formed at least $50\%$ of their stars within the last
$\sim 2\,$Gyr of the time of observation. They predicted that
these \emph{Late Bloomers} , account for about $30\%$ of the Milky-Way-sized
galaxies ($M_{\star} \sim 10^{10.5}\,$M$_{\odot}$) at $z = 0.7$, but their
frequency declines after $z = 0.3$, and they effectively disappear at the
present epoch.

Motivated by the debate on the possible VYG
nature of very low metallicity galaxies such as \izw,
we \citep[][hereafter
Paper~I]{Tweed+18} independently predicted the frequency of
VYGs in the local Universe using analytical and semi-analytic models of
galaxy formation. We found that the predicted fraction of VYGs
depends on galaxy stellar mass, as well as on the model of galaxy formation
adopted. The fraction of galaxies with $10^8 < M_{\star} < 10^{10}\,$\Msun\
that are VYGs ranges between $ \sim 0.4\%$ up to $4\%$ depending
on the model, but the fraction of massive galaxies ($M_{\star} \gtrsim
10^{11}\,$\Msun)  that are VYGs is less than $\sim 0.01\%$ for all models.

In a second article (\citealp{Mamon+20}, hereafter Paper~II), we used the
galaxy SFRs inferred from the Sloan Digital Sky Survey - Data Release 12
(SDSS-DR12) spectra to identify the VYGs and computed the fraction of these
systems in the local Universe. We found that the observed VYG fractions
decreases more gradually with stellar mass when compared to the predictions
from Paper~I. We also found that the fractions of VYGs strongly depends on
the spectral model used in the spectral fitting procedure, with differences
up to $1\,$dex between different models, and that the number of VYGs in the SDSS can be as
high as a few tens of thousands, depending on the model.
Finally, Paper~II discusses the possibility that old stellar populations can 
dominate the mass despite 1) the spectral fitting and 2) the conservative choice of selecting
VYGs that have blue colour gradients.

In the present work, we explore in detail the properties of a subsample of the large sample of VYG candidates identified in Paper~II, as well as the environment where they reside. 
To minimize the dependency of SFHs on the spectral model used and select a
reliable sample of VYGs, we require that each galaxy in the subsample has
more than half of its stellar mass younger than $1\,$Gyr, according to each of
three different spectral models. 

Our article is organized as follows: 
in Sect.~\ref{sec:data}, we describe how we select the VYGs and the control sample galaxies (CSGs). In the following sections, we describe the properties (Sect.~\ref{sec:properties}), gas content (Sect.~\ref{sec:gas}) and  environment (Sect.~\ref{sec:environment}) of the VYGs, and compare them with those of the CSGs.
In Sect.~\ref{sec:discuss}, we summarise how the VYGs differ from other galaxies and discuss our results.
Finally, our conclusions are given in Sec.~\ref{sec:concl}.
Throughout the paper, we adopt the seven-year Wilkinson Microwave Anisotropy Probe cosmological parameters of a 
flat Lambda Cold Dark Matter Universe with $\Omega_{\rm m} = 0.275$, $\Omega_{\Lambda} = 0.725$ and 
$H_0 = 70.2\,$km$\, {\rm s}^{-1}\,$Mpc$^{-1}$ \citep{Komatsu.etal:2011}.

\section{Data and Sample Selection}
\label{sec:data}

\subsection{Ages from galaxy spectra}
\label{sec:data:sfh}

Galaxy SFHs and ages were derived from the SDSS-DR12 spectra using stellar population synthesis analysis.
Because they have a fairly high spectral resolution ($R \approx 2000$), a good signal to noise 
(${\rm S/N} \gtrsim 10$), 
cover a wide spectral range (from below $3800\,$\AA\ to $9200\,$\AA) and are flux calibrated, 
the SDSS spectra allow the derivation of the SFH of each galaxy, by matching the 
observed spectral energy distribution with a linear combination of single
stellar population (SSP) spectra with non-negative coefficients.

As described in Paper~II, we considered two non-parametric algorithms to estimate the SFH:
the STARLIGHT \citep{CidFernandes+05} algorithm, 
and the VESPA database \citep{Tojeiro+09}.
The two algorithms have been run using the
\citeauthor{Bruzual&Charlot03} (\citeyear{Bruzual&Charlot03}, hereafter BC03) 
model,  calculated with Padova 1994 
stellar evolution tracks \citep{Bressan+93,Fagotto+94a,Fagotto+94b,Girardi+96} and
assuming the \cite{Chabrier03} initial mass function (IMF).
The BC03 model employs the STELIB stellar library \citep{LeBorgne+03}.
VESPA has also been run using the SSPs of \citet[][M05]{Maraston05} 
based on the \cite{Kroupa01} IMF. We have also run STARLIGHT with
the Medium resolution Isaac Newton Telescope Library of Empirical Spectra
(MILES, \citealp{Sanchez-Blazquez+06}), using the updated version 10.0
(\citealp{Vazdekis+15}, hereafter V15) of the code presented in
\cite{Vazdekis+10}.
The V15 models were computed  with the \cite{Kroupa01} IMF, and
stellar evolution tracks from BaSTI 
(Bag of Stellar Tracks and Isochrones, \citealp{Pietrinferni+04,Pietrinferni+06}).
While STARLIGHT was run assuming a screen dust model and \citet*{Cardelli.etal:1989} reddening curve, VESPA assumed either a mixed
slab interstellar dust model \citep{Charlot&Fall00} or combined it with
extra dust around young stars (also from \citeauthor{Charlot&Fall00}).

We ran STARLIGHT considering 15 bins of ages, ranging from 30~Myr up to 13.5~Gyr,
and 6 metallicity bins between [M/H]$\,= -1.3$ and $+0.4$. 
The SFHs from VESPA were obtained considering 16 bins of ages (from 20~Myr to 14~Gyr) and 4 (for BC03 models) or 5 (for M05 models) bins of metallicity. 
For each galaxy, and for each one of the models,
we compute the \emph{median} age
when the stellar mass was half of its final value by interpolating the
cumulative fractional mass as a function of the logarithm of age.

In summary, we have 6 different SFH estimates for each galaxy: those determined with VESPA using BC03 and M05 models, and assuming two 
different dust modelling in each case; and those obtained with STARLIGHT using BC03 and V15 models. We have investigated how these 6 SFHs correlate with other
indicators of recent SF activity, such as the H$\alpha$ equivalent width. We found that, 
among all models, 
 three (V15 with STARLIGHT, and the 2-component dust BC03 and M05 models with VESPA) lead to a tighter correlation
between $f_{\rm young}$ and H$\alpha$ EW (see Paper~II for details). Therefore, we have selected these three models as the benchmark for the rest of our analysis.
Moreover, because individual SFHs are not fully reliable,
we classify a galaxy as a VYG only if all these three models agree on the youth of the galaxy (see Sect.~\ref{sec:data:samples}). 

\subsection{Handling aperture effects}
\label{sec:data:aperture}

One of the main issues when dealing with SFHs derived from SDSS spectra is the finite aperture of the SDSS fibres (3 arcsec diameter). These spectra sample only 
the inner regions of most galaxies. 
To account for the stars lying outside of the fibre aperture, the stellar masses obtained through the SPS analysis of the SDSS spectra, as described in Sect.~\ref{sec:data:sfh}, are corrected by a factor 
${\rm dex}[0.4 (m_{{\rm fiber}, z} - m_{{\rm model}, z}]$,
where $m_{{\rm fiber}, z}$ and $m_{{\rm model}, z}$ are the fiber and model magnitudes in the $z$ band. This correction assumes that the SFH is homogeneous throughout the galaxy. We use the $z$-band magnitude because it traces the stellar mass of the galaxy better than bluer filters.

Although the value of the total stellar mass is not strongly affected by assuming the same SFH throughout the galaxy, the finite aperture of the SDSS fibres has a more important consequence on the selection of the VYGs. Many galaxies with
blue nuclei have redder envelopes. In fact, 17 per cent of all galaxies in our sample have redder global colours than their fibre colour.
Therefore, using spectra within fibres that probe only the nuclei
would make us classify mistakenly such galaxies as VYGs, even though the bulk of their
stellar mass (in the outer regions) is old.

As discussed in Paper~II, this aperture effect is potentially serious, since the median fraction of galaxy light subtended by the fibre, $f_{L}$, is only 26 per cent, and only 3 per cent of our sample have fibres collecting more than half of the galaxy light. Therefore, instead of limiting our sample to galaxies with high $f_L$ values, we follow the approach of Paper~II and require a VYG galaxy to be one where the $g-i$ colour of the global galaxy (using model magnitudes) is bluer than the corresponding fibre colour. 

\subsection{Selection of VYG and control samples}
\label{sec:data:samples}

\subsubsection{Selection of very young galaxies}

For our comparison of the properties of VYGs to that of a control sample, we
started with the sample of 404\,931 SDSS MGS galaxies with $0.005 < z < 0.12$
that had good spectra and colours (the {\tt clean} sample
described in Sect.~2 of Paper~II).\footnote{The table
with the properties of the galaxies of the {\tt clean} sample is available
at \url{http://vizier.u-strasbg.fr/viz-bin/VizieR?-source=J/MNRAS/492/1791}.}
We then assembled a sample of VYGs that satisfy the following six additional conditions:

\begin{itemize}
\itemsep 0.25\baselineskip
\itemindent 0pt
\item[]{\sc i}) their median ages are younger than 1 Gyr according to all 3 spectral models (V15 with STARLIGHT, and the 2-component dust BC03 and M05 models with VESPA; 1214 galaxies); 

\item[]{\sc ii}) their stellar masses derived with STARLIGHT using V15 models are $\geq 10^8\,$M$_{\odot}$ (838 galaxies);

\item[]{\sc iii}) the signal to noise of their SDSS spectra satisfy $\rm S/N \geq 10$, where S/N is computed within a window of $50\,$\AA\ centered at $4755\,$\AA\ (rest frame; 654 galaxies);

\item[]{\sc iv}) their specific SF rates (sSFRs) are available in the MPA/JHU {\tt SpecLineExtra} table of the SDSS database (634 galaxies);  

\item[]{\sc v}) they do not lie in the AGN region of the BPT diagram \citep*{BPT81} -- we adopted the relation by \citet{Kewley+01} to separate the star-forming and AGN galaxies (633 galaxies);

\item[]{\sc vi}) they show blue colour gradients: 
$\Delta (g-i) = (g-i)_{\rm model} - (g-i)_{\rm fiber}  \leq 0$ (207 galaxies).
\end{itemize}

The number of galaxies that remain in the sample after each cut is indicated in parentheses. The list of criteria above is very conservative, and yield a relatively small sample of 207 VYGs. As shown in Paper~II, the number of VYGs in the SDSS can be as high as a few tens of thousands, but to avoid possible contaminants in the sample, we opted for the conservative approach. 
The images of these 207 VYGs are shown in Figs.~\ref{fig:vyg_images1} to \ref{fig:vyg_images4}.

\subsubsection{Control sample}


%
\begin{figure*}
\centering
\includegraphics[width=0.9\hsize]{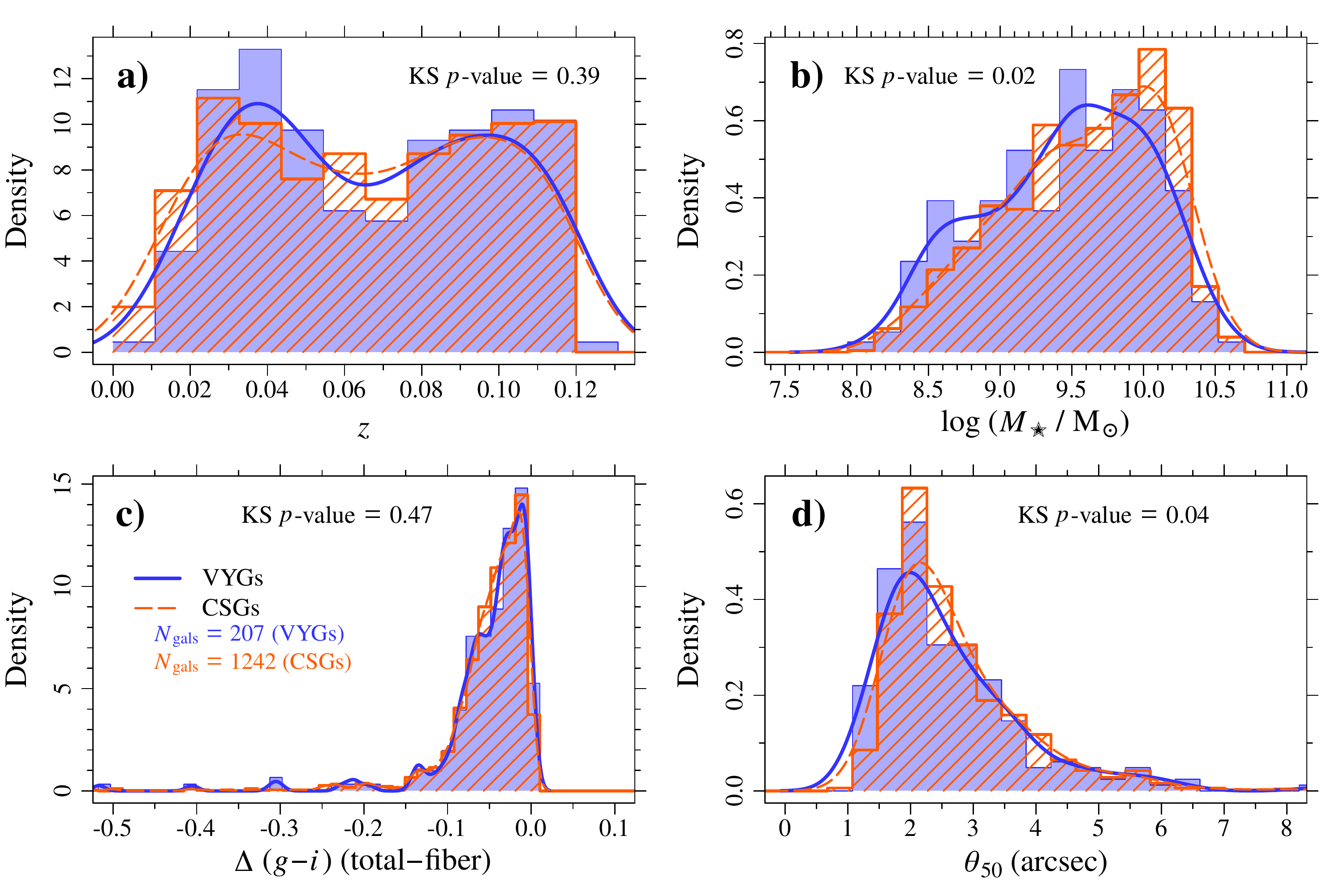} 
\caption{Redshifts ({\bf a}), stellar masses ({\bf b}), difference between
the total and the fiber $g-i$ colours ({\bf c}), and angular radius
containing 50\% of the Petrosian flux, $\theta_{50}$ ({\bf
d}). The \emph{blue histograms} show the distributions of $z$,
$\log\,(M_{\star}/\msun)$ (from STARLIGHT V15), $\Delta\,(g-i)$, and $\theta_{50}$ of the VYG sample.
The control sample (CSG), drawn from the general sample of galaxies by applying the PSM technique (see the text for details), is shown as the orange histogram. The curves are obtained
by smoothing the positions of the data points (not the histograms) using a
Gaussian kernel with the standard deviation equal to one third of the standard
deviation of the data points. In each panel, we indicate the $p$-value of the
Kolmogorov-Smirnov test.
}
\label{fig:afterPSM}
\end{figure*}

The control sample galaxies (CSGs) were also selected from the \emph{clean} sample after removing the AGNs using the same criteria adopted for the VYG sample (item~\emph{v} above). The CSGs were chosen to be older than $1\,$Gyr according to all the three models.
We matched the CSG galaxies to have the same distributions of redshift,
stellar mass (according to STARLIGHT ran with V15), angular size and the total-fibre colour difference, $\Delta (g-i)$ (see item~\emph{vi} above). Matching the  radii
$\theta_{50}$ containing $50\%$ of the Petrosian flux in the $r-$band
ensures that the aperture effects are similar for both samples. The redshifts were included in the matching procedure to 
ensure that VYGs and CSGs potentially equally populate the same features of the large-scale distribution of galaxies.

We applied the Propensity Score Matching (PSM) technique \citep{PSM:1983,deSouza+16} to build the control sample.
We used the  {\sc MatchIt} package \citep{MatchIt:2011}, written in 
R\footnote{\url{https://cran.r-project.org/}} \citep{R:2015}.  This
technique allows us to select from the sample of normal galaxies (all minus VYGs)
a control sample in which the distribution of observed
properties is as close as possible to that of the VYGs.  
We adopted the Mahalanobis distance approach \citep{Mahalanobis36, Bishop:2006} and the nearest-neighbour
method to perform the matching.

This procedure yielded a sample of 1242 CSGs, i.e. 6 times the size of the VYG sample. 
We have also tried control samples with different sizes, spanning the range from 1 to 10 times the size of the VYG sample. We found that the results on the comparison between the samples are
not affected. The control sample with 6 times the size of the VYG sample was chosen to provide good statistics while keeping the distribution of stellar masses within the same range for both samples.

%
\begin{figure*}
\centering
\includegraphics[width=0.8\hsize]{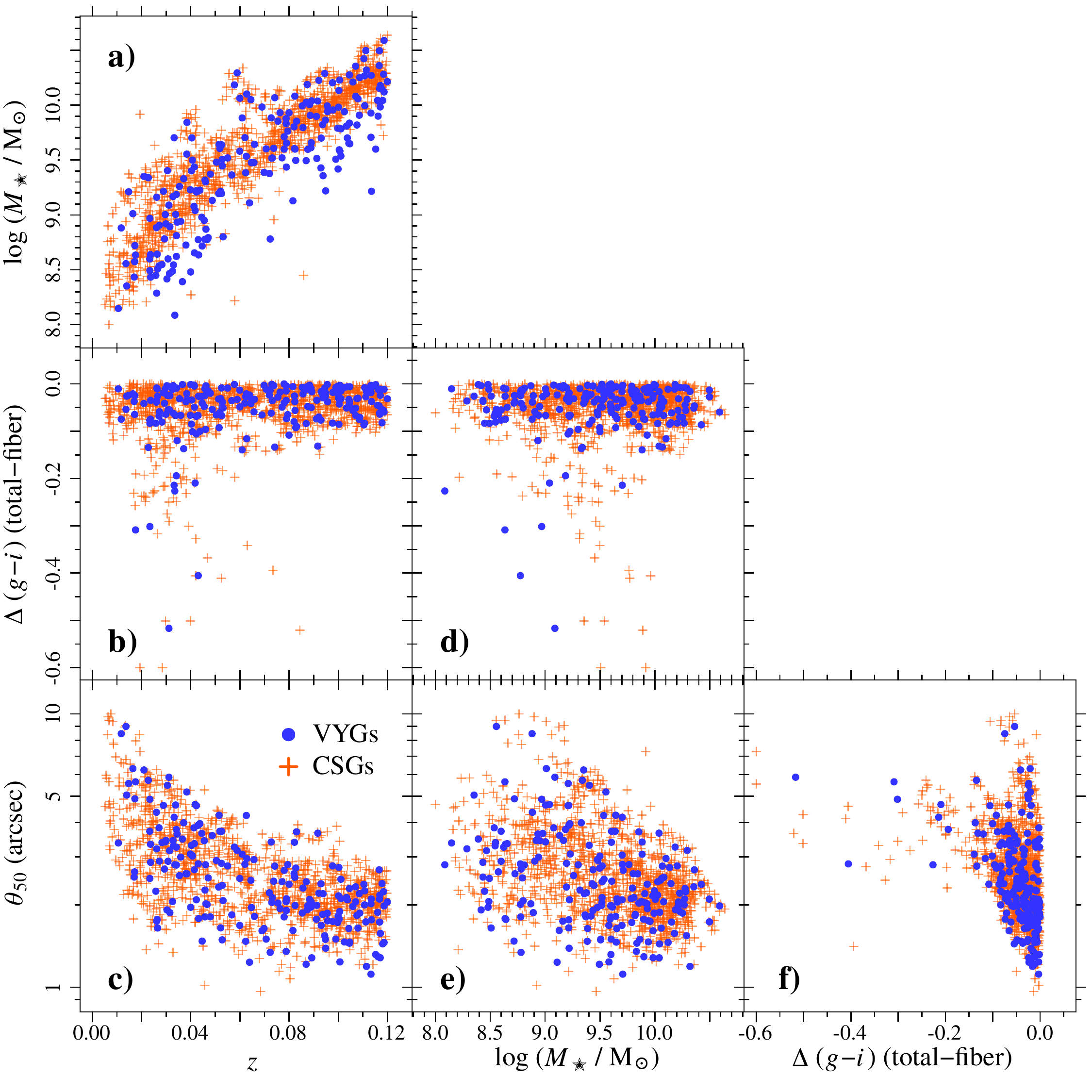} 
\caption{Relations between redshift, stellar mass, difference between the
total and the fiber $g-i$ colours, and angular radius containing 50\% of the
Petrosian flux for the VYGs (\emph{blue symbols}) and for the CSGs
(\emph{orange}).}
\label{fig:afterPSM_b}
\end{figure*}

The distributions of redshifts, stellar masses, $\Delta (g-i)$, and $\theta_{50}$ of the VYGs and CSGs are shown in Fig.~\ref{fig:afterPSM}. 
Also shown are the $p$-values derived from the Kolmogorov-Smirnov (KS)
test. The VYG and CSG distributions of redshifts, $\Delta (g-i)$ and
$\theta_{50}$ (Fig.~\ref{fig:afterPSM}a,c,d) are very similar. On the other
hand, we can see in Fig.~\ref{fig:afterPSM}b that the VYG sample contains
more low-mass galaxies than the CSG sample. It is particularly difficult to
draw a control sample of SDSS galaxies with masses below $\sim
10^9\,$M$_{\odot}$, since only $\sim 3.5\%$ of them are in this mass range. 
Besides, $95\%$ of them are at $z \lesssim 0.02$ , the
redshift completeness cut-off limit of the SDSS sample defined in Paper~II
for VYGs within this mass range. This makes it very difficult to select CSGs
with the same stellar mass vs. redshift distribution as that of the
VYGs. This is illustrated in Fig.~\ref{fig:afterPSM_b}a, where we see that,
at a given mass, the VYGs tend to lie at higher redshifts compared to the CSGs. We have tried to overcome this issue by changing the PSM parameters and the size of the control sample, but we always end up with a similar $M_{\star}$ vs. $z$ distribution. 
In any event, the results and conclusions presented in this work are not significantly affected by the difference between the VYG and CSG $M_{\star}$ vs. $z$ relations, as we discuss in Sect.~\ref{sec:caveats:Mz}.
The other panels in Fig.~\ref{fig:afterPSM_b} will be discussed in later sections.

\section{Properties of very young galaxies}
\label{sec:properties}

In this Section, we investigate some basic properties of the VYGs by comparing their sSFRs, colours, positions in the BPT diagram \citep{BPT81}, 
and morphologies with those of the CSGs.

%
\begin{figure*}
\centering
\includegraphics[width=0.95\hsize]{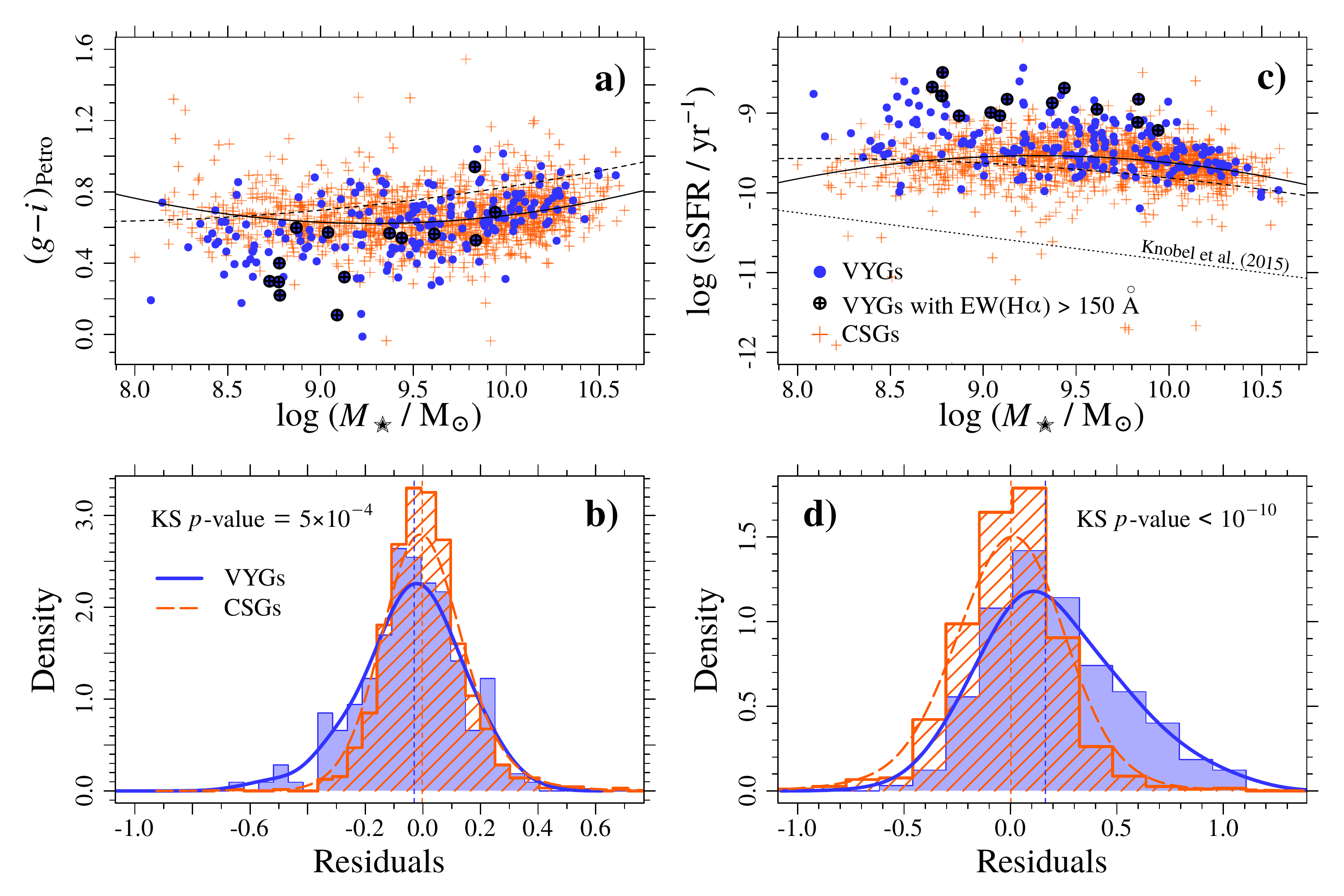} 
\caption{{\bf Top:} Petrosian $g-i$ colours ({\bf a}) and specific star
formation rates ({\bf b}, from the MPA/JHU table in SDSS) 
of VYGs (\emph{blue circles}) and normal galaxies (\emph{orange symbols}). The VYGs with EW(H$\alpha$)$\, \geq 150\,$\AA\ are indicated by the \emph{black crossed circles}.
The \emph{solid lines} in both panels correspond to a second-order polynomial
fit to the colour and $\log {\rm sSFR}$ vs. log stellar mass relations for galaxies in the control sample. 
The \emph{dashed lines} shows the best-fit relations for a sample of galaxies selected using the same approach adopted to build the control sample, but without including the redshift in the PSM procedure (see text for details). The difference between these two curves illustrates the effect of requiring the CSGs to be at similar distances as the VYGs. 
The line separating the star-forming and passive galaxies determined by \citet{Knobel.etal:2015} is indicated by the \emph{dotted line}. {\bf Bottom:} Residuals from the second-order polynomial fits shown for the VYG (\emph{blue histograms}) and control (\emph{orange}) samples. In each panel, we indicate the KS test $p$-values.}
\label{fig:sSFR_gi}
\end{figure*}

\subsection{Star formation rates and colours}
We extracted from the SDSS-DR12 database the optical magnitudes as well as the sSFRs ({\tt specsfr\_tot\_p50} from the {\tt galSpecExtra} table, obtained with the algorithm of \citealp{Brinchmann+04}). 
To compute the $g-i$ colours, we used the extinction- and $k$-corrected Petrosian magnitudes.
The $k$-correction was obtained with the {\tt kcorrect} code (version 4\_2) of \citet{Blanton+03_kcorrect}, choosing as reference the median redshift of the SDSS sample ($z = 0.1$).

In Figure~\ref{fig:sSFR_gi}, we compare the sSFRs and colours of the VYGs
with those of normal galaxies. Although we allow passive galaxies to be
included in the control sample, the CSGs are naturally more likely to be
star-forming for two reasons. 
First, $\sim 90\%$ of the galaxies with stellar
masses in the range of those of the VYGs are star-forming
systems (the line separating the star-forming and passive galaxies determined by
\citealp{Knobel.etal:2015} is shown in Fig.~\ref{fig:sSFR_gi}c).
Second, requiring CSGs to be at similar redshifts and to have similar stellar
masses as those of VYGs favours the selection of star-forming systems, since
they are brighter and more likely to be seen at distances where VYGs are
observed.
As described in Sect.~\ref{sec:data}, the VYG sample contains slightly more low-mass galaxies than the control sample, even after the PSM, since the SDSS-MGS is highly incomplete for stellar masses below $\sim 10^9$. To minimize the effects of this difference on our comparison results, we have fitted the relations between the galaxy properties and the stellar mass (using only the galaxies in the control sample) and analysed the residuals from these fits. 

The curvature that we observe in the best fits to the $g-i$ and $\log {\rm sSFR}$ vs. stellar mass relations (shown as solid lines in Fig.~\ref{fig:sSFR_gi}) is, again, a consequence of requiring the CSGs to have a similar redshift distribution as that of the VYGs. If we build a control sample without including the redshift in the PSM procedure, i.e., using only $\log M_{\star}$, $\theta_{50}$, and $\Delta(g-i)$, the $g-i$ colours (sSFRs) of the selected galaxies systematically increase (decrease) with stellar mass, and the best-fit relations (indicated by the dashed curves in Fig.~\ref{fig:sSFR_gi}) do not show the curvature. 

Not surprisingly, the VYGs, which have made more than half of their stellar mass in the last Gyr, show higher sSFRs than the CSGs. The effect is especially evident for the low-mass galaxies ($\log (M_{\star}/{\rm M}_{\odot}) < 9.0$). The statistical significance of the difference between the 2 samples is $p < 10^{-10}$ (Fig.~\ref{fig:sSFR_gi}a,b). 
As expected, the VYGs are bluer than normal galaxies, and those with high
H$\alpha$ equivalent widths have more extreme sSFRs.
Again, the differences are more significant for low-mass galaxies.

\begin{figure}
\centering
\includegraphics[width=0.9\hsize]{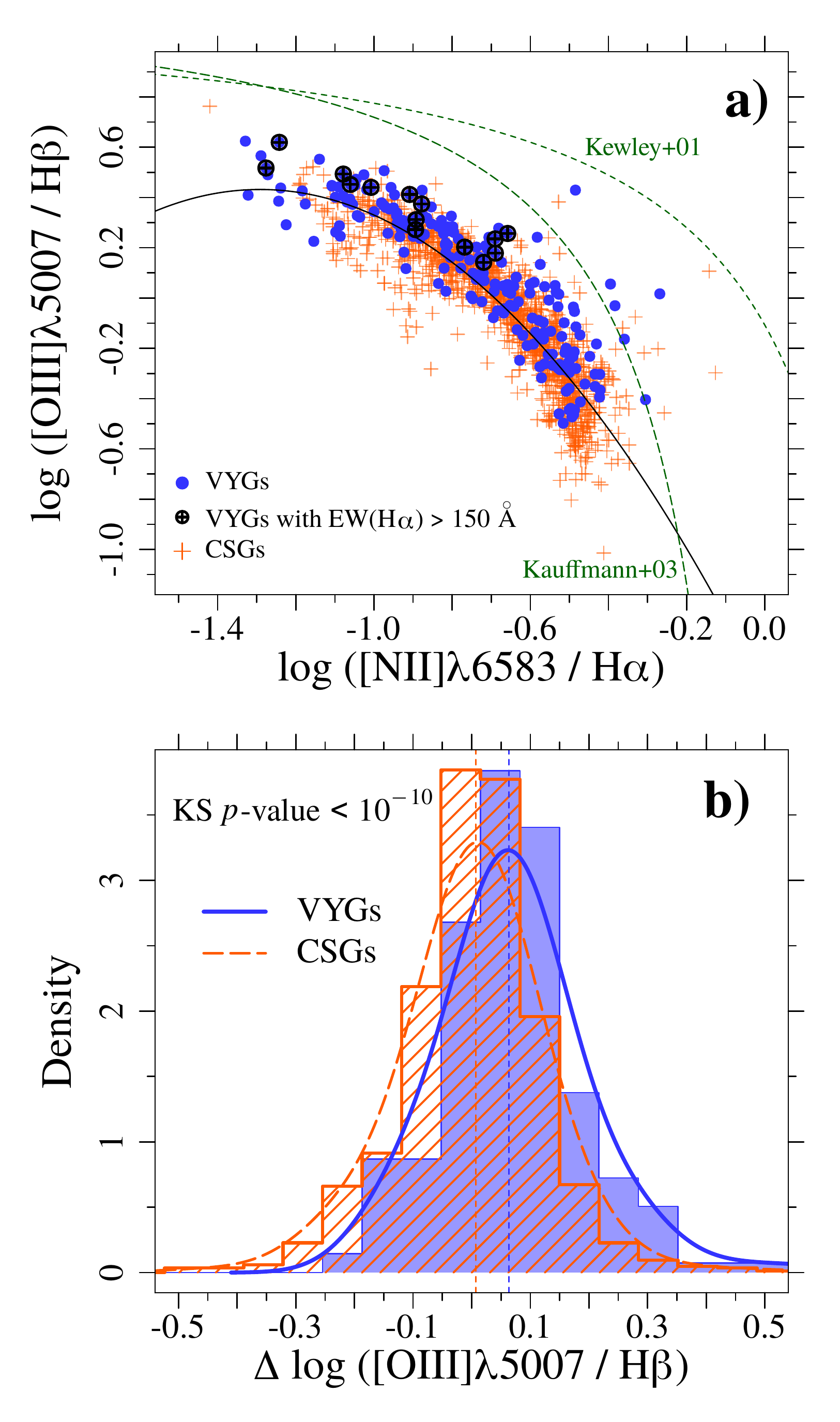} 
\caption{{\bf Top:} BPT diagram showing the VYG (\emph{blue circles}) and control (\emph{orange symbols}) samples. The VYGs with EW(H$\alpha$)$\, \geq 150\,$\AA\ are indicated by the \emph{black crossed circles}. The relations by \citet{Kewley+01} and \citet{Kauffmann+03} separating star-forming galaxies and AGNs are indicated as short- and long-dashed lines, respectively. The relation between the flux ratios of the CSGs was fitted with a second-order polynomial, as indicated as the \emph{black solid line}. {\bf Bottom:} The \emph{blue} and \emph{orange} histograms show the distribution of residuals from the polynomial fit for the VYGs and CSGs, respectively.
\label{fig:BPT}
}
\end{figure}

The BPT diagram \citep{BPT81} of the VYG and control samples is shown in Figure~\ref{fig:BPT}, where the emission-line fluxes are taken from MPA-JHU catalogue (table {\tt galSpecLine} based on the SDSS DR12, \citealp{Brinchmann+04,Tremonti+04}).
We find a higher fraction of VYGs in the composite region of the diagram compared to the normal galaxies (4.5 times higher).
By construction of the VYG and CSG samples, there are no galaxies in the AGN region of the diagram.
We used the CSGs to derive a fit for the $\log ([$O\,{\sc iii}$]/{\rm H}\beta)$ vs.  $\log ([$N\,{\sc ii}$]/{\rm H}\alpha)$ relation, and computed the residuals (Fig.~\ref{fig:BPT}b). The distribution of the residuals for the VYGs is shifted towards higher ionization levels compared to the CSGs, with a KS test indicating that the difference is  statistically significant ($p$-value$\, < 10^{-10}$).

%

\subsection{Galaxy morphologies and structural parameters}
\label{sec:properties:mfmtk}

To characterise the VYG and CSG morphologies, we have used the morphological classification from the Galaxy Zoo 1 project (GZ1, \citealp{Lintott.etal:2011}) and that by \citet[][hereafter DS18]{DominguezSanchez.etal:2018}. The GZ1 catalogue is based on simple visually-inspected classifications by hundreds of thousands of volunteers, and for each galaxy, it provides the debiased fraction of votes for each morphological type and classifies the galaxies in three categories: elliptical, spiral and uncertain.
The DS18 catalogue was obtained with Deep Learning algorithms, using Convolutional Neural Networks and visual classification catalogues from Galaxy Zoo 2 project (GZ2, \citealp{Willett.etal:2013}) and by \citet{Nair.Abraham:2010} as training sets. Besides providing classifications following a scheme similar to that of the GZ2, it also computes the T-type of each galaxy \citep{deVaucouleurs:1963}.

\begin{figure}
\centering
\begin{tabular}{ccc}
 \includegraphics[width=0.9\hsize]{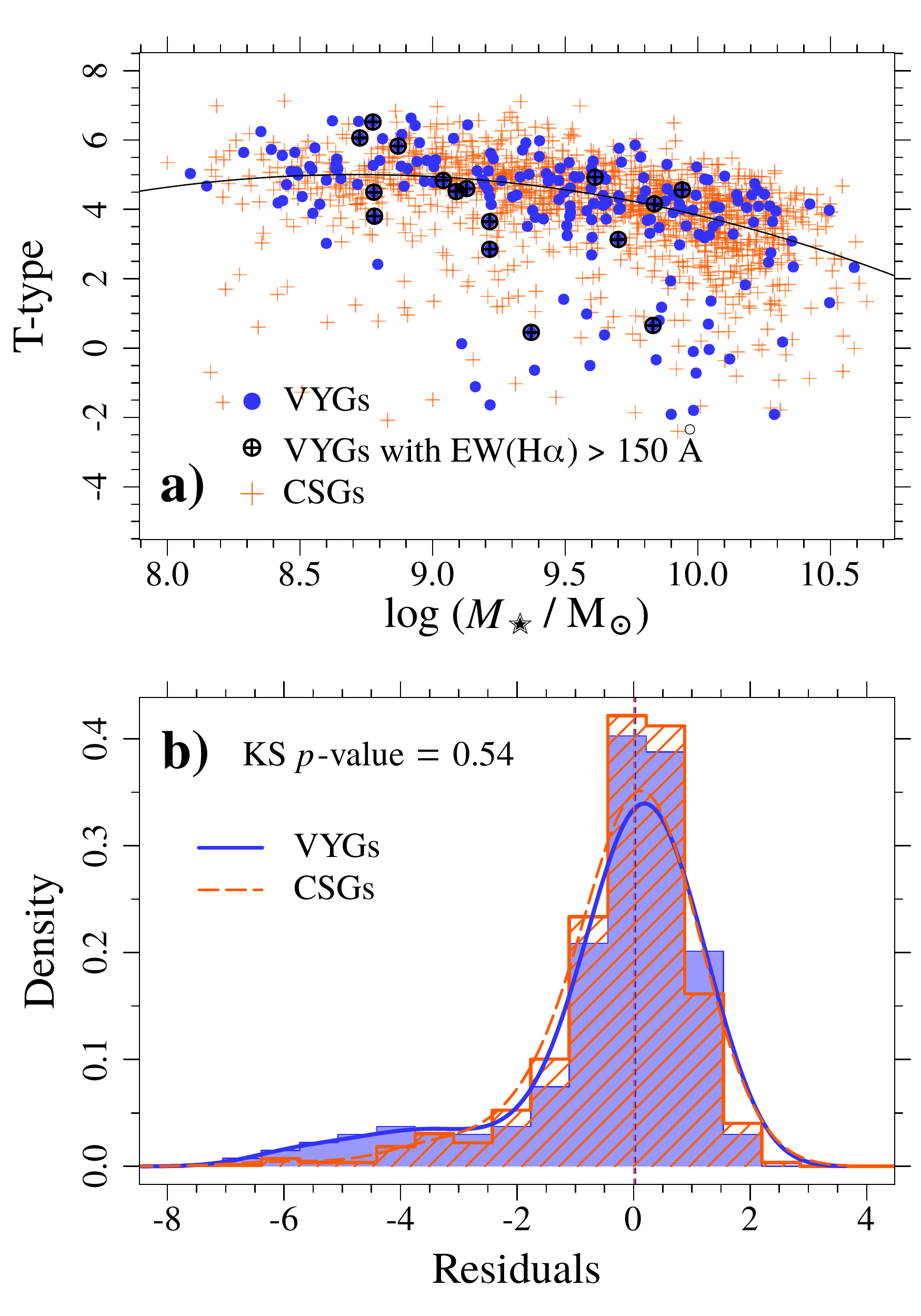} 
\end{tabular}
\caption{{\bf Top:} Morphological T-types from \citet{DominguezSanchez.etal:2018} as a function of stellar mass. The best second-order polynomial fit to the CSG T-type vs. $\log (M_{\star} / {\rm M}_{\odot})$ relation is indicated by the \emph{black solid line}. {\bf Bottom:} Distribution of VYG and CSG residuals from the second-order polynomial fit. The notations in both panels are the same as in Fig.~\ref{fig:sSFR_gi}.
\label{fig:Ttype}
}
\end{figure}

We cross-matched our samples with the GZ1 and DS18 catalogues, and found
morphological data for nearly all our galaxies: 1384
of our galaxies are in GZ1 (205 VYGs, 1179 CSGs) and 1441 in DS18 (203 VYGs,
1238 CSGs).
Most of the VYGs and CSGs are classified as \emph{uncertain}
according to GZ1 (76\% and 74\%, respectively); all the other VYGs and CSGs are classified as \emph{spirals}, except for one CSG that is an \emph{elliptical}.
We also compared the distributions of
probabilities of being disk- ({\tt P\_CS\_DEBIASED} and $p_{\rm disk}$ in the
GZ1 and DS18 catalogues, respectively) or bulge-dominated galaxies ({\tt
P\_EL\_DEBIASED}, $p_{\rm bulge}$), and found no significant difference
between the VYGs and CSGs. Finally, the VYG and CSG distributions of T-types
from the DS18 catalogue are also very similar. Fig.~\ref{fig:Ttype} shows that most of the galaxies ($97.6\%$) have T-type$\geq 0$ (94.1\% VYGS, 98.1\% CSGs), and a KS test confirms the low statistical differences between the two sample (KS $p$-value$\, = 0.54$). We find an excess of VYGs with low T-types compared to the CSGs. The number of VYGs and CSGs that lie below 3$\,\sigma$ of the T-type vs. stellar mass relation is 15 (7.4\%) and 27 (2.2\%), respectively. 

%

%
\begin{figure}
\centering
\begin{tabular}{ccc}
 \includegraphics[width=0.95\hsize]{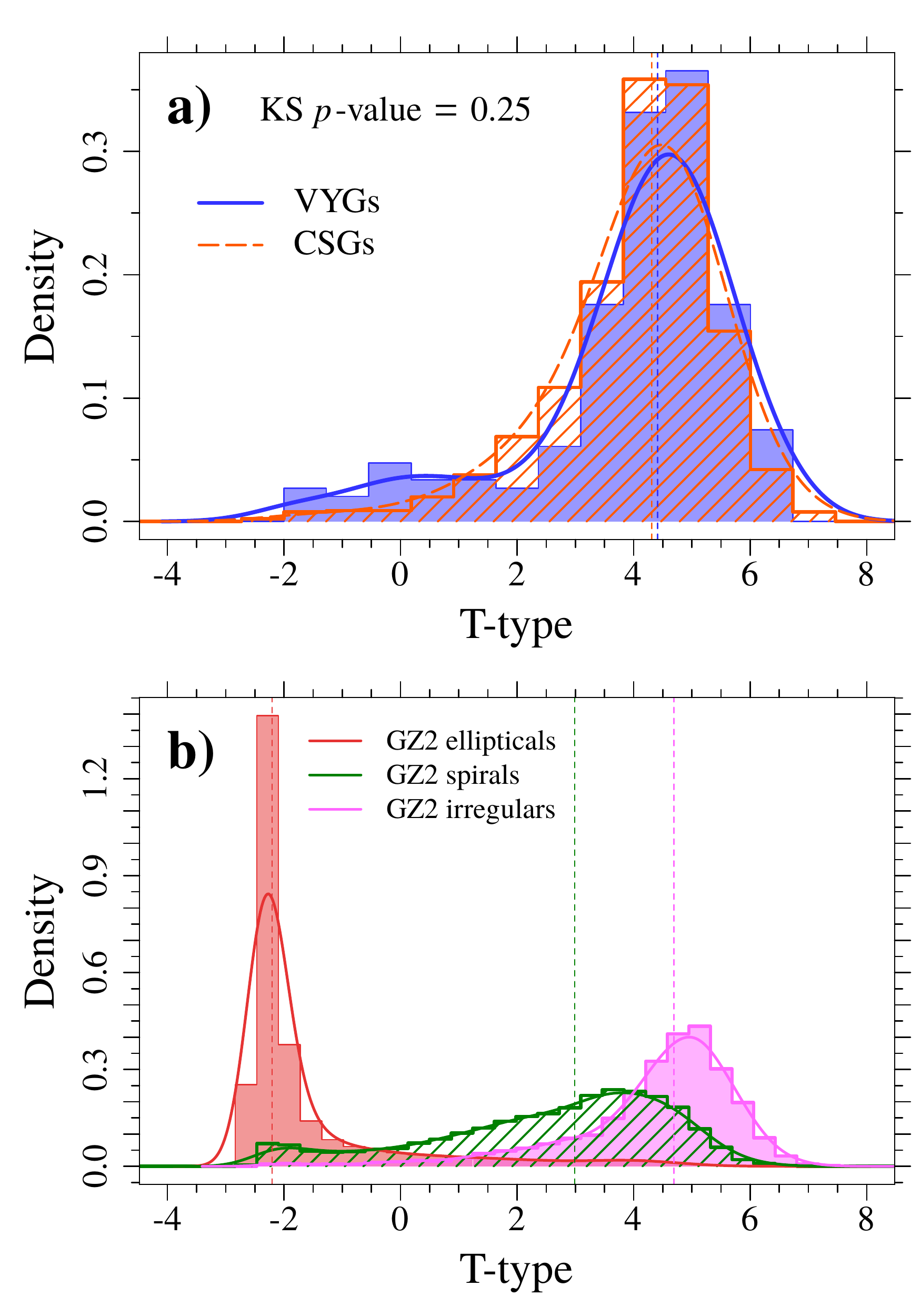} 
\end{tabular}
\caption{{\bf Top:} Distributions of T-types of the VYGs (\emph{blue histogram}) and CSGs (\emph{orange}). {\bf Bottom:} Distribution of T-types of galaxies classified as ellipticals (\emph{red}), spirals (\emph{green}) and irregulars (\emph{magenta}) in Galaxy Zoo 2 (GZ2). 
\label{fig:Ttype_gz2}
}
\end{figure}

The high fraction of \emph{uncertain} classifications in GZ1 indicate that most of our galaxies cannot be simply classified as ellipticals or spirals. Even when discarding galaxies with small angular sizes, the fractions of VYGs and CSGs that are classified as \emph{uncertain} is very high (61\% and 50\% of VYGs and CSGs with $\theta_{50} > 3''$, respectively).
The morphological classification by DS18 also points to this conclusion, since only a small fraction of VYGs and CSGs could be reliably classified as one of these two morphological types (15\% of spirals with $p_{\rm disk} \ge 0.8$ and 0.5\% of ellipticals with $p_{\rm bulge} \ge 0.8$).
This is confirmed by examination of
Figs.~\ref{fig:vyg_images1}--\ref{fig:vyg_images4}, which display the VYG
images. On the other hand, the distribution of T-types (Fig.~\ref{fig:Ttype})
is not compatible with irregular morphology, which is characterized by high
T-type values ($9- 10$). However, the range of T-types values in the DS18
catalogue goes from $\sim -3$ to $\sim 6$, and only 0.4\% of the galaxies
have T-type values greater than 6 (2861 out of 670\,722 galaxies).

To understand what morphological types these T-type values correspond to, we used the parameter {\tt gz2\_class} from GZ2 to select elliptical, spiral and irregular galaxies and analysed the T-type distributions of these three morphological classes. 
The {\tt gz2\_class} parameter is a string that indicates the most common consensus classification for the galaxy, and is composed by the letter E (galaxies that are smooth) or S (galaxies with disks and/or features), followed by other letters indicating several galaxy features (see the Appendix in \citealp{Willett.etal:2013} for details). 
We assumed to be ellipticals all galaxies with {\tt gz2\_class}$\, =\, $Er or Ei, where the letters `r' and `i' indicate that the galaxy is round or in-between round and cigar-shaped. 
We used the classification for bulge prominence as an indication for spiral morphology, and to avoid selecting lenticulars, we selected galaxies with no bulge or systems for which the bulge is just noticeable ({\tt gz2\_class}$\, = $SBc+, SBd+, Sc+, or Sd+, where `+' indicates other features). The same selection was made for the irregulars, but {\tt gz2\_class} contains the identifier for irregular morphology `(i)', i.e., {\tt gz2class} = Sc+(i), SBc+(i), Sd+(i), SBd+(i).

Fig.~\ref{fig:Ttype_gz2} shows that the VYG and CSG T-type distributions
(Fig.~\ref{fig:Ttype_gz2}a) are similar to that of irregulars, as seen in
Fig.~\ref{fig:Ttype_gz2}b.
The irregulars have median T-type = 4.7, and the 10th and 90th percentiles are 2.4 and 5.8. The spirals have lower T-types, with median 3.0, and 10th and 90th percentiles of -0.7 and 4.8. 
The median of the VYG and CSG T-type values are 4.4 and 4.3, respectively, with 10th and 90th percentiles of 1.2 and 5.6 (VYGs) and 2.2 and 5.5 (CSGs). 
As already indicated in the results shown in Fig.~\ref{fig:Ttype}, the we find an excess of VYGs with small T-types, and the number of VYGs with negative values is 3.2 times higher than the number of CSGs with T-type$\, < 0$ (5.9\% of the VYGs and 1.9\% of the CSGs). Fisher's \citep{Fisher:1935} and Barnard's \citep{Barnard:1945} tests, performed using the 
R packages {\tt stats} \citep{R:2015} and {\tt Barnard} \citep{Erguler:2016}, indicate that this difference between the VYG and CSG early-type fractions is statistically significant, with $p$-values$\,=0.002$ and $0.03$, respectively.
The inspection of the images of these galaxies reveals that both VYGs and CSGs with negative T-types have spheroid-like morphology, but VYGs spheroids are bluer than the CSGs spheroids, with median $g-i$ colours of 0.7 (VYGs) and 0.9 (CSGs).


Parametric models of bulges and/or disks may not give a good representation of these systems. So, to obtain more information and characterize the galaxy morphology, we
have adopted a non-parametric approach. We used the popular CAS system (concentration, asymmetry and clumpiness) presented in \citet{Abraham+94,Abraham+96} and \citet*{Conselice+00}. 

We measured 
the concentration as the ratio of radii containing 90 and 50 per cent of the
Petrosian flux in the $r$ band,
$\theta_{90}/\theta_{50}$, which we retrieved from the SDSS-DR12 database as
{\tt petroR90\_r} and {\tt petroR50\_r}, respectively.\footnote{SDSS does not
provide radii at less than half the flux, which are more sensitive to
variations in the point spread function.}

We also compute the galaxy surface brightness as
\begin{equation}
    \mu_{50} = m_r - k_{0.1} + 2.5  \log (2 \pi \theta_{50}^2) - 10 \log (1 + z)
    \label{eq:mu50}
\end{equation}
\noindent where $m_r$ is the extinction-corrected Petrosian magnitude in the $r$ band, $k_{0.1}$ is the $k$-correction, and $\theta_{50}$ is the radius (in arcsec) containing 50\% of the Petrosian flux in the $r$ band. The $k$-correction was obtained with the code {\tt kcorrect} with the SDSS filters shifted to $z = 0.1$

The galaxy asymmetry and clumpiness were measured using 
\textsc{ Morfometryka}\footnote{\url{http://morfometryka.ferrari.pro.br}} \citep{Ferrari.etal:2015}, which is a code to perform structural and morphometric measurements on galaxy images. 
The asymmetry coefficient is determined by comparing the galaxy image with a rotated version of itself. Three different asymmetry estimates are computed by the code. Here we adopt the standard one defined by \citet{Abraham+96}, $A_1$, with the difference that we do not subtract the background asymmetry, since this procedure leads to an unstable estimate of the coefficient $A_1$ and makes it dependent of the region selected to measure the sky asymmetry. Instead, only the region within the Petrosian radius is used. 

Similarly, \textsc{Morfometryka} also computes three different clumpiness coefficients (confusingly denoted ``smoothness'' and denoted by $S_1$ and $S_2$, even though higher $S_i$ values correspond to more clumpy distributions) by comparing the galaxy image with a smoothed version of itself. In this work we adopt $S_1$ defined by \citet*{Lotz+04} using a Hamming window \citep{Hamming:1998} with size $\theta_{\rm Petro}/4$, where $\theta_{\rm Petro}$ is the galaxy Petrosian radius.

Although the \textsc{Morfometryka} code returns many other parameters, 
including modified versions of asymmetry and clumpiness as described in detail in \citet{Ferrari.etal:2015},
we show only $A_1$ and $S_1$ because they are less dependent on the S/N of the images.
In particular, the Gini and M20 parameters, which are an extension of the CAS
system \citep{Lotz+04}, are also measured by  \textsc{Morfometryka}. However, we
found them to be very dependent on the S/N of the image. Besides, it is not straightforward to physically interpret the Gini and M20 parameters.

All the \textsc{Morfometryka} fits were visually inspected, and we excluded 27 (13\%) VYGs and 132 (11\%) CSGs with fits affected by foreground stars or very close objects (distances $\lesssim 5''$). 

%
\begin{figure*}
\centering
\begin{tabular}{cc}
 \includegraphics[width=0.45\hsize]{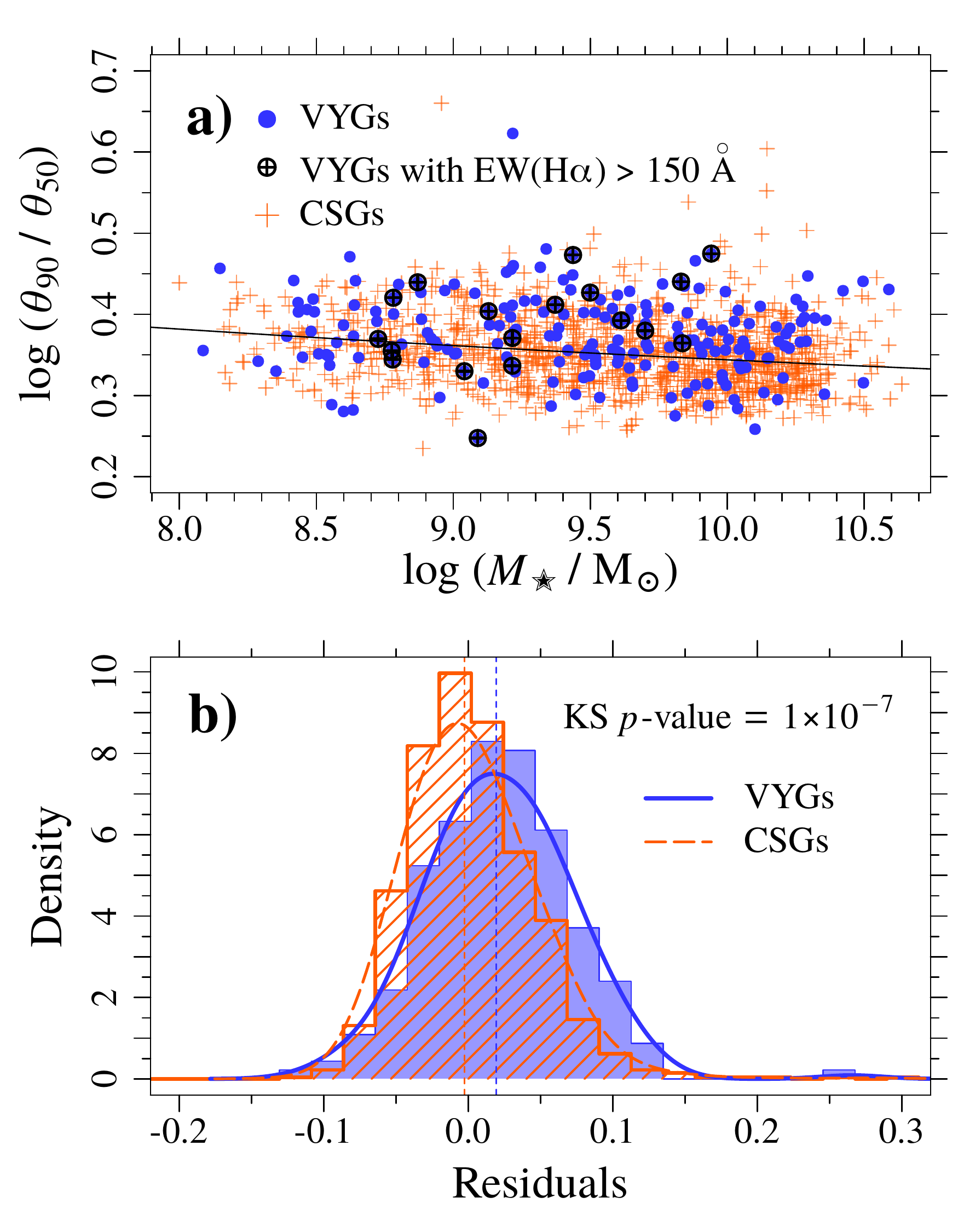} & 
 \includegraphics[width=0.45\hsize]{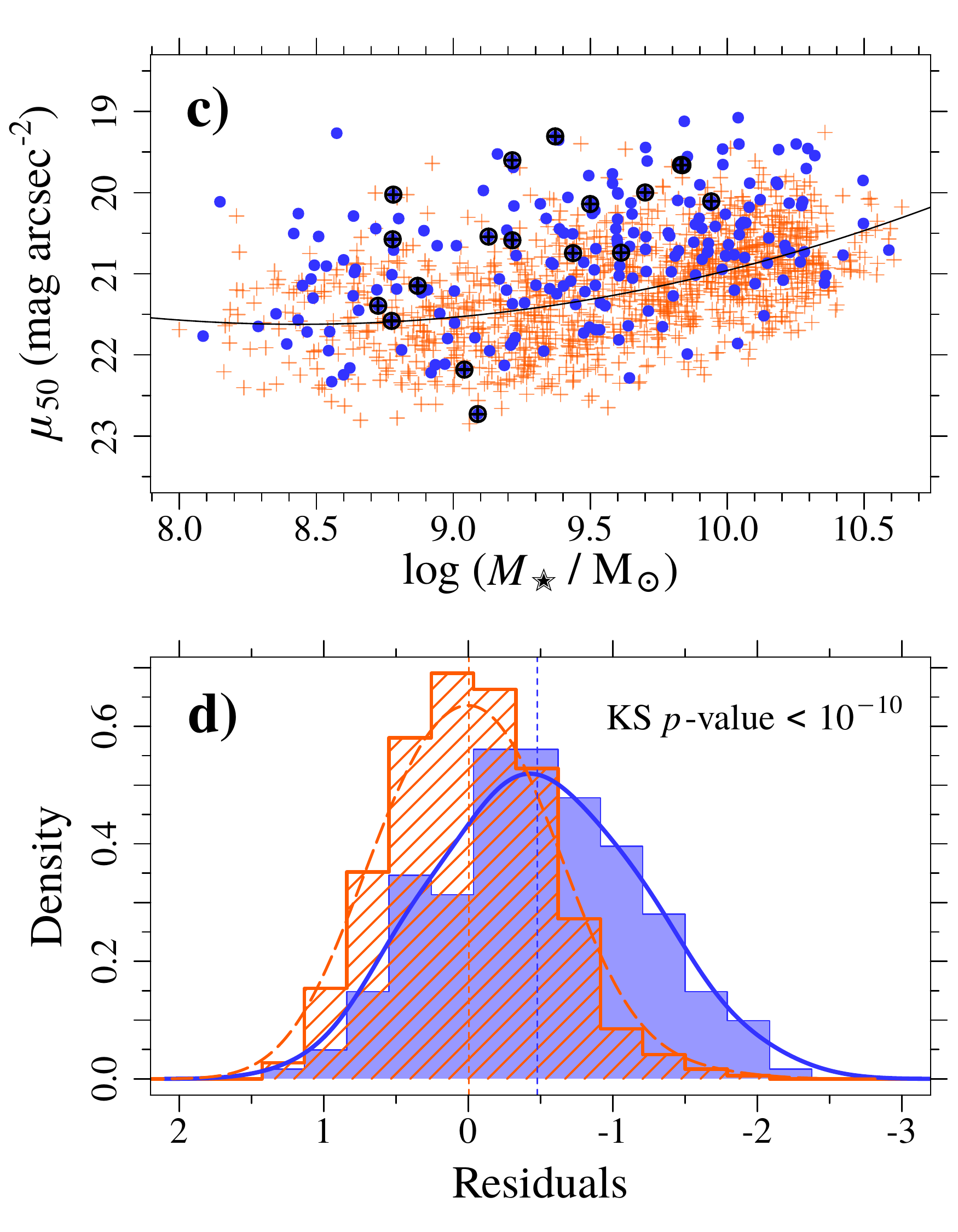} \\ 
\end{tabular}
\caption{{\bf Top:} Concentration (\emph{left}) and surface brightness (\emph{right}) of 
VYGs (\emph{blue circles}) and normal galaxies (\emph{orange symbols}). The VYGs with EW(H$\alpha$)$\, \geq 150\,$\AA\ are indicated by the \emph{black crossed circles}.
The solid lines in both panels correspond to a second-order polynomial fit to the concentration and $\mu_{50}$ vs. stellar mass relations for galaxies in the control sample. {\bf Bottom:}
Distribution of residuals from the second-order polynomial fit for the VYG (\emph{blue histograms}) and control (\emph{orange}) samples. In each panel, we indicate the $p$-value of KS tests.}
\label{fig:PSM_C_mu}
\end{figure*}

%
\begin{figure*}
\centering
\begin{tabular}{cc}
 \includegraphics[width=0.45\hsize]{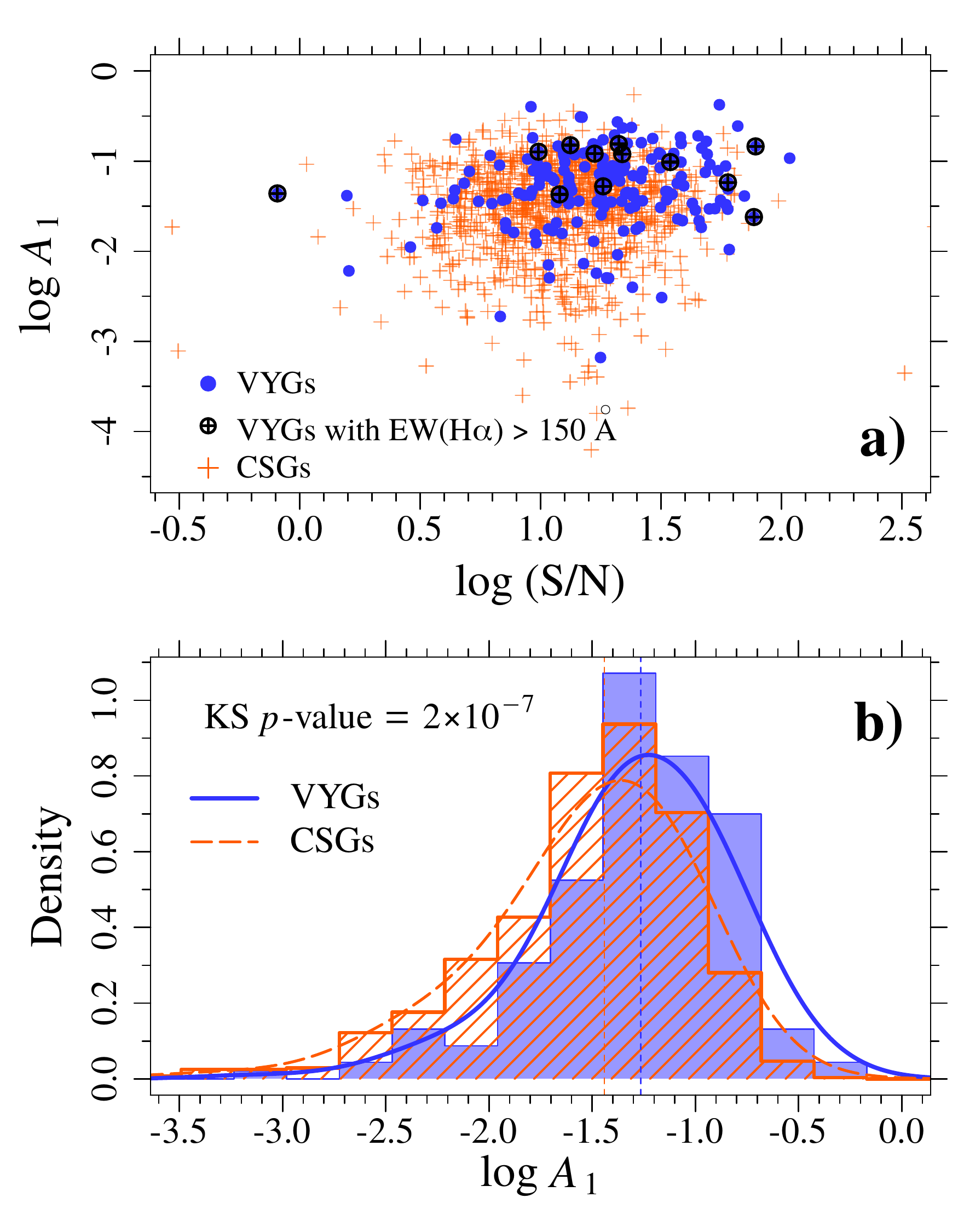}  & 
 \includegraphics[width=0.45\hsize]{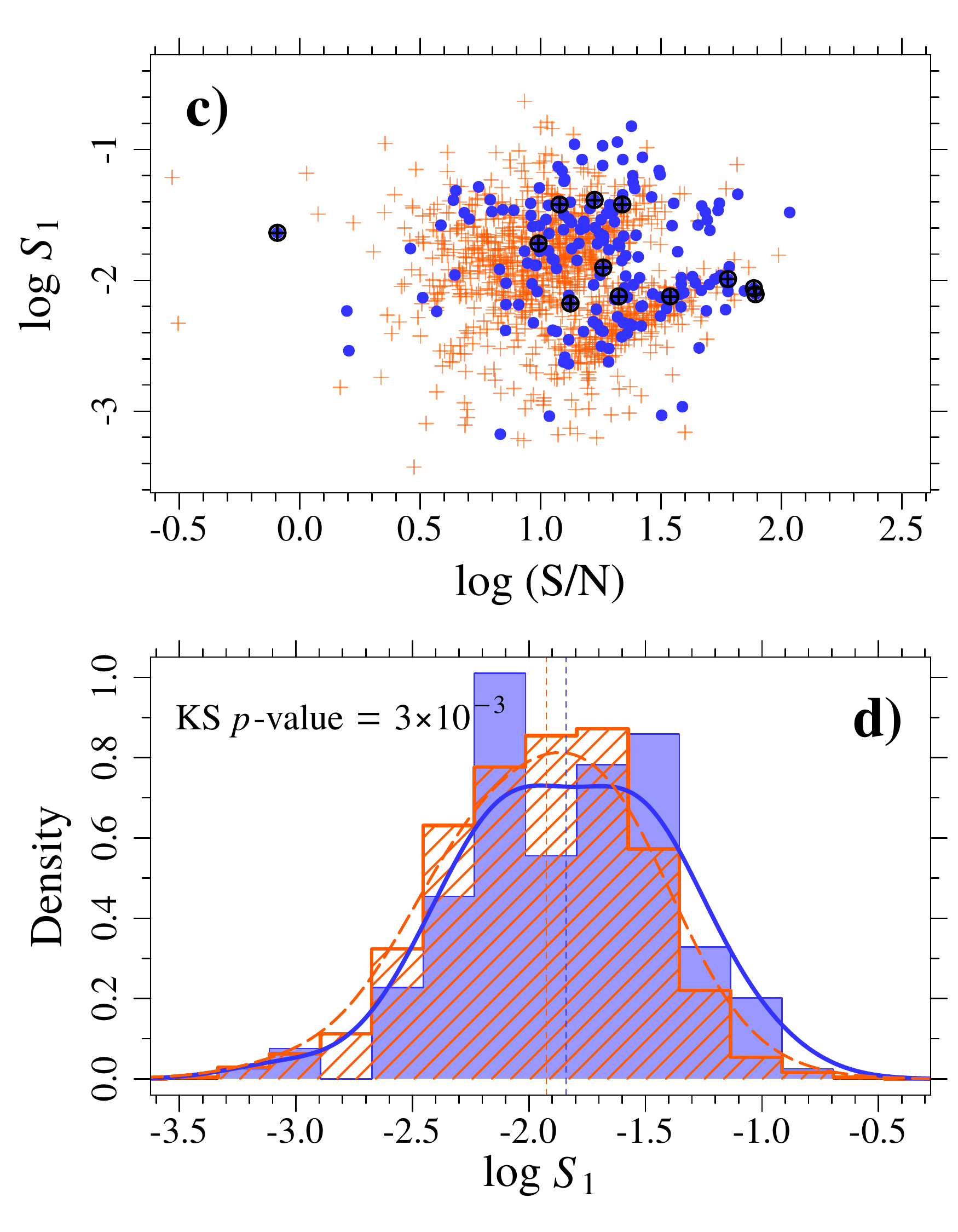} \\
\end{tabular}
\caption{{\bf Top:} Asymmetry ($A_1$) and clumpiness ($S_1$) parameters as a function of signal-to-noise ratio.
The notation is the same as in Fig.~\ref{fig:sSFR_gi}. {\bf Bottom:}
Distribution of $\log A_1$ and $\log S_1$ for the VYG (\emph{blue
histograms}) and control (\emph{orange}) samples. In each panel, we show the
KS test $p$-values.
\label{fig:PSM_A1_S1}
}
\end{figure*}

In Fig.~\ref{fig:PSM_C_mu}, we show the galaxy concentration (panel a) and surface brightness (panel c) as a function of the stellar mass. The residuals derived from fitting the data show that the VYGs are more concentrated and have higher surface brightness than normal galaxies (Figs.~\ref{fig:PSM_C_mu}b,d). The KS tests indicate that these results have high statistical significance, with $p = 10^{-7}$ (concentration) and $p < 10^{-10}$ 
(surface brightness).

Since the structural parameters, such as asymmetry and clumpiness, may be sensitive to the S/N of the images, we show in Fig.~\ref{fig:PSM_A1_S1}a,c the $A_1$ and $S_1$ parameters as a function of the signal-to-noise ratios instead of stellar masses. We confirm the weak dependency of of $A_1$ with S/N, with Kendall and Spearman correlation coefficients $\tau = 0.02$ and $\rho = 0.03$. On the other hand, $S_1$ shows an anti-correlation with S/N ($\tau = -0.11$ and $p$-value$ = 2 \times 10^{-9}$; 
$\rho = -0.17$ and $p$-value$ = 8 \times 10^{-11}$). 
In summary, the comparison between the $S_1$ distributions of the VYGs and control samples should be considered with caution. 
The distributions of these parameters are shown in panels b and d. We can see that the VYGs tend to be more asymmetric and clumpy when compared to normal galaxies, with high statistical significance ($p = 2 \times 10^{-7}$ for the asymmetry parameter and $p = 3 \times 10^{-3}$ for the clumpiness). In Fig.~\ref{fig:PSM_A1_S1} we show only galaxies with good \textsc{Morfometryka} fits, i.e., galaxies with nearby objects (in projection) affecting the estimate of the structural parameters are excluded from the plots (27 VYGs and 132 CSGs, corresponding to 13\% and 11\%, respectively). 

To obtain meaningful structural parameter measurements, the angular size of the galaxy must be greater than the resolution of the SDSS image and the atmospheric seeing. It can be seen in Fig.~\ref{fig:afterPSM}d that many of our galaxies have small angular sizes and might be unresolved. However, we note that, since the VYG and CSG $\theta_{50}$ distributions are similar by construction, differences between the VYG and CSG structural parameters are unlikely to arise from differences in the angular sizes of galaxies in these two samples.

\section{Ionized gas, neutral gas and dust content}
\label{sec:gas}

\subsection{Oxygen abundances of the ionized gas}
\label{sec:properties:oh}

%
\begin{figure}
\centering
\includegraphics[width=\hsize]{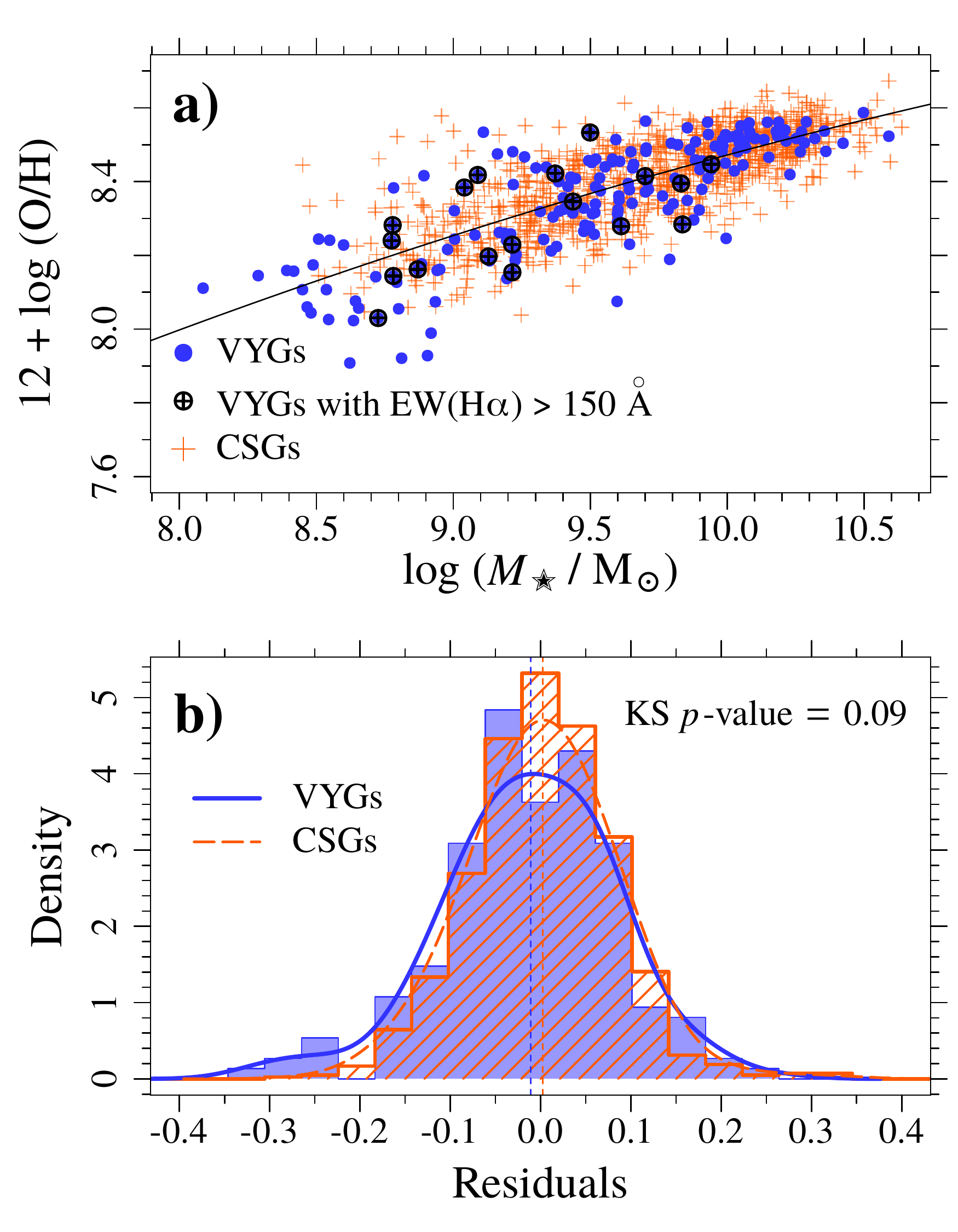} 
\caption{Ionized gas oxygen abundances as a function of stellar mass. The abundances were derived using emission line fluxes from the SDSS table {\tt GalSpecLine} (MPA-JHU). 
The notation is the same as in Fig.~\ref{fig:sSFR_gi}. The bottom panel shows the distribution of residuals from the second-order polynomial fit for the VYG (\emph{blue histograms}) and control (\emph{orange}) samples. 
\label{fig:PSM_OH}
}
\end{figure}

The gas oxygen abundance, an indicator of metallicity, was derived from the emission-line fluxes retrieved from the SDSS database. 
Two sets of measurements are available: one given by the MPA-JHU catalogue (table {\tt galSpecLine} in SDSS DR12, \citealp{Brinchmann+04,Tremonti+04}) and the other by the catalogue of the Portsmouth group (table {\tt emissionLinesPort} in SDSS DR12). The first one employs the \citet{Bruzual&Charlot03} models to fit the stellar continuum, while models from \citet{Maraston2011} and \citet*{Thomas2011} are used in the Portsmouth measurements.
The use of line measurements from two different catalogues allows to estimate the uncertainties in the abundances of the
target galaxies due to uncertainties in the line flux measurements.

We extracted from those catalogues the measurements of the 
[O\,{\sc ii}]$\lambda$3727, 
[O\,{\sc ii}]$\lambda$3729, 
H$\beta$,  
[O\,{\sc iii}]$\lambda$4959, 
[O\,{\sc iii}]$\lambda$5007,
[N\,{\sc ii}]$\lambda$6548,
H$\alpha$, 
[N\,{\sc ii}]$\lambda$6584, 
[S\,{\sc ii}]$\lambda$6717, and 
[S\,{\sc ii}]$\lambda$6731 
emission lines in the spectra.
We restricted our analysis to galaxies whose emission lines were measured with S/N$\, > 3$ for all lines.
We corrected the emission line fluxes for interstellar reddening using 
the observed H$\alpha$/H$\beta$ ratio and the reddening function
from \citet*{Cardelli.etal:1989} for 
$R_V = 3.1$, which leads to  $A_{{\rm H}{\beta}} = 1.164\,A_{V}$.

The ``direct $T_{e}$'' method \citep[e.g.,][]{Dinerstein1990} is believed to 
provide the most reliable abundance determinations in SF 
regions from the emission lines in their spectra. 
However, it requires high-precision spectroscopy
to detect weak auroral lines such as [O\,{\sc iii}]$\lambda$4363
and [N\,{\sc ii}]$\lambda$5755.  Unfortunately, these auroral lines
are usually not detected or they are measured with a large uncertainty in the SDSS spectra. 
The abundances in SDSS objects with emission line spectra 
are thus usually estimated using "the strong line method" 
of \citet{Alloin1979} and \citet{Pagel1979}, using 
 intensity ratios of the strong emission lines in the spectra. 

Two families of calibrations are widely used. The family of the calibrations 
following \citet{Pagel1979} is based on the oxygen lines. Both 1D
(e.g. \citealt*{Zaritsky1994}; \citealt{Tremonti+04}) and 2D (or parametric, $p$-method)
calibrations \citep{Pilyugin2000,Pilyugin2001,Pilyugin2005} have been suggested.
The family of the calibrations following \citet{Alloin1979} is based on the nitrogen lines.
There are only 1D calibrations of this kind for abundance determinations
in high metallicity objects \citep[e.g.][]{Pettini2004,Marino2013}.
There is no
unique relation applicable across the whole range of
metallicities of H\,{\sc ii} regions. Instead, 
distinct calibration relations are constructed for high (upper branch) 
and low metallicities (lower branch).  
But these branches are not applicable to objects lying in the transition zone 
\citep[e.g.][and discussion there]{Pilyugin2005}. Moreover, to choose the relevant calibration relation, one has to know 
{\em a priori} to which interval of metallicity the H\,{\sc ii} region belongs. A
wrong choice would lead to 
a wrong abundance. 
Many of our galaxies may have moderate to low metallicities, 
i.e. they belong to the transition zone and to the lower branch, making   
abundance determinations for our objects particularly difficult. 

We used the simple 3D calibration relations proposed by 
\citet{Pilyugin2016}.  The oxygen abundances (O/H)$_{R}$ are
determined using the $R$ calibration, i.e., (O/H)$_{R}$ = $f$(R$_2$, R$_3$, N$_2$), 
where the oxygen R$_2$, R$_3$ and the nitrogen N$_2$ lines are given by  
\begin{eqnarray*}
R_2  & = & I_{\rm [O\,II] \lambda 3727+ \lambda 3729} /I_{{\rm H}\beta },  \\
R_3  & = & I_{{\rm [O\,III]} \lambda 4959+ \lambda 5007} /I_{{\rm H}\beta },\ {\rm and}  \\
N_2  & = & I_{\rm [N\,II] \lambda 6548+ \lambda 6584} /I_{{\rm H}\beta }.
\end{eqnarray*}

One advantage of this calibration is that it is applicable over the entire metallicity range
of H\,{\sc ii} regions. Although distinct relations for high- and
low-metallicity objects are constructed, the separation between these
two can be simply obtained from the intensity of the $N_{2}$ line.
Moreover, the applicability ranges of the high- and low-metallicity
relations overlap, 
making the transition zone disappear. 

Since the [O\,{\sc iii}]$\lambda$5007 and $\lambda$4959 lines originate 
from transitions from the same energy level, their flux ratio is determined 
only by the transition probability ratio, close to 3 
\citep{Storey2000}. Therefore, the value of $R_3$ can be estimated as
$R_3  = I_{{\rm [O\,III]} \lambda 4959+ \lambda 5007} /I_{{\rm H}\beta }$ or  
$R_3  = 1.33I_{{\rm [O\,III]} \lambda 5007} /I_{{\rm H}\beta }$. 
The stronger line [O\,{\sc iii}]$\lambda$5007 is usually measured with 
higher precision than the weaker line [O\,{\sc iii}]$\lambda$4959. 
We thus used the latter expression to calculate $R_3$, 
instead of the sum of the [O\,{\sc iii}] line fluxes.
The same applies to 
the nitrogen lines [N\,{\sc ii}]$\lambda$6584 and $\lambda$6548. They  
also originate from transitions from the same energy level and the 
transition probability ratio for those lines is again
close to 3 \citep{Storey2000}. We again calculate $N_2$ as  
$N_2 = 1.33$~[N\,{\sc ii}]$\lambda$6584/H$\beta$, instead of using the sum of the 
[N\,{\sc ii}] lines. 

The radiation of the diffuse ionized gas is believed to
contribute significantly to some fibre SDSS  spectra,
and may increase the strength of the low-ionization lines [N\,{\sc ii}], [O\,{\sc
ii}], and [S\,{\sc ii}] relative to the Balmer lines.
If this is the case, then the abundances derived from the SDSS spectra by 
strong-line methods may have large errors
\citep{Belfiore2015,Belfiore2017,Zhang2017,Sanders2017}.  
\citet{Pilyugin2018}. 
found that the mean increase of R$_{2}$ and
N$_{2}$ is less than a factor of $\sim 1.3$, 
and the $R$ calibration produces reliable abundances. 

For each galaxy in our sample with an available oxygen line R$_{2}$ and a S/N greater than 3 for the emission lines of interest, we have determined two values of the oxygen abundance, one based on the MPA-JHU line measurements and the other on the Portsmouth ones.  
We have found that the two sets of abundances 
are in excellent agreement ($\Delta = 0.005 \pm 0.02\, $dex). Hence, 
we only present here the results based on the MPA-JHU catalogue. 

In Fig.~\ref{fig:PSM_OH}a, we show how the VYG and CSG oxygen abundances vary with galaxy stellar mass. Only galaxies with good MPA-JHU emission-line measurements (i.e., S/N$\,\ge 3$) are shown (183 VYGs and 1031 CSGs, corresponding to $88\%$ and $83\%$ of the galaxies in these samples, respectively). We fitted the $12+\log ({\rm O/H})$ vs. $\log (M_{\star}/{\rm M}_{\odot})$ relation of the CSG, and show the distribution of residuals for both samples in Fig.~\ref{fig:PSM_OH}b.
The residuals from the fit indicate that there is no significant difference between the oxygen abundances of the VYGs and the CSGs as a function of stellar mass (KS test $p$-value$\,=0.09$). We also compared the best-fit to the VYG $12+\log ({\rm O/H})$ vs. $\log (M_{\star}/{\rm M}_{\odot})$ relation to that of the CSGs, and find that the difference between the two fits does not exceed 0.015 dex within the range $8.0 \leq \log (M_{\star}/{\rm M}_{\odot}) \leq 10.6$. Using a first-order polynomial to fit the VYG and CSG relations, we also get very similar slopes for both samples (0.23 and 0.22 for the VYGs and CSGs, respectively).

On the other hand, when we limited our analysis to galaxies with  $M_{\star} \leq 10^9\, {\rm M}_{\odot}$,
we found that low-mass VYGs  have lower oxygen abundances by 0.07\,dex, on average, compared to those of the CSGs, with a KS test indicating marginal statistical significance ($p$-value$\,=0.02$). But a large fraction of the low-mass CSGs are excluded from this analysis when we require emission-line measurements with  S/N$\,\ge 3$. Among the 210 CSGs with $M_{\star} \leq 10^9\, {\rm M}_{\odot}$, only 78 ($37\%$) pass the S/N criteria, while $73\%$ of the VYGs within the same mass range have good line-emission measurements (36 out of 49). The higher sSFRs of low-mass VYGs compared to the those of the CSGs might explain why a larger fraction of low-mass CSGs is excluded from our analysis of the oxygen abundances. Galaxies with higher sSFRs have stronger emission lines which result in flux measurements with higher S/N ratios. Therefore, the selection effects introduced by this cut in S/N affects the low-mass VYG and CSG samples differently, and the comparison between their oxygen abundances must be seen with caution.

Finally, there are 6 low-metallicity VYGs with oxygen abundances $3\, \sigma$ below the $\log({\rm O/H}+12)$ vs. stellar mass relation: they have lower metallicities for their stellar masses. However, 
this result is not statistically significant.

\subsection{Neutral hydrogen content}
\label{sec:gas:neutral}

%
\begin{figure*}
\centering
\begin{tabular}{cc}
 \includegraphics[width=0.45\hsize]{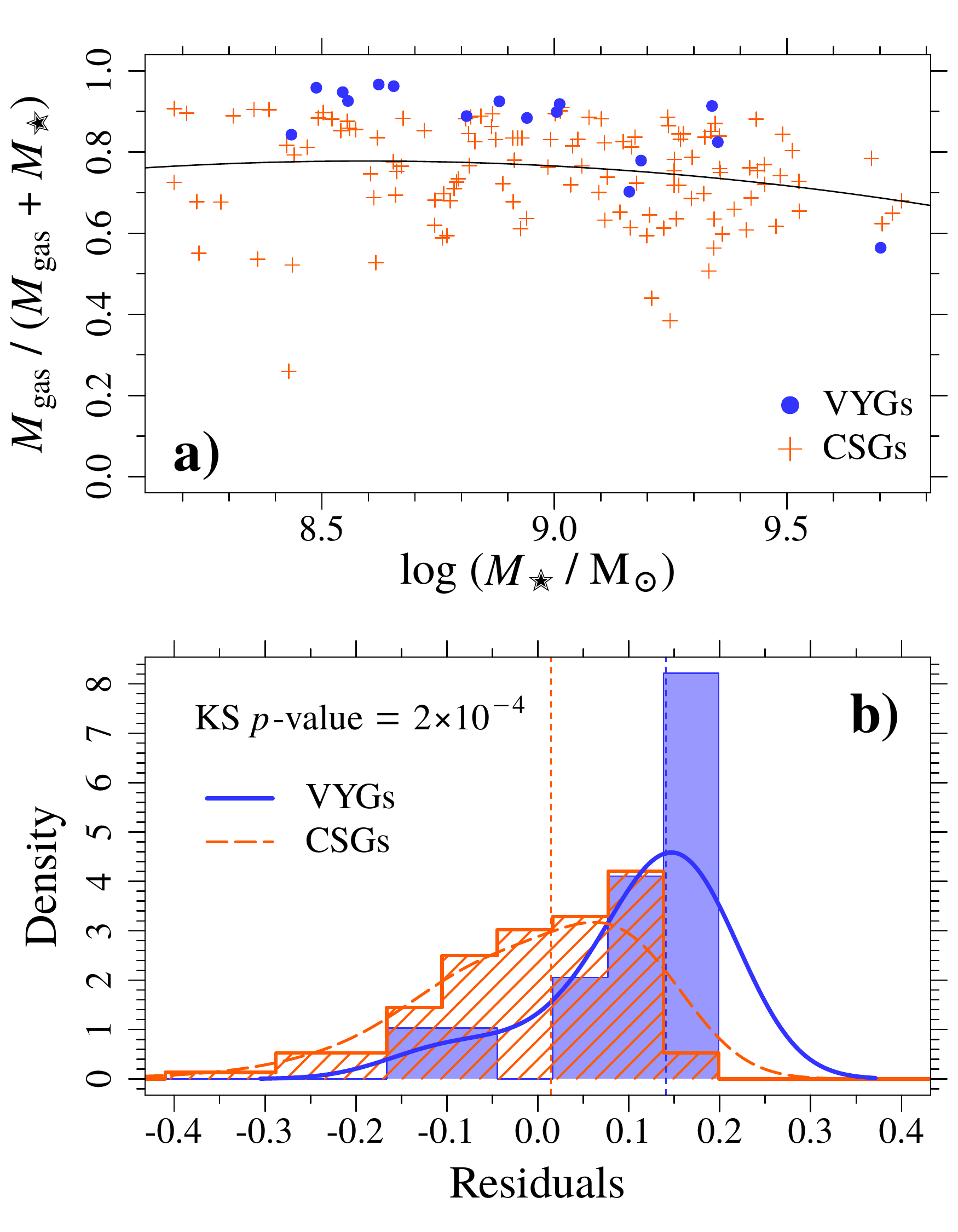}  & 
 \includegraphics[width=0.45\hsize]{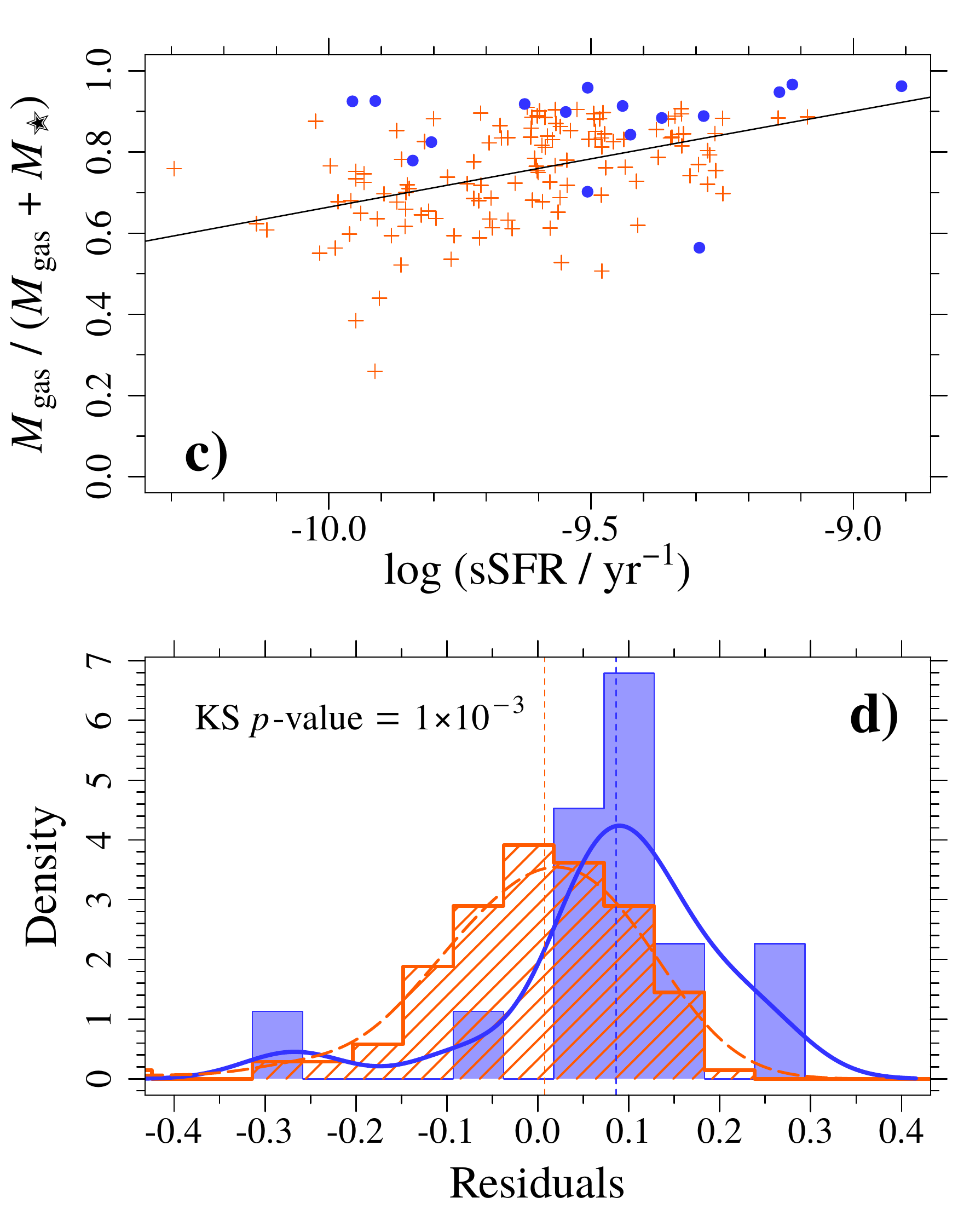} \\
\end{tabular}
\caption{Fraction of atomic gas in galaxies with H\,{\sc i} detections in the ALFALFA survey as a function of galaxy stellar mass (\emph{upper left}) and $\log ({\rm sSFR})$ (\emph{upper right}). The residuals from the linear fits to the CSG $f_{\rm HI}$ vs. stellar mass and vs. $\log ({\rm sSFR})$ relations are shown in the \emph{bottom panels}.
The \emph{blue} and \emph{orange} histograms correspond to the VYG and CSGs,
respectively and, in each panel, we indicate the $p$-values of the KS tests.
\label{fig:HIgas}
}
\end{figure*}

The atomic gas content of our galaxies was investigated using  the data from the Arecibo Legacy Fast ALFA (ALFALFA) survey \citep{Haynes.etal:2011, Haynes.etal:2018}.
The ALFALFA H\,{\sc i} source catalogue contains $31\,502$ detections, and
approximately half of our objects lies in the ALFALFA-SDSS overlap region
($\sim 4000\,$deg$^2$). The coordinates of the most probable optical
counterpart (OC) of each H\,{\sc i} detection are available in the catalogue.
We cross-matched our VYG and CSG samples by adopting a maximum angular
separation of $5''$ between the OC and the SDSS galaxy. We found 149 galaxies
with H\,{\sc i} detections, among which 19 are VYGs and 130 are CSGs,
corresponding to $9.2\%$ and $10.5\%$ of these samples, respectively. For
each galaxy, we computed the atomic gas mass fraction $f_{\rm gas}$, i.e.,
the gas mass divided by the sum of the gas and stellar masses, $f_{\rm gas} =
M_{\rm gas} / (M_{\rm gas} + M_{\star})$.
We accounted for the He gas by assuming that the total gas mass satisfies 
the relation $M_{\rm gas}
= 1.33 \,M_{\rm H\textsc{i}}$. The stellar masses were obtained from our SPS
analysis, using the V15 spectral model with STARLIGHT. 

Measurements of H\,{\sc i} masses may be contaminated by the neutral gas of
nearby galaxies because of the large Arecibo beam size
(FWHM$\,=3.5\,$arcmin). We checked for such neighbour contamination by  using
the SDSS spectroscopic catalogue to identify the VYGs and CSGs that have
other galaxies within a radius of $1.8\,$arcmin ($\sim 90$\% of the beam
flux) and $\Delta z \leq 2\, W_{20}/c$, where $W_{20}$ is the velocity width
of the H\,{\sc i} line profile in km\,s$^{-1}$, measured at the 20\% level of
of each of the two peaks on the low- and high-velocity horns of the profile
(see \citealp{Springob.etal:2005}) and $c$ is the speed of light.
Although this procedure misses galaxies fainter than the magnitude limit of
the SDSS spectroscopic observations ($m_r \leq 17.77$), we use the spectra to
estimate the total stellar mass within the Arecibo beam.
The stellar masses of the nearby galaxies were computed using the STARLIGHT
code with the V15 spectral models as described in Sect.~\ref{sec:data}.

We found that 3 VYGs and 5 CSGs ($15.8\%$ and $3.8\%$, respectively) have other galaxies at distances $\leq 1.8\,$arcmin and within the $\Delta z$ limit described above. The fraction of VYGs with nearby galaxies is 4.1 times higher than that of the CSGs. However, the stellar masses of the galaxies around the VYGs are lower than those of galaxies in the CSG surroundings. The total stellar mass within the beam is $0.26\,$dex higher (median) than the mass of the VYGs, while for the CSGs, the total mass is $\sim 0.9\,$dex higher (median). 
The median gas fraction computed with the total stellar mass within the beam, $M_{\rm gas} / (M_{\rm gas} + M_{\star, \rm beam})$, is 0.88 and 0.66 for these VYGs and CSGs, respectively. When comparing these values with the median gas mass fractions of VYGs  and CSGs with no nearby galaxies within $1.8\,$arcmin ($\langle f_{\rm gas} \rangle = 0.91$ and $0.76$, respectively) , we see that the CSG gas fractions are more affected by nearby galaxies. Therefore, we conclude that the presence of other galaxies does not lead to higher VYGs gas fractions compared to the CSGs.

In Fig.~\ref{fig:HIgas} we show how $f_{\rm gas}$ varies with galaxy stellar
mass (Fig.~\ref{fig:HIgas}a) and sSFR (Fig.~\ref{fig:HIgas}c). The 3 VYGs and
5 CSGs that have nearby galaxies within the Arecibo beam, identified as
described above, are excluded from the plot. For a given stellar mass and
sSFR, the fraction of atomic gas in VYGs is systematically larger compared to
the one in CSGs. About $80\%$ of the VYGs with H\,{\sc i} detections and no
nearby galaxies (13 out of 16 VYGs) have $f_{\rm gas} > 0.8$, while only $42\%$ of the CSGs have such high gas mass fractions (52 out of 125 CSGs). We fitted the CSG $f_{\rm gas}$ vs. stellar mass and vs. $\log ({\rm sSFR / yr^{-1}})$ relations, and found that 14 out of 16 VYGs ($87.5\%$) lie above the best-fit relation. The KS tests applied to the distributions of the residuals provide $p$-values$\,= 2\times10^{-4}$ and $1\times10^{-3}$ (Figs.~\ref{fig:HIgas}b,d).

%
\begin{figure}
\centering
\begin{tabular}{c}
 \includegraphics[width=0.9\hsize]{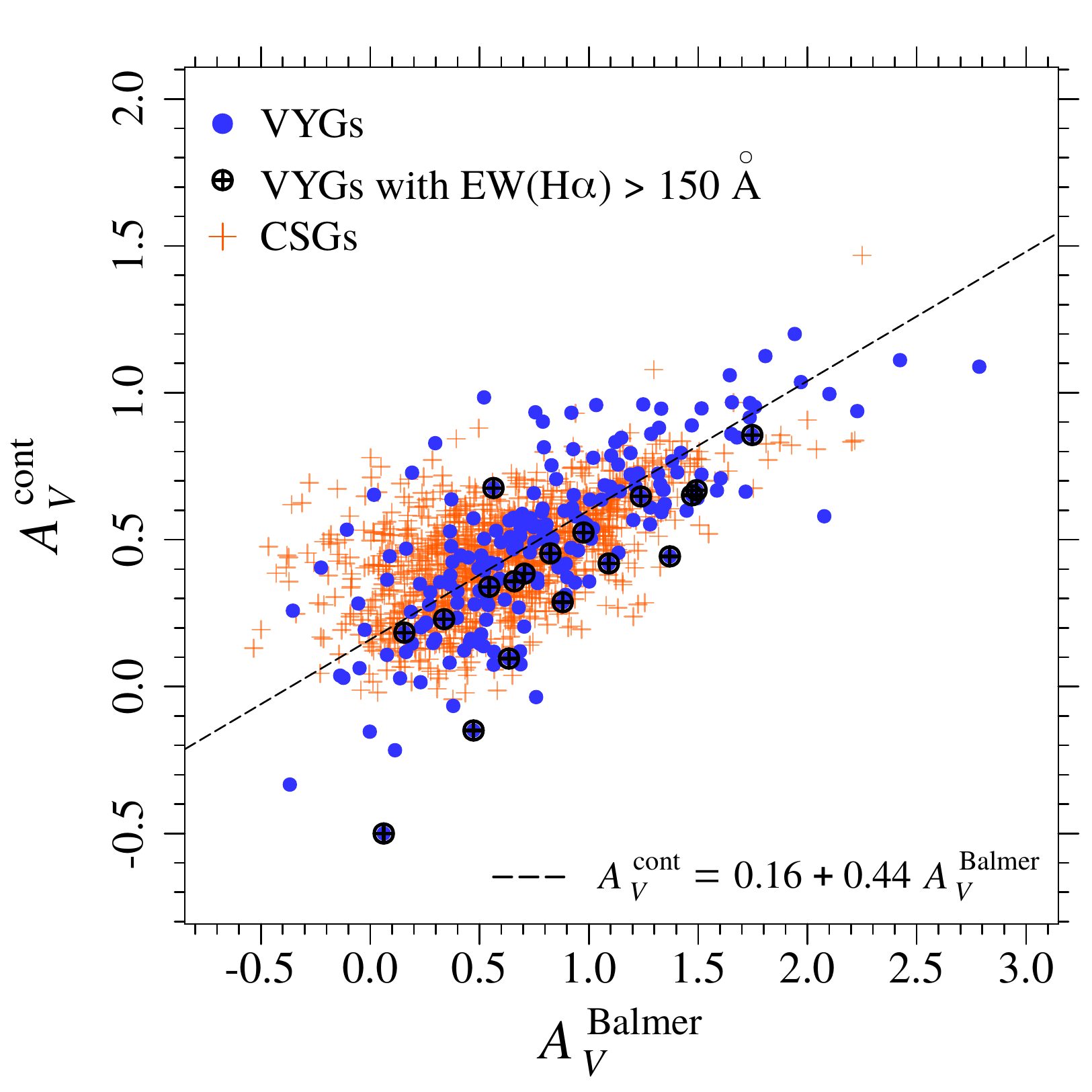}  
\end{tabular}
\caption{Internal extinction estimated from spectral fitting analysis using STARLIGHT vs. extinction obtained from the Balmer decrement. The notation is the same as in Fig.~\ref{fig:sSFR_gi}. The \emph{dashed line} indicates the relation $A_V^{\rm cont} = 0.16 + 0.44 \, A_V^{\rm Balmer}$, which is the relation found by \citet{Calzetti+2000} plus an offset of $0.16$ magnitudes.
\label{fig:Av_Av}
}
\end{figure}

%
\begin{figure*}
\centering
\begin{tabular}{cc}
 \includegraphics[width=0.45\hsize]{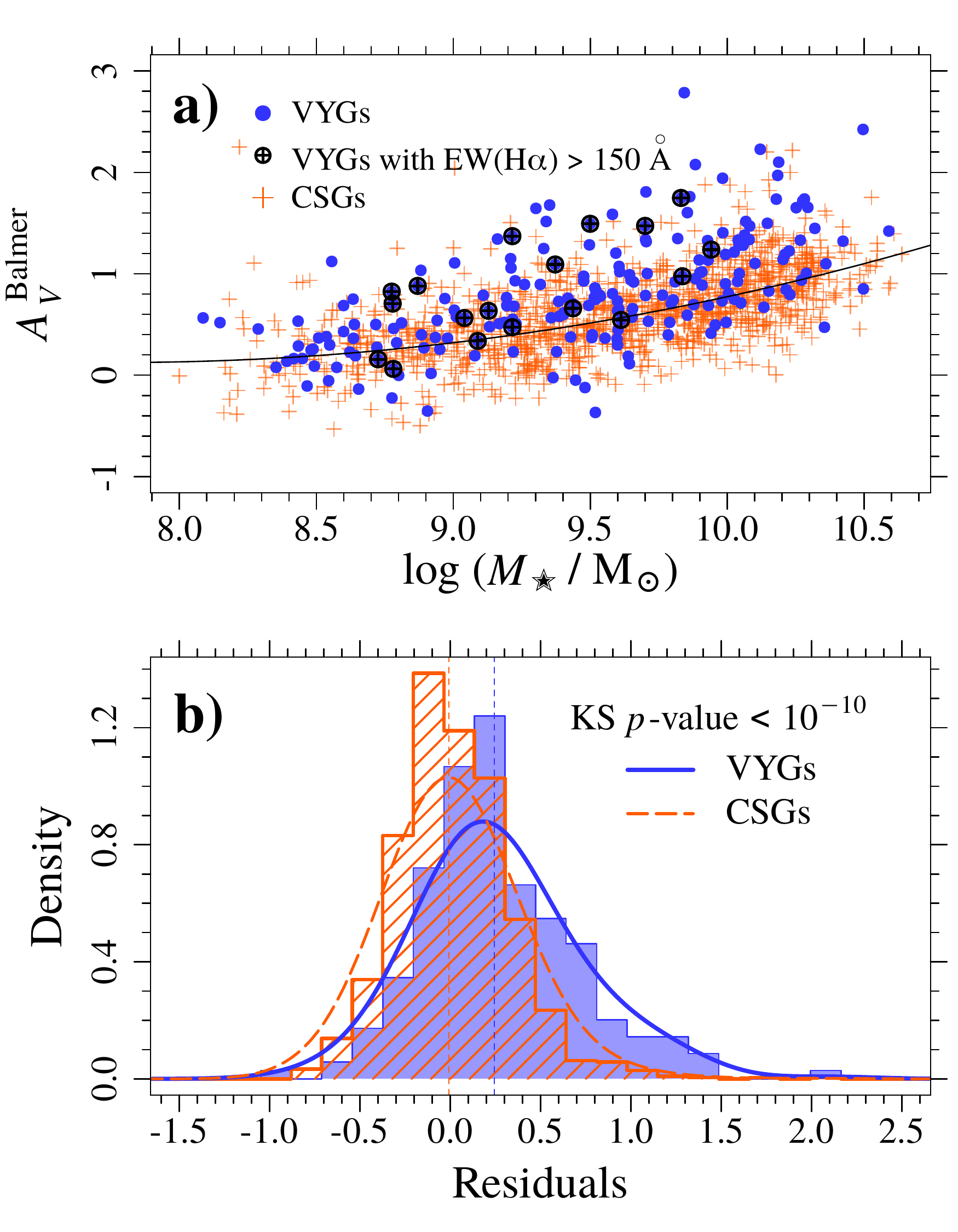}  & 
 \includegraphics[width=0.45\hsize]{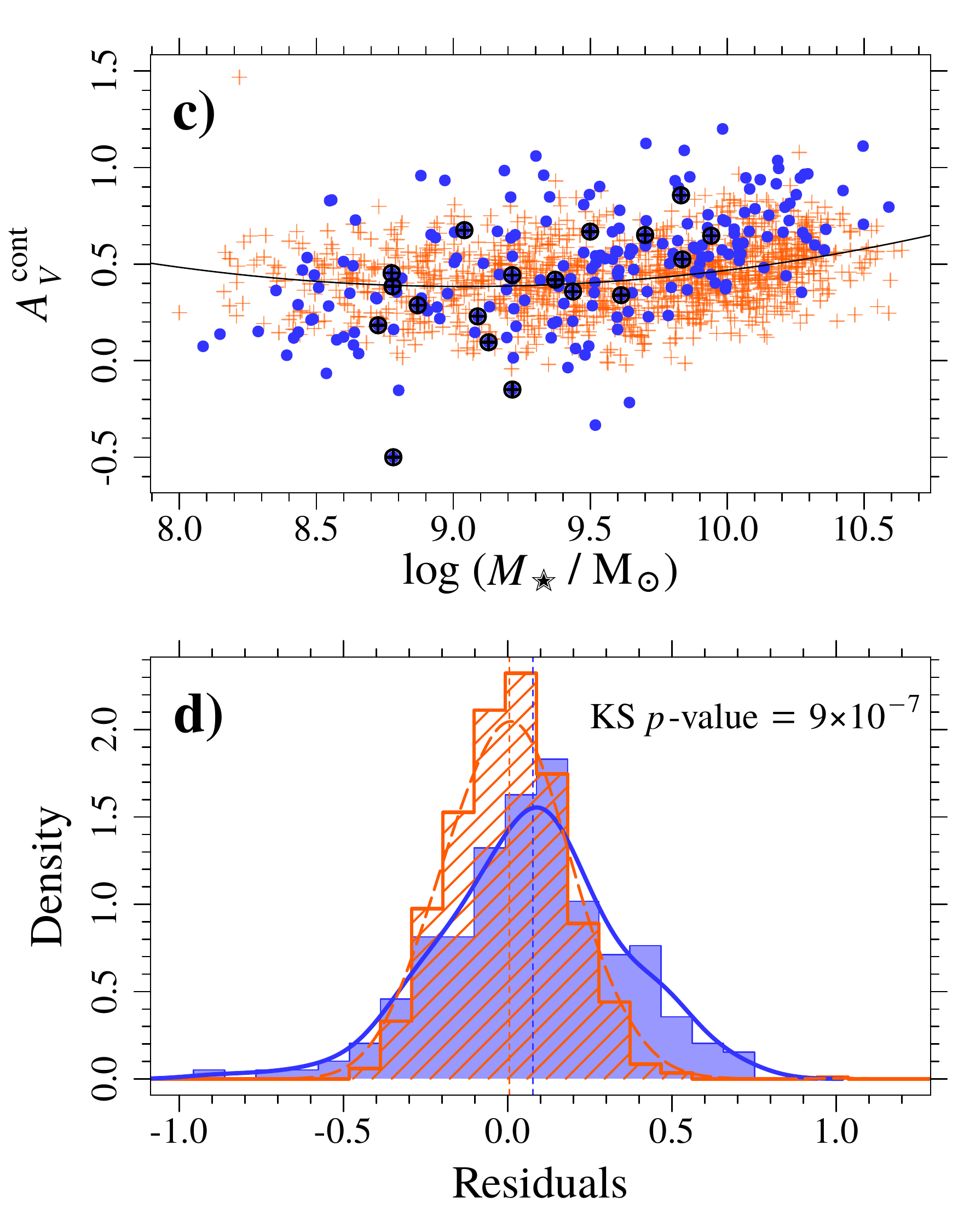} \\
\end{tabular}
\caption{{\bf Top:} Internal extinction estimated from the Balmer decrement
 (\emph{left}) and from spectral fitting analysis using STARLIGHT
 (\emph{right}), both as a function of stellar mass. The solid lines in both panels indicate the second-order polynomial fits to the $A_V^{\rm Balmer}$  and $A_V^{\rm cont}$ vs. stellar mass relations. The notation is the same as in Fig.~\ref{fig:sSFR_gi}. {\bf Bottom:} Residuals from the second-order polynomial fit for the VYG (\emph{blue histograms}) and control (\emph{orange}) samples. 
\label{fig:dust}
}
\end{figure*}

The gas mass fractions of VYGs are $\sim\,$20\% higher than those in CSGs. These large values might be due to the fact that VYGs have not managed to convert the neutral gas into stars before the last Gyr.  We explore in the next section whether environmental effects can be the cause of that delayed SF.

It is important to note that the gas mass fractions and the properties inferred from the SDSS spectra are measured within different volumes of the galaxy, since the ALFALFA beam is $\sim 70$ times the size of the SDSS fibre.

%
\begin{figure*}
\centering
\begin{tabular}{cc}
 \includegraphics[width=0.45\hsize]{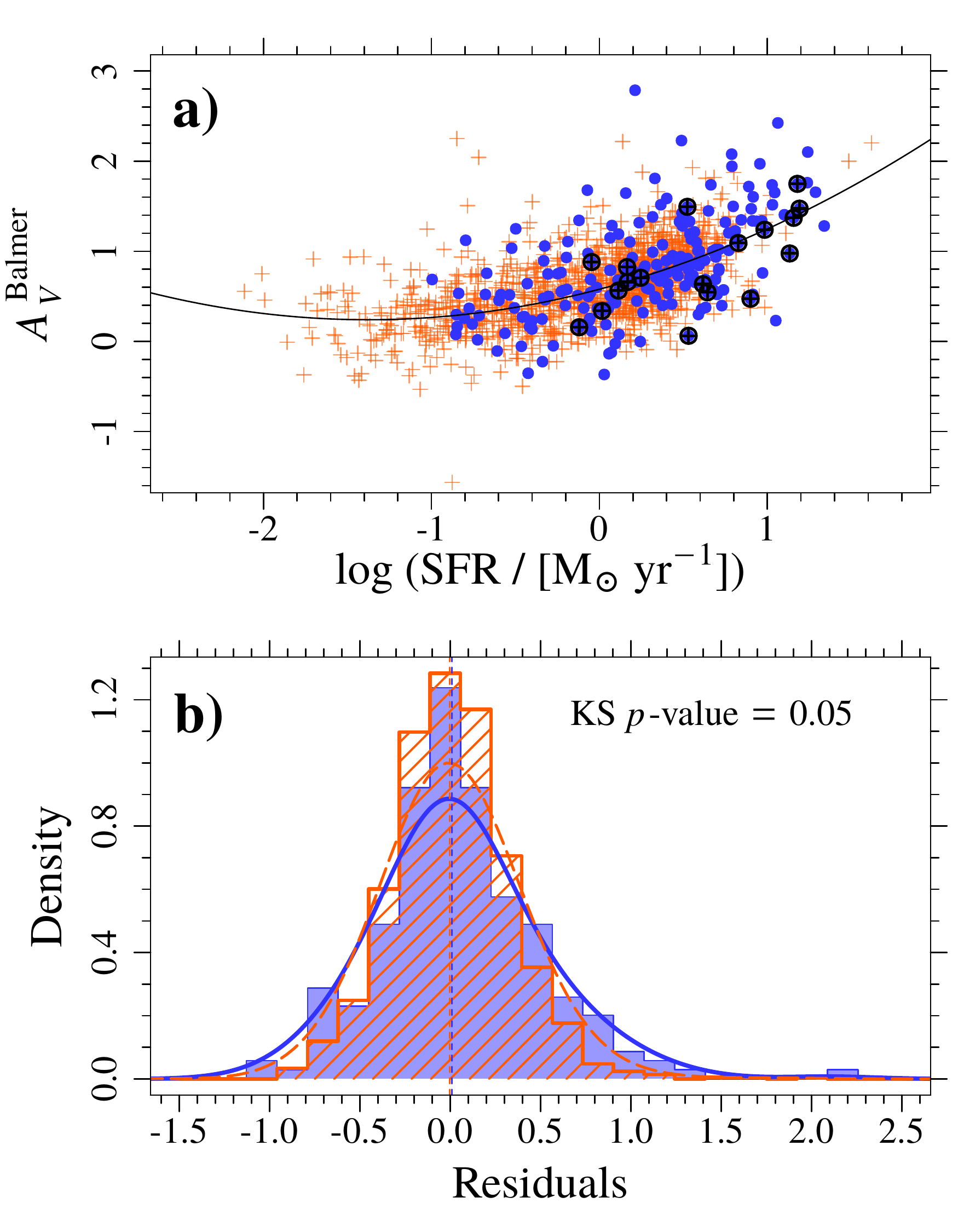}  & 
 \includegraphics[width=0.45\hsize]{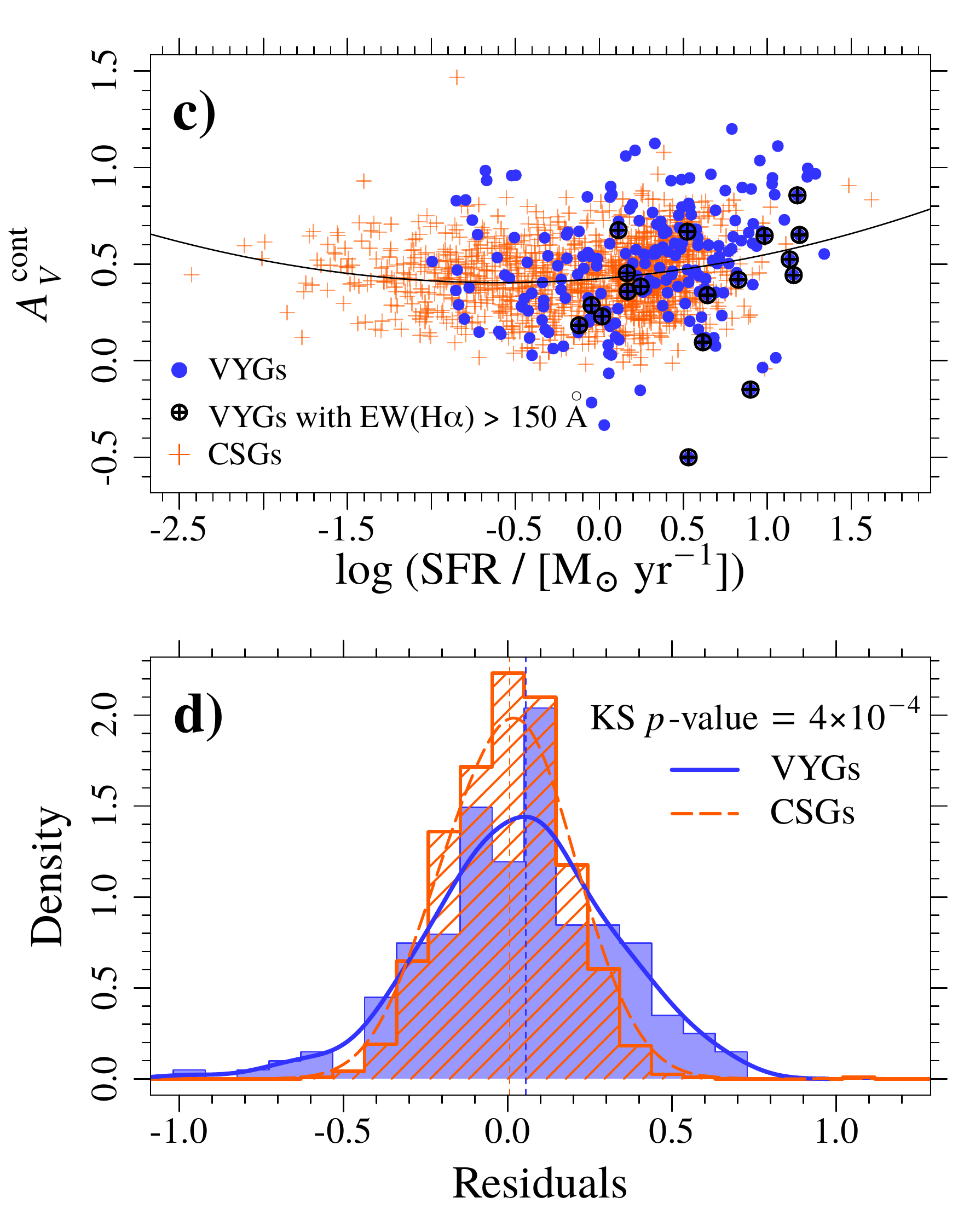} \\
\end{tabular}
\caption{{\bf Top:} Internal extinction estimated from the Balmer decrement
 (\emph{left}) and from spectral fitting analysis using STARLIGHT
 (\emph{right}), both as a function of the star formation rate. The solid lines in both panels indicate the second-order polynomial fits to the $A_V^{\rm Balmer}$  and $A_V^{\rm cont}$ vs. $\log {\rm SFR}$ relations. The notation is the same as in Fig.~\ref{fig:sSFR_gi}. {\bf Bottom:} Residuals from the second-order polynomial fit for the VYG (\emph{blue histograms}) and control (\emph{orange}) samples. 
\label{fig:dust_SFR}
}
\end{figure*}

\subsection{Dust content and internal extinction}

We estimated the amount of dust in our galaxies using two measures. 
We first derived the extinction in the $V$ band, $A_V^{\rm cont}$, from the SPS analysis
using the V15 spectral model with STARLIGHT and assuming the \citet*{Cardelli.etal:1989} reddening law.
We also measured the dust extinction from the Balmer decrement,  $A_V^{\rm Balmer}$, 
which is inferred from the ratio of the H$\alpha$ and
H$\beta$ emission-line fluxes $F_{{\rm H}\alpha}/F_{{\rm H}\beta}$. This ratio
should be 2.87 in the case of no extinction and a temperature $T=10\,000\,\rm K$ \citep{Savage&Mathis79}.
To derive $A_V^{\rm Balmer}$, we first compute the
the colour excess $E(B - V)$ given by 
(see, e.g., \citealp{Momcheva+2013}; and \citealp{Dominguez+13} for details)
$$
E(B - V) = \frac{2.5}{k(\lambda_{{\rm H}\beta}) - k(\lambda_{{\rm H}\alpha})} \log\left[ \frac{F_{{\rm H}\alpha}}{2.87\,F_{{\rm H}\beta}}\right]\ .
$$
\noindent where $k(\lambda_{{\rm H}\beta})$ and $k(\lambda_{{\rm H}\alpha})$ are the values of the
reddening curves at H$\beta$ and H$\alpha$ wavelengths, respectively. The factor $2.5 / [k(\lambda_{{\rm H}\beta}) - k(\lambda_{{\rm H}\alpha})] = 2.33$ 
for the \citealp*{Cardelli.etal:1989} reddening law.
The extinction in the $V$ band is then computed as $A_V^{\rm Balmer} = R_V \,E(B - V)$, 
with $R_V = 3.1$. 

We compared the two extinction estimates in Fig.~\ref{fig:Av_Av}, and it can be seen that $A_V^{\rm Balmer}$ tend to be greater than $A_V^{\rm cont}$. This result has been already found in previous studies (e.g., \citealp{Calzetti+2000}; \citealp{Asari+2007}; \citealp{Kreckel+2013}; \citealp{Florido+2015}) and suggests that $A_V^{\rm Balmer}$ and $A_V^{\rm cont}$ are probing dust in different components of the galaxy ISM. The $A_V^{\rm Balmer}$ value is dominated by dust absorption within molecular clouds in active star-forming regions, while the reddening of the stellar continuum is mainly due to absorption by dust distributed more homogeneously through the galaxy ISM. The slope of the relation between $A_V^{\rm Balmer}$ and $A_V^{\rm cont}$ found by \citet{Calzetti+2000} for starburst galaxies is consistent with our measurements, as indicated by the line shown in Fig.~\ref{fig:Av_Av}. However, we find an offset of $0.16$ magnitudes relative to the \citeauthor{Calzetti+2000} relation, which might be due to more diffuse dust in our galaxies compared to the starburst galaxies studied by these authors. 

In Figs.~\ref{fig:dust}a,c, we show $A_V^{\rm Balmer}$ and $A_V^{\rm cont}$ as a
function of stellar mass for the VYGs and CSGs. We fitted the
$A_V^{\rm Balmer}$ and $A_V^{\rm cont}$ vs. $\log M_{\star}$ relations for the CSGs
and the residuals from these fits are shown in Figs.~\ref{fig:dust}b,d. 
The VYGs have systematically higher internal extinctions, with KS tests
indicating high statistical significance ($p$-values $< 10^{-6}$).
However, Figs.~\ref{fig:dust}a,c suggest that the higher dust content of VYGs
relative to CSGs is most pronounced at high galaxy masses and is nonexistent
($A_V^{\rm Balmer}$) or reversed ($A_V^{\rm cont}$) at low masses. We will discuss 
the differences between the VYGs and CSGs in the high- and low-mass regimes in Sect.~\ref{sec:discuss:mass}.

The amount of light absorbed in the interstellar medium depends on the size,
geometry and inclination of the galaxy. By construction, the distribution of
the CSG physical sizes is close to that of the VYGs, since the angular sizes
and redshifts were included in the matching procedure to construct the control
sample, leading to similar $\theta_{50}$ vs. $z$ distributions shown in
Fig.~\ref{fig:afterPSM_b}c. In any event, we compared the VYG and CSG
$A_V^{\rm Balmer}$ and $A_V^{\rm cont}$, both normalized by galaxy radius:
$A_V^{\rm Balmer}/R_{50}$ and $10^{-0.4\,A_V}/R_{50}$. After fitting the
normalized quantities vs. stellar mass relations and analysing the residuals,
we still find significantly higher extinction in VYGs, with KS
$p$-values$\,=2\times10^{-8}$ (for $A_V^{\rm Balmer}/R_{50}$) and
$2\times10^{-7}$ (for $10^{-0.4\,A_V}/R_{50}$).

We assessed the dependency of the internal extinction on the galaxy geometry
and inclination by using the minor-to-major axis ratio $b/a$ as a proxy for these properties. Admittedly, this is a very simplified approach, given the complexity of the ISM structure and geometry, but the detailed characterization of the dust content of the ISM is beyond the scope of this paper. 
We find that $A_V^{\rm cont}$ increases with decreasing $b/a$, as expected if $b/a$
indicates the inclination of disk galaxies. The Kendall and Spearman tests
show that this anti-correlation is statistically significant, with
coefficients $\tau = -0.15$ and $\rho = -0.22$, and $p$-values$\,< 10^{10}$
for both tests. On the other hand, $A_V^{\rm Balmer}$
does not show any dependency with the axis ratio, with Kendall $\tau = 0.03$ 
and Spearman $\rho = 0.04$ ($p$-values$\,=0.12$
and $0.13$). The difference in the behaviour of $A_V^{\rm cont}$ and $A_V^{\rm Balmer}$ with galaxy inclination reinforces the idea that they are probing dust in different components of the galaxy ISM.

After taking into account the dependency of $A_V^{\rm cont}$ with $b/a$ by fitting the relation for the CSGs and analysing the residuals, the VYGs still have more extinction, with KS $p$-values$\,=1\times10^{-5}$ (for $A_V^{\rm Balmer}$) and $5\times10^{-6}$ (for $A_V^{\rm cont}$). Therefore, the differences between $A_V^{\rm Balmer}$ and $A_V^{\rm cont}$ of VYGs and CSGs do not appear to be due to distinct sizes, geometries and/or inclination angles, but must be truly related to the amount of dust within these systems, which is significantly higher in VYGs compared to the general population of galaxies.


We extracted the IR magnitudes from the Wide-field Infrared Survey Explorer (WISE, \citealt{WISE10})
 database, and computed the $r-W1$ and $r-W2$ colours for the VYGs and CSGs, where $W1$ and $W2$ are the WISE bands at 3.4 and 4.6\,$\mu$m, respectively. We fitted the $r-W1$ and $r-W2$ vs. stellar mass relations, and the residuals show that VYGs tend to be redder than the CSGs, which is consistent with an excess of dust in VYGs.

The higher amount of dust in VYGs compared to the CSGs may be a consequence of their higher star formation rates (SFR), since there is a correlation between the dust mass and SFR \citep[e.g.]{daCunha+2010, Hjorth+2014}. In Fig.~\ref{fig:dust_SFR}, we show $A_V^{\rm Balmer}$ and $A_V^{\rm cont}$ as a function of $\log {\rm SFR}$. Both $A_V^{\rm Balmer}$ and $A_V^{\rm cont}$ increase with increasing SFR, but the correlation is much stronger for $A_V^{\rm Balmer}$. Kendall and Spearman correlation tests confirm the tighter relation between $A_V^{\rm Balmer}$ and SFR compared to the $A_V^{\rm cont}-\log {\rm SFR}$ relation. We obtained the following coefficients and $p$-values: $\tau = 0.45$, $\rho = 0.63$, and 
$p$-value$\, < 10^{-15}$ ($A_V^{\rm Balmer}-\log {\rm SFR}$ relation);  
and $\tau = 0.08$, $\rho = 0.13$, and $p$-value$\, < 10 ^{-5}$ 
($A_V^{\rm cont}-\log {\rm SFR}$ relation).
The different behaviour between $A_V^{\rm Balmer}$ and $A_V^{\rm cont}$ suggests, again, that they trace the absorption by dust of different ISM components. 

As shown in Fig.~\ref{fig:dust_SFR}b, after removing the dependency with SFR, we see only a marginally-significant difference between the VYG and CSG $A_V^{\rm Balmer}$. This result shows that the amount of dust in VYGs is higher due to their higher SFRs compared to those of CSGs. On the other hand, after taking the  $A_V^{\rm cont}-\log {\rm SFR}$ relation into account, VYGs have higher $A_V^{\rm cont}$ values than CSGs (Fig.~\ref{fig:dust_SFR}d). 
These results suggest that, while the dust mass vs. SFR relations for H{\sc II} regions in VYGs and CSGs are similar, the amount of diffuse ISM dust in VYGs is higher compared to that of CSGs. Given that VYGs have a significant fraction of $\lesssim 1$\,Gyr stellar populations, they are expected to have a large number of TP-AGB stars polluting the ISM with dust. Another possibility is that the timescales for the destruction of the ISM dust by sputtering or other processes are longer than 1\,Gyr. 
We will discuss the VYG and CSG internal extinction and their amount of dust in Sect.~\ref{sec:discuss:dust}.

\section{The environment of very young galaxies}
\label{sec:environment}

If physical processes external to the VYGs are responsible for the recent SF activity, their environment is expected to be different from that of the CSGs. 
To investigate this possibility, we analyse successively the \emph{local}, \emph{group} and \emph{large-scale} environments of our sample galaxies.
We identified possible local effects due to interactions with other galaxies in the immediate surroundings, 
by searching for neighbours within 200~kpc from each of our sample galaxies. We also visually inspected the galaxy images and 
classified them according to the presence of tidal features indicating interactions with neighbours or recent mergers.
We then characterized the group environment of a galaxy by the halo mass of the group nearest to it and its position within that group, i.e., its 
distance relative to the group centre. Lastly, we determine the large-scale environment by the position of the galaxies relative to voids and filaments.
We describe  our approach in detail below.

\subsection{Local environment}
\label{sec:environment:local}

%
\begin{figure*}
\centering
\begin{tabular}{cc}
 \includegraphics[width=0.45\hsize]{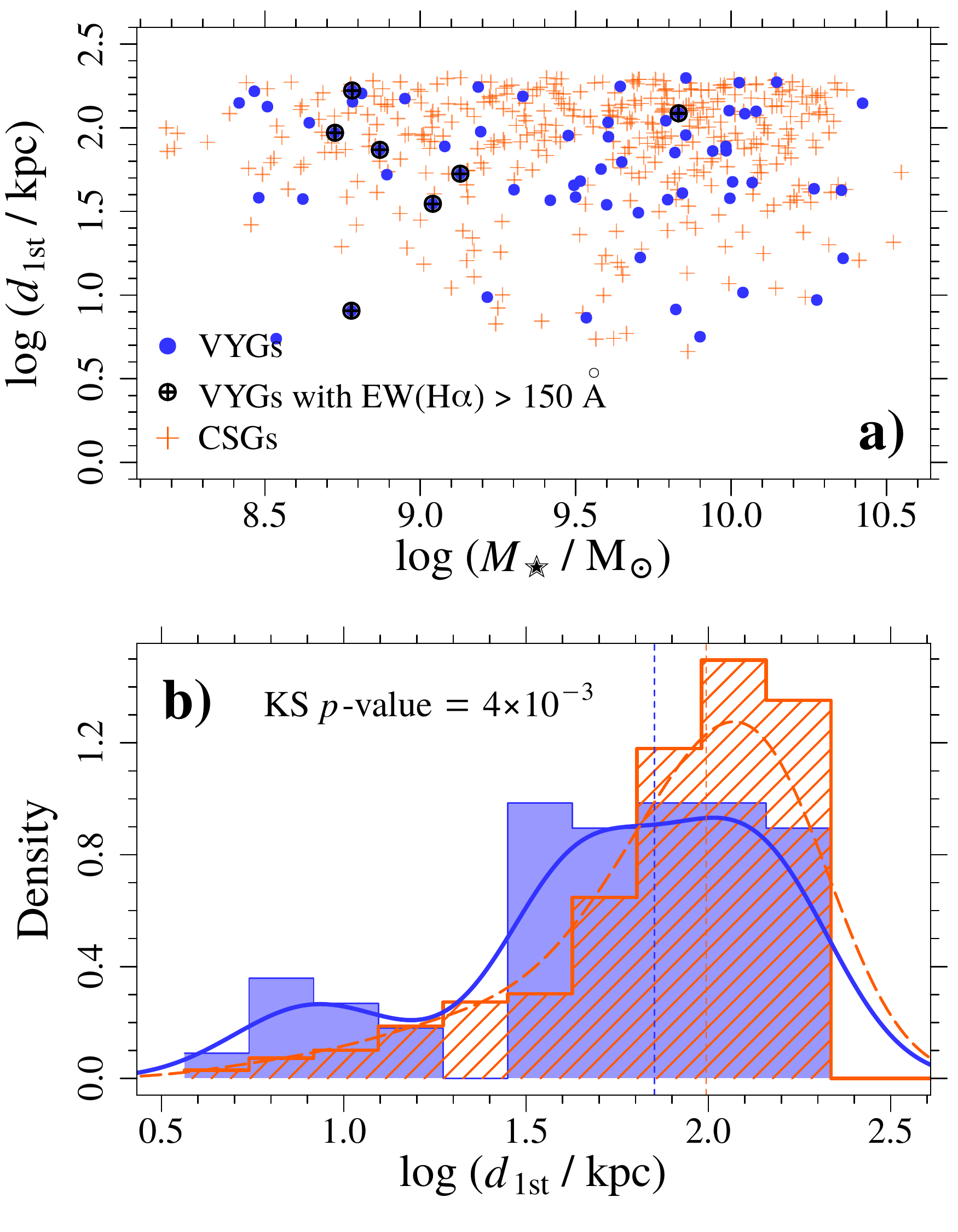}  & 
 \includegraphics[width=0.45\hsize]{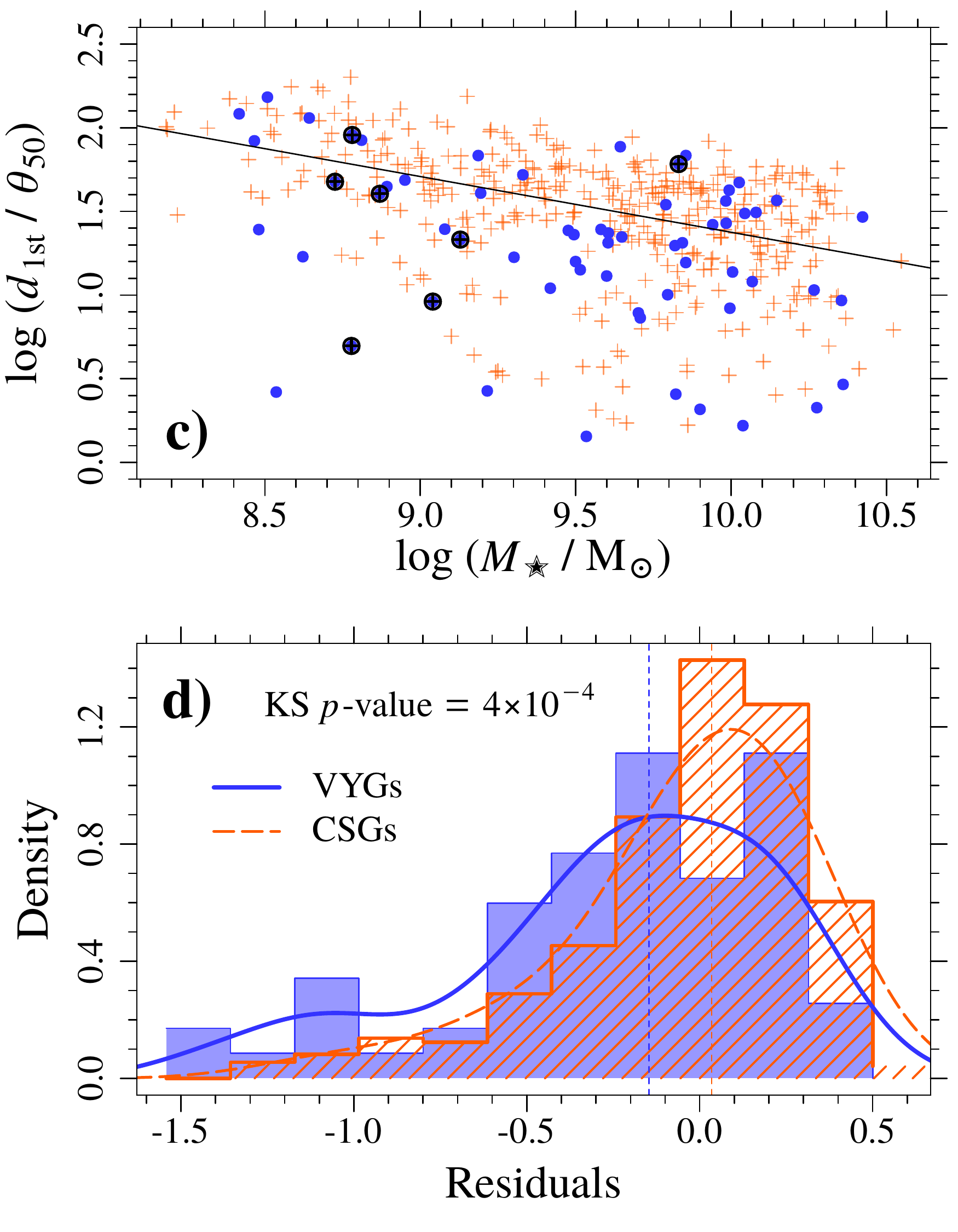} \\
\end{tabular}
\caption{{\bf Top:} Distance to the closest neighbour galaxy with $m_r + 1$, where $m_r$ is the $r$-band  Petrosian magnitude of our galaxies. We show the distances in kpc (\emph{left}) and normalised by the galaxy radius containing 50\% of the Petrosian flux, $\theta_{50}$ (\emph{right}). We show only galaxies that have a neighbour within $200\,$kpc and that are far ($> 200\,$kpc) from bright stars and the borders of the SDSS coverage area (63 VYGs and 392 CSGs).
The notation is the same as in Fig.~\ref{fig:sSFR_gi}. {\bf Bottom:} Distribution of distances in kpc (\emph{left}) and the residuals from the linear fit to the normalised distance vs. stellar mass relation of the CSGs (\emph{right}).
The \emph{blue} and \emph{orange} histograms correspond to the VYG and control sample, respectively and, in each panel, we indicate the $p$-values of the KS tests.
\label{fig:distance_1st_neighbour}
}
\end{figure*}

To investigate the local environment, we have computed 
for each galaxy in the VYG and control samples the distance to its closest neighbour, as follows.
We retrieved from the SDSS-DR12 photometric catalog all galaxies within $200\,$kpc from each of our sample galaxies, not requiring that the neighbour have spectroscopic observations. We adopt this approach because the great majority of the neighbour galaxies do not have listed redshifts in the SDSS. We selected objects with magnitudes $m_r + 1$, where $m_r$ is the extinction-corrected Petrosian magnitude in the $r$ band of our sample galaxy. 

We ensured not missing galaxies by requesting 
that at least 95 per cent of the region within $200\,$kpc from each sample galaxy lies within the SDSS coverage area. For this purpose, we adopted the SDSS-DR7 spectroscopic angular selection function mask\footnote{We used the file {\tt sdss\_dr72safe0\_res6d.pol}, which can be downloaded from \url{https://space.mit.edu/~molly/mangle/download/data.html}.} provided by the NYU Value-Added Galaxy Catalog team \citep{Blanton+05_NYUVAGC}, and assembled with the package {\tt MANGLE 2.1}
\citep{Hamilton.Tegmark:2004, Swanson.etal:2008}. 
After excluding the galaxies close to bright stars or to the borders of the survey, we get a sample of 883 galaxies, among which 117 are VYGs and 766 are CSGs. 

To minimize the contamination by background galaxies, we discard objects with $g-i$ colours $2\, \sigma$ above the red sequence, at the redshift of our sample galaxies. The method used to identify the red sequence at different redshifts is described in Appendix~\ref{ap:red_seq}. 

We found that 428 (out of 883) galaxies do not have any neighbour with magnitude $m_r + 1$ within $200\,$kpc (54 VYGs and 374 CSGs, corresponding to $46.2$\% and $48.8$\%).
Figure~\ref{fig:distance_1st_neighbour} shows the distribution of distances to the closest neighbour with magnitude $m_r + 1$ for the 455 galaxies that have a nearby object at distances $\leq 200\,$kpc. The median distances are $71$ and $98.6\,$kpc for the VYG and control sample, respectively.

We find that 9 out of 117 VYGs (7.7\%) have a close companion at distances smaller than 5 times $R_{50}$, while the number for normal galaxies is 22 out of 766 (2.9\%). A Barnard's test indicates that this difference is statistically significant, with a $p$-value$\,=0.02$. These results indicate that the VYGs are more likely to be interacting or merging with nearby galaxies than the control galaxies. 

\subsection{Interactions and mergers with neighbour galaxies}
\label{sec:environment:mergers}

%
\begin{table}
 \centering
  \caption{Fractions of VYGs and CSGs that show signs of interactions}
\tabcolsep=3pt
 \begin{tabular}{lccll}
 \hline
 Signs of interaction    & VYGs & Control & \multicolumn{2}{c}{$p$-value} \\
                         \cline{4-5} 
  &  &  & \multicolumn{1}{c}{Fisher} & \multicolumn{1}{c}{Barnard} \\
  \hline
  \multicolumn{1}{c}{(1)} & (2) & (3) & \multicolumn{1}{c}{(4)} & \multicolumn{1}{c}{(5)} \\
  \hline\hline
  with neighbours or not       & 40.6$\pm$3.0\% & 23.2$\pm$3.8\% & 0.0002 & 0.0002 \\ 
  galaxy and neighbour         & 12.6$\pm$2.0\% & \ \,6.3$\pm$1.2\% & 0.04 & 0.04 \\ 
  galaxy and close neighbour   & 12.1$\pm$1.8\% & \ \,6.3$\pm$1.2\% & 0.06 & 0.05 \\
  with no neighbour            & \ \,8.2$\pm$1.3\% & \ \,3.9$\pm$2.5\% & 0.10 & 0.09 \\
  \hline
 \end{tabular}
\label{tab:interacting}
\end{table}

Besides identifying the closest neighbour, as described above, we have also looked for signs of interactions or features indicating recent merger events. To identify merging systems, we use the classification from the Galaxy Zoo 1 project \citep{Lintott.etal:2011}. The method consists in converting a set of visually-inspected classifications by hundreds of thousands of volunteers into a single parameter, $p_{\rm merger}$, which corresponds to the weighted fraction of votes in the ``merger'' category. 
\citet{Darg.etal:2010} have shown that galaxies with $p_{\rm merger} \gtrsim 0.4$ are, in fact, true mergers. 
Although $p_{\rm merger}$ is certainly related to the probability that a galaxy is part of an ongoing merger, adopting a single critical $p_{\rm merger}$ threshold to classify a galaxy as merging may be over-simplistic. 

\begin{figure}
    \centering
 \includegraphics[width=0.98\hsize]{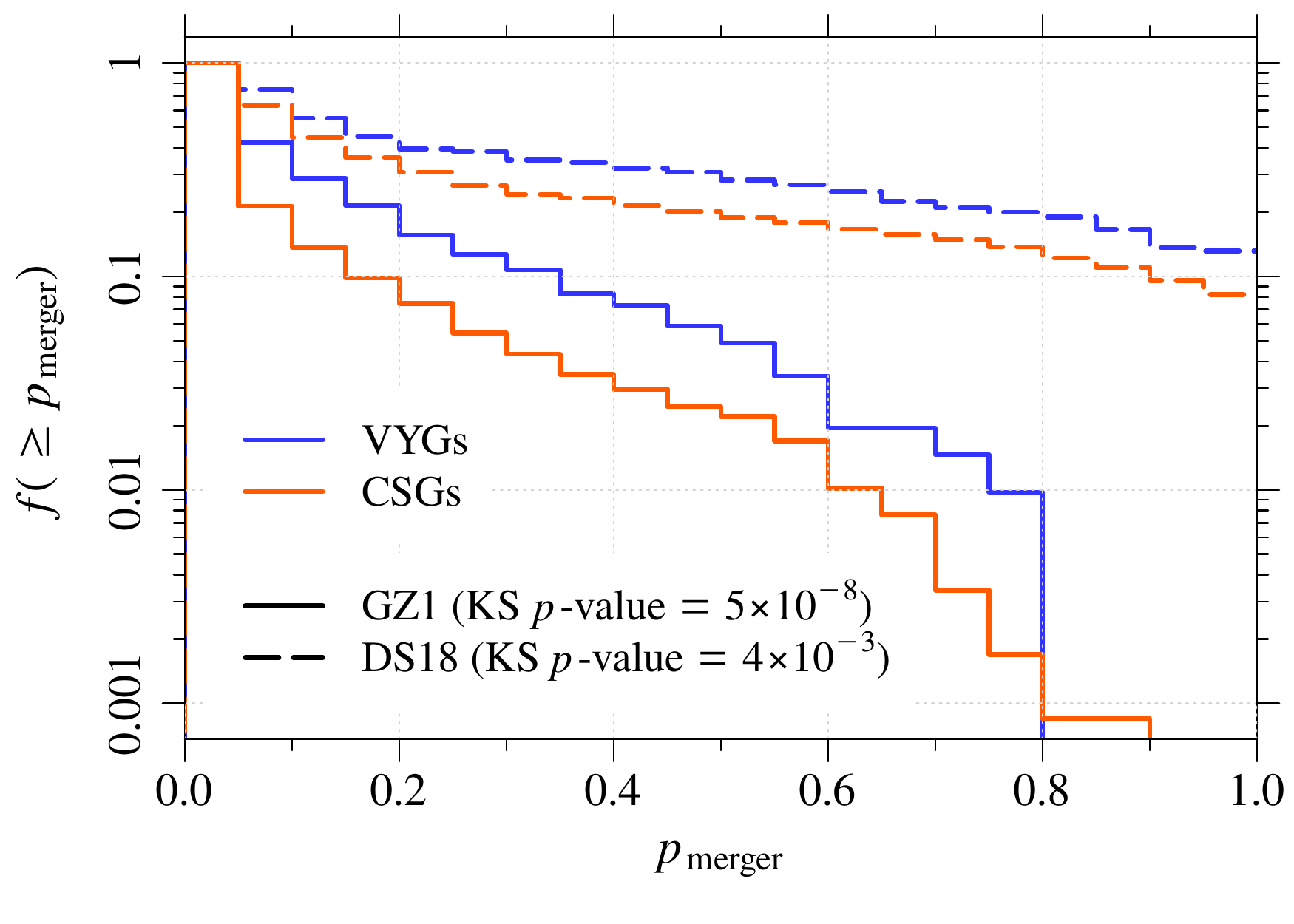}  
    \caption{Fraction of galaxies that are classified as mergers in Galaxy
 Zoo (GZ1, \emph{solid lines}) and by \citet{DominguezSanchez.etal:2018}
 (DS18, \emph{dashed lines}) for different threshold values of $p_{\rm
 merger}$. The \emph{blue} and \emph{orange lines} correspond to the VYGs and
 CSGs, respectively.
 }
    \label{fig:p_mergers}
\end{figure}

To avoid choosing a specific threshold, we computed the fraction of VYGs and
CSGs that could be classified as merging systems for different $p_{\rm
merger}$ threshold values, as shown in Fig.~\ref{fig:p_mergers}. Regardless
of the threshold adopted, the fraction of VYGs that are merging is $\sim 2$
times higher than that of the CSGs (KS $p$-value$\,= 5 \times 10^{-8}$).
In the same figure, we also show the merger fraction according to the morphological classification by \cite{DominguezSanchez.etal:2018}. The fractions are higher than those based on Galaxy Zoo, and the difference between the VYGs and CSGs is smaller ($1.5$ times higher), but still with a high statistical significance (KS $p$-value$\,= 0.004$). 

However, the Galaxy Zoo classification is biased towards a specific phase of the merger, i.e., when the two galaxies are close enough to be classified as merging. Interacting systems in the early phases of the merger and those which may never merge will be missed. Besides, the features indicating the post-merger phase might be only identifiable by expert astronomers. 

%

Therefore, we performed our own visual classification based on the presence
of tidal features indicating interactions with neighbour galaxies. The
classification was performed independently by five members of our team. Each
participant classified 414 galaxies (207 VYGs and 207 CSGs for comparison),
and was asked to identify tidal features and signs of interactions in the
test galaxies as well as their neighbours. To classify the galaxies, each
vote received a value $v$, depending on how visible the interaction features
are. The participants could answer \texttt{yes}, if they clearly see
interaction signs ($v_{\tt yes} = 1$), \texttt{maybe} if they are not sure
($v_{\tt mb} = 0.5$) or \texttt{no} if they see none ($v_{\tt no} = 0$). For
the neighbours, the vote for \texttt{no neighbours} received $v_0 = -1$. We
then computed the mean of the values for the galaxy, $p_{\rm i,g}$, and its
neighbour, $p_{\rm i,n}$,  as follows:
\begin{eqnarray} 
p_{\rm i,g} &\!\!\!\!=\!\!\!\!& \frac{v_{\rm yes} \, N_{\tt yes,g} + v_{\tt mb} \, N_{\tt
mb,g}}{N_{\rm votes}} \ , \nonumber \\
p_{\rm i,n} &\!\!\!\!=\!\!\!\!& \frac{v_{\rm yes} \, N_{\tt yes,n} + v_{\tt
mb} \, N_{\tt mb,n} + v_{\rm 0} \, N_{\rm 0,n}}{N_{\rm votes}} \ , \nonumber
\end{eqnarray}
\noindent where $N_{\tt yes,g}$ and $N_{\tt mb,g}$ are the number of {\tt
yes} and {\tt maybe} votes for the galaxy and $N_{\tt yes,n}$, $N_{\tt mb,n}$
and $N_{\rm 0,n}$ are the number of {\tt yes}, {\tt maybe} and {\tt no-neighbour} votes for the galaxy neighbour. If the mean of the votes is $p_{\rm i} \ge 0.75$, the galaxy is classified as ``interacting''.
More details of our interaction classification scheme are given in Appendix~\ref{ap:classification}.

The advantage of our classification scheme of interactions is that it also
considers the neighbouring galaxies. In Table~\ref{tab:interacting}, we
compare the fractions of interacting VYGs and CSGs regardless of whether
there is a neighbour, and also those of galaxies that have neighbours
classified as interacting. We also computed the fractions of interacting
galaxies that have no neighbour or those that have neighbours with no sign of
interaction. In all three cases, the fraction of VYGs classified as
interacting is roughly twice that of CSGs, in agreement with the results obtained with the Galaxy Zoo classification. 

\begin{figure}
    \centering
 \includegraphics[width=0.95\hsize]{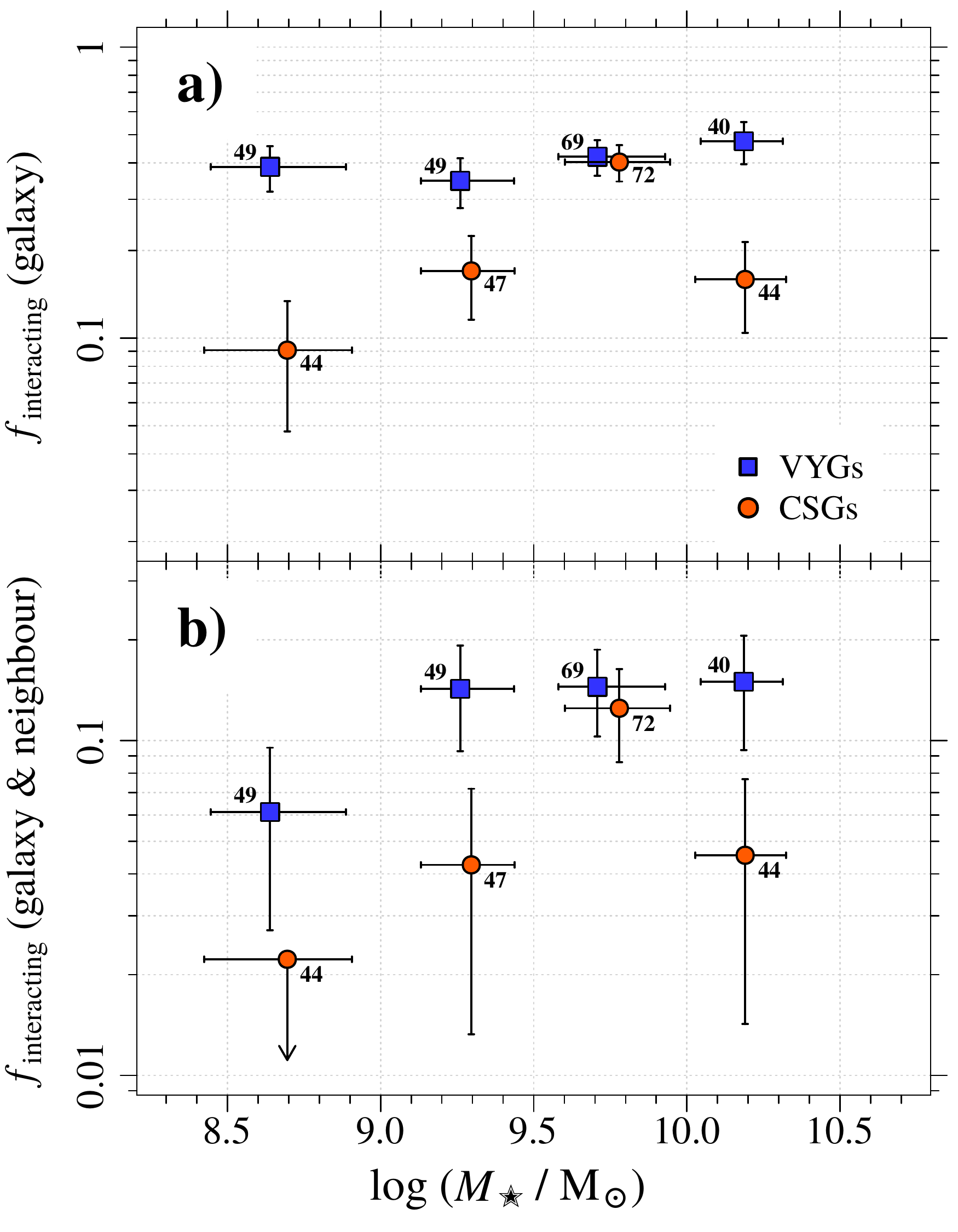}  
 \caption{Fraction of galaxies with visually detected signs of interaction (from our 5-expert analysis)  among 
 VYGs (\emph{blue squares}) and CSGs (\emph{orange
 circles}), as a function of their stellar mass.
 In
 the \emph{upper panel}, we include all galaxies classified as interacting,
 regardless of neighbours. In the \emph{bottom panel}, we show the fractions of
 systems where the neighbour galaxy also show signs of interactions. 
 The \emph{error bars} indicate the $1\sigma$ binomial errors for $f_{\rm interacting}$ and the 16th
 and 84th percentiles of $\log M_{\star}$.
 }
    \label{fig:frac_interaction_logM}
\end{figure}

In Fig.~\ref{fig:frac_interaction_logM}, we show how the fractions of
galaxies, classified as interacting according to our scheme, vary as a
function of galaxy stellar mass. In the upper panel, we show the fractions of
all interacting galaxies regardless whether there is a neighbour or not. The
fraction of VYGs that are interacting is roughly independent of stellar mass at $\sim
30-40\%$, while the corresponding fractions of CSGs increases with $M_{\star}$ up to $M_{\star} \sim 10^{10}\,$M$_{\odot}$, and decreases again for $M_{\star} \gtrsim 10^{10}\,$M$_{\odot}$.
The difference between these two samples is larger for low-mass galaxies:  $38.8\%$ of the VYGs with $M_{\odot} < 10^9\,$M$_{\odot}$ are classified as interacting, a fraction that is $8.5$ times higher than that of CSGs in the same mass range ($f_{\rm interacting}= 4.5\%$). 

The fraction of interacting galaxies with a companion also classified as interacting is shown in the lower panel of Fig.~\ref{fig:frac_interaction_logM}. For the VYG sample, $f_{\rm interacting} = 6.1\%$ for galaxies with $M_{\odot} < 10^9\,$M$_{\odot}$ and varies between $10$ and $15\%$ for more massive objects ($M_{\odot} > 10^9\,$M$_{\odot}$). These fractions are $> 6.7$ times higher when compared to those of the CSGs, except for the mass bin $9.5 < \log (M_{\star} / {\rm M}_{\odot}) < 10$, where the $f_{\rm interacting}$ are similar for the two samples. 

\begin{figure}
    \centering
 \includegraphics[width=0.95\hsize]{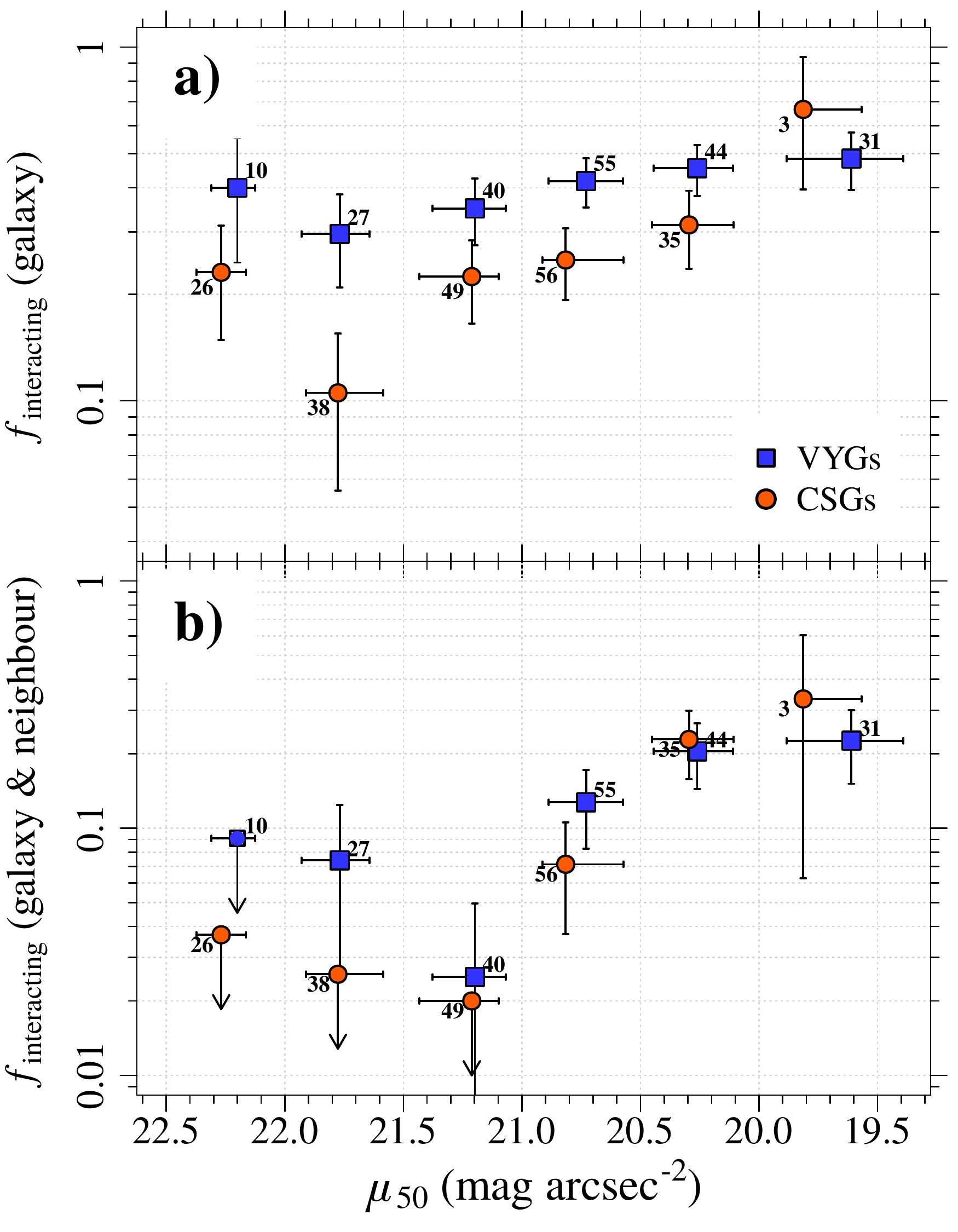}  
 \caption{Fraction of VYGs (\emph{blue squares}) and CSGs (\emph{orange circles}) that show signs of interactions in bins of galaxy surface brightness. The notation is the same as in Fig.~\ref{fig:frac_interaction_logM}. }
    \label{fig:frac_interaction_mur}
\end{figure}

Could this higher fraction of interacting galaxies among VYGs be due to a surface brightness effect? 
The VYGs have higher surface brightness compared to the CSGs, as shown in Fig.~\ref{fig:PSM_C_mu}.
This could naturally lead to higher fractions of galaxies classified as
interacting, since it would be easier to identify structures and features in
the galaxy images. To test this hypothesis, we compared $f_{\rm interacting}$
in bins of $\mu_{50}$. As we can see in Fig.~\ref{fig:frac_interaction_mur}a,
the fraction of interacting VYGs is always higher than that of CSGs with
similar $\mu_{50}$, with differences ranging from $1.5$ to $5.2$ times
higher, except for the brightest $\mu_{50}$ bin. But the difference is within
the errors and might be a result of poor statistics, since there are only 3
CSGs in this bin. We conclude that the higher fraction of interacting galaxies
among VYGs is not caused by surface brightness effects. 

Finally, in Fig.~\ref{fig:frac_interaction_mur}b, we show $f_{\rm interacting}$ when both the galaxy and its neighbour(s) show signs of interactions. For the faintest $\mu_{50}$ bin ($\mu_{50} > 22$), no systems are classified as interacting. The CSG fractions are lower than those of the VYGs for galaxies with $20.5 < \mu_{50} < 22$, but they are similar for brighter objects ($\mu_{50} < 20.5$). However, the interpretation of these results is not straightforward, since these fractions also depend on the surface brightness of the neighbours that can be fainter than that of our galaxies. 

In summary,  VYGs exhibit  a significantly higher fraction of mergers and
interactions than do CSGs.   

%

%

\subsection{Group environment}
\label{sec:environment:group}

We now analyze the group environment of the VYGs and CSGs.
We selected groups and clusters from the updated version
of the catalogue compiled by \citet{Yang+07}. The 
catalogue contains $473\,482$ groups drawn from a sample of
$601\,751$ galaxies, mostly from the SDSS-DR7 \citep{Abazajian+09}.
Only groups with $\log (M_{\rm halo}/{\rm M}_{\odot}) \geq 12.3$ were selected, and we assigned our sample galaxies 
to the nearest group, following the method described in \citet*{Trevisan+17_Let}. 
A galaxy is assigned to the group that gravitationally attracts it the most, 
i.e. the group with the lowest distance, in units of virial radius, $r_{\rm vir}$. 

We define $r_{\rm vir}$ of a group as the radius, $r_{100}$, of a sphere that
is $\Delta_{\rm v} = 100$ times denser than the critical density of the
Universe. We obtained $r_{\rm vir}$ by first deducing $r_{200,{\rm
m}}$ (of spheres that are 200 times denser than the \emph{mean} density of
the Universe
from the $M_{200,{\rm m}}$ masses given in the \citeauthor{Yang+07}
catalogue (which are based on abundance matching with the group luminosities).
We then calculated the $r_{\rm vir}$, following appendix~A of \citet*{Trevisan+17} for 
the conversion from quantities relative to the mean density to those relative to the critical density,
and the corresponding virial masses, $M_{\rm vir} = (\Delta_{\rm v}/2)\,
H^2(z)\,r_{\rm vir}^3/G$.

To assign the galaxies to their nearest group, we compute the distances, $d$, between the galaxies and the group centres, assuming two regimes. For galaxies far away from the group, $d$ is given by the redshift-space distance. 
For a galaxy close to a group, the strong redshift distortions are taken into account by 
using the overdensity in projected phase space introduced by \cite{Yang+05,Yang+07}. 
We convert this overdensity to an equivalent redshift-space distance by
joining the two estimators at a given radius, $R_{\rm n}$,  
which marks the transition between the non-linear and the linear regimes. We adopt $R_{\rm n} = 2\, r_{\rm vir}$, and galaxies can be assigned to distances up to $20\, r_{\rm vir}$ from a group.
A small fraction ($< 4\%$) of the galaxies was not assigned to any group (7 VYGs and 48 CSGs). Among these galaxies, 22 of them were not assigned because they lie at $z < 0.01$, which is the lower redshift limit of the group catalogue. The other 33 galaxies lie close to the borders of the SDSS Survey or a bright star, and the group catalogue may be incomplete in these regions. Using the package {\tt MANGLE 2.1}, as described in Sect.~\ref{sec:environment:local}, we estimate that more than 20\% of the region within $500\,$kpc from these galaxies lies outside the SDSS coverage area.

\begin{figure}
    \centering
 \includegraphics[width=0.95\hsize]{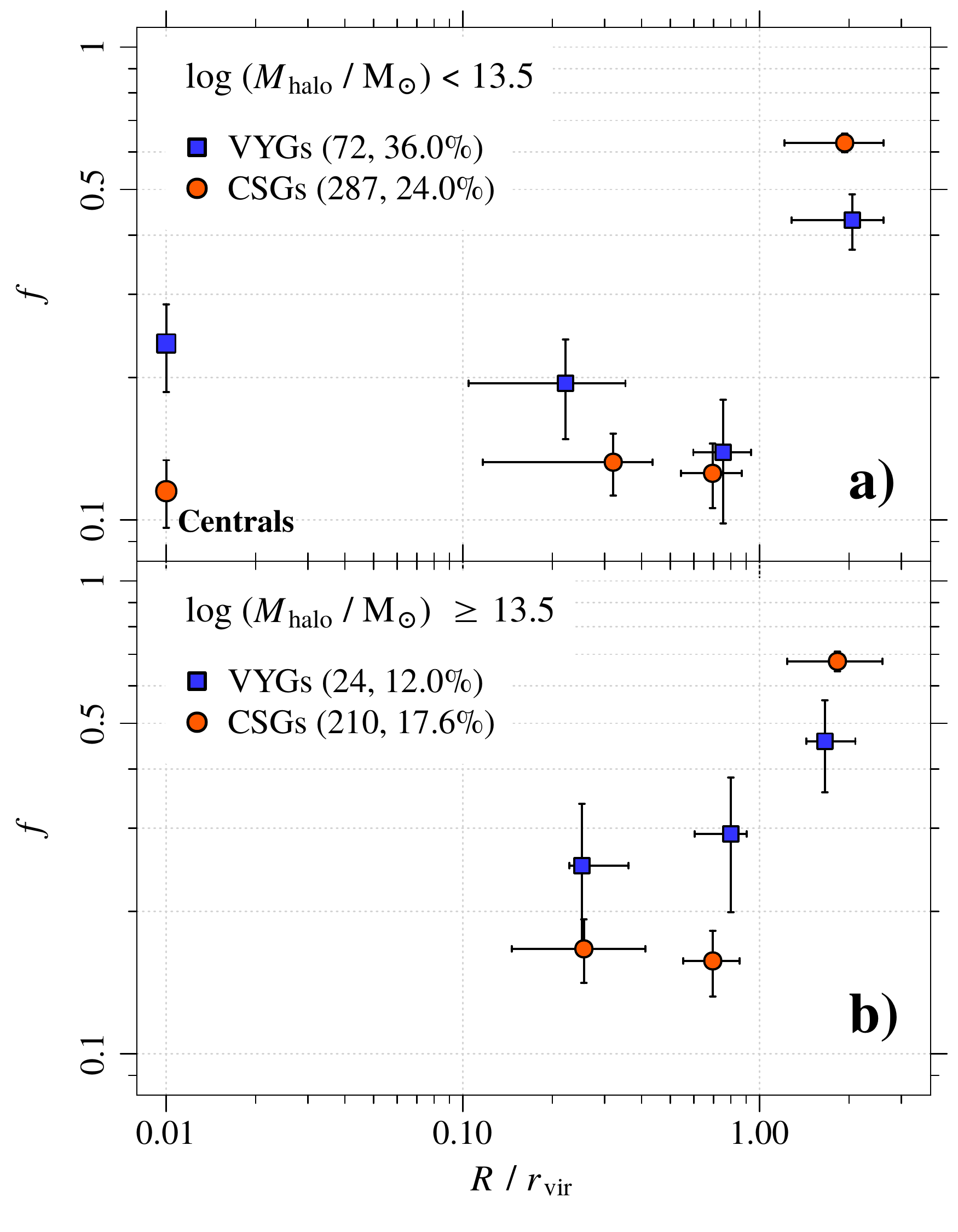}  
 \caption{Distribution of radial positions relative to nearest group among VYGs (\emph{blue squares}) and CSGs (\emph{orange
 circles}).
 The \emph{upper} and \emph{lower panels} show the radial distributions of galaxies relative to low and high mass groups, respectively  
 (there are no centrals among VYGs and CSGs in the higher mass groups, because nearly all centrals of high-mass groups are passive galaxies). 
 The error bars indicate the $1\sigma$ binomial errors for $f$ and the 16th
 and 84th percentiles of $R/r_{\rm vir}$. 
} 
    \label{fig:frac_groups}
\end{figure}

We computed the fractions of VYGs and CSGs at different distances from the
group/cluster centre. 
Table~\ref{tab:frac_groups} indicates that roughly half of galaxies are found at distances 
$R > 3\,r_{\rm vir}$ from the group centres (52\% and 58.4\% of the VYGs and CSGs, respectively).
Analysing the galaxies within groups, we see that VYGs
reside preferentially in low-mass haloes compared to CSGs. As shown in
Fig.~\ref{fig:frac_groups} and Table~\ref{tab:frac_groups}, the fraction of
VYGs that reside in groups with $\log (M_{\rm halo}/{\rm M}_{\odot}) < 13.5$
is $\sim$ 50 per cent higher than the corresponding fraction of CSGs. Fisher
and Barnard tests indicate a high statistical significance for this result ($p$-values $= 0.0006$ and $0.002$, respectively).

The way that the VYGs are distributed within and around the group haloes is also different from that of the CSGs. 
By comparison to CSGs, VYGs are more likely to lie within groups.\footnote{Of course, projection effects limit us to a cylindrical view of groups, preventing us from knowing which galaxies are within or outside groups defined as spheres. These projection effects are much more severe for star-forming galaxies: nearly half of those that are inside the virial cylinder actually lie outside the virial sphere \citep{MMR11}. However, both VYGs and CSGs are star-forming systems, and it is unlikely that projection effects would affect these two samples differently.}
As shown in Fig.~\ref{fig:frac_groups} and Table~\ref{tab:frac_groups}, $~63\%$ of the CSGs in low-mass haloes are found between $1$ and $3\,r_{\rm vir}$, while only $43\%$ of the VYGs reside in this region. The fraction of VYGs that are centrals is double that of CSGs, with statistical tests indicating a marginal significance ($p$-values $\sim 0.02$). 

Only a small fraction of VYGs and CSGs are in more massive haloes with $\log [M_{\rm halo}/{\rm M}_{\odot}] \geq 13.5$ ($12\%$ of the VYGs and $17.6\%$ of the control sample). But we find that VYGs are also more likely to be found in the inner parts of the these high-mass groups, with $54.2\%$ of the VYGs lying at distances $< 1\,r_{\rm vir}$. For the CSGs, this fraction is only $32.4\%$, with statistical tests indicating that this difference is marginally significant ($p$-values $\sim 0.04$). 

From the results shown in Fig.~\ref{fig:frac_groups} and Table~\ref{tab:frac_groups}, we see that the VYGs are more likely to be found in the inner parts of low mass groups when compared to the CSGs. We find 41 VYGs that lie within $1\,r_{\rm vir}$ from the centre of low mass groups. These number corresponds to 20.5\% of all 200 VYGs that were assigned to a halo. This fraction is much higher than that of CSGs residing in the inner regions of low-mass groups: only 9\% (107 out of 1194) of the CSGs are at distances $<1\,r_{\rm vir}$ from the centre of haloes with $\log (M_{\rm halo}/{\rm M}_{\odot}) < 13.5$. Fisher's and Barnard's tests indicate high statistical significance, with $p$-values of $6\times10^{-6}$ and $6\times10^{-5}$, respectively.


We checked if the definition of the group and cluster centres could affect our results. In our group assignment scheme, we assume the position of the brightest group galaxy to be the centre of the group. Since VYGs have young stellar populations, they are expected to be brighter than non-VYGs with similar masses, and this could lead to higher fractions of VYGs that are centrals compared to normal galaxies. Therefore, we repeated our assignment procedure using the most massive galaxy as the centre of the group haloes. We find that the number of VYGs that are central galaxies in haloes with $\log (M_{\rm halo}/{\rm M}_{\odot}) < 13.5$ remains unchanged (17 VYGs). On the other hand, the number of centrals in the control sample decreases from 33 ($11.5\%$) to 27 ($9.2\%$), and the statistical significance of the result of a larger fraction of centrals in VYGs increases ($p$-values$ = 0.002 - 0.003$). 

%
\renewcommand{\arraystretch}{1.5}
\begin{table*}
 \centering
 \caption{Kolmogorov-Smirnov $p$-values for different comparisons between
 VYGs and CSGs}
 \footnotesize
 \begin{tabular}{c|cccc|ccc|c|c}
   \hline
     & \multicolumn{4}{c|}{\large $12.3 \leq \log M_{\rm halo} < 13.5$}  & 
     \multicolumn{3}{c|}{\large $13.5 \leq \log M_{\rm halo}$} & 
     {\large Field} & {\large Total} \\

     & Centrals & $0  - 0.5\, r_{\rm vir}$ & $0.5 - 1\, r_{\rm vir}$ &
     $1  - 3\, r_{\rm vir}$ & 
     $0  - 0.5\, r_{\rm vir}$ & $0.5 - 1\, r_{\rm vir}$ &
     $1  - 3\, r_{\rm vir}$ & 
     $> 3\, r_{\rm vir}$ & \\
     \hline \hline

     \multirow{2}{*}{VYGs} & {\bf 17 (23.6\%)} & 14 (19.4\%) & 10 (13.9\%) & {\bf 31 (43.1\%)} &  
     6 (25\%) & 7 (29.2\%) & {\bf 11 (45.8\%)} & & \\

& \multicolumn{4}{c|}{\cellcolor{lightgrey}\bf 72 (36\%)}
     &  \multicolumn{3}{c|}{\cellcolor{lightgrey}24 (12\%)} &  
     \multicolumn{1}{c|}{\cellcolor{lightgrey}104 (52\%)} & \multicolumn{1}{c}{\cellcolor{lightgrey}200} \\

     \hline

     \multirow{2}{*}{Control} & {\bf 33 (11.5\%)} & 38 (13.2\%) & 36 (12.5\%) & {\bf 180 (62.7\%)} &  
    35 (16.7\%) & 33 (15.7\%) & {\bf 142 (67.6\%)} &  & \\
    
     & \multicolumn{4}{c|}{\cellcolor{lightgrey}\bf 287 (24\%)}
     &  \multicolumn{3}{c|}{\cellcolor{lightgrey}210 (17.6\%)} &  
     \multicolumn{1}{c|}{\cellcolor{lightgrey}697 (58.4\%)} & \multicolumn{1}{c}{\cellcolor{lightgrey}1194}   \\
    \hline

     \multicolumn{10}{c}{\bf \large Fisher's tests} \\
     \hline    \hline
     \multirow{2}{*}{$p$-value} & {\bf 0.013} & 0.192 & 0.843 & {\bf 0.003} &  
    0.392 & 0.146 & {\bf 0.042} &  & \\
    
     & \multicolumn{4}{c|}{\cellcolor{lightgrey}\bf 0.0006}
     &  \multicolumn{3}{c|}{\cellcolor{lightgrey}0.052} &  
     \multicolumn{1}{c|}{\cellcolor{lightgrey}0.104} & \multicolumn{1}{c}{\cellcolor{lightgrey}}   \\
    \hline
    
         \multicolumn{10}{c}{\bf \large Barnard's tests} \\
     \hline    \hline
     \multirow{2}{*}{$p$-value} & {\bf 0.017} & 0.108 & 0.535 & {\bf 0.005} &  
    0.213 & 0.095 & {\bf 0.044} &  & \\
    
     & \multicolumn{4}{c|}{\cellcolor{lightgrey}\bf 0.002}
     &  \multicolumn{3}{c|}{\cellcolor{lightgrey}0.065} &  
     \multicolumn{1}{c|}{\cellcolor{lightgrey}0.065} & \multicolumn{1}{c}{\cellcolor{lightgrey}}   \\
    \hline
    \end{tabular}
    \\
    \parbox{\hsize}{Notes:
    The number of galaxies in groups are divided in bins of projected distance to the group centre, $R$. The number of central galaxies is shown only for low-mass groups, since none of our VYG and CSGs are found in the centre of high-mass groups. The last lines show the $p$-values of Fisher's and Barnard's tests comparing the fractions of VYGs and CSGs in each bin of $R $ and in the field. Results with $p$-values$< 0.05$ are highlighted in boldface.}
    \label{tab:frac_groups}
\end{table*}
\renewcommand{\arraystretch}{1}

%

\subsection{Large-scale environment: filaments and voids}
\label{sec:environment:largescale}

The positions of our galaxies with respect to the large-scale filamentary
structure were obtained from the catalogue of filaments established
by \citet{Tempel.etal:2014}. For each SDSS-DR7 galaxy, the catalogue provides the distance to the closest filament, $D_{\rm fil}$. Since the catalogue is based on SDSS-DR7 (while our samples were drawn from SDSS-DR12) and contains only galaxies in the main survey area ($6.5 \lesssim {\rm RA} \lesssim 18\,$h), $D_{\rm fil}$ is available for only 185 VYGs and 1126 CSGs.

There is 
no statistical difference between the VYG and CSG distributions of distances to the nearest filament (KS test $p$-value is$\, =0.22$), and similar fractions of VYGs and CSGs are within $D_{\rm fil} \le D_{\rm max} = 1\,$Mpc: $35.7\%$ of the VYGs and $31.6\%$ of the CSGs ($p$-values$\, = 0.27$ and $0.31$ for Fisher's and Barnard's tests). In addition, the luminosities of the filaments containing VYGs and CSGs within 1\,Mpc are similar. We used the luminosities $L_{0.5}$ and $L_{1.0}$ computed by \citet{Tempel.etal:2014} as the the sum of luminosities of observed galaxies that are closer than 0.5 and 1.0 $h^{-1}\,$Mpc from the filament axis. 
After removing the dependency of $L_{0.5}$ and $L_{1.0}$ on the stellar mass
of the VYGs and CSGs, we see no significant differences between the
distribution of luminosities of filaments containing VYGs and those containing CSGs (KS test
$p$-values$\,=0.22$ and 0.08 for $L_{0.5}$ and $L_{1.0}$, respectively).
Since an roughly half of VYGs and CSGs are within the virial
spheres of groups, we investigated if the statistics of distance to nearest filaments are
blurred by these galaxies.  We repeated the analysis using only galaxies that are at distances $R > 3 r_{\rm vir}$ from all groups, and we still do not see any significant differences between the VYG and CSG distributions of $D_{\rm fil}$, $L_{0.5}$ and $L_{1.0}$.

We repeated the analysis using different maximum distances to the filaments ($D_{\rm max} =\,$0.2 up to 3\,Mpc), but we find no significant evidence for higher (or lower) fractions of VYGs lying along the filaments compared to those of CSGs. The distributions of luminosities of filaments containing VYGs within $D_{\rm max}$ are also very similar to those of filaments containing CSGs, with KS tests indicating low statistical significance regardless of the $D_{\rm max}$ value adopted. 

We also determined the position of our galaxies relative to voids by using the void catalogue by \citet{Sutter.etal:2012}\footnote{\url{http://www.cosmicvoids.net/}}, which is also based on SDSS-DR7. To cover the redshift range of our sample, we also used three catalogues of voids that were identified using different samples of SDSS galaxies: {\tt dim1} (for galaxies at $0 \leq z < 0.05$), {\tt dim2} ($0.05 \leq z < 0.1$) and  {\tt bright1} ($0.1 \leq z < 0.15$). We used the subcatalogues called ``\emph{centrals}'' by \citet{Sutter.etal:2012}. To avoid the biases introduced by the survey boundaries and masks, the \emph{centrals} subcatalogues exclude voids that, when rotated in any direction about its barycentre, intersects a boundary galaxy.  

For each VYG and CSG, we computed the distance to the centre of the closest void in units of the void radius, $D_{\rm void}$, assuming that the voids are spherical. The distances $D_{\rm void}$ are given by
$$
D_{\rm void} = \frac{1}{r_{\rm void}\, (1 + z_{\rm void})} \sqrt{\left[D_{\rm c}(z) - D_{\rm c}(z_{\rm void})\right]^2 + \left[\theta\, \overline{D_{\rm c}}\right]^2} \ ,
$$
\noindent where $z$ and $z_{\rm void}$ are the redshifts of the galaxy and the centre of the void, respectively; $\theta$ is the angular separation between the void centre and the galaxy; $D_{\rm c}(z)$ and $D_{\rm c}(z_{\rm void})$ are the comoving distances to the galaxy and to the void centre; $\overline{D_{\rm c}}$ is the mean comoving distance, $\overline{D_{\rm c}} = \left[D_{\rm c}(z) + D_{\rm c}(z_{\rm void})\right]/2$; and $r_{\rm void}$ is the void radius in Mpc.
Since the catalogue is restricted to voids in the main survey area, we compute the distances for our galaxies within $6.5 \lesssim {\rm RA} \lesssim 18\,$h only (187 VYGs and 1155 CSGs). 

Almost all of our galaxies (both VYGs and CSGs) are located far from the centre of the voids. Only $8.4\%$ of the galaxies (113 out of 1342) are at distances $D_{\rm void} \le 1$, and the $p$-values of statistical tests do not provide any evidence that the VYGs prefer (or avoid) these low-density environments. We obtain $p$-values$\, = 0.48$ and $0.67$ when applying Fisher's and Barnard's tests to the fractions of VYGs and CSGs that are at distances $D_{\rm void} \le 1$ (18 VYGs and 95 CSGs, corresponding to $9.6\%$ and $8.2\%$, respectively). 
Both samples have similar $D_{\rm void}$ distributions, with median and
standard deviations $D_{\rm void} = 2.52 \pm 1.38$ (VYGs) and $2.54 \pm 1.27$
(CSGs). We applied a KS test to compare the distributions and obtained
$p$-value$\, = 0.22$.
In summary, the large-scale environments of VYGs and CSGs are similar.
\section{Discussion}
\label{sec:discuss}

\subsection{How do very young galaxies differ from others?}
\label{sec:discuss:differ}

Our comparison of the properties of VYGs and CSGs
reveals that the VYGs are different from the general population of galaxies in many aspects. The results presented in Sections~\ref{sec:properties} to \ref{sec:environment} 
can be summarized as follows:

\begin{itemize}
\itemsep 0.5\baselineskip
\item \emph{VYGs are bluer and have higher sSFRs than the CSGs.} 
(Fig.~\ref{fig:sSFR_gi}), which confirms that our sample of VYGs indeed have
younger stellar populations. 
 
  \item \emph{In VYGs, the gas has higher ionization ratios}
  (Fig.~\ref{fig:BPT}), which might be simply a consequence of higher sSFRs. 
 
  \item \emph{VYGs contain a higher fraction of  spheroidal systems compared
  to CSGs} (Figs.~\ref{fig:Ttype} and \ref{fig:Ttype_gz2}).
  These VYG spheroids correspond to $\sim 6\%$ of sample and they are bluer than the CSG spheroids. On the other hand, we did not find significant differences between the overall distributions of VYG and CSG T-types, but most of our galaxies appear to be irregulars, so these catalogues may not provide a good description of their morphologies.
 
 \item \emph{VYGs have higher concentrations and surface brightness.} (Fig.~\ref{fig:PSM_C_mu}), indicating that the SF activity in VYGs is occurring in the inner parts of the galaxy. However, we cannot determine how the size of the VYGs compares to the general population of galaxies, since our control sample was defined by using the redshifts and angular effective radii
 in the PSM procedure; therefore, the VYG and CSG distributions of physical radii are similar by construction. 
  
 \item \emph{VYGs are more asymmetric and more clumpy than CSGs.}
 (Fig.~\ref{fig:PSM_A1_S1}), which may be a consequence of interactions with neighbour galaxies. 
 
 \item Among galaxies detected in H\,{\sc i}, \emph{VYGs have significantly
 higher fractions of atomic gas than CSGs} (Fig.~\ref{fig:HIgas}). Around $80\%$ of the VYGs with H\,{\sc i} detections have $f_{\rm gas} > 0.8$, while only $42\%$ of the CSGs have such high amounts of H\,{\sc i} gas. 

 \item \emph{The internal extinction in VYGs is higher than in the CSGs} (Fig.~\ref{fig:dust}), indicating that these young systems have a higher amount of dust compared to the general population of star-forming galaxies. We discuss the VYG internal extinction and dust content in Sect.~\ref{sec:discuss:dust}.

 \item \emph{Compared to the CSGs, the VYGs are at smaller distances from their nearest companions and are more likely to be interacting/merging with a neighbour galaxy}.  
 These results are shown in Figs.~\ref{fig:distance_1st_neighbour} to \ref{fig:frac_interaction_mur} and Table~\ref{tab:interacting}, and will be discussed in Sect.~\ref{sec:discuss:mergers}.
 
 \item Roughly half of the VYGs lie outside groups; but compared to CSG, \emph{VYGs are more likely to be found in the inner parts of low-mass groups} (Fig.~\ref{fig:frac_groups} and Table~\ref{tab:frac_groups}). We discuss this result in Sect.~\ref{sec:discuss:mergers}.
\end{itemize}

Our sample of VYGs includes some starburst galaxies with very large H$\alpha$
equivalent widths, which are expected to have different properties. So, one
could argue that the differences that we find between the VYG and control
samples could be due to these objects. However, even when excluding these
galaxies from our sample, the differences between the VYGs and the CSGs are
still statistically significant, as indicated by the $p$-values in columns~3
and~4 of Table~\ref{tab:p-values}.

Although VYGs differ from the general population of galaxies in all the properties listed above, the VYGs are very similar to the CSGs in the following aspects: 

\begin{itemize}
\itemsep 0.5\baselineskip
 \item \emph{VYGs have the same gas metallicity as the CSGs}
 (Fig.~\ref{fig:PSM_OH}), except for
 hints of lower gas metallicity of VYGs at the low-mass end. We discuss in
 Sect.~\ref{sec:discuss:met} the implications of the lack of  differences in
 the gas metallicities of VYGs vs. CSG.
 
 \item \emph{The distribution of VYGs relative to cosmic filaments and voids is very similar to that of CSGs.} 
\end{itemize}

\subsection{The VYG metallicities and relation with other young galaxies in the local Universe}
\label{sec:discuss:met}

The similar gas metallicities of VYGs and CSGs suggests that SF in VYGs is not being fuelled by infalling metal-poor gas, but by gas that was already enriched by previous generations of stars. 
This result has a consequence when
comparing the VYGs with other populations of very young galaxies. As already
discussed in Sect.~\ref{sec:introduction}, a few low-mass star-forming
galaxies with extremely low metallicities in the local Universe, such as the blue compact dwarf galaxies
I\,Zw\,18 and SBS\,0335--052, are
strong VYG candidates.\footnote{\izw\ and SBS\,0335--052 are too close to lie
in the redshift range of the parent {\tt clean} galaxy sample.}
As shown by \citet{Izotov+19}, these objects strongly deviate from the oxygen abundance vs. $M_{\star}$ relation defined by the bulk of star-forming galaxies.
Therefore, the normal metallicities of VYGs at given mass
suggests that VYGs do not resemble objetcs like
I\,Zw\,18. 

Although we find that the gas mass fractions of VYGs with H{\sc i} detections is very high,   I\,Zw\,18 has an even more extreme $f_{\rm gas}$ value. The H{\sc i} mass of I\,Zw\,18 is 50 times higher than its stellar mass, and the total neutral gas mass fraction is $f_{\rm gas} = 0.98$.\footnote{To compute the gas mass fraction of I\,Zw\,18, we adopted $M_{\rm H\textsc{i}} = 1.25 \times 10^8\,$M$_{\odot}$ \citep{Engelbracht+08, Thuan+16} and $M_{\star} = 2.6 \times 10^{6}\,$M$_{\odot}$ \citep{Izotov+2014_SDSS_SFing}.} The most gas-rich VYG in our ALFALFA sample has $f_{\rm gas} = 0.97$, and only 3 out of 16 VYGs with H{\sc i} detection have $f_{\rm gas} \geq 0.95$. But the stellar mass of I\,Zw\,18 is 40 times lower than the lowest masses of our VYG sample, so it is difficult to make a meaningful comparison between this galaxy and the VYGs.

Furthermore, while our VYGs tend to have close companions, 
I\,Zw\,18 appears to be a very isolated system. Indeed, using the same approach
described in Sect.~\ref{sec:environment:local} to investigate the local
environment of our VYG sample, and assuming $m_{r,{\rm I\,Zw\,18}}=16.4$ as
the extinction-corrected Petrosian magnitude in the $r$ band of I\,Zw\,18, we
find that there is no other galaxy brighter than $m_r = m_{r,{\rm I\,Zw\,18}}
+ 1$ up to $200\,$kpc from I\,Zw\,18. 
This suggests that interactions with
nearby galaxies cannot be the mechanism triggering the burst of SF in
I\,Zw\,18, and other processes must be invoked to explain the SF
activity in this galaxy.  But one cannot rule out the possibility that \izw\
has very recently grown by gas infall and/or mergers, as evidenced by its irregular morphology
and the complex kinematics of its atomic gas \citep{vanZee+98}.

In any event, it is difficult to compare this population of extremely metal-deficient dwarf galaxies to the VYGs studied here. First, while these dwarfs have stellar masses $\sim 10^{5.5}-10^{8.0}$ M$_{\odot}$ \citep{Izotov+2014_SDSS_SFing, Izotov+19},
our sample is restricted to more massive objects with  $M_{\star} > 10^8\,$M$_{\odot}$. Moreover, the techniques employed to study the stellar populations of nearby dwarfs are different from those used here. Our ages were determined through SPS analysis of the integrated galaxy spectra, while the age of objects like I\,Zw\,18 are inferred from colour-magnitude diagrams of resolved stellar populations. In Paper~II, we determined the age of I\,Zw\,18 using the same method (with STARLIGHT using the \citealp{Vazdekis+15} spectral model) and found that 100\% of its stellar mass was formed in the last $100\,$Myr, so this galaxy easily meets the VYG classification.

Since these metal-poor dwarfs are very nearby objects, with distances
$\lesssim 20\,$Mpc, it is very difficult to investigate their global
environment. Catalogues of groups and clusters are not reliable at very low
redshifts due to uncertainties introduced by peculiar velocities of
galaxies. In addition, images of close objects contain more spatial
information than those of galaxies at higher redshifts. Hence, 
morphological classification and morphometry of these galaxies (e.g. with
 \textsc{Morfometryka}) depend on galaxy distance, and correcting for
this dependence to compare I\,Zw\,18 analogues with our VYGs is beyond the
scope of this paper.

%
\begin{table}
\centering
\caption{$p$-values of KS tests applied to the control and VYG samples \label{tab:p-values}}
\begin{tabular}{lccc}
\hline
Property & \multicolumn{3}{c}{KS $p$-values} \\
\cline{2-4} 
         &        All &                      VYGs with &                       VYGs with \\
         &       VYGs & EW(H$\alpha$)$\, \le 150\,$\AA &  EW(H$\alpha$)$\, \le 100\,$\AA \\
\hline
\multicolumn{1}{c}{(1)} &        (2) &                            (3) &        (4) \\
\hline\hline
$\log {\rm sSFR}$       &       $< 10^{-10}$ & $9 \times 10^{-10}$ & $2 \times 10^{-5}$ \\  
($g - i$)$_{\rm Petro}$ & $5 \times 10^{-4}$ &  $2 \times 10^{-3}$ &               0.01 \\  
BPT diagram             &       $< 10^{-10}$ &  $7 \times 10^{-8}$ & $3 \times 10^{-5}$ \\  
$\log ({\rm O/H})$      &               0.09 &                0.18 &               0.55 \\  
$f_{\rm gas}$           & $2 \times 10^{-4}$ &  $2 \times 10^{-4}$ & $7 \times 10^{-5}$ \\  
$A_V^{\rm Balmer}$      &       $< 10^{-10}$ &        $< 10^{-10}$ &$2 \times 10^{-10}$ \\  
$A_V^{\rm cont}$             & $9 \times 10^{-7}$ &  $3 \times 10^{-7}$ & $3 \times 10^{-8}$ \\  
$C_1$                   & $1 \times 10^{-7}$ &  $3 \times 10^{-6}$ & $2 \times 10^{-6}$ \\  
$\mu_{50}$              &       $< 10^{-10}$ &        $< 10^{-10}$ & $6 \times 10^{-8}$ \\  
$A_1$                   & $2 \times 10^{-7}$ &  $3 \times 10^{-5}$ & $2 \times 10^{-5}$ \\  
$S_1$                   & $3 \times 10^{-3}$ &                0.02 & $4 \times 10^{-3}$ \\  
$d_{\rm 1st}/$kpc        & $4 \times 10^{-3}$ &                0.01 &              0.03 \\  
$d_{\rm 1st}/\theta_{50}$& $4 \times 10^{-4}$ &  $5 \times 10^{-3}$ &              0.03 \\  
\hline
\end{tabular}
\end{table}

\subsection{VYGs versus late bloomer galaxies at intermediate redshifts}
\label{sec:discuss:dressler}

A sample of young systems at intermediate redshifts was recently studied by \citet[][hereafter DKA18]{Dressler.etal:2018}. They derived the SFHs of galaxies at $0.45 < z < 0.75$ and with $M_{\star} > 10^{10}\,$M$_{\odot}$, and identified a galaxy population of \emph{late bloomers} (LB), i.e., galaxies that formed at least 50\% of their stellar mass within 2\,Gyr of the epoch of observations. Their SFHs were inferred from the Carnegie-\emph{Spitzer-IMACS} Survey photometry. 

DKA18 found that LBs account for $\sim 20\%$ of galaxies at $z \sim 0.6$, and their fractions systematically decrease with decreasing redshift, and the most massive LBs ($M_{\star} \gtrsim 10^{11}\,$M$_{\odot}$) pratically disappear at $z \sim 0$.
In Paper~II, we showed that the fractions of VYGs in the local Universe are consistent with the extrapolation of the fractions of LBs vs. redshift determined by DKA18 (see Fig.~14 in Paper~II).

\emph{Are the properties of VYGs similar to those of LBs?} It is difficult to answer this question, given the stellar mass ranges of the VYG and LB samples; while LBs are, by definition, more massive than $M_{\star} > 10^{10}\,$M$_{\odot}$, most of our VYGs have $M_{\star} \lesssim 10^{10}\,$M$_{\odot}$, and it is well known that galaxy properties correlate with stellar mass. A significant fraction of LBs appear to be spiral galaxies, but DKA18 also find some LBs with early-type morphology, with SEDs that are consistent with that of a post-starburst galaxy. Although VYGs appear to be mostly irregular, it is interesting that we also find an excess of spheroidal systems, suggesting that some common mechanism is producing young spheroidal systems at different redshifts. 

One very interesting result obtained by DKA18 is that LBs avoid lying close to galaxies that are not LBs up to distances of $8\,$Mpc, indicating that galaxy SFHs can trace local and large-scale environmental histories. In  Sect.~\ref{sec:environment:largescale}, we found 
no difference between the VYG and CSG positions relative to filaments and voids, and the luminosities of the filaments close to VYGs and CSGs are also similar. 

\begin{figure}
    \centering
 \includegraphics[width=0.98\hsize]{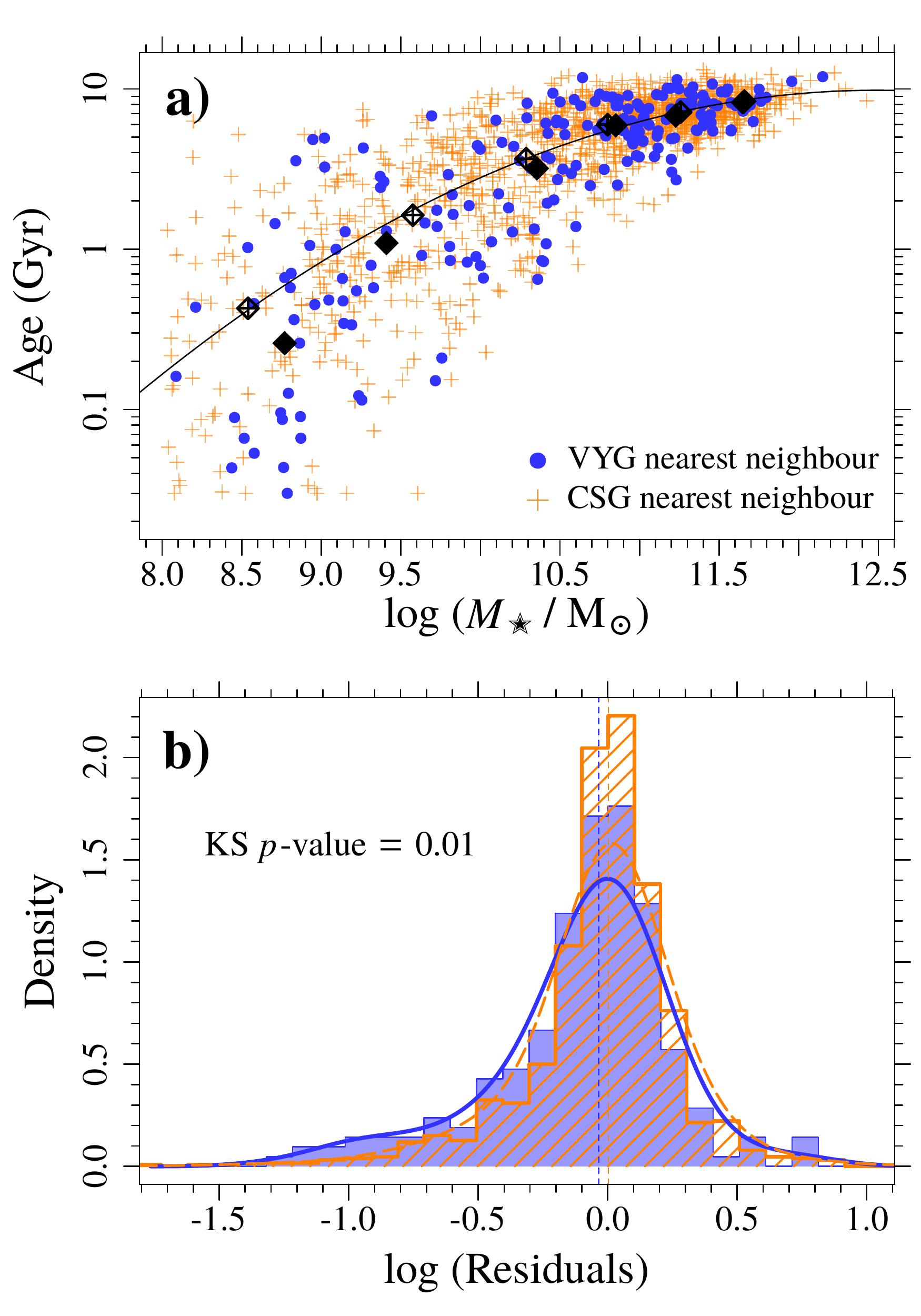}  
    \caption{{Top:} Mass-weighted age of the VYG and CSG nearest SDSS spectroscopic galaxy (\emph{blue} and \emph{orange}
    symbols, respectively). The median ages in bins of stellar mass are indicated by the \emph{black diamonds}, and the best-fit to the age-mass relation for the CSG neighbours is shown as the \emph{black solid line}. {\bf Bottom:} Distribution of the logarithm of the residuals from the best-fit relation shown in the top panel. The \emph{blue} and \emph{orange histograms} corresponds to the VYG and CSG nearest neighbours.}
    \label{fig:age_neighbour}
\end{figure}

However, the distributions of VYGs and CSGs relative to the large-scale structures being similar does not mean that the properties of galaxies within these structures are also similar. To investigate this, we used the SDSS spectroscopic catalogue to identify the VYG and CSG nearest galaxies (in comoving units). 
As shown in Fig.~\ref{fig:age_neighbour}, the nearest neighbours of VYGs tend to be younger compared to those that are close to CSGs, in agreement with DKA18. The difference is more pronounced for low-mass neighbours; VYG neighbours with $M_{\star} \leq 10^{10}\,$M$_{\odot}$ are $\sim 0.17\,$dex, on average, younger than CSG neighbours in the same stellar mass range ($p$-value$ = 0.02$).  
Since VYGs tend to have more close companions than CSGs, we checked if the neighbours in the immediate surroundings are leading to this result by comparing the ages of closest galaxies at distances $> 1\,$Mpc only. We confirm that we see the effect in both small ($< 1\,$Mpc, $p$-value$\, = 0.02$) and large ($> 1\,$Mpc, $p$-value$\, = 0.04$) scales.
This is also in agreement with DKA18, who found  that LBs avoid non-LBs up to $\sim 8\,$Mpc, which means that the ages of the LB nearest galaxies are younger compared to the general population of galaxies.

%

\subsection{VYGs and CSGs in the low- and high-mass regimes}
\label{sec:discuss:mass}

%
\begin{table}
    \centering
    \caption{Comparison of VYGs to control galaxies, splitting by mass}
    \begin{tabular}{lccccc}
\hline
 & \multicolumn{2}{c}{$\log (M_{\star}/{\rm M}_{\odot}) \leq 9.5$} & &
\multicolumn{2}{c}{$\log (M_{\star}/{\rm M}_{\odot}) > 9.5$} \\
\cline{2-3}\cline{5-6} 
\multirow{2}{*}{Property} & VYG & \multirow{2}{*}{KS $p$-value} & & VYG & \multirow{2}{*}{KS $p$-value} \\
& \emph{median} & & & \emph{median} &  \\

\hline
\multicolumn{1}{c}{(1)} & (2) & (3) & & (4) & (5) \\
\hline \hline
sSFR                   & higher & $ < 10^{-10}$      & & higher & $2 \times 10^{-3}$ \\
$(g - i)_{\rm Petro}$  & {\bf lower} & $1 \times 10^{-9}$ & & {\bf same}   & 0.10 \\
BPT diagram            & higher & $2 \times 10^{-4}$ & & higher & $6 \times 10^{-8}$ \\
log(O/H)               & {\bf lower}  & 0.06               & & {\bf same}   & 0.86 \\
$f_{\rm gas}$          & higher & $2 \times 10^{-4}$ & & --     & -- \\
$A_V^{\rm Balmer}$     & higher & $6 \times 10^{-6}$ & & higher & $< 10^{-10}$ \\
$A_V^{\rm cont}$       & {\bf lower}  &  0.01              & & {\bf higher} & $1 \times 10^{-7}$ \\
$C_1$                  & higher & $3 \times 10^{-3}$ & & higher & $3 \times 10^{-5}$\\
$\mu_{50}$             & higher & $1 \times 10^{-6}$ & & higher & $< 10^{-10}$\\
$A_1$                  & higher & $4 \times 10^{-5}$ & & higher & $3 \times 10^{-3}$ \\
$S_1$                  & {\bf higher} & $0.08$             & & {\bf same}   & 0.05 \\
$d_{\rm 1st} / {\rm kpc}$    & lower & 0.06          & & lower  & 0.02 \\
$d_{\rm 1st} / \theta_{50}$  & lower & 0.05          & & lower  & 0.01 \\
\hline
\end{tabular} 
\parbox{\hsize}{Notes: 
For each property, columns 2 and 4 indicate how the \emph{median} VYG value compares with that of the CSGs, while columns 3 and 5 show the respective $p$-values for that property in the corresponding mass bin. Changes in the trend between low and high mass are indicated in {\bf bold}. 
}
    \label{tab:low_high_mass}
\end{table}

Our method of comparing the residuals to the CSG trend with mass of the properties of VYGs and CSGs
may hide how differences between  VYGs and CSGs vary from low to high stellar masses.
To address this question, we repeated our analysis, separating the galaxies into two mass bins, above and below $\log (M_{\star}/{\rm M}_{\odot}) = 9.5$, which is close to the median of the VYG sample.
Table~\ref{tab:low_high_mass} shows the results.

For several properties, we see similar trends for the VYGs compared to the CSGs in the low- and high-stellar mass regimes. On the other hand, differences between the VYG and CSG colours and asymmetries is seen only for low-mass galaxies. Furthermore, as already mentioned in Sect.~\ref{sec:discuss:met}, low-mass VYGs appear to be more metal-poor than CSGs with similar masses, and this is not observed for the high-mass VYGs. 
According to our merger diagnostics, merger events appear to be an important mechanism for VYGs of all masses, but the low gas metallicity in less massive VYG suggest both mergers and gas infall from the cosmic web are operating in the low-mass regime.\footnote{But in a recent study, \cite{Tiwari+20} find that galaxies in filaments are enriched relative to field galaxies.}
This qualitatively agrees with our understanding that only high mass galaxies \emph{typically} grow by mergers, which are too infrequent in low mass ones, which thus must \emph{usually} grow by gas infall \citep{Cattaneo+11}.
Following this argument, one would not expect that low-mass VYGs also have more neighbours, in contrast with the marginal indication of more neighbours around low-mass VYGs compared to low-mass CSGs shown in Table~\ref{tab:low_high_mass}.
Besides, all of the galaxies in our VYG and CSG samples have lower mass than the critical $z$\,=\,0 mass of $M_\star = 2\times 10^{11}\,\msun$ that \citeauthor{Cattaneo+11} found for the growth by mergers (see Fig.~\ref{fig:afterPSM}). Nevertheless, one can reason that VYGs recently formed their stars by mergers, but will subsequently live a more quiet period of gas infall.

From low- to high-mass galaxies, we see a reversal of the trend of the VYG 
$A_V^{\rm cont}$ with respect to that of the CSGs, i.e., low-mass VYGs have lower
$A_V^{\rm cont}$ values than the CSG values, while for high-mass VYGs we see the opposite. 
These high $A_V^{\rm cont}$ values might be responsible for the lack of bluer colours among high-mass VYGs. The behaviour of $A_V^{\rm cont}$ with galaxy mass, and the larger scatter that we observe in the $A_V^{\rm cont}-M_{\star}$ and $A_V^{\rm cont}-{\rm SFR}$ relations compared to the $\tau_{\alpha \beta}$ relations suggest that the amount of extinction of the stellar continuum might be determined by different -- and maybe more -- factors than the amount of extinction of the gas emission. We will further discuss the  dust content in Sect.~\ref{sec:discuss:dust}.

\subsection{How can a galaxy retain so much gas for so long?}
\label{sec:discuss:gas}

As shown in Fig.~\ref{fig:HIgas}, among galaxies whose atomic gas is detected
with ALFALFA, VYGs have very high atomic gas mass
fractions, falling in the range 0.8--1, the same as found by \citet{Thuan+16}
for extremely metal-deficient blue compact dwarf starburst galaxies. It is not clear if
the ALFALFA sample is representative of the whole VYG population, 
but it seems that at least the low-mass VYGs with H{\sc i} detections 
have the amount of gas necessary to fuel the the
intense SF activity required to form so many stars in such a small timescale. 
\emph{But how can a galaxy retain this gas until recent times?}


A possible scenario comes from a recent study
by \citet{Zhang.etal:2019}, who, using data from SDSS, ALFALFA,
GASS \citep{Catinella+10} and COLD GASS \citep{Saintonge+11}, found that massive quiescent
central disk galaxies contain as much atomic gas as corresponding
star-forming galaxies  ($M_{\rm HI}/M_{\star} > 0.1$).
Moreover, both galaxy classes have identical H\,{\sc i} spectra indicating regularly
rotating H\,{\sc i} disks with radius $\sim 30\,$kpc and little kinematic
perturbations. Following \cite{Renzini20}, the H\,{\sc i} gas in quenched disks can be stored in an
outer ring with high angular momentum and large infall timescales. These
galaxies are quenched because of the reduced molecular gas content and lower
SF efficiency. However, perturbations could drive the outer atomic gas
inwards and trigger SF.

This scenario could be extended to the lower masses of 
our VYGs. However, \citeauthor{Zhang.etal:2019}  only considered
central galaxies, and only one out of the 17 VYGs that are centrals in
low-mass groups ($12.3 \leq \log [M_{\rm halo} / {\rm M}_{\odot}] < 13.5$)
has H\,{\sc i} detection in the ALFALFA survey. All the other central VYGs are
beyond the $z = 0.05$ redshift limit of ALFALFA. We thus cannot determine if
our central VYGs also have high fractions of atomic gas. 

\begin{figure}
\centering
\begin{tabular}{c}
 \includegraphics[width=0.9\hsize]{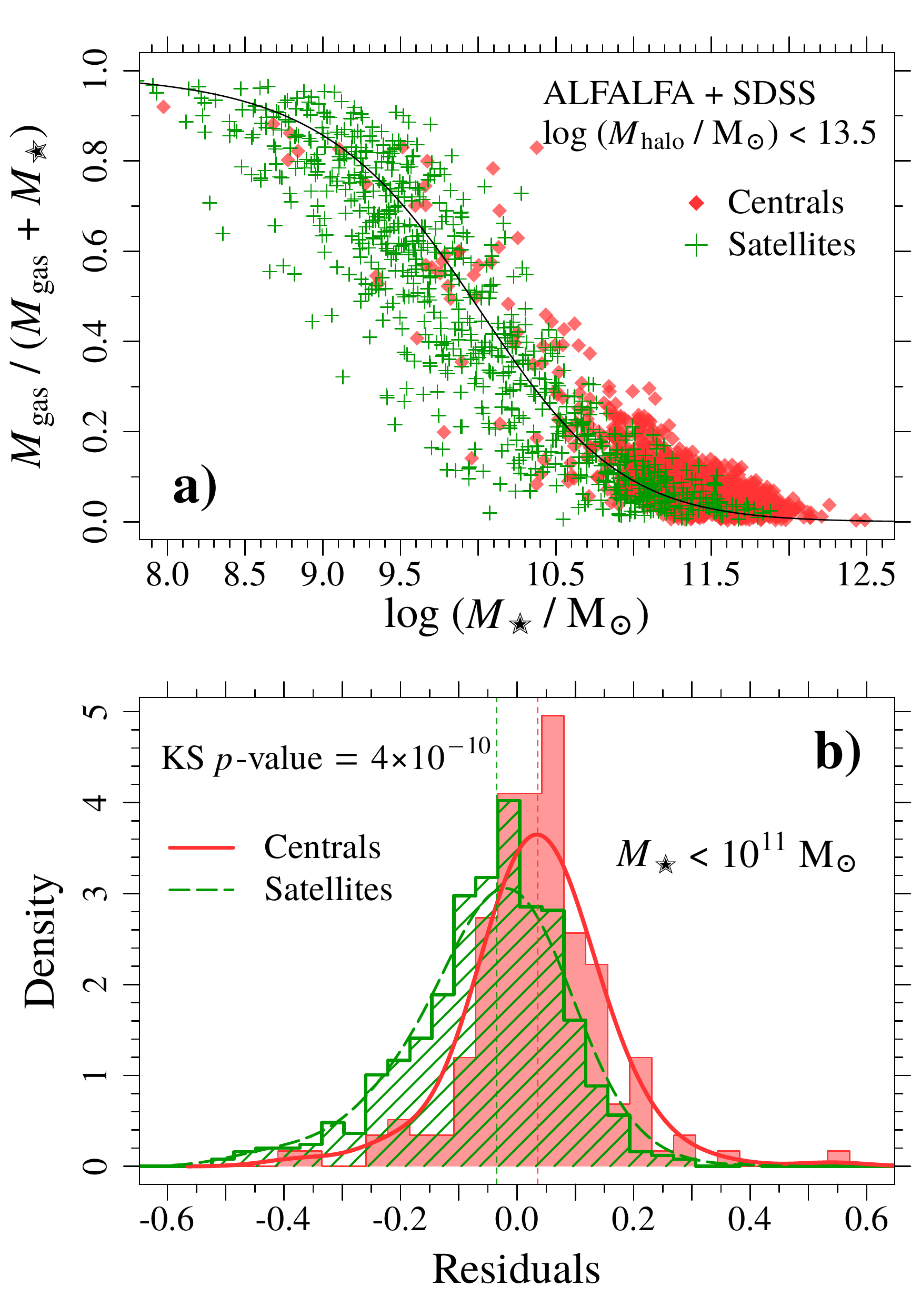}  
\end{tabular}
\caption{{\bf Top:} Fraction of atomic gas in SDSS galaxies with H\,{\sc i}
detections in the ALFALFA survey as a function of galaxy stellar
mass. The \emph{red} and \emph{green} symbols indicate the central and
satellite galaxies in low-mass groups ($\log [M_{\rm halo} / {\rm
M}_{\odot}] \leq 13.5$). The \emph{black solid line} is the best-fit to the
satellite $f_{\rm gas}$ vs. $\log M_{\star}$ relation (see text for details).   
{\bf Bottom:} The residuals from the best-fit indicated in panel \textbf{a}
for galaxies with $M_{\star} < 10^{11} {\rm M}_{\odot}$. The \emph{red}
and \emph{green histograms} correspond to low-mass centrals and satellites,
respectively. }
\label{fig:HIgas_sdss}
\end{figure}

To investigate the gas fraction of low-mass central galaxies, we computed $f_{\rm gas}$ for a general sample of SDSS galaxies that have a counterpart in the ALFALFA catalogue. To avoid measurements of H\,{\sc i} masses that may be contaminated by the neutral gas of nearby galaxies because of the large Arecibo beam size, we excluded galaxies that have companions within a radius of $1.8\,$arcmin and $\Delta z \leq 2\, W_{20}/c$, i.e., we adopt the same approach as described in Sect.~\ref{sec:gas:neutral}. Fig.~\ref{fig:HIgas_sdss}a shows $f_{\rm gas}$ vs. stellar mass of centrals and satellite galaxies in  groups with $12.3 \leq \log [M_{\rm halo} / {\rm M}_{\odot}] < 13.5$, the same halo mass range of the central VYG haloes. We fitted the satellite $f_{\rm gas}$ vs. $\log M_{\star}$ relation assuming that $f_{\rm gas}$ can be described as
$$
f_{\rm gas} = \frac{1}{1 + \exp\, \left\{\sum_{i = 0}^{4} \, a_i \, \left[\log (M_{\star}/{\rm M}_{\odot})\right]^i \right\}}
$$

Fig.~\ref{fig:HIgas_sdss}b shows the distributions of the residuals from the
best-fit relation for the centrals and satellite galaxies with $M_{\star} <
10^{11} {\rm M}_{\odot}$, which is the mass range of our VYG sample. Central galaxies tend to have higher fractions of atomic gas than
satellites. This suggests that the higher gas fractions of VYGs may be related to their excess of centrals (Fig.~\ref{fig:frac_groups}).

Instead of retaining their gas, 
some VYGs may be experiencing gas infall from their surroundings.
It could be pristine gas
from the cosmic web, but, as already mentioned in
Sect.~\ref{sec:discuss:met}, 
this only affects low-mass VYGs, at best, since high-mass VYGs have similar metallicities as CSGs of similar mass, while low-mass VYGs have a marginal indication of lower metallicity.
Since a large
fraction of the VYGs have companions, the VYG gas can be replenished by
the gas from the merging neighbour galaxies. This is more likely than coming
from the cosmic web of gas,
since the gas brought in by the neighbouring galaxies is already somewhat enriched.

Although we find that VYGs appear to have more atomic gas, differences in the fraction
of atomic (and molecular) gas alone cannot account for the observed
distribution of objects above the main sequence of star-forming
galaxies. Indeed, using data from the ALFALFA, GASS
and COLD GASS surveys to quantify how the mean atomic and molecular gas mass
fractions vary in the SFR-$M_{\star}$ plane, \citet{Saintonge+16} showed that galaxies with
very high sSFR, such as those that are necessary to produce a VYG, not only
have large gas fractions, but also have high SF efficiencies. It is
known that mergers and interactions might enhance the efficiency in which the
gas is converted into stars. We explore this scenario in the following
Section.

%

\subsection{Are VYGs the products of mergers?}
\label{sec:discuss:mergers}

Many studies have shown that mergers and interactions induce an intense
SF activity in galaxies \citep{Joseph&Wright85}, with a recent
exception \citep{Pearson+19}.
Using hydrodynamical
simulations, \citet{DiMatteo+2008} found that the SF activity of starbursts
triggered by mergers is enhanced by a factor of 3, on average, compared to
starbursts in normal galaxies, and is greater than 5 in about 15\% of major
galaxy interactions and mergers. Moreover, multiple bursts can occur, with
the bulk of SF among merging galaxies occurring at the 2nd passage, even
though there is continuous SF, as well as a first burst at the first
pericentre.

Our results suggest that mergers and interactions are important processes
triggering recent SF activity. Compared to the control sample, the VYGs are
twice as likely to have close companions at distances $< 5\, \theta_{50}$
(Fig.~\ref{fig:distance_1st_neighbour}). Moreover, the fraction of VYGs
showing signs of interactions is double that of CSGs. Using the parameters
$p_{\rm merger}$ from the DS18 and GZ1 catalogues, we have shown that the fraction
of VYGs classified as mergers is always twice the corresponding fraction for
CSGs, regardless of the threshold adopted for $p_{\rm merger}$
(Fig.~\ref{fig:p_mergers}). Our visual inspection of the SDSS images also
shows that the fraction of VYGs showing sings of interaction is twice that of
CSGs
(Fig.~\ref{fig:frac_interaction_logM} and Table~\ref{tab:interacting}), and
this result does not depend on the galaxy surface brightness
(Fig.~\ref{fig:frac_interaction_mur}).

The  preference of VYGs to lie in the inner parts of low-mass groups (Fig.~\ref{fig:frac_groups} and Table~\ref{tab:frac_groups}) also supports that scenario, since interactions and mergers are more likely to occur in this environment, where encounter velocities are slower, hence the encounters are more efficient (e.g., \citealp{Mamon92, Mamon00_IAP}). 

The VYG morphologies also point to the merging nature of a significant fraction of these systems, and may be related to different phases of mergers. VYGs are more clumpy and asymmetric than CSGs (Fig.~\ref{fig:PSM_A1_S1}), which may be signs of ongoing interactions with neighbour galaxies.
In addition, VYGs have higher concentrations than CSGs (Fig.~\ref{fig:PSM_C_mu}), indicating central SF activity. Hydrodynamical simulations show that global torques induced by mergers drive gas inwards, leading to more compact systems \citep{Hopkins+2010, Hopkins+2013}. 

These compact galaxies are also seen as an excess of VYGs with low T-type
values compared to those of the CSGs (Figs.~\ref{fig:Ttype}
and \ref{fig:Ttype_gz2}). These early-type VYGs are bluer than CSGs with
T-type$\,\leq 0$, and through a visual inspection of their images, we see
that these systems are blue compact spheroids.
These very compact VYGs could
represent recent post-merger galaxies.
Indeed, post-merger galaxies have 3 times higher gas fractions than control
galaxies \citep{Ellison+18}. However, high gas fractions -- as in our VYGs --
reduce the loss of the gas angular momentum and increases the changes of disk
survival \citep{Hopkins+2013}. Therefore, the lack of disks  in these
galaxies suggests that the progenitor galaxies already had low angular
momentum, such as irregular galaxies, or the orbital angular momentum of the merging system
is low (i.e., head-on merging encounters).
Otherwise, other mechanisms should be invoked to account for the formation of these blue compact VYGs.

\subsubsection{Internal extinction, dust mass and relation with mergers}
\label{sec:discuss:dust}

We found that VYGs have more internal extinction than CSGs, and this may be
related to a merging process. As discussed in Sect.~\ref{sec:gas}, at a fixed stellar mass, 
VYGs have higher $\tau_{\alpha \beta}$ (which probes the extinction in H{\sc ii} regions) and $A_V^{\rm cont}$ (which is more related to dust in the diffuse ISM component) values compared to those of the CSGs (Fig.~\ref{fig:dust}). While $A_V^{\rm cont}$ is inferred from 
the stellar continuum through SPS analysis,  $\tau_{\alpha \beta}$ 
is obtained from H$\alpha$ and H$\beta$ emission
lines. Since both parameters indicate that VYGs have higher internal
extinction than CSGs, it is unlikely that this result is a consequence of
biases and uncertainties in the SPS analysis or in the measurements of the
emission line fluxes.  Therefore, our results might be truly related to a
higher amount of dust in VYGs compared to the general population of
star-forming galaxies.

The high VYG $\tau_{\alpha \beta}$ values appear to be related to their high SFRs (Fig.~\ref{fig:dust_SFR}a,b), since VYGs and CSGs follow a similar $\tau_{\alpha \beta}-\log {\rm SFR}$ relation. On the other hand, differences between the 
VYG and CSG $A_V^{\rm cont}$ do not appear to be fully accounted by
differences in the galaxy sizes, morphology, inclination or SFRs.
Differences between VYG and CSG gas metallicities cannot explain the higher
amount of dust in VYGs either, since VYGs and CSGs have similar oxygen abundances
(Fig.~\ref{fig:PSM_OH}). Other factors can also contribute to higher internal 
extinction in VYGs, such as the HI gas mass \citep{Casasola+20}. However, the VYGs have higher $A_V^{\rm cont}$ and $\tau_{\alpha \beta}$ values compared to those of the CSGs at a fixed $f_{\rm gas}$. 
\emph{So, why do VYGs have more dust in their diffuse ISM component than CSGs?}
Ultra Luminous Infrared Galaxies (ULIRGs), whose very strong infrared emission
is caused by dust, are observed to have one \citep{Armus+87} or
several \citep{Borne+00} companions, implying that mergers triggered the
extreme starburst leading to extreme amounts of dust.

\emph{Does the connection between star formation rates, dust and mergers extend to low-mass
galaxies?} To address this question, we analysed a sample of SDSS galaxies with very 
high SFRs and used the parameter $p_{\rm merg}$
from the catalogue of morphology by (\citealp{DominguezSanchez.etal:2018};
see Sect.~\ref{sec:environment:mergers}) to access the probability of these systems of being mergers.
A significant fraction of VYGs in our sample have $\log [{\rm SFR/(M_{\odot}\,yr}^{-1})]  \geq 0.8$ (14.5\%, i.e., 30 VYGs), and these galaxies are the ones associated with high internal extinction (see Fig.~\ref{fig:dust_SFR}a). However, in a general sample of $\sim 323\,087$ SDSS galaxies at $z \leq 0.1$, only $4\,107$ ($\sim 1.3\%$) have these extreme SFR values, and they are typically more massive than $10^{10}\,$M$_{\odot}$. We analysed the internal extinction and merger probabilities of these $4\,107$ with very high SFRs by comparing them to a sample of galaxies with similar stellar masses and $\log [{\rm SFR/(M_{\odot}\,yr}^{-1})] < 0.8$. 

As expected, systems with high SFRs have larger $\tau_{\alpha \beta}$ and $A_V^{\rm cont}$ values compared to low-SFR galaxies with similar masses.
In addition, we find that the distribution of $p_{\rm merg}$ is shifted towards higher values compared to those of the low-SFR galaxies, and the fraction of high-SFR galaxies with $p_{\rm merg} > 0.5$ is almost two times larger than that of low-SFR objects ($26\%$ and $15\%$, respectively).
The same results are observed when we consider only galaxies with $\log (M_{\star} / {\rm M}_{\odot}) < 10.5$; i.e., at lower stellar masses, galaxies with high SFRs also have more internal extinction and higher probabilities of being merging systems than low-SFR galaxies.
Besides, in the low-mass regime, the differences between $p_{\rm merg}$ of high- and low-SFR galaxies is more pronounced when compared to the results obtained for all galaxies. The fraction of low-SFR systems with $p_{\rm merg} > 0.5$ remains unchanged (15\%), but this fraction increases to $42\%$ among the high-SFR low-mass galaxies. These results suggest that the relation between SFR, dust and merger events can be, at least at some level, extended to the low stellar mass regime, and the high SFRs and internal extinction that we find for our VYGs can be an indication of mergers and interations.

The internal extinction $\tau_{\alpha \beta}$ inferred from emission lines appear to be more related to the SFR and merger events than $A_V^{\rm cont}$. Once we take the $\tau_{\alpha \beta}$-SFR relation into account, we see no differences between the VYGs and CSGs. On the other hand, differences are still seen for $A_V^{\rm cont}$. Looking at Table~\ref{tab:low_high_mass}, we see that the trend of VYG $A_V^{\rm cont}$ values compared to those of CSGs is reverted from low- to high-mass. Besides, the scatter in the $A_V^{\rm cont}-M_{\star}$ and $A_V^{\rm cont}-{\rm SFR}$ relations is larger compared to the relations for $\tau_{\alpha \beta}$. All these results suggest that dust production in the diffuse ISM is a complex process with several variables. While the properties of the molecular clouds in the star-forming regions of the galaxy appear to be dominant for $\tau_{\alpha \beta}$, $A_V^{\rm cont}$, tracing a more diffuse dust component, should depend on other galaxy properties, such as galaxy size, inclination, and the balance between the production and destruction of dust grains in the ISM.  In other words, while the small-scale physics within molecular clouds appear to be roughly independent of the galaxy mass and properties, the processes regulating the production and destruction of dust in the ISM might more dependent of the galaxy integrated properties.

\subsubsection{Predictions from models of galaxy formation}

In Paper~I \citep{Tweed+18}, we used analytical models of galaxy formation,
which were applied to high-mass resolution Monte Carlo halo merger trees, to
predict the fraction of VYGs in the local Universe. Four models were
considered: one physically motivated model by \citet{Cattaneo+11}, including
a more realistic cut off at low halo masses due to reionization
feedback \citep[][C+G]{Gnedin00}; two empirical models
by \citet*[][MNW]{Moster+13} and \citet*[][BWC]{Behroozi+13}, that use
abundance matching to link the stellar-mass and the halo-mass functions; and
an empirical model by \citet*[][MCP]{Mutch+13}, where the stellar mass growth
rate is proportional to the halo mass growth rate times a function of halo
mass and redshift. The models were run with either the bursty or quiet halo merging schemes.

The effects of the bursty and quiet schemes on the predictions of VYG fractions depend on the model used. The merging scheme has virtually no effect on the VYG fractions versus stellar mass in the MNW and MCP models (fig.~10 of \citeauthor{Tweed+18}). On the other hand, the VYG fractions in the quiet merging scheme are reduced by over one order of magnitude in the C+G model at intermediate mass  and in the BWC model at low mass. \citeauthor{Tweed+18} also showed that, with the bursty galaxy merging, VYGs are associated with a recent major halo merger, in contrast to the quiet merging scheme.

In Paper~II \citep{Mamon+20}, we compared the predictions from these models
to the observed fractions of VYGs. We found that the observations support the bursty scheme for C+G and BWC models, but little can be said about the bursty and quiet MNW and MCP models due to the small differences between these two schemes for these two models.  

Since these models do not provide any prediction on the properties of the
VYGs, no direct comparison between our results and these models is
possible. However, we do present new constraints in favour of the bursty merging
scheme, as our results indicate that, in comparison to CSGs, VYGs 
are more likely to be associated with gas-rich mergers.

\subsubsection{Are most or all VYGs associated with mergers?}

While we find many indications that mergers and interactions are
responsible for triggering an intense SF activity in VYGs, it
is not clear if these processes  account for most or all of the VYGs. Our visual
inspection of the VYG images revealed that at least half of our galaxies do
not show any signs of interactions. However, the SDSS images are not deep
enough, and we could be missing faint tidal features.
So, future studies using deeper imaging data could help
understanding the nature of those VYGs for which no
signs of interactions are observed. In addition, since mergers leave
signatures in the gas and stellar kinematics, spatially-resolved
spectroscopic observations of VYGs might also help constraining this
scenario, in particular to identify the isolated VYGs that are in a
post-merger phase.

\subsection{Caveats}
\subsubsection{Colour gradients and outer halo of old stellar populations}
\label{sec:caveats:gradients}
As described in Sect.~\ref{sec:data:aperture},  to minimize aperture effects
due to the limited size of the SDSS fibre, we required that the total $g-i$ VYG
colour be bluer than the corresponding fibre colour. We showed in
Paper~II that the requirement of a blue gradient does not rule out the possibility of a
hidden old stellar population in the VYG, but at least it mitigates
the aperture effect. But one must keep in mind that $i$) a blue gradient
in a galaxy can also be a consequence of a negative metallicity gradient due
to the age-metallicity degeneracy; and $ii$) this selection criteria can
produce biased samples, as we discuss below.

It is known that different galaxy formation scenarios and physical mechanisms leave typical imprints on how the stellar population properties vary with the distance to the galaxy centre \citep{Kobayashi:2004,Hirschmann+15, Ferreras+19}. 
Therefore, when imposing a blue gradient to select our galaxies, we favour
physical mechanisms and formation scenarios that leave negative age and/or
metallicity gradients. Several studies \citep[e.g.][]{Dimatteo+09,LaBarbera.etal:2010,LaBarbera.etal:2012,Bernardi+19,Zhuang+19,Zibetti+20} have shown that connecting observed galaxy gradients to the physical processes that produce them is not a straightforward task, so it is difficult to determine how this selection criteria affects our results and conclusions. Although the CSGs were selected to have negative (blue) colour gradients as for the VYGs, we cannot claim that our results and conclusions can be extended to VYGs that have positive (red) colour gradients.


\subsubsection{Uncertainties in the spectral  analysis}

Young stellar populations are much more luminous than old ones. So, even within the fibre, the young population can outshine the old stars, making it hard to be detected through SPS analysis of the SDSS spectra and possibly leading to the overestimation of the fraction of mass in populations younger than 1\,Gyr, $f_{M_\star,{\rm y}}$.

To investigate the uncertainties in the determination of $f_{M_\star,{\rm
y}}$ using the STARLIGHT code, we used the \citet{Vazdekis+15} spectral
models to create a set of simulated spectra assuming two bursts of SF, one
occurring within the last 1\,Gyr and the other 12\,Gyr ago. The age 
of the young population was randomly selected among 21 ages between 0.03 and 1\,Gyr,
while the old population has a fixed age of 12\,Gyr.  We used as input
$f^{\rm  in}_{M_\star,{\rm y}} = 0.1, 0.2, 0.3$ and 0.4, and created 1000 simulated
spectra for each $f^{\rm in}_{M_\star,{\rm y}}$. The internal extinction and
S/N of the simulated spectra were selected from the distributions of $A_V^{\rm cont}$ and S/N
of the VYG sample.

%
\begin{table}
\centering
\caption{False positive VYGs from simulated spectra}
\def\arraystretch{1.5}
\begin{tabular}{ccc}
\hline
     \multirow{2}{*}{$f^{\rm in}_{M_\star,{\rm y}}$} & \multirow{2}{*}{median $f^{\rm out}_{M_\star,{\rm y}}$} & fraction of runs \\
        &     & with $f^{\rm out}_{M_\star,{\rm y}} \ge 0.5$ \\  
    (1) & (2) & (3) \\
     \hline
     0.10 & $0.10^{+0.04}_{-0.07}$ & \ \,0.1\% \\ 
     0.20 & $0.19^{+0.11}_{-0.12}$ & \ \,6.9\% \\  
     0.30 & $0.31^{+0.38}_{-0.18}$ & 20.7\% \\
     0.40 & $0.46^{+0.48}_{-0.29}$ & 44.0\%\ \ \ \\
     \hline
\end{tabular}
\parbox{\hsize}{Notes: Columns are {  (1)} Input mass fraction of the
        young stellar population (age $< 1\,$Gyr); 1000 simulated spectra
        were created for each $f^{\rm in}_{M_\star,{\rm y}}$ value. { 
        (2)} Median of the output $f^{\rm out}_{M_\star,{\rm y}}$ values; the
        upper and lower errors were obtained from the 84th and 16th
        percentiles of the  $f^{\rm out}_{M_\star,{\rm y}}$
        distributions. {  (3)} Fraction of runs with $f^{\rm
        out}_{M_\star,{\rm y}} \ge 0.5$.
        }
\label{tab_starlight_sims}
\end{table}

We ran STARLIGHT with the same configuration used in the SDSS spectral
analysis of Paper~II, and computed the fraction of runs with output $f^{\rm out}_{M_\star,{\rm y}} \ge 0.5$, i.e., the fraction of simulated galaxies that would be wrongly classified as VYGs. We show the results in Table~\ref{tab_starlight_sims}, where we also present the median of the $f^{\rm out}_{M_\star,{\rm y}}$ values obtained. 

We find a good agreement between the median estimated fraction of young mass and the input fraction. 
The fraction of incorrect VYG classifications rises very rapidly with the fraction of young stars: from rare with 10 per cent of young mass in the input spectra to 44 per cent with 40 per cent of young mass in the spectrum.
This indicates that the reliability of the VYG classification is not too low.

It is difficult to estimate the number of false VYGs in our sample, because
we do not know neither the true distribution of $f_{M_\star,{\rm y}}$ nor the
age and duration of the past and recent bursts of SF. Besides, our 2-burst simulations with
are very simplistic.
On the other hand, they represent a
`worst' possible scenario of a very old and low-luminosity population
(12 Gyr) mixed with a very recent burst (< 1\,Gyr). Many studies show that
the SFHs of real low-mass galaxies are more extended \citep{Thomas+10,
DeLucia+06, Trevisan.etal:2012}, and their old ($>\,1\,$Gyr) stellar
component would be brighter and thus easier to detect than the population in our simulations, which was formed in single burst 12\,Gyr ago.  
Nevertheless, we argued in Paper~II that
a conservative lower limit on the reliability of the VYG classification
combined with the observed cosmic SFH implies that present-day galaxies
with $M_{\star} > 10^8\,$M$_{\odot}$ experienced, on average, at most four major
starbursts over cosmic time.

In summary, our simulations show that it is possible that the fraction of
mass in populations younger than 1\,Gyr in some of our VYGs is, in fact, less
than 50\%. But our results suggest that, even if $f_{M_\star,{\rm y}}$ is
less than 50\%, the young population still accounts for a significant
fraction of the total stellar mass of the galaxy ($f_{M_\star,{\rm
y}} \gtrsim 30\%$). In addition, the results from 3 different spectral
models have been combined to increase the reliability of our VYG sample.

Finally, one may think that the SPS analysis is not the proper method to
derive the fraction of mass of young populations in galaxies, due to the
large uncertainties associated with this method, in particular for galaxies
with few absorption features in their spectra (such as galaxies with a very
recent starburst). However, the differences in the properties of VYGs
relative to our control sample confirms that these systems are different,
i.e. younger. These differences cannot be attributed to the presence of starburst
galaxies with large equivalent widths of H$\alpha$ 
in the VYG sample, since our conclusions remain unchanged when we exclude these
galaxies from the VYG sample (see Table~\ref{tab:p-values}). Besides, we found
differences in properties that were derived through methods that are
independent of the SPS analysis, such as  morphology, 
environment, the $\tau_{\rm \alpha\beta}$ measure of interstellar dust,
and the gas mass fraction.

\subsubsection{Imperfect control sample}
\label{sec:caveats:Mz}

As discussed in Sect.~\ref{sec:data} and shown in Fig.~\ref{fig:afterPSM_b}a, the $\log M_{\star}$ vs. $z$ distribution of the CSGs is slightly different from that of the VYGs. To investigate if it affects our results, we repeated all the analysis presented in this work after excluding the galaxies that are more than $2\sigma$ away from the CSG $\log M_{\star}$ vs. $z$ relation (29 VYGs and 53 CSGs). We used a 2nd-order polynomial to fit the relation. 
All the results presented in Table~\ref{tab:p-values} remain statistically significant with $p$-values below $< 5\times10^{-3}$, except for the parameters ($g - i$)$_{\rm Petro}$, $S_1$, and $d_{\rm 1st}/$kpc, for which KS tests indicate marginal statistical significance with $p$-values$\,= 0.04$, 0.01 and 0.03, respectively. Therefore, we conclude that the outliers in the $\log M_{\star}$ vs. $z$ relation do not affect our results and conclusions.

\subsubsection{Nebular continuum emission}

The SPS fitting of the spectra with STARLIGHT and VESPA both ignored the
nebular contribution to the continuum, which becomes  important in strongly
star-forming galaxies. The ratio between the ionized gas and the 
continuum emissions increases with wavelength, ranging from $\sim 20\%$ at
$\sim 4000\,$\AA\ to more than $40\%$ around $7000\,$\AA\ in galaxies with
very high sSFRs \citep{Izotov+11}.  As discussed in Paper~II, the neglect of
this red nebular component causes the SPS analysis to overestimate the ages
of starburst galaxies. 
This was confirmed by \citet*{Cardoso+19}, who found that the FADO SPS code,
which includes the nebular continuum emission in the fit, obtains lower ages
than does STARLIGHT. Therefore, including the nebular continuum emission in our analysis would lead to even higher fractions of stellar mass formed within the last 1\,Gyr, and would not affect the results and conclusions presented in this work.

\section{Summary and Conclusions}
\label{sec:concl}

In this work, we investigated the properties and environments of very young
galaxies. The SFHs of galaxies were inferred from SDSS spectra through SPS
analysis, and we selected as VYGs those galaxies with over half  their
stellar mass formed within the last $1\,$Gyr according to each of three different stellar population models. We built a control sample of normal galaxies with similar stellar masses, redshifts, radii and colour gradients, and the comparison between these two samples revealed that VYGs have different properties and reside in different environments. In Sect.~\ref{sec:discuss:differ}, we summarise all the VYGs properties that we investigated and discuss how they compare to those of CSGs. From our results, we conclude that:

\begin{itemize}
  \item \emph{VYGs and other young systems like I\,Zw\,18 are likely to be different in nature}, as discussed in Sect.~\ref{sec:discuss:gas}. I\,Zw\,18 analogues are low-mass very metal-poor systems that deviate from the mass-metallicity relation of normal galaxies, while VYGs follow it. \\
  
  \item \emph{Star formation in the more massive VYGs is not being fed by infalling metal-poor gas from the cosmic web}, since these galaxies have similar gas oxygen abundances as the control galaxies.
  \emph{But gas infall and mergers may both be important triggers of star formation in the VYGs at the low-mass end}, given hints of their lower metallicities and their excess of near neighbours.
  \\
  
  \item \emph{Gas-rich mergers and interactions are important mechanisms for
  producing very young systems in the local Universe}
  (Sect.~\ref{sec:discuss:mergers}). Differences in the morphology (VYGs are
  more clumpy, more asymmetric and a larger fraction show tidal features and
  other signs of interactions compared to the CSGs) and the environment (VYGs
  are more likely to have companions and to be found in the inner parts of
  low-mass groups) of the VYGs support this scenario, in particular for the massive VYGs. These results 
  agree with the predictions by \citet[][Paper~I]{Tweed+18} for their
  subset of  analytical models of galaxy formation where halo
  mergers are associated with starbursts. 
\end{itemize}

Since mergers leave typical signatures in the gas and stellar kinematics of the galaxy, future studies of VYGs using spatially-resolved spectroscopic data can help constraining the merger scenario. Moreover, deep imaging data might help understanding the nature of the VYGs for which there is no evident sign of interactions in the SDSS images. We have also obtained GMRT and VLA HI radio observations to investigate the gas distribution and kinematics in and around VYGs.

Finally, in the present study, we showed that the VYGs differ from normal galaxies in various ways, but \emph{did their progenitors have peculiar properties before becoming VYGs? Or does any galaxy can become a VYG?} In a future study, we will address the nature of the VYG progenitors by identifying these systems in hydrodynamical simulations and tracing back the assembly history of these systems. 

We provide an electronic table with observed and derived properties of our VYGs and CSGs. The description of the table columns are given in appendix~\ref{ap:images}.

\section*{Acknowledgments}
The authors thank the referee Daniel Kelson for his excellent comments and suggestions, which led to an improved version of the manuscript. We also thank Mojtaba Raouf for useful suggestions.
MT thanks the support of CNPq (process \#307675/2018-1), the Institut Lagrange de Paris and 
the program L'Or\'eal UNESCO ABC \emph{Para Mulheres na Ci\^encia}.
M.T. and T.X.T. are grateful for the hospitality of the Institut d'Astrophysique de Paris 
where part of this work was carried out. 
G.A.M. acknowledges the Brazilian CNPq (grant \#451451/2019-8) awarded to M.T. and thanks the Universidade Federal do Rio Grande do Sul for its hospitality.
L.S.P acknowledges support of the program of the NAS of
Ukraine for the development of priority fields of scientific research
(CPCEL 6541230).
We acknowledge Roberto Cid Fernandes for making his STARLIGHT code
public and Rita Tojeiro for making the VESPA output publicly available. 
We acknowledge the use of SDSS data (\url{http://www.sdss.org/collaboration/credits.html}), 
{\tt TOPCAT} Table/VOTable Processing Software 
\citep[][\url{http://www.star.bris.ac.uk/mbt/topcat/}]{Taylor:2005}, and R language and environment for statistical computing \citep{R:2015}.
This research has made use of the NASA/IPAC Extragalactic Database (NED),
which is operated by the Jet Propulsion Laboratory, California Institute of
Technology, under contract with the National Aeronautics and Space
Administration.

\section*{Data availability}
The data underlying this article are available in the article and in its online supplementary material.


\bibliography{master}

\begin{thebibliography}{}
\makeatletter
\relax
\def\mn@urlcharsother{\let\do\@makeother \do\$\do\&\do\#\do\^\do\_\do\%\do\~}
\def\mn@doi{\begingroup\mn@urlcharsother \@ifnextchar [ {\mn@doi@}
  {\mn@doi@[]}}
\def\mn@doi@[#1]#2{\def\@tempa{#1}\ifx\@tempa\@empty \href
  {http://dx.doi.org/#2} {doi:#2}\else \href {http://dx.doi.org/#2} {#1}\fi
  \endgroup}
\def\mn@eprint#1#2{\mn@eprint@#1:#2::\@nil}
\def\mn@eprint@arXiv#1{\href {http://arxiv.org/abs/#1} {{\tt arXiv:#1}}}
\def\mn@eprint@dblp#1{\href {http://dblp.uni-trier.de/rec/bibtex/#1.xml}
  {dblp:#1}}
\def\mn@eprint@#1:#2:#3:#4\@nil{\def\@tempa {#1}\def\@tempb {#2}\def\@tempc
  {#3}\ifx \@tempc \@empty \let \@tempc \@tempb \let \@tempb \@tempa \fi \ifx
  \@tempb \@empty \def\@tempb {arXiv}\fi \@ifundefined
  {mn@eprint@\@tempb}{\@tempb:\@tempc}{\expandafter \expandafter \csname
  mn@eprint@\@tempb\endcsname \expandafter{\@tempc}}}

\bibitem[\protect\citeauthoryear{{Abazajian} et~al.,}{{Abazajian}
  et~al.}{2009}]{Abazajian+09}
{Abazajian} K.~N.,  et~al., 2009, \mn@doi [\apjs]
  {10.1088/0067-0049/182/2/543}, \href
  {https://ui.adsabs.harvard.edu/abs/2009ApJS..182..543A} {182, 543}

\bibitem[\protect\citeauthoryear{{Abraham}, {Valdes}, {Yee}  \& {van den
  Bergh}}{{Abraham} et~al.}{1994}]{Abraham+94}
{Abraham} R.~G.,  {Valdes} F.,  {Yee} H.~K.~C.,   {van den Bergh} S.,  1994,
  \mn@doi [\apj] {10.1086/174550}, \href
  {http://cdsads.u-strasbg.fr/abs/1994ApJ...432...75A} {432, 75}

\bibitem[\protect\citeauthoryear{{Abraham}, {van den Bergh}, {Glazebrook},
  {Ellis}, {Santiago}, {Surma}  \& {Griffiths}}{{Abraham}
  et~al.}{1996}]{Abraham+96}
{Abraham} R.~G.,  {van den Bergh} S.,  {Glazebrook} K.,  {Ellis} R.~S.,
  {Santiago} B.~X.,  {Surma} P.,   {Griffiths} R.~E.,  1996, \mn@doi [\apjs]
  {10.1086/192352}, \href {http://cdsads.u-strasbg.fr/abs/1996ApJS..107....1A}
  {107, 1}

\bibitem[\protect\citeauthoryear{{Alloin}, {Collin-Souffrin}, {Joly}  \&
  {Vigroux}}{{Alloin} et~al.}{1979}]{Alloin1979}
{Alloin} D.,  {Collin-Souffrin} S.,  {Joly} M.,   {Vigroux} L.,  1979, \aap,
  \href {http://adsabs.harvard.edu/abs/1979A%26A....78..200A} {78, 200}

\bibitem[\protect\citeauthoryear{{Aloisi} et~al.,}{{Aloisi}
  et~al.}{2007}]{Aloisi.etal:2007}
{Aloisi} A.,  et~al., 2007, \mn@doi [\apjl] {10.1086/522368}, \href
  {https://ui.adsabs.harvard.edu/abs/2007ApJ...667L.151A} {667, L151}

\bibitem[\protect\citeauthoryear{{Armus}, {Heckman}  \& {Miley}}{{Armus}
  et~al.}{1987}]{Armus+87}
{Armus} L.,  {Heckman} T.,   {Miley} G.,  1987, \mn@doi [\aj] {10.1086/114517},
  \href {https://ui.adsabs.harvard.edu/abs/1987AJ.....94..831A} {94, 831}

\bibitem[\protect\citeauthoryear{{Asari}, {Cid Fernandes}, {Stasi{\'n}ska},
  {Torres-Papaqui}, {Mateus}, {Sodr{\'e}}, {Schoenell}  \& {Gomes}}{{Asari}
  et~al.}{2007}]{Asari+2007}
{Asari} N.~V.,  {Cid Fernandes} R.,  {Stasi{\'n}ska} G.,  {Torres-Papaqui}
  J.~P.,  {Mateus} A.,  {Sodr{\'e}} L.,  {Schoenell} W.,   {Gomes} J.~M.,
  2007, \mn@doi [\mnras] {10.1111/j.1365-2966.2007.12255.x}, \href
  {https://ui.adsabs.harvard.edu/abs/2007MNRAS.381..263A} {381, 263}

\bibitem[\protect\citeauthoryear{{Baldwin}, {Phillips}  \&
  {Terlevich}}{{Baldwin} et~al.}{1981}]{BPT81}
{Baldwin} J.~A.,  {Phillips} M.~M.,   {Terlevich} R.,  1981, \mn@doi [\pasp]
  {10.1086/130766}, \href
  {https://ui.adsabs.harvard.edu/abs/1981PASP...93....5B} {93, 5}

\bibitem[\protect\citeauthoryear{{Balogh} et~al.,}{{Balogh}
  et~al.}{2009}]{Balogh.etal:2009}
{Balogh} M.~L.,  et~al., 2009, \mn@doi [MNRAS]
  {10.1111/j.1365-2966.2009.15193.x}, 398, 754

\bibitem[\protect\citeauthoryear{{Barnard}}{{Barnard}}{1945}]{Barnard:1945}
{Barnard} G.~A.,  1945, \mn@doi [\nat] {10.1038/156783b0}, \href
  {http://adsabs.harvard.edu/abs/1945Natur.156..783B} {156, 783}

\bibitem[\protect\citeauthoryear{{Behroozi}, {Wechsler}  \&
  {Conroy}}{{Behroozi} et~al.}{2013}]{Behroozi+13}
{Behroozi} P.~S.,  {Wechsler} R.~H.,   {Conroy} C.,  2013, \mn@doi [\apj]
  {10.1088/0004-637X/770/1/57}, \href
  {http://adsabs.harvard.edu/abs/2013ApJ...770...57B} {770, 57}

\bibitem[\protect\citeauthoryear{{Belfiore} et~al.,}{{Belfiore}
  et~al.}{2015}]{Belfiore2015}
{Belfiore} F.,  et~al., 2015, \mn@doi [\mnras] {10.1093/mnras/stv296}, \href
  {http://adsabs.harvard.edu/abs/2015MNRAS.449..867B} {449, 867}

\bibitem[\protect\citeauthoryear{{Belfiore} et~al.,}{{Belfiore}
  et~al.}{2017}]{Belfiore2017}
{Belfiore} F.,  et~al., 2017, \mn@doi [\mnras] {10.1093/mnras/stx789}, \href
  {http://adsabs.harvard.edu/abs/2017MNRAS.469..151B} {469, 151}

\bibitem[\protect\citeauthoryear{{Bernardi}, {Dom{\'\i}nguez S{\'a}nchez},
  {Brownstein}, {Drory}  \& {Sheth}}{{Bernardi} et~al.}{2019}]{Bernardi+19}
{Bernardi} M.,  {Dom{\'\i}nguez S{\'a}nchez} H.,  {Brownstein} J.~R.,  {Drory}
  N.,   {Sheth} R.~K.,  2019, \mn@doi [\mnras] {10.1093/mnras/stz2413}, \href
  {https://ui.adsabs.harvard.edu/abs/2019MNRAS.489.5633B} {489, 5633}

\bibitem[\protect\citeauthoryear{Bishop}{Bishop}{2006}]{Bishop:2006}
Bishop C.~M.,  2006, Pattern Recognition and Machine Learning.
Springer, \url {http://research.microsoft.com/en-us/um/people/cmbishop/prml/}

\bibitem[\protect\citeauthoryear{{Blanton} et~al.,}{{Blanton}
  et~al.}{2003}]{Blanton+03_kcorrect}
{Blanton} M.~R.,  et~al., 2003, \mn@doi [\aj] {10.1086/342935}, \href
  {http://adsabs.harvard.edu/abs/2003AJ....125.2348B} {125, 2348}

\bibitem[\protect\citeauthoryear{{Blanton} et~al.,}{{Blanton}
  et~al.}{2005}]{Blanton+05_NYUVAGC}
{Blanton} M.~R.,  et~al., 2005, \mn@doi [\aj] {10.1086/429803}, \href
  {https://ui.adsabs.harvard.edu/abs/2005AJ....129.2562B} {129, 2562}

\bibitem[\protect\citeauthoryear{{Borne}, {Bushouse}, {Lucas}  \&
  {Colina}}{{Borne} et~al.}{2000}]{Borne+00}
{Borne} K.~D.,  {Bushouse} H.,  {Lucas} R.~A.,   {Colina} L.,  2000, \mn@doi
  [\apjl] {10.1086/312461}, \href
  {https://ui.adsabs.harvard.edu/abs/2000ApJ...529L..77B} {529, L77}

\bibitem[\protect\citeauthoryear{{Bressan}, {Fagotto}, {Bertelli}  \&
  {Chiosi}}{{Bressan} et~al.}{1993}]{Bressan+93}
{Bressan} A.,  {Fagotto} F.,  {Bertelli} G.,   {Chiosi} C.,  1993, \aaps, \href
  {http://cdsads.u-strasbg.fr/abs/1993A%26AS..100..647B} {100, 647}

\bibitem[\protect\citeauthoryear{{Brinchmann}, {Charlot}, {White}, {Tremonti},
  {Kauffmann}, {Heckman}  \& {Brinkmann}}{{Brinchmann}
  et~al.}{2004}]{Brinchmann+04}
{Brinchmann} J.,  {Charlot} S.,  {White} S.~D.~M.,  {Tremonti} C.,  {Kauffmann}
  G.,  {Heckman} T.,   {Brinkmann} J.,  2004, \mn@doi [\mnras]
  {10.1111/j.1365-2966.2004.07881.x}, \href
  {http://adsabs.harvard.edu/abs/2004MNRAS.351.1151B} {351, 1151}

\bibitem[\protect\citeauthoryear{{Bruzual} \& {Charlot}}{{Bruzual} \&
  {Charlot}}{2003}]{Bruzual&Charlot03}
{Bruzual} G.,  {Charlot} S.,  2003, \mn@doi [\mnras]
  {10.1046/j.1365-8711.2003.06897.x}, \href
  {http://adsabs.harvard.edu/abs/2003MNRAS.344.1000B} {344, 1000}

\bibitem[\protect\citeauthoryear{{Calzetti}, {Armus}, {Bohlin}, {Kinney},
  {Koornneef}  \& {Storchi-Bergmann}}{{Calzetti} et~al.}{2000}]{Calzetti+2000}
{Calzetti} D.,  {Armus} L.,  {Bohlin} R.~C.,  {Kinney} A.~L.,  {Koornneef} J.,
   {Storchi-Bergmann} T.,  2000, \mn@doi [\apj] {10.1086/308692}, \href
  {https://ui.adsabs.harvard.edu/abs/2000ApJ...533..682C} {533, 682}

\bibitem[\protect\citeauthoryear{{Cardelli}, {Clayton}  \& {Mathis}}{{Cardelli}
  et~al.}{1989}]{Cardelli.etal:1989}
{Cardelli} J.~A.,  {Clayton} G.~C.,   {Mathis} J.~S.,  1989, \mn@doi [ApJ]
  {10.1086/167900}, \href {http://adsabs.harvard.edu/abs/1989ApJ...345..245C}
  {345, 245}

\bibitem[\protect\citeauthoryear{{Cardoso}, {Gomes}  \& {Papaderos}}{{Cardoso}
  et~al.}{2019}]{Cardoso+19}
{Cardoso} L. S.~M.,  {Gomes} J.~M.,   {Papaderos} P.,  2019, \mn@doi [\aap]
  {10.1051/0004-6361/201833438}, \href
  {https://ui.adsabs.harvard.edu/abs/2019A&A...622A..56C} {622, A56}

\bibitem[\protect\citeauthoryear{{Casasola} et~al.,}{{Casasola}
  et~al.}{2020}]{Casasola+20}
{Casasola} V.,  et~al., 2020, \mn@doi [\aap] {10.1051/0004-6361/201936665},
  \href {https://ui.adsabs.harvard.edu/abs/2020A&A...633A.100C} {633, A100}

\bibitem[\protect\citeauthoryear{{Catinella} et~al.,}{{Catinella}
  et~al.}{2010}]{Catinella+10}
{Catinella} B.,  et~al., 2010, \mn@doi [\mnras]
  {10.1111/j.1365-2966.2009.16180.x}, \href
  {https://ui.adsabs.harvard.edu/abs/2010MNRAS.403..683C} {403, 683}

\bibitem[\protect\citeauthoryear{{Cattaneo}, {Mamon}, {Warnick}  \&
  {Knebe}}{{Cattaneo} et~al.}{2011}]{Cattaneo+11}
{Cattaneo} A.,  {Mamon} G.~A.,  {Warnick} K.,   {Knebe} A.,  2011, \mn@doi
  [\aap] {10.1051/0004-6361/201015780}, \href
  {http://adsabs.harvard.edu/abs/2011A%26A...533A...5C} {533, A5}

\bibitem[\protect\citeauthoryear{{Chabrier}}{{Chabrier}}{2003}]{Chabrier03}
{Chabrier} G.,  2003, \mn@doi [\pasp] {10.1086/376392}, \href
  {http://adsabs.harvard.edu/abs/2003PASP..115..763C} {115, 763}

\bibitem[\protect\citeauthoryear{{Charlot} \& {Fall}}{{Charlot} \&
  {Fall}}{2000}]{Charlot&Fall00}
{Charlot} S.,  {Fall} S.~M.,  2000, \mn@doi [\apj] {10.1086/309250}, \href
  {http://adsabs.harvard.edu/abs/2000ApJ...539..718C} {539, 718}

\bibitem[\protect\citeauthoryear{{Cid Fernandes}, {Mateus}, {Sodr{\'e}},
  {Stasi{\'n}ska}  \& {Gomes}}{{Cid Fernandes} et~al.}{2005}]{CidFernandes+05}
{Cid Fernandes} R.,  {Mateus} A.,  {Sodr{\'e}} L.,  {Stasi{\'n}ska} G.,
  {Gomes} J.~M.,  2005, \mn@doi [\mnras] {10.1111/j.1365-2966.2005.08752.x},
  \href {http://cdsads.u-strasbg.fr/abs/2005MNRAS.358..363C} {358, 363}

\bibitem[\protect\citeauthoryear{{Conselice}, {Bershady}  \&
  {Jangren}}{{Conselice} et~al.}{2000}]{Conselice+00}
{Conselice} C.~J.,  {Bershady} M.~A.,   {Jangren} A.,  2000, \mn@doi [\apj]
  {10.1086/308300}, \href {http://cdsads.u-strasbg.fr/abs/2000ApJ...529..886C}
  {529, 886}

\bibitem[\protect\citeauthoryear{{Contreras Ramos} et~al.,}{{Contreras Ramos}
  et~al.}{2011}]{ContrerasRamos.etal:2011}
{Contreras Ramos} R.,  et~al., 2011, \mn@doi [\apj]
  {10.1088/0004-637X/739/2/74}, \href
  {https://ui.adsabs.harvard.edu/abs/2011ApJ...739...74C} {739, 74}

\bibitem[\protect\citeauthoryear{{Darg} et~al.,}{{Darg}
  et~al.}{2010}]{Darg.etal:2010}
{Darg} D.~W.,  et~al., 2010, \mn@doi [\mnras]
  {10.1111/j.1365-2966.2009.15686.x}, \href
  {https://ui.adsabs.harvard.edu/abs/2010MNRAS.401.1043D} {401, 1043}

\bibitem[\protect\citeauthoryear{{De Lucia}, {Springel}, {White}, {Croton}  \&
  {Kauffmann}}{{De Lucia} et~al.}{2006}]{DeLucia+06}
{De Lucia} G.,  {Springel} V.,  {White} S. D.~M.,  {Croton} D.,   {Kauffmann}
  G.,  2006, \mn@doi [\mnras] {10.1111/j.1365-2966.2005.09879.x}, \href
  {https://ui.adsabs.harvard.edu/abs/2006MNRAS.366..499D} {366, 499}

\bibitem[\protect\citeauthoryear{{Di Matteo}, {Bournaud}, {Martig}, {Combes},
  {Melchior}  \& {Semelin}}{{Di Matteo} et~al.}{2008}]{DiMatteo+2008}
{Di Matteo} P.,  {Bournaud} F.,  {Martig} M.,  {Combes} F.,  {Melchior} A.~L.,
   {Semelin} B.,  2008, \mn@doi [\aap] {10.1051/0004-6361:200809480}, \href
  {https://ui.adsabs.harvard.edu/abs/2008A&A...492...31D} {492, 31}

\bibitem[\protect\citeauthoryear{{Di Matteo}, {Pipino}, {Lehnert}, {Combes}  \&
  {Semelin}}{{Di Matteo} et~al.}{2009}]{Dimatteo+09}
{Di Matteo} P.,  {Pipino} A.,  {Lehnert} M.~D.,  {Combes} F.,   {Semelin} B.,
  2009, \mn@doi [\aap] {10.1051/0004-6361/200911715}, \href
  {https://ui.adsabs.harvard.edu/abs/2009A&A...499..427D} {499, 427}

\bibitem[\protect\citeauthoryear{{Dinerstein}}{{Dinerstein}}{1990}]{Dinerstein1990}
{Dinerstein} H.~L.,  1990, in {Thronson} Jr. H.~A.,  {Shull} J.~M.,  eds,
  Astrophysics and Space Science Library Vol. 161, The Interstellar Medium in
  Galaxies. pp 257--285, \mn@doi{10.1007/978-94-009-0595-5_10}

\bibitem[\protect\citeauthoryear{{Dom{\'\i}nguez S{\'a}nchez},
  {Huertas-Company}, {Bernardi}, {Tuccillo}  \& {Fischer}}{{Dom{\'\i}nguez
  S{\'a}nchez} et~al.}{2018}]{DominguezSanchez.etal:2018}
{Dom{\'\i}nguez S{\'a}nchez} H.,  {Huertas-Company} M.,  {Bernardi} M.,
  {Tuccillo} D.,   {Fischer} J.~L.,  2018, \mn@doi [\mnras]
  {10.1093/mnras/sty338}, \href
  {https://ui.adsabs.harvard.edu/abs/2018MNRAS.476.3661D} {476, 3661}

\bibitem[\protect\citeauthoryear{{Dom{\'\i}nguez} et~al.,}{{Dom{\'\i}nguez}
  et~al.}{2013}]{Dominguez+13}
{Dom{\'\i}nguez} A.,  et~al., 2013, \mn@doi [\apj]
  {10.1088/0004-637X/763/2/145}, \href
  {https://ui.adsabs.harvard.edu/abs/2013ApJ...763..145D} {763, 145}

\bibitem[\protect\citeauthoryear{{Dressler}, {Kelson}  \&
  {Abramson}}{{Dressler} et~al.}{2018}]{Dressler.etal:2018}
{Dressler} A.,  {Kelson} D.~D.,   {Abramson} L.~E.,  2018, \mn@doi [\apj]
  {10.3847/1538-4357/aaedbe}, \href
  {https://ui.adsabs.harvard.edu/abs/2018ApJ...869..152D} {869, 152}

\bibitem[\protect\citeauthoryear{{Ellison}, {Catinella}  \&
  {Cortese}}{{Ellison} et~al.}{2018}]{Ellison+18}
{Ellison} S.~L.,  {Catinella} B.,   {Cortese} L.,  2018, \mn@doi [\mnras]
  {10.1093/mnras/sty1247}, \href
  {https://ui.adsabs.harvard.edu/abs/2018MNRAS.478.3447E} {478, 3447}

\bibitem[\protect\citeauthoryear{{Engelbracht}, {Rieke}, {Gordon}, {Smith},
  {Werner}, {Moustakas}, {Willmer}  \& {Vanzi}}{{Engelbracht}
  et~al.}{2008}]{Engelbracht+08}
{Engelbracht} C.~W.,  {Rieke} G.~H.,  {Gordon} K.~D.,  {Smith} J. D.~T.,
  {Werner} M.~W.,  {Moustakas} J.,  {Willmer} C.~N.~A.,   {Vanzi} L.,  2008,
  \mn@doi [\apj] {10.1086/529513}, \href
  {https://ui.adsabs.harvard.edu/abs/2008ApJ...678..804E} {678, 804}

\bibitem[\protect\citeauthoryear{Erguler}{Erguler}{2016}]{Erguler:2016}
Erguler K.,  2016, Barnard: Barnard's Unconditional Test.
\url {https://CRAN.R-project.org/package=Barnard}

\bibitem[\protect\citeauthoryear{{Fagotto}, {Bressan}, {Bertelli}  \&
  {Chiosi}}{{Fagotto} et~al.}{1994a}]{Fagotto+94a}
{Fagotto} F.,  {Bressan} A.,  {Bertelli} G.,   {Chiosi} C.,  1994a, \aaps,
  \href {http://cdsads.u-strasbg.fr/abs/1994A%26AS..104..365F} {104, 365}

\bibitem[\protect\citeauthoryear{{Fagotto}, {Bressan}, {Bertelli}  \&
  {Chiosi}}{{Fagotto} et~al.}{1994b}]{Fagotto+94b}
{Fagotto} F.,  {Bressan} A.,  {Bertelli} G.,   {Chiosi} C.,  1994b, \aaps,
  \href {http://cdsads.u-strasbg.fr/abs/1994A%26AS..105...29F} {105, 29}

\bibitem[\protect\citeauthoryear{{Ferrari}, {de Carvalho}  \&
  {Trevisan}}{{Ferrari} et~al.}{2015}]{Ferrari.etal:2015}
{Ferrari} F.,  {de Carvalho} R.~R.,   {Trevisan} M.,  2015, \mn@doi [ApJ]
  {10.1088/0004-637X/814/1/55}, \href
  {http://adsabs.harvard.edu/abs/2015ApJ...814...55F} {814, 55}

\bibitem[\protect\citeauthoryear{{Ferreras} et~al.,}{{Ferreras}
  et~al.}{2019}]{Ferreras+19}
{Ferreras} I.,  et~al., 2019, \mn@doi [\mnras] {10.1093/mnras/stz2095}, \href
  {https://ui.adsabs.harvard.edu/abs/2019MNRAS.489..608F} {489, 608}

\bibitem[\protect\citeauthoryear{{Fisher}}{{Fisher}}{1935}]{Fisher:1935}
{Fisher} R.~A.,  1935, \mn@doi [Journal of the Royal Statistical Society]
  {10.2307/2342435}, 98, 39–

\bibitem[\protect\citeauthoryear{{Florido}, {Zurita}, {P{\'e}rez},
  {P{\'e}rez-Montero}, {Coelho}  \& {Gadotti}}{{Florido}
  et~al.}{2015}]{Florido+2015}
{Florido} E.,  {Zurita} A.,  {P{\'e}rez} I.,  {P{\'e}rez-Montero} E.,  {Coelho}
  P.~R.~T.,   {Gadotti} D.~A.,  2015, \mn@doi [\aap]
  {10.1051/0004-6361/201526191}, \href
  {https://ui.adsabs.harvard.edu/abs/2015A&A...584A..88F} {584, A88}

\bibitem[\protect\citeauthoryear{{Fraley} \& {Raftery}}{{Fraley} \&
  {Raftery}}{2002}]{mclust:2002}
{Fraley} C.,  {Raftery} A.~E.,  2002, \mn@doi [Journal of the American
  Statistical Association] {10.1198/016214502760047131}, 97, 611

\bibitem[\protect\citeauthoryear{{Fraley}, {Raftery}, {Murphy}  \&
  {Scrucca}}{{Fraley} et~al.}{2012}]{mclust:2012}
{Fraley} C.,  {Raftery} A.~E.,  {Murphy} T.~B.,   {Scrucca} L.,  2012, mclust
  Version 4 for R: Normal Mixture Modeling for Model-Based Clustering,
  Classification, and Density Estimation

\bibitem[\protect\citeauthoryear{{Girardi}, {Bressan}, {Chiosi}, {Bertelli}  \&
  {Nasi}}{{Girardi} et~al.}{1996}]{Girardi+96}
{Girardi} L.,  {Bressan} A.,  {Chiosi} C.,  {Bertelli} G.,   {Nasi} E.,  1996,
  \aaps, \href {http://adsabs.harvard.edu/abs/1996A%26AS..117..113G} {117, 113}

\bibitem[\protect\citeauthoryear{{Gnedin}}{{Gnedin}}{2000}]{Gnedin00}
{Gnedin} N.~Y.,  2000, \mn@doi [\apj] {10.1086/317042}, \href
  {http://adsabs.harvard.edu/abs/2000ApJ...542..535G} {542, 535}

\bibitem[\protect\citeauthoryear{{Hamilton} \& {Tegmark}}{{Hamilton} \&
  {Tegmark}}{2004}]{Hamilton.Tegmark:2004}
{Hamilton} A.~J.~S.,  {Tegmark} M.,  2004, \mn@doi [MNRAS]
  {10.1111/j.1365-2966.2004.07490.x}, \href
  {http://adsabs.harvard.edu/abs/2004MNRAS.349..115H} {349, 115}

\bibitem[\protect\citeauthoryear{Hamming}{Hamming}{1998}]{Hamming:1998}
Hamming R.~W.,  1998, Digital filters, 3 edn.
Dover, Mineola, N.Y.

\bibitem[\protect\citeauthoryear{{Haynes} et~al.,}{{Haynes}
  et~al.}{2011}]{Haynes.etal:2011}
{Haynes} M.~P.,  et~al., 2011, \mn@doi [\aj] {10.1088/0004-6256/142/5/170},
  \href {https://ui.adsabs.harvard.edu/abs/2011AJ....142..170H} {142, 170}

\bibitem[\protect\citeauthoryear{{Haynes} et~al.,}{{Haynes}
  et~al.}{2018}]{Haynes.etal:2018}
{Haynes} M.~P.,  et~al., 2018, \mn@doi [\apj] {10.3847/1538-4357/aac956}, \href
  {https://ui.adsabs.harvard.edu/abs/2018ApJ...861...49H} {861, 49}

\bibitem[\protect\citeauthoryear{{Hirschmann}, {Naab}, {Ostriker}, {Forbes},
  {Duc}, {Dav{\'e}}, {Oser}  \& {Karabal}}{{Hirschmann}
  et~al.}{2015}]{Hirschmann+15}
{Hirschmann} M.,  {Naab} T.,  {Ostriker} J.~P.,  {Forbes} D.~A.,  {Duc} P.-A.,
  {Dav{\'e}} R.,  {Oser} L.,   {Karabal} E.,  2015, \mn@doi [\mnras]
  {10.1093/mnras/stv274}, \href
  {https://ui.adsabs.harvard.edu/abs/2015MNRAS.449..528H} {449, 528}

\bibitem[\protect\citeauthoryear{{Hjorth}, {Gall}  \&
  {Micha{\l}owski}}{{Hjorth} et~al.}{2014}]{Hjorth+2014}
{Hjorth} J.,  {Gall} C.,   {Micha{\l}owski} M.~J.,  2014, \mn@doi [\apjl]
  {10.1088/2041-8205/782/2/L23}, \href
  {https://ui.adsabs.harvard.edu/abs/2014ApJ...782L..23H} {782, L23}

\bibitem[\protect\citeauthoryear{Ho, Imai, King  \& Stuart}{Ho
  et~al.}{2011}]{MatchIt:2011}
Ho D.~E.,  Imai K.,  King G.,   Stuart E.~A.,  2011, Journal of Statistical
  Software, 42, 1

\bibitem[\protect\citeauthoryear{{Hopkins} et~al.,}{{Hopkins}
  et~al.}{2010}]{Hopkins+2010}
{Hopkins} P.~F.,  et~al., 2010, \mn@doi [\apj] {10.1088/0004-637X/715/1/202},
  \href {https://ui.adsabs.harvard.edu/abs/2010ApJ...715..202H} {715, 202}

\bibitem[\protect\citeauthoryear{{Hopkins}, {Cox}, {Hernquist}, {Narayanan},
  {Hayward}  \& {Murray}}{{Hopkins} et~al.}{2013}]{Hopkins+2013}
{Hopkins} P.~F.,  {Cox} T.~J.,  {Hernquist} L.,  {Narayanan} D.,  {Hayward}
  C.~C.,   {Murray} N.,  2013, \mn@doi [\mnras] {10.1093/mnras/stt017}, \href
  {https://ui.adsabs.harvard.edu/abs/2013MNRAS.430.1901H} {430, 1901}

\bibitem[\protect\citeauthoryear{{Izotov} \& {Thuan}}{{Izotov} \&
  {Thuan}}{1998}]{Izotov.Thuan:1998}
{Izotov} Y.~I.,  {Thuan} T.~X.,  1998, \mn@doi [\apj] {10.1086/305440}, \href
  {https://ui.adsabs.harvard.edu/abs/1998ApJ...497..227I} {497, 227}

\bibitem[\protect\citeauthoryear{{Izotov} \& {Thuan}}{{Izotov} \&
  {Thuan}}{2004}]{Izotov.Thuan:2004}
{Izotov} Y.~I.,  {Thuan} T.~X.,  2004, \mn@doi [\apj] {10.1086/424990}, \href
  {https://ui.adsabs.harvard.edu/abs/2004ApJ...616..768I} {616, 768}

\bibitem[\protect\citeauthoryear{{Izotov}, {Guseva}  \& {Thuan}}{{Izotov}
  et~al.}{2011}]{Izotov+11}
{Izotov} Y.~I.,  {Guseva} N.~G.,   {Thuan} T.~X.,  2011, \mn@doi [\apj]
  {10.1088/0004-637X/728/2/161}, \href
  {https://ui.adsabs.harvard.edu/abs/2011ApJ...728..161I} {728, 161}

\bibitem[\protect\citeauthoryear{{Izotov}, {Guseva}, {Fricke}  \&
  {Henkel}}{{Izotov} et~al.}{2014}]{Izotov+2014_SDSS_SFing}
{Izotov} Y.~I.,  {Guseva} N.~G.,  {Fricke} K.~J.,   {Henkel} C.,  2014, \mn@doi
  [\aap] {10.1051/0004-6361/201322338}, \href
  {https://ui.adsabs.harvard.edu/abs/2014A&A...561A..33I} {561, A33}

\bibitem[\protect\citeauthoryear{{Izotov}, {Thuan}, {Guseva}  \&
  {Liss}}{{Izotov} et~al.}{2018}]{Izotov.etal:2018}
{Izotov} Y.~I.,  {Thuan} T.~X.,  {Guseva} N.~G.,   {Liss} S.~E.,  2018, \mn@doi
  [\mnras] {10.1093/mnras/stx2478}, \href
  {https://ui.adsabs.harvard.edu/abs/2018MNRAS.473.1956I} {473, 1956}

\bibitem[\protect\citeauthoryear{{Izotov}, {Thuan}  \& {Guseva}}{{Izotov}
  et~al.}{2019}]{Izotov+19}
{Izotov} Y.~I.,  {Thuan} T.~X.,   {Guseva} N.~G.,  2019, \mn@doi [\mnras]
  {10.1093/mnras/sty3472}, \href
  {https://ui.adsabs.harvard.edu/abs/2019MNRAS.483.5491I} {483, 5491}

\bibitem[\protect\citeauthoryear{{Joseph} \& {Wright}}{{Joseph} \&
  {Wright}}{1985}]{Joseph&Wright85}
{Joseph} R.~D.,  {Wright} G.~S.,  1985, \mn@doi [\mnras]
  {10.1093/mnras/214.2.87}, \href
  {https://ui.adsabs.harvard.edu/abs/1985MNRAS.214...87J} {214, 87}

\bibitem[\protect\citeauthoryear{{Kauffmann} et~al.,}{{Kauffmann}
  et~al.}{2003}]{Kauffmann+03}
{Kauffmann} G.,  et~al., 2003, \mn@doi [\mnras]
  {10.1046/j.1365-8711.2003.06291.x}, \href
  {https://ui.adsabs.harvard.edu/abs/2003MNRAS.341...33K} {341, 33}

\bibitem[\protect\citeauthoryear{{Kewley}, {Dopita}, {Sutherland}, {Heisler}
  \& {Trevena}}{{Kewley} et~al.}{2001}]{Kewley+01}
{Kewley} L.~J.,  {Dopita} M.~A.,  {Sutherland} R.~S.,  {Heisler} C.~A.,
  {Trevena} J.,  2001, \mn@doi [\apj] {10.1086/321545}, \href
  {http://adsabs.harvard.edu/abs/2001ApJ...556..121K} {556, 121}

\bibitem[\protect\citeauthoryear{{Knobel}, {Lilly}, {Woo}  \&
  {Kova{\v{c}}}}{{Knobel} et~al.}{2015}]{Knobel.etal:2015}
{Knobel} C.,  {Lilly} S.~J.,  {Woo} J.,   {Kova{\v{c}}} K.,  2015, \mn@doi
  [\apj] {10.1088/0004-637X/800/1/24}, \href
  {https://ui.adsabs.harvard.edu/abs/2015ApJ...800...24K} {800, 24}

\bibitem[\protect\citeauthoryear{{Kobayashi}}{{Kobayashi}}{2004}]{Kobayashi:2004}
{Kobayashi} C.,  2004, \mn@doi [MNRAS] {10.1111/j.1365-2966.2004.07258.x},
  \href {http://adsabs.harvard.edu/abs/2004MNRAS.347..740K} {347, 740}

\bibitem[\protect\citeauthoryear{{Komatsu} et~al.,}{{Komatsu}
  et~al.}{2011}]{Komatsu.etal:2011}
{Komatsu} E.,  et~al., 2011, \mn@doi [ApJS] {10.1088/0067-0049/192/2/18}, \href
  {http://adsabs.harvard.edu/abs/2011ApJS..192...18K} {192, 18}

\bibitem[\protect\citeauthoryear{{Kreckel} et~al.,}{{Kreckel}
  et~al.}{2013}]{Kreckel+2013}
{Kreckel} K.,  et~al., 2013, \mn@doi [\apj] {10.1088/0004-637X/771/1/62}, \href
  {https://ui.adsabs.harvard.edu/abs/2013ApJ...771...62K} {771, 62}

\bibitem[\protect\citeauthoryear{{Kroupa}}{{Kroupa}}{2001}]{Kroupa01}
{Kroupa} P.,  2001, \mn@doi [\mnras] {10.1046/j.1365-8711.2001.04022.x}, \href
  {http://cdsads.u-strasbg.fr/abs/2001MNRAS.322..231K} {322, 231}

\bibitem[\protect\citeauthoryear{{La Barbera}, {De Carvalho}, {De La Rosa},
  {Gal}, {Swindle}  \& {Lopes}}{{La Barbera}
  et~al.}{2010}]{LaBarbera.etal:2010}
{La Barbera} F.,  {De Carvalho} R.~R.,  {De La Rosa} I.~G.,  {Gal} R.~R.,
  {Swindle} R.,   {Lopes} P.~A.~A.,  2010, \mn@doi [AJ]
  {10.1088/0004-6256/140/5/1528}, \href
  {http://adsabs.harvard.edu/abs/2010AJ....140.1528L} {140, 1528}

\bibitem[\protect\citeauthoryear{{La Barbera}, {Ferreras}, {de Carvalho},
  {Bruzual}, {Charlot}, {Pasquali}  \& {Merlin}}{{La Barbera}
  et~al.}{2012}]{LaBarbera.etal:2012}
{La Barbera} F.,  {Ferreras} I.,  {de Carvalho} R.~R.,  {Bruzual} G.,
  {Charlot} S.,  {Pasquali} A.,   {Merlin} E.,  2012, \mn@doi [MNRAS]
  {10.1111/j.1365-2966.2012.21848.x}, \href
  {http://adsabs.harvard.edu/abs/2012MNRAS.426.2300L} {426, 2300}

\bibitem[\protect\citeauthoryear{{Le Borgne} et~al.,}{{Le Borgne}
  et~al.}{2003}]{LeBorgne+03}
{Le Borgne} J.-F.,  et~al., 2003, \mn@doi [\aap] {10.1051/0004-6361:20030243},
  \href {http://cdsads.u-strasbg.fr/abs/2003A%26A...402..433L} {402, 433}

\bibitem[\protect\citeauthoryear{{Lintott} et~al.,}{{Lintott}
  et~al.}{2011}]{Lintott.etal:2011}
{Lintott} C.,  et~al., 2011, \mn@doi [MNRAS]
  {10.1111/j.1365-2966.2010.17432.x}, \href
  {http://adsabs.harvard.edu/abs/2011MNRAS.410..166L} {410, 166}

\bibitem[\protect\citeauthoryear{{Lotz}, {Primack}  \& {Madau}}{{Lotz}
  et~al.}{2004}]{Lotz+04}
{Lotz} J.~M.,  {Primack} J.,   {Madau} P.,  2004, \mn@doi [\aj]
  {10.1086/421849}, \href {http://cdsads.u-strasbg.fr/abs/2004AJ....128..163L}
  {128, 163}

\bibitem[\protect\citeauthoryear{{Madau} \& {Dickinson}}{{Madau} \&
  {Dickinson}}{2014}]{Madau.Dickinson:2014}
{Madau} P.,  {Dickinson} M.,  2014, \mn@doi [\araa]
  {10.1146/annurev-astro-081811-125615}, \href
  {https://ui.adsabs.harvard.edu/abs/2014ARA&A..52..415M} {52, 415}

\bibitem[\protect\citeauthoryear{{Mahajan}, {Mamon}  \&
  {Raychaudhury}}{{Mahajan} et~al.}{2011}]{MMR11}
{Mahajan} S.,  {Mamon} G.~A.,   {Raychaudhury} S.,  2011, \mn@doi [\mnras]
  {10.1111/j.1365-2966.2011.19236.x}, 416, 2882

\bibitem[\protect\citeauthoryear{Mahalanobis}{Mahalanobis}{1936}]{Mahalanobis36}
Mahalanobis P.~C.,  1936, Proceedings of the National Institute of Sciences
  (Calcutta), 2, 49

\bibitem[\protect\citeauthoryear{{Mamon}}{{Mamon}}{1992}]{Mamon92}
{Mamon} G.~A.,  1992, \mn@doi [\apjl] {10.1086/186656}, \href
  {http://adsabs.harvard.edu/abs/1992ApJ...401L...3M} {401, L3}

\bibitem[\protect\citeauthoryear{{Mamon}}{{Mamon}}{2000}]{Mamon00_IAP}
{Mamon} G.~A.,  2000, in {Combes} F.,  {Mamon} G.~A.,   {Charmandaris} V.,
  eds,  Astronomical Society of the Pacific Conference Series Vol. 197,
  Dynamics of Galaxies: from the Early Universe to the Present. p.~377
  (\mn@eprint {arXiv} {astro-ph/9911333})

\bibitem[\protect\citeauthoryear{{Mamon}, {Trevisan}, {Thuan}, {Gallazzi}  \&
  {Dav{\'e}}}{{Mamon} et~al.}{2020}]{Mamon+20}
{Mamon} G.~A.,  {Trevisan} M.,  {Thuan} T.~X.,  {Gallazzi} A.,   {Dav{\'e}} R.,
   2020, \mn@doi [\mnras] {10.1093/mnras/stz3556}, \href
  {https://ui.adsabs.harvard.edu/abs/2020MNRAS.492.1791M} {492, 1791
  (Paper~II)}

\bibitem[\protect\citeauthoryear{{Maraston}}{{Maraston}}{2005}]{Maraston05}
{Maraston} C.,  2005, \mn@doi [\mnras] {10.1111/j.1365-2966.2005.09270.x},
  \href {http://cdsads.u-strasbg.fr/abs/2005MNRAS.362..799M} {362, 799}

\bibitem[\protect\citeauthoryear{{Maraston} \& {Str{\"o}mb{\"a}ck}}{{Maraston}
  \& {Str{\"o}mb{\"a}ck}}{2011}]{Maraston2011}
{Maraston} C.,  {Str{\"o}mb{\"a}ck} G.,  2011, \mn@doi [\mnras]
  {10.1111/j.1365-2966.2011.19738.x}, \href
  {http://adsabs.harvard.edu/abs/2011MNRAS.418.2785M} {418, 2785}

\bibitem[\protect\citeauthoryear{{Marino} et~al.,}{{Marino}
  et~al.}{2013}]{Marino2013}
{Marino} R.~A.,  et~al., 2013, \mn@doi [\aap] {10.1051/0004-6361/201321956},
  \href {http://adsabs.harvard.edu/abs/2013A%26A...559A.114M} {559, A114}

\bibitem[\protect\citeauthoryear{{McGee}, {Balogh}, {Wilman}, {Bower},
  {Mulchaey}, {Parker}  \& {Oemler}}{{McGee} et~al.}{2011}]{McGee.etal:2011}
{McGee} S.~L.,  {Balogh} M.~L.,  {Wilman} D.~J.,  {Bower} R.~G.,  {Mulchaey}
  J.~S.,  {Parker} L.~C.,   {Oemler} A.,  2011, \mn@doi [MNRAS]
  {10.1111/j.1365-2966.2010.18189.x}, 413, 996

\bibitem[\protect\citeauthoryear{{Momcheva}, {Lee}, {Ly}, {Salim}, {Dale},
  {Ouchi}, {Finn}  \& {Ono}}{{Momcheva} et~al.}{2013}]{Momcheva+2013}
{Momcheva} I.~G.,  {Lee} J.~C.,  {Ly} C.,  {Salim} S.,  {Dale} D.~A.,  {Ouchi}
  M.,  {Finn} R.,   {Ono} Y.,  2013, \mn@doi [\aj]
  {10.1088/0004-6256/145/2/47}, \href
  {https://ui.adsabs.harvard.edu/abs/2013AJ....145...47M} {145, 47}

\bibitem[\protect\citeauthoryear{{Moster}, {Naab}  \& {White}}{{Moster}
  et~al.}{2013}]{Moster+13}
{Moster} B.~P.,  {Naab} T.,   {White} S.~D.~M.,  2013, \mn@doi [\mnras]
  {10.1093/mnras/sts261}, \href
  {http://adsabs.harvard.edu/abs/2013MNRAS.428.3121M} {428, 3121}

\bibitem[\protect\citeauthoryear{{Mutch}, {Croton}  \& {Poole}}{{Mutch}
  et~al.}{2013}]{Mutch+13}
{Mutch} S.~J.,  {Croton} D.~J.,   {Poole} G.~B.,  2013, \mn@doi [\mnras]
  {10.1093/mnras/stt1453}, \href
  {http://cdsads.u-strasbg.fr/abs/2013MNRAS.435.2445M} {435, 2445}

\bibitem[\protect\citeauthoryear{{Nair} \& {Abraham}}{{Nair} \&
  {Abraham}}{2010}]{Nair.Abraham:2010}
{Nair} P.~B.,  {Abraham} R.~G.,  2010, \mn@doi [\apjs]
  {10.1088/0067-0049/186/2/427}, \href
  {https://ui.adsabs.harvard.edu/abs/2010ApJS..186..427N} {186, 427}

\bibitem[\protect\citeauthoryear{{Pagel}, {Edmunds}, {Blackwell}, {Chun}  \&
  {Smith}}{{Pagel} et~al.}{1979}]{Pagel1979}
{Pagel} B.~E.~J.,  {Edmunds} M.~G.,  {Blackwell} D.~E.,  {Chun} M.~S.,
  {Smith} G.,  1979, \mn@doi [\mnras] {10.1093/mnras/189.1.95}, \href
  {http://adsabs.harvard.edu/abs/1979MNRAS.189...95P} {189, 95}

\bibitem[\protect\citeauthoryear{{Pearson} et~al.,}{{Pearson}
  et~al.}{2019}]{Pearson+19}
{Pearson} W.~J.,  et~al., 2019, \mn@doi [\aap] {10.1051/0004-6361/201936337},
  \href {https://ui.adsabs.harvard.edu/abs/2019A&A...631A..51P} {631, A51}

\bibitem[\protect\citeauthoryear{{Peng} et~al.,}{{Peng} et~al.}{2010}]{Peng+10}
{Peng} Y.-j.,  et~al., 2010, \mn@doi [\apj] {10.1088/0004-637X/721/1/193}, 721,
  193

\bibitem[\protect\citeauthoryear{{Pettini} \& {Pagel}}{{Pettini} \&
  {Pagel}}{2004}]{Pettini2004}
{Pettini} M.,  {Pagel} B.~E.~J.,  2004, \mn@doi [\mnras]
  {10.1111/j.1365-2966.2004.07591.x}, \href
  {http://adsabs.harvard.edu/abs/2004MNRAS.348L..59P} {348, L59}

\bibitem[\protect\citeauthoryear{{Pietrinferni}, {Cassisi}, {Salaris}  \&
  {Castelli}}{{Pietrinferni} et~al.}{2004}]{Pietrinferni+04}
{Pietrinferni} A.,  {Cassisi} S.,  {Salaris} M.,   {Castelli} F.,  2004,
  \mn@doi [\apj] {10.1086/422498}, \href
  {http://cdsads.u-strasbg.fr/abs/2004ApJ...612..168P} {612, 168}

\bibitem[\protect\citeauthoryear{{Pietrinferni}, {Cassisi}, {Salaris}  \&
  {Castelli}}{{Pietrinferni} et~al.}{2006}]{Pietrinferni+06}
{Pietrinferni} A.,  {Cassisi} S.,  {Salaris} M.,   {Castelli} F.,  2006,
  \mn@doi [\apj] {10.1086/501344}, \href
  {http://cdsads.u-strasbg.fr/abs/2006ApJ...642..797P} {642, 797}

\bibitem[\protect\citeauthoryear{{Pilyugin}}{{Pilyugin}}{2000}]{Pilyugin2000}
{Pilyugin} L.~S.,  2000, \aap, \href
  {http://adsabs.harvard.edu/abs/2000A%26A...362..325P} {362, 325}

\bibitem[\protect\citeauthoryear{{Pilyugin}}{{Pilyugin}}{2001}]{Pilyugin2001}
{Pilyugin} L.~S.,  2001, \mn@doi [\aap] {10.1051/0004-6361:20010079}, \href
  {http://adsabs.harvard.edu/abs/2001A%26A...369..594P} {369, 594}

\bibitem[\protect\citeauthoryear{{Pilyugin} \& {Grebel}}{{Pilyugin} \&
  {Grebel}}{2016}]{Pilyugin2016}
{Pilyugin} L.~S.,  {Grebel} E.~K.,  2016, \mn@doi [\mnras]
  {10.1093/mnras/stw238}, \href
  {http://adsabs.harvard.edu/abs/2016MNRAS.457.3678P} {457, 3678}

\bibitem[\protect\citeauthoryear{{Pilyugin} \& {Thuan}}{{Pilyugin} \&
  {Thuan}}{2005}]{Pilyugin2005}
{Pilyugin} L.~S.,  {Thuan} T.~X.,  2005, \mn@doi [\apj] {10.1086/432408}, \href
  {http://adsabs.harvard.edu/abs/2005ApJ...631..231P} {631, 231}

\bibitem[\protect\citeauthoryear{{Pilyugin}, {Grebel}, {Zinchenko}, {Nefedyev},
  {Shulga}, {Wei}  \& {Berczik}}{{Pilyugin} et~al.}{2018}]{Pilyugin2018}
{Pilyugin} L.~S.,  {Grebel} E.~K.,  {Zinchenko} I.~A.,  {Nefedyev} Y.~A.,
  {Shulga} V.~M.,  {Wei} H.,   {Berczik} P.~P.,  2018, \mn@doi [\aap]
  {10.1051/0004-6361/201732185}, \href
  {https://ui.adsabs.harvard.edu/abs/2018A&A...613A...1P} {613, A1}

\bibitem[\protect\citeauthoryear{{R Core Team}}{{R Core Team}}{2015}]{R:2015}
{R Core Team} 2015, R: A Language and Environment for Statistical Computing.
R Foundation for Statistical Computing, Vienna, Austria, \url
  {https://www.R-project.org}

\bibitem[\protect\citeauthoryear{{Renzini}}{{Renzini}}{2020}]{Renzini20}
{Renzini} A.,  2020, \mn@doi [\mnras] {10.1093/mnrasl/slaa054}, \href
  {https://ui.adsabs.harvard.edu/abs/2020MNRAS.495L..42R} {495, L42}

\bibitem[\protect\citeauthoryear{{Rosenbaum} \& {Rubin}}{{Rosenbaum} \&
  {Rubin}}{1983}]{PSM:1983}
{Rosenbaum} P.~R.,  {Rubin} D.~B.,  1983, \mn@doi [Biometrika]
  {10.1093/biomet/70.1.41}, 70, 41

\bibitem[\protect\citeauthoryear{{Saintonge} et~al.,}{{Saintonge}
  et~al.}{2011}]{Saintonge+11}
{Saintonge} A.,  et~al., 2011, \mn@doi [\mnras]
  {10.1111/j.1365-2966.2011.18823.x}, \href
  {https://ui.adsabs.harvard.edu/abs/2011MNRAS.415...61S} {415, 61}

\bibitem[\protect\citeauthoryear{{Saintonge} et~al.,}{{Saintonge}
  et~al.}{2016}]{Saintonge+16}
{Saintonge} A.,  et~al., 2016, \mn@doi [\mnras] {10.1093/mnras/stw1715}, \href
  {https://ui.adsabs.harvard.edu/abs/2016MNRAS.462.1749S} {462, 1749}

\bibitem[\protect\citeauthoryear{{S{\'a}nchez-Bl{\'a}zquez}
  et~al.,}{{S{\'a}nchez-Bl{\'a}zquez} et~al.}{2006}]{Sanchez-Blazquez+06}
{S{\'a}nchez-Bl{\'a}zquez} P.,  et~al., 2006, \mn@doi [\mnras]
  {10.1111/j.1365-2966.2006.10699.x}, \href
  {http://cdsads.u-strasbg.fr/abs/2006MNRAS.371..703S} {371, 703}

\bibitem[\protect\citeauthoryear{{Sanders}, {Shapley}, {Zhang}  \&
  {Yan}}{{Sanders} et~al.}{2017}]{Sanders2017}
{Sanders} R.~L.,  {Shapley} A.~E.,  {Zhang} K.,   {Yan} R.,  2017, \mn@doi
  [\apj] {10.3847/1538-4357/aa93e4}, \href
  {http://adsabs.harvard.edu/abs/2017ApJ...850..136S} {850, 136}

\bibitem[\protect\citeauthoryear{{Savage} \& {Mathis}}{{Savage} \&
  {Mathis}}{1979}]{Savage&Mathis79}
{Savage} B.~D.,  {Mathis} J.~S.,  1979, \mn@doi [\araa]
  {10.1146/annurev.aa.17.090179.000445}, \href
  {https://ui.adsabs.harvard.edu/abs/1979ARA&A..17...73S} {17, 73}

\bibitem[\protect\citeauthoryear{{Skillman} \& {Kennicutt}}{{Skillman} \&
  {Kennicutt}}{1993}]{Skillman.Kennicutt:1993}
{Skillman} E.~D.,  {Kennicutt} Jr. R.~C.,  1993, \mn@doi [\apj]
  {10.1086/172868}, \href
  {https://ui.adsabs.harvard.edu/abs/1993ApJ...411..655S} {411, 655}

\bibitem[\protect\citeauthoryear{{Springob}, {Haynes}, {Giovanelli}  \&
  {Kent}}{{Springob} et~al.}{2005}]{Springob.etal:2005}
{Springob} C.~M.,  {Haynes} M.~P.,  {Giovanelli} R.,   {Kent} B.~R.,  2005,
  \mn@doi [\apjs] {10.1086/431550}, \href
  {https://ui.adsabs.harvard.edu/abs/2005ApJS..160..149S} {160, 149}

\bibitem[\protect\citeauthoryear{{Storey} \& {Zeippen}}{{Storey} \&
  {Zeippen}}{2000}]{Storey2000}
{Storey} P.~J.,  {Zeippen} C.~J.,  2000, \mn@doi [\mnras]
  {10.1046/j.1365-8711.2000.03184.x}, \href
  {http://adsabs.harvard.edu/abs/2000MNRAS.312..813S} {312, 813}

\bibitem[\protect\citeauthoryear{{Sutter}, {Lavaux}, {Wandelt}  \&
  {Weinberg}}{{Sutter} et~al.}{2012}]{Sutter.etal:2012}
{Sutter} P.~M.,  {Lavaux} G.,  {Wandelt} B.~D.,   {Weinberg} D.~H.,  2012,
  \mn@doi [\apj] {10.1088/0004-637X/761/1/44}, \href
  {https://ui.adsabs.harvard.edu/abs/2012ApJ...761...44S} {761, 44}

\bibitem[\protect\citeauthoryear{{Swanson}, {Tegmark}, {Hamilton}  \&
  {Hill}}{{Swanson} et~al.}{2008}]{Swanson.etal:2008}
{Swanson} M.~E.~C.,  {Tegmark} M.,  {Hamilton} A.~J.~S.,   {Hill} J.~C.,  2008,
  \mn@doi [\mnras] {10.1111/j.1365-2966.2008.13296.x}, \href
  {http://adsabs.harvard.edu/abs/2008MNRAS.387.1391S} {387, 1391}

\bibitem[\protect\citeauthoryear{{Taylor}}{{Taylor}}{2005}]{Taylor:2005}
{Taylor} M.~B.,  2005, in {Shopbell} P.,  {Britton} M.,   {Ebert} R.,  eds,
  Astronomical Society of the Pacific Conference Series Vol. 347, Astronomical
  Data Analysis Software and Systems XIV. p.~29

\bibitem[\protect\citeauthoryear{{Tempel}, {Stoica}, {Mart{\'\i}nez},
  {Liivam{\"a}gi}, {Castellan}  \& {Saar}}{{Tempel}
  et~al.}{2014}]{Tempel.etal:2014}
{Tempel} E.,  {Stoica} R.~S.,  {Mart{\'\i}nez} V.~J.,  {Liivam{\"a}gi} L.~J.,
  {Castellan} G.,   {Saar} E.,  2014, \mn@doi [\mnras] {10.1093/mnras/stt2454},
  \href {https://ui.adsabs.harvard.edu/abs/2014MNRAS.438.3465T} {438, 3465}

\bibitem[\protect\citeauthoryear{{Thomas}, {Maraston}, {Schawinski}, {Sarzi}
  \& {Silk}}{{Thomas} et~al.}{2010}]{Thomas+10}
{Thomas} D.,  {Maraston} C.,  {Schawinski} K.,  {Sarzi} M.,   {Silk} J.,  2010,
  \mn@doi [\mnras] {10.1111/j.1365-2966.2010.16427.x}, 404, 1775

\bibitem[\protect\citeauthoryear{{Thomas}, {Maraston}  \& {Johansson}}{{Thomas}
  et~al.}{2011}]{Thomas2011}
{Thomas} D.,  {Maraston} C.,   {Johansson} J.,  2011, \mn@doi [\mnras]
  {10.1111/j.1365-2966.2010.18049.x}, \href
  {http://adsabs.harvard.edu/abs/2011MNRAS.412.2183T} {412, 2183}

\bibitem[\protect\citeauthoryear{{Thuan}, {Goehring}, {Hibbard}, {Izotov}  \&
  {Hunt}}{{Thuan} et~al.}{2016}]{Thuan+16}
{Thuan} T.~X.,  {Goehring} K.~M.,  {Hibbard} J.~E.,  {Izotov} Y.~I.,   {Hunt}
  L.~K.,  2016, \mn@doi [\mnras] {10.1093/mnras/stw2259}, \href
  {https://ui.adsabs.harvard.edu/abs/2016MNRAS.463.4268T} {463, 4268}

\bibitem[\protect\citeauthoryear{{Tiwari}, {Mahajan}  \& {Singh}}{{Tiwari}
  et~al.}{2020}]{Tiwari+20}
{Tiwari} J.,  {Mahajan} S.,   {Singh} K.~P.,  2020, \mn@doi [\na]
  {10.1016/j.newast.2020.101417}, \href
  {https://ui.adsabs.harvard.edu/abs/2020NewA...8101417T} {81, 101417}

\bibitem[\protect\citeauthoryear{{Tojeiro}, {Wilkins}, {Heavens}, {Panter}  \&
  {Jimenez}}{{Tojeiro} et~al.}{2009}]{Tojeiro+09}
{Tojeiro} R.,  {Wilkins} S.,  {Heavens} A.~F.,  {Panter} B.,   {Jimenez} R.,
  2009, \mn@doi [\apjs] {10.1088/0067-0049/185/1/1}, \href
  {http://cdsads.u-strasbg.fr/abs/2009ApJS..185....1T} {185, 1}

\bibitem[\protect\citeauthoryear{{Tremonti} et~al.,}{{Tremonti}
  et~al.}{2004}]{Tremonti+04}
{Tremonti} C.~A.,  et~al., 2004, \mn@doi [\apj] {10.1086/423264}, \href
  {http://adsabs.harvard.edu/abs/2004ApJ...613..898T} {613, 898}

\bibitem[\protect\citeauthoryear{{Trevisan}, {Ferreras}, {de La Rosa}, {La
  Barbera}  \& {de Carvalho}}{{Trevisan} et~al.}{2012}]{Trevisan.etal:2012}
{Trevisan} M.,  {Ferreras} I.,  {de La Rosa} I.~G.,  {La Barbera} F.,   {de
  Carvalho} R.~R.,  2012, \mn@doi [ApJ] {10.1088/2041-8205/752/2/L27}, \href
  {http://adsabs.harvard.edu/abs/2012ApJ...752L..27T} {752, L27}

\bibitem[\protect\citeauthoryear{{Trevisan}, {Mamon}  \&
  {Khosroshahi}}{{Trevisan} et~al.}{2017a}]{Trevisan+17}
{Trevisan} M.,  {Mamon} G.~A.,   {Khosroshahi} H.~G.,  2017a, \mn@doi [\mnras]
  {10.1093/mnras/stw2588}, \href
  {http://adsabs.harvard.edu/abs/2017MNRAS.464.4593T} {464, 4593}

\bibitem[\protect\citeauthoryear{{Trevisan}, {Mamon}  \& {Stalder}}{{Trevisan}
  et~al.}{2017b}]{Trevisan+17_Let}
{Trevisan} M.,  {Mamon} G.~A.,   {Stalder} D.~H.,  2017b, \mn@doi [\mnras]
  {10.1093/mnrasl/slx092}, \href
  {http://adsabs.harvard.edu/abs/2017MNRAS.471L..47T} {471, L47}

\bibitem[\protect\citeauthoryear{{Tweed}, {Mamon}, {Thuan}, {Cattaneo},
  {Dekel}, {Menci}, {Calura}  \& {Silk}}{{Tweed} et~al.}{2018}]{Tweed+18}
{Tweed} D.~P.,  {Mamon} G.~A.,  {Thuan} T.~X.,  {Cattaneo} A.,  {Dekel} A.,
  {Menci} N.,  {Calura} F.,   {Silk} J.,  2018, \mn@doi [\mnras]
  {10.1093/mnras/sty507}, \href
  {http://adsabs.harvard.edu/abs/2018MNRAS.477.1427T} {477, 1427 (Paper~I)}

\bibitem[\protect\citeauthoryear{{Vazdekis}, {S{\'a}nchez-Bl{\'a}zquez},
  {Falc{\'o}n-Barroso}, {Cenarro}, {Beasley}, {Cardiel}, {Gorgas}  \&
  {Peletier}}{{Vazdekis} et~al.}{2010}]{Vazdekis+10}
{Vazdekis} A.,  {S{\'a}nchez-Bl{\'a}zquez} P.,  {Falc{\'o}n-Barroso} J.,
  {Cenarro} A.~J.,  {Beasley} M.~A.,  {Cardiel} N.,  {Gorgas} J.,   {Peletier}
  R.~F.,  2010, \mn@doi [\mnras] {10.1111/j.1365-2966.2010.16407.x}, \href
  {http://cdsads.u-strasbg.fr/abs/2010MNRAS.404.1639V} {404, 1639}

\bibitem[\protect\citeauthoryear{{Vazdekis} et~al.,}{{Vazdekis}
  et~al.}{2015}]{Vazdekis+15}
{Vazdekis} A.,  et~al., 2015, \mn@doi [\mnras] {10.1093/mnras/stv151}, \href
  {http://cdsads.u-strasbg.fr/abs/2015MNRAS.449.1177V} {449, 1177}

\bibitem[\protect\citeauthoryear{{Weinmann}, {van den Bosch}, {Yang}  \&
  {Mo}}{{Weinmann} et~al.}{2006}]{Weinmann+06}
{Weinmann} S.~M.,  {van den Bosch} F.~C.,  {Yang} X.,   {Mo} H.~J.,  2006,
  \mn@doi [\mnras] {10.1111/j.1365-2966.2005.09865.x}, \href
  {http://adsabs.harvard.edu/abs/2006MNRAS.366....2W} {366, 2}

\bibitem[\protect\citeauthoryear{{Willett} et~al.,}{{Willett}
  et~al.}{2013}]{Willett.etal:2013}
{Willett} K.~W.,  et~al., 2013, \mn@doi [\mnras] {10.1093/mnras/stt1458}, \href
  {https://ui.adsabs.harvard.edu/abs/2013MNRAS.435.2835W} {435, 2835}

\bibitem[\protect\citeauthoryear{{Woo} et~al.,}{{Woo} et~al.}{2013}]{Woo+13}
{Woo} J.,  et~al., 2013, \mn@doi [\mnras] {10.1093/mnras/sts274}, \href
  {http://adsabs.harvard.edu/abs/2013MNRAS.428.3306W} {428, 3306}

\bibitem[\protect\citeauthoryear{{Wright} et~al.,}{{Wright}
  et~al.}{2010}]{WISE10}
{Wright} E.~L.,  et~al., 2010, \mn@doi [\aj] {10.1088/0004-6256/140/6/1868},
  \href {https://ui.adsabs.harvard.edu/abs/2010AJ....140.1868W} {140, 1868}

\bibitem[\protect\citeauthoryear{{Yang}, {Mo}, {van den Bosch}  \&
  {Jing}}{{Yang} et~al.}{2005}]{Yang+05}
{Yang} X.,  {Mo} H.~J.,  {van den Bosch} F.~C.,   {Jing} Y.~P.,  2005, \mn@doi
  [\mnras] {10.1111/j.1365-2966.2005.08560.x}, \href
  {https://ui.adsabs.harvard.edu/abs/2005MNRAS.356.1293Y} {356, 1293}

\bibitem[\protect\citeauthoryear{{Yang}, {Mo}, {van den Bosch}, {Pasquali},
  {Li}  \& {Barden}}{{Yang} et~al.}{2007}]{Yang+07}
{Yang} X.,  {Mo} H.~J.,  {van den Bosch} F.~C.,  {Pasquali} A.,  {Li} C.,
  {Barden} M.,  2007, \mn@doi [\apj] {10.1086/522027}, \href
  {http://adsabs.harvard.edu/abs/2007ApJ...671..153Y} {671, 153}

\bibitem[\protect\citeauthoryear{{Zaritsky}, {Kennicutt}  \&
  {Huchra}}{{Zaritsky} et~al.}{1994}]{Zaritsky1994}
{Zaritsky} D.,  {Kennicutt} Jr. R.~C.,   {Huchra} J.~P.,  1994, \mn@doi [\apj]
  {10.1086/173544}, \href {http://adsabs.harvard.edu/abs/1994ApJ...420...87Z}
  {420, 87}

\bibitem[\protect\citeauthoryear{{Zhang} et~al.,}{{Zhang}
  et~al.}{2017}]{Zhang2017}
{Zhang} K.,  et~al., 2017, \mn@doi [\mnras] {10.1093/mnras/stw3308}, \href
  {http://adsabs.harvard.edu/abs/2017MNRAS.466.3217Z} {466, 3217}

\bibitem[\protect\citeauthoryear{{Zhang} et~al.,}{{Zhang}
  et~al.}{2019}]{Zhang.etal:2019}
{Zhang} C.,  et~al., 2019, \mn@doi [\apjl] {10.3847/2041-8213/ab4ae4}, \href
  {https://ui.adsabs.harvard.edu/abs/2019ApJ...884L..52Z} {884, L52}

\bibitem[\protect\citeauthoryear{{Zhuang}, {Leaman}, {van de Ven}, {Zibetti},
  {Gallazzi}, {Zhu}, {Falc{\'o}n-Barroso}  \& {Lyubenova}}{{Zhuang}
  et~al.}{2019}]{Zhuang+19}
{Zhuang} Y.,  {Leaman} R.,  {van de Ven} G.,  {Zibetti} S.,  {Gallazzi} A.,
  {Zhu} L.,  {Falc{\'o}n-Barroso} J.,   {Lyubenova} M.,  2019, \mn@doi [\mnras]
  {10.1093/mnras/sty2916}, \href
  {https://ui.adsabs.harvard.edu/abs/2019MNRAS.483.1862Z} {483, 1862}

\bibitem[\protect\citeauthoryear{{Zibetti}, {Gallazzi}, {Hirschmann},
  {Consolandi}, {Falc{\'o}n-Barroso}, {van de Ven}  \& {Lyubenova}}{{Zibetti}
  et~al.}{2020}]{Zibetti+20}
{Zibetti} S.,  {Gallazzi} A.~R.,  {Hirschmann} M.,  {Consolandi} G.,
  {Falc{\'o}n-Barroso} J.,  {van de Ven} G.,   {Lyubenova} M.,  2020, \mn@doi
  [\mnras] {10.1093/mnras/stz3205}, \href
  {https://ui.adsabs.harvard.edu/abs/2020MNRAS.491.3562Z} {491, 3562}

\bibitem[\protect\citeauthoryear{{da Cunha}, {Eminian}, {Charlot}  \&
  {Blaizot}}{{da Cunha} et~al.}{2010}]{daCunha+2010}
{da Cunha} E.,  {Eminian} C.,  {Charlot} S.,   {Blaizot} J.,  2010, \mn@doi
  [\mnras] {10.1111/j.1365-2966.2010.16344.x}, \href
  {https://ui.adsabs.harvard.edu/abs/2010MNRAS.403.1894D} {403, 1894}

\bibitem[\protect\citeauthoryear{{de Souza} et~al.,}{{de Souza}
  et~al.}{2016}]{deSouza+16}
{de Souza} R.~S.,  et~al., 2016, \mn@doi [\mnras] {10.1093/mnras/stw1459},
  \href {https://ui.adsabs.harvard.edu/abs/2016MNRAS.461.2115D} {461, 2115}

\bibitem[\protect\citeauthoryear{{de Vaucouleurs}}{{de
  Vaucouleurs}}{1963}]{deVaucouleurs:1963}
{de Vaucouleurs} G.,  1963, \mn@doi [\apjs] {10.1086/190084}, \href
  {https://ui.adsabs.harvard.edu/abs/1963ApJS....8...31D} {8, 31}

\bibitem[\protect\citeauthoryear{{van Zee}, {Westpfahl}, {Haynes}  \&
  {Salzer}}{{van Zee} et~al.}{1998}]{vanZee+98}
{van Zee} L.,  {Westpfahl} D.,  {Haynes} M.~P.,   {Salzer} J.~J.,  1998,
  \mn@doi [\aj] {10.1086/300251}, \href
  {http://cdsads.u-strasbg.fr/abs/1998AJ....115.1000V} {115, 1000}

\bibitem[\protect\citeauthoryear{{von der Linden}, {Wild}, {Kauffmann}, {White}
   \& {Weinmann}}{{von der Linden} et~al.}{2010}]{vonderLinden+10}
{von der Linden} A.,  {Wild} V.,  {Kauffmann} G.,  {White} S.~D.~M.,
  {Weinmann} S.,  2010, \mn@doi [\mnras] {10.1111/j.1365-2966.2010.16375.x},
  \href {http://adsabs.harvard.edu/abs/2010MNRAS.404.1231V} {404, 1231}

\makeatother
\end{thebibliography}

\appendix

\section{Determination of the Red Sequence at different redshifts}
\label{ap:red_seq}

%
\begin{figure}
\centering
\begin{tabular}{c}
 \resizebox{0.95\hsize}{!}{\includegraphics{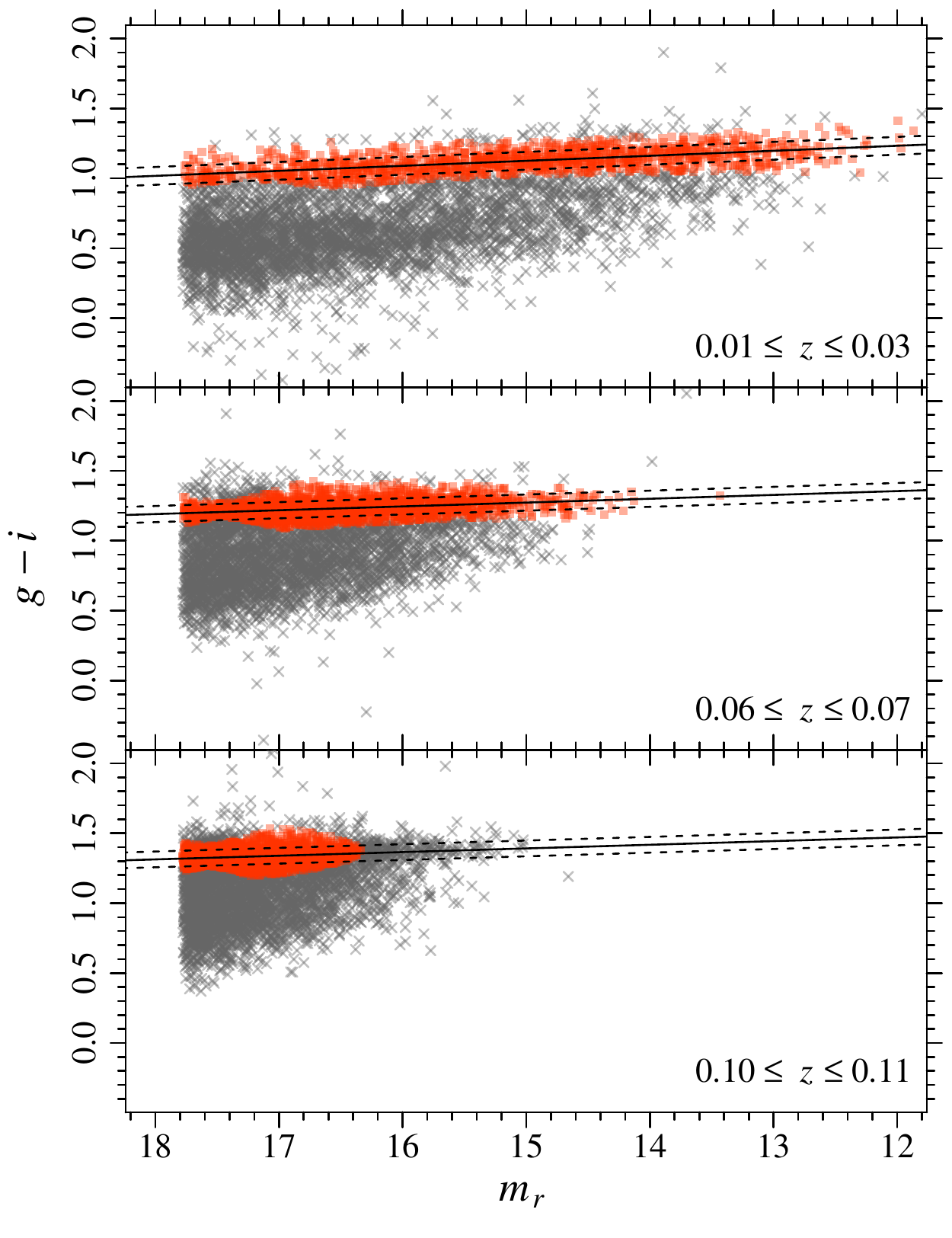}} 
\end{tabular}
\caption {Identification of  Red Sequence galaxies in the colour-magnitude
diagram for three different bins of redshifts, using {\tt mclust} (see
text).
}
\label{Fig_rs_zbins}
\end{figure}

%
\begin{figure}
\centering
\begin{tabular}{c}
 \resizebox{0.95\hsize}{!}{\includegraphics{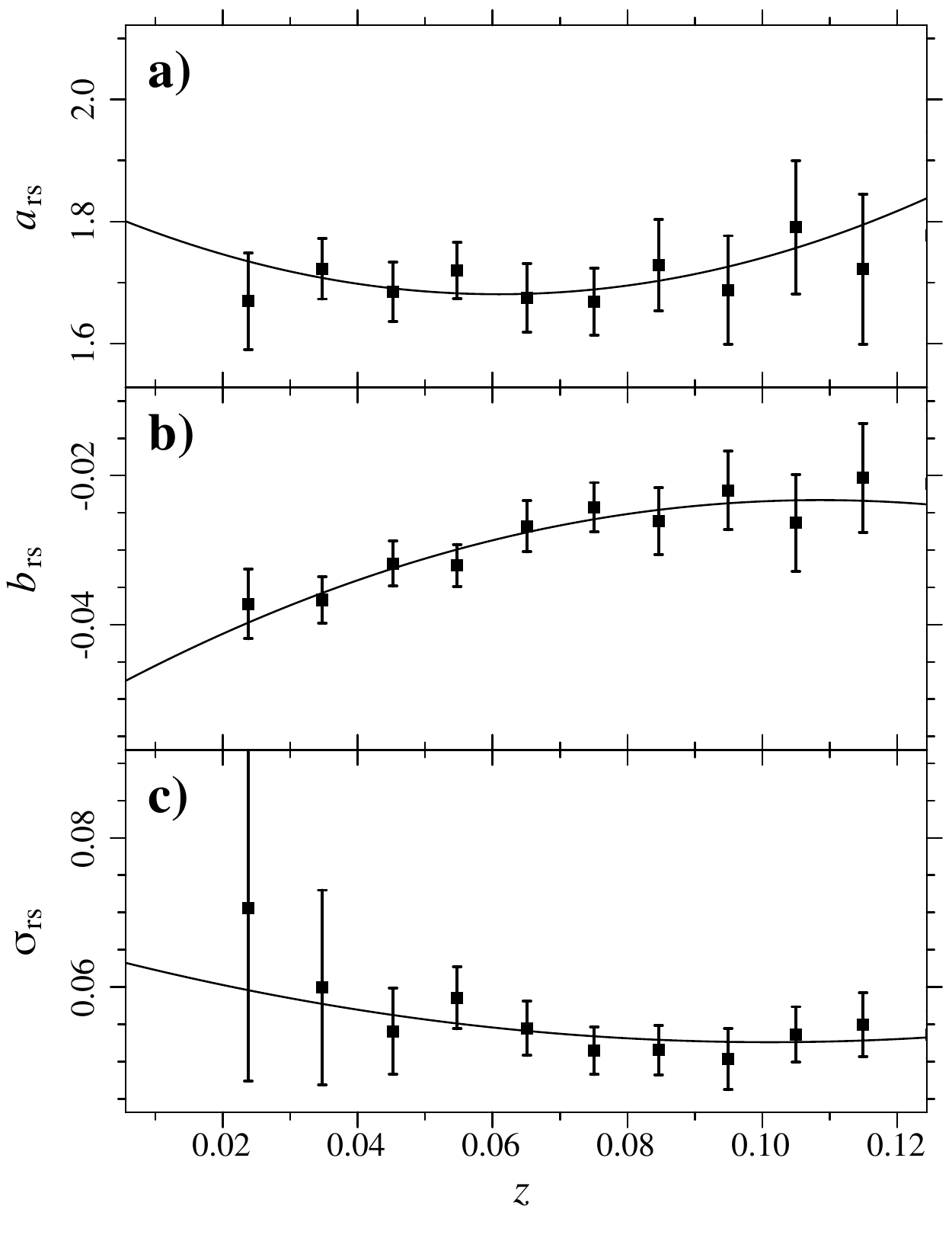}} \\
\end{tabular}
\caption {Parameters of the fit to the Red Sequence galaxies at different
redshifts, where $(g-i)(m_r, z) = a_{\rm RS}(z) + b_{\rm RS}(z) \, m_r$ with
scatter $\sigma_{\rm RS}(z)$.}
\label{Fig_fit_rs_coeffs}

\end{figure}

As described in Sect.~\ref{sec:environment}, we relied on the photometric
SDSS data to identify neighbouring galaxies and determine the local
environment of our galaxies.  This approach allows  the identification of companions
that are fainter than the MGS flux limit of $m_r = 17.77$, which is the
magnitude limit below which the spectroscopic catalogue is 95\%
complete. This also avoids the spectroscopic incompleteness due to the SDSS fiber
collision issue.  However, by using the photometry only, our determinations
of the local environment will be affected by projection effects. Therefore,
we used the spectroscopic catalogue to identify the Red Sequence (RS) at different
redshifts, and used these relations to mitigate the contamination by
background galaxies by excluding objects that are $2\, \sigma$ above the Red
Sequence.

We identified the position of the RS in the colour-magnitude
diagram at different redshifts, as follows. We retrieved galaxies from the
SDSS spectroscopic catalogue within $0.01 \leq z \leq 0.2$, and divided the
sample in 18 bins of redshift in steps of $\Delta z=0.01$ (except the first
bin, which includes galaxies in the range $0.01 \leq z < 0.03$). For each
bin, we identified the galaxies belonging to the RS using
{\tt mclust} \citep{mclust:2012, mclust:2002}, which is  an R package
for model-based clustering, classification, and density estimation, based on
Gaussian finite mixture modelling via expectation-maximization. We
ran {\tt mclust} on 100 bootstrap re-samplings for each redshift bin,
selecting 5000 galaxies in each realization. Once the RS is
identified, we then fit the $(g - i)_{\rm rs} = a_{\rm RS} + b_{\rm RS}\,
m_r$ relation and its scatter, $\sigma_{\rm RS}$. Figure~\ref{Fig_rs_zbins}
illustrates the procedure showing three different redshift bins.

Fig.~\ref{Fig_fit_rs_coeffs}  shows how $a_{\rm RS}$, $b_{\rm RS}$, and
$\sigma_{\rm RS}$  vary with $z$. We fitted these relations with a 2nd-order
polynomial, weighting the data points by the inverse of the squared  error, obtaining:
\begin{eqnarray}
 a_{\rm RS}(z) & = & 1.826 - 4.759\,z + 39.054\,z^2\ , \nonumber \\
 b_{\rm RS}(z) & = & -0.05024 + 0.49631\,z - 2.28545\,z^2\ , \nonumber \\
 \sigma_{\rm RS}(z) & = & 0.06452 - 0.23577\,z + 1.16574\,z^2\ . 
\end{eqnarray}

\section{Visual classification of interacting and merging galaxies}
\label{ap:classification}

%
\begin{figure}
\centering
\begin{tabular}{c}
 \resizebox{0.95\hsize}{!}{\includegraphics{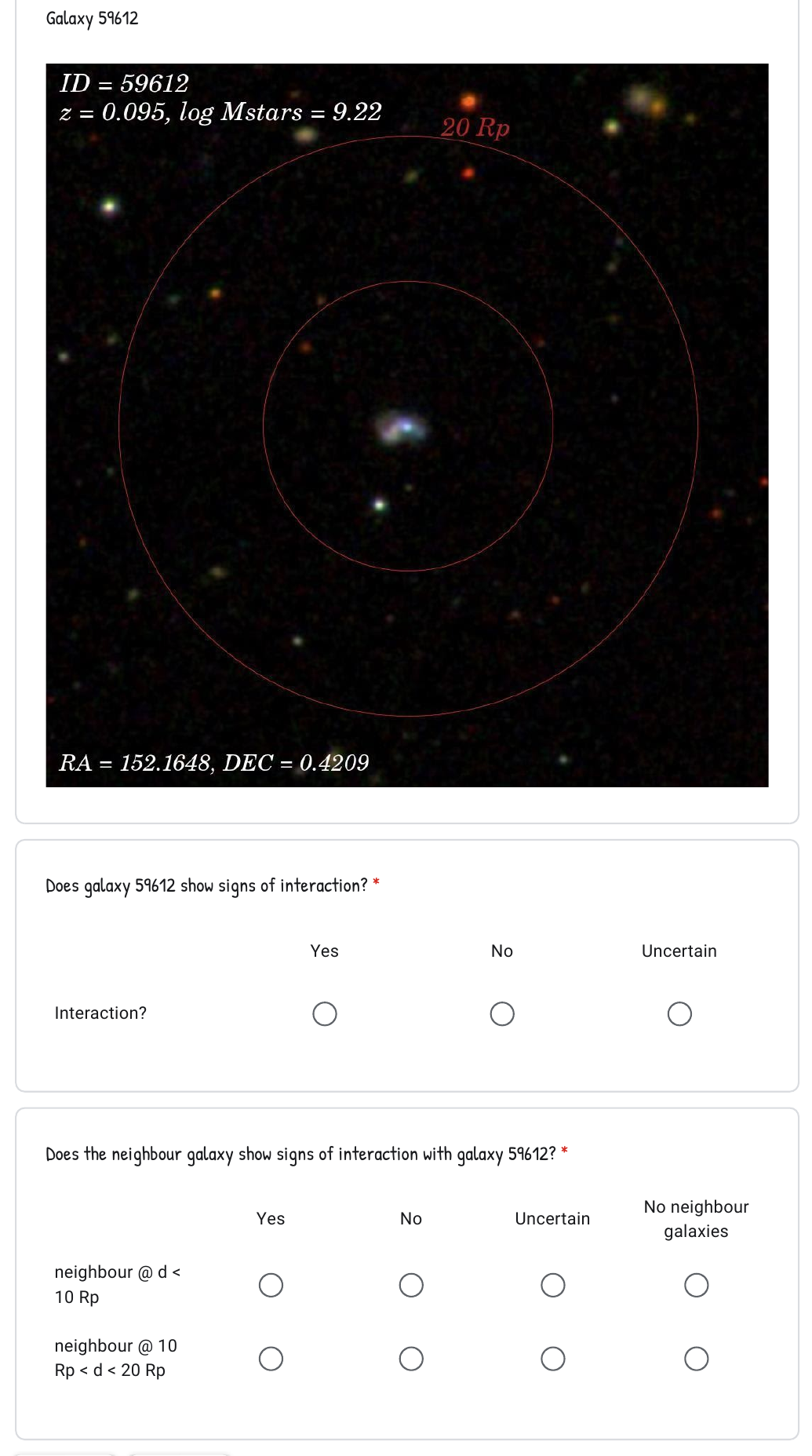}} \\
\end{tabular}
\caption {Example illustrating our form for visual classification of merging
and interacting galaxies. This particular galaxy was classified as
interacting system with both galaxy and neighbour showing signs of
interaction.}
\label{Fig_form}
\end{figure}

We visually inspected our galaxies to identify features indicating
interactions and recent mergers. While this approach is subjective, our eyes
can still catch some features that the morphometry of galaxies is not able
to. Besides the 207 VYGs, we selected 207 randomly chosen CSGs for
comparison, and these 414 galaxies were classified by 5 participants, all
astronomers. 

We aimed to identify the following types of objects:
\begin{description}
  \item[1.] \emph{Interacting / merging systems}: the galaxy has one or several companions, at least one of which shows tidal features and/or other signs of interactions;
  \item[2.] \emph{Pairs / multiple systems}: the galaxy has one or several  close companions, but no clear sign of interaction between them (we might miss the tidal features because the SDSS images are not deep enough -- or the pair is just a projection effect);
  \item[3.] \emph{Post-mergers}: we see signs of recent interaction, i.e., tidal tails, but no neighbour galaxy;
  \item[4.] \emph{Non-interacting, isolated systems}: no signs of interactions and no neighbour galaxies.
\end{description}

For this classification, we created a form with the galaxy images and the following questions, as shown in Fig.~\ref{Fig_form}. The following instructions were given to the participants:

\begin{itemize}
  \item[1.] \emph{Does the galaxy show signs of interaction, like tidal distortions and tails?}

If there are tidal features, the participants were instructed to answer \emph{yes} to this question even if there are no neighbour galaxies, since we also want to identify post-mergers.

  \item[2.] \emph{Does the neighbour galaxy show signs of interaction with the test galaxy?}

This question is divided in two parts, one for neighbour galaxies within $10\,R_{\rm p}$, where $R_{\rm p}$ is the Petrosian radius in the $r$ band of our test galaxy, or for neighbour galaxies between $10$ and $20\,R_{\rm p}$. In each image, the $10-$ and $20-R_{\rm p}$ regions are indicated by red circles, as indicated in Fig.~\ref{Fig_form}.
\end{itemize}

\section{VYG images and data}
\label{ap:images}

\begin{figure*}
\setlength{\tabcolsep}{1pt}
\begin{tabular}{ccccccc}
 \includegraphics[width=0.13\hsize]{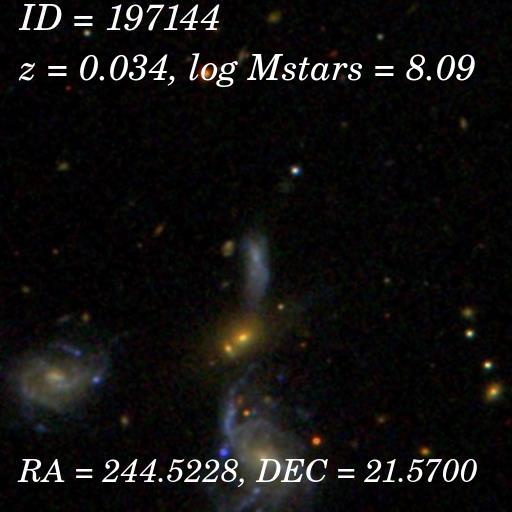} &
 \includegraphics[width=0.13\hsize]{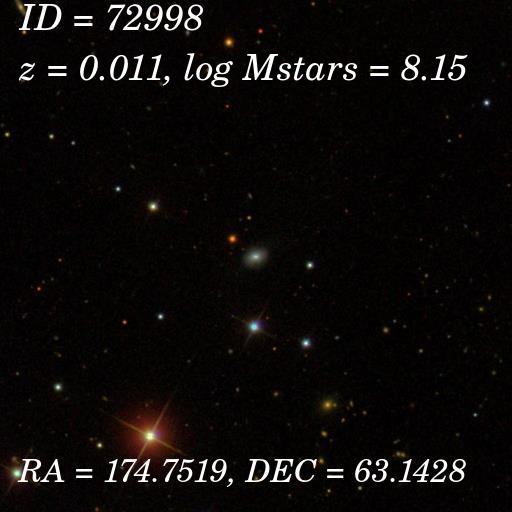} &
 \includegraphics[width=0.13\hsize]{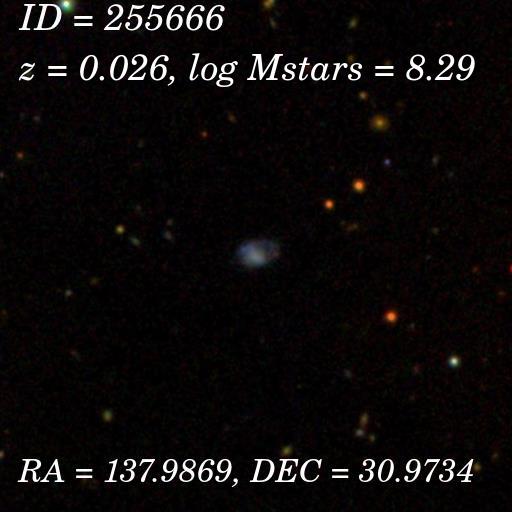} &
 \includegraphics[width=0.13\hsize]{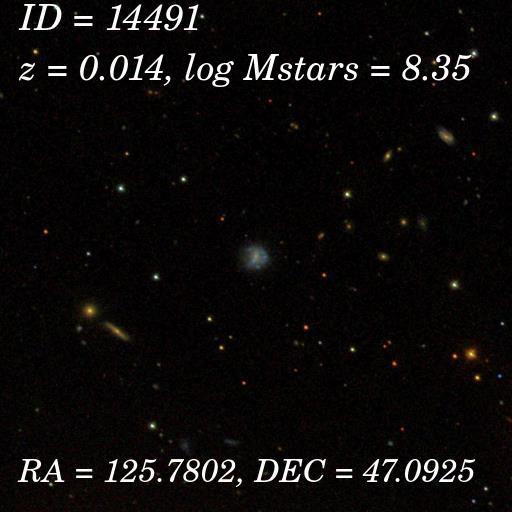} &
 \includegraphics[width=0.13\hsize]{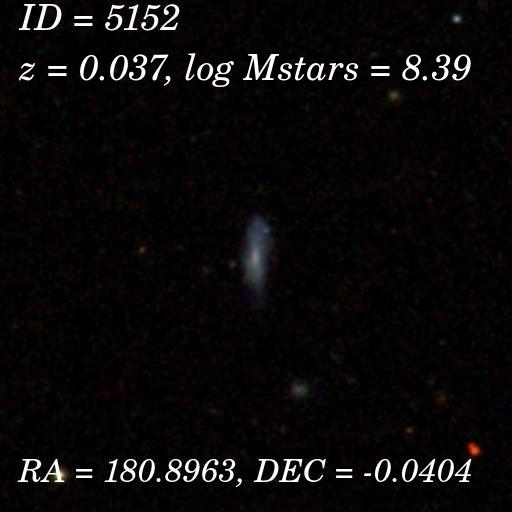} &
 \includegraphics[width=0.13\hsize]{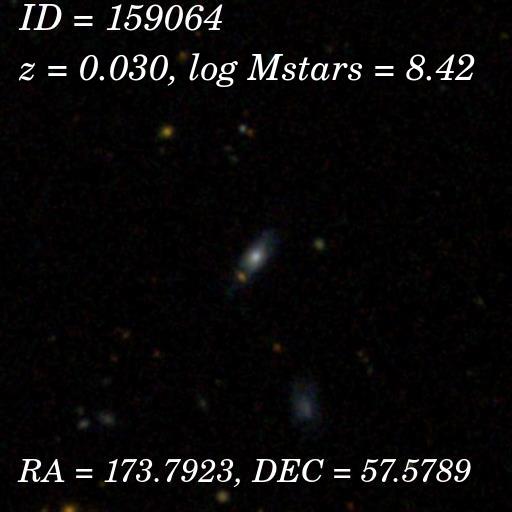} &
 \includegraphics[width=0.13\hsize]{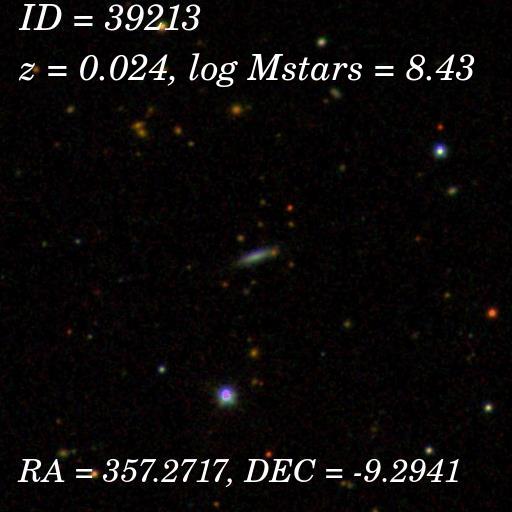} \\
 
 \includegraphics[width=0.13\hsize]{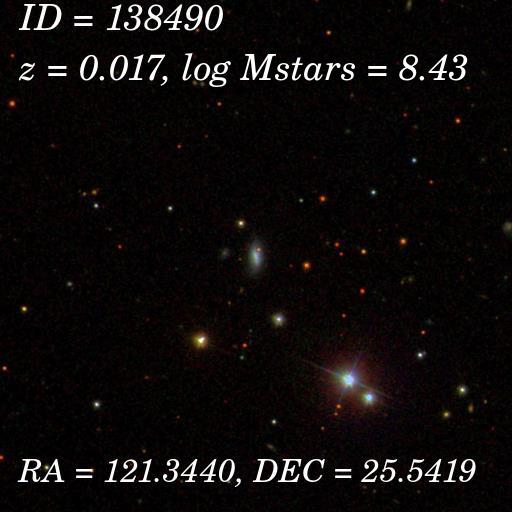} &
 \includegraphics[width=0.13\hsize]{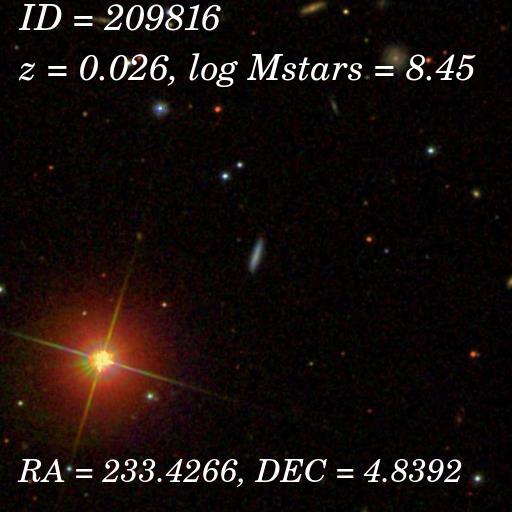} &
 \includegraphics[width=0.13\hsize]{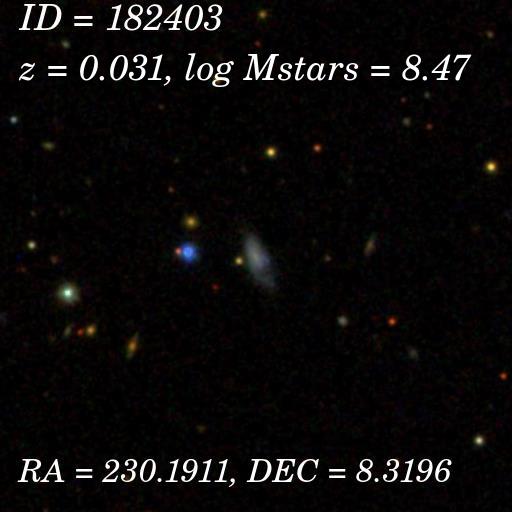} &
 \includegraphics[width=0.13\hsize]{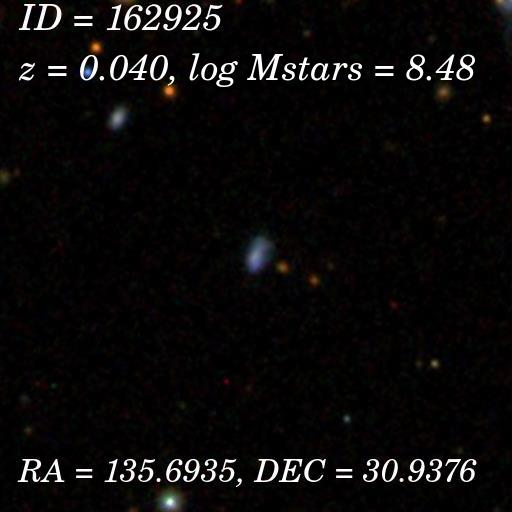} &
 \includegraphics[width=0.13\hsize]{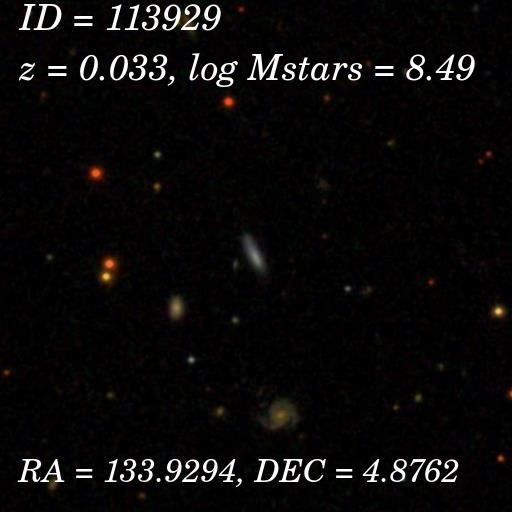} &
 \includegraphics[width=0.13\hsize]{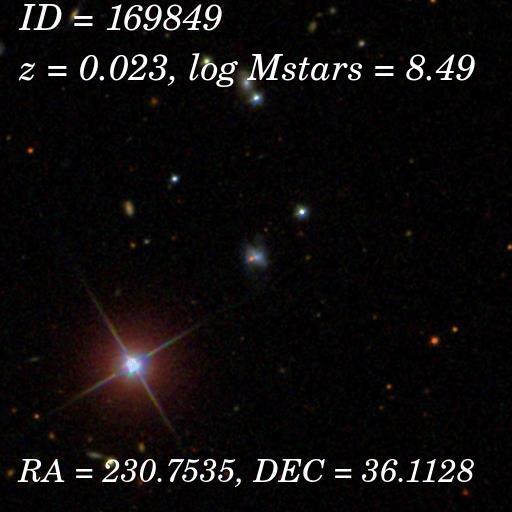} &
 \includegraphics[width=0.13\hsize]{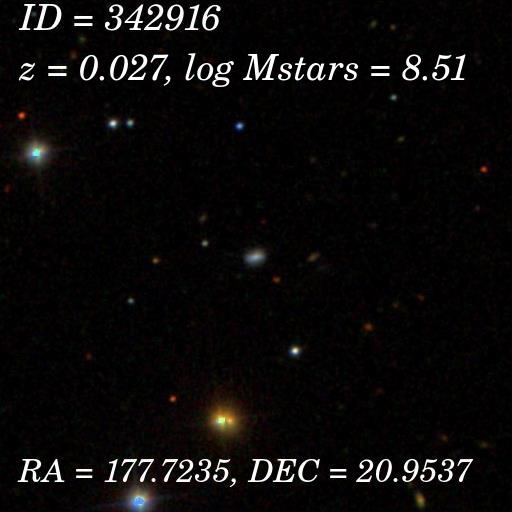} \\
 
 \includegraphics[width=0.13\hsize]{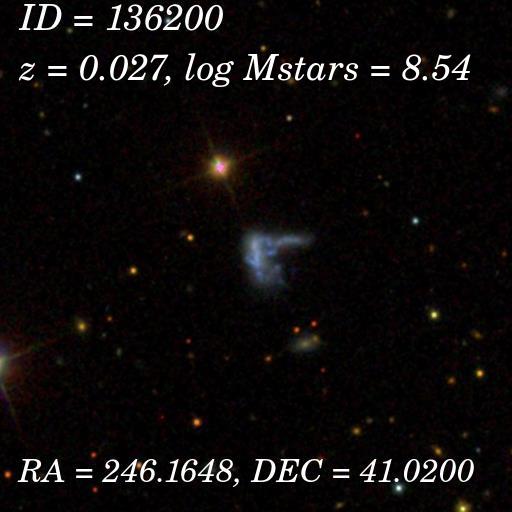} &
 \includegraphics[width=0.13\hsize]{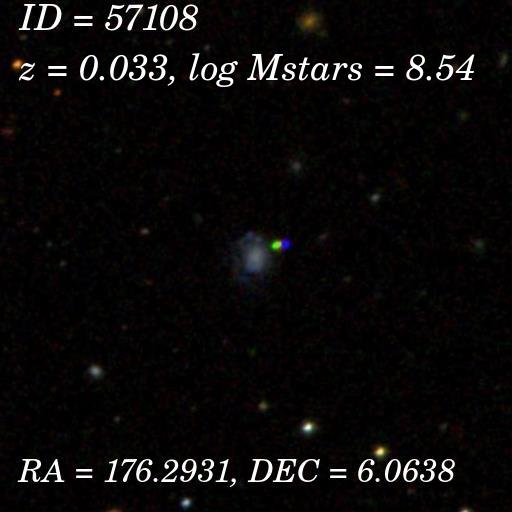} &
 \includegraphics[width=0.13\hsize]{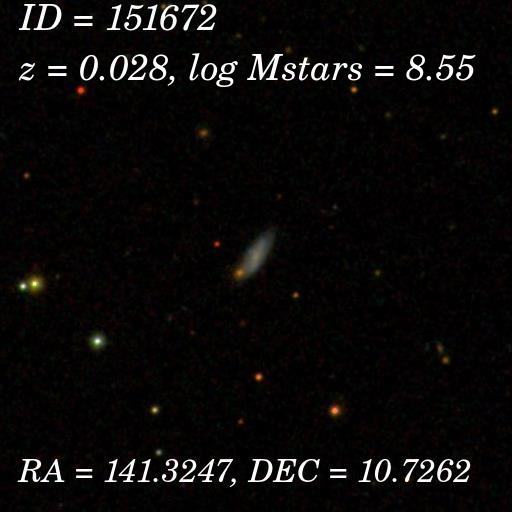} &
 \includegraphics[width=0.13\hsize]{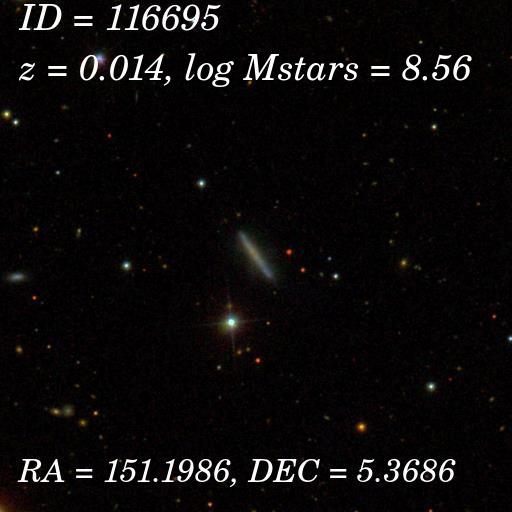} &
 \includegraphics[width=0.13\hsize]{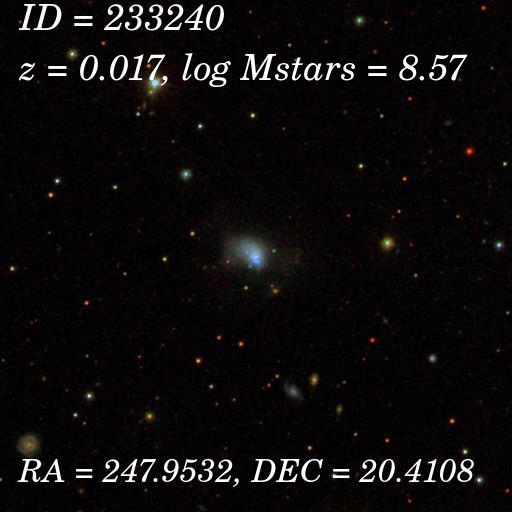} &
 \includegraphics[width=0.13\hsize]{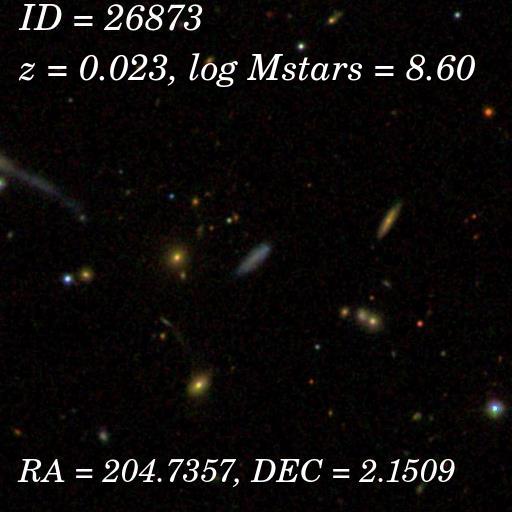} &
 \includegraphics[width=0.13\hsize]{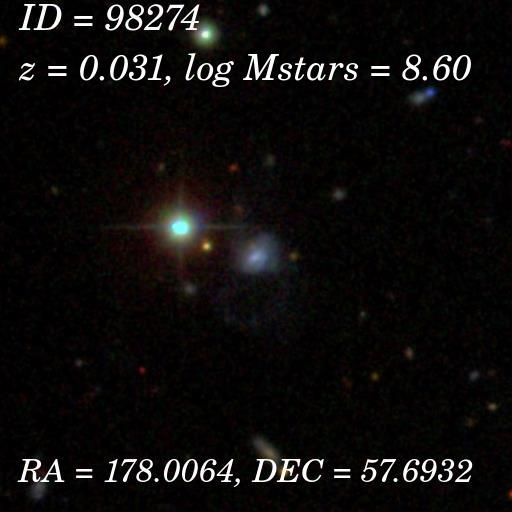} \\
 
 \includegraphics[width=0.13\hsize]{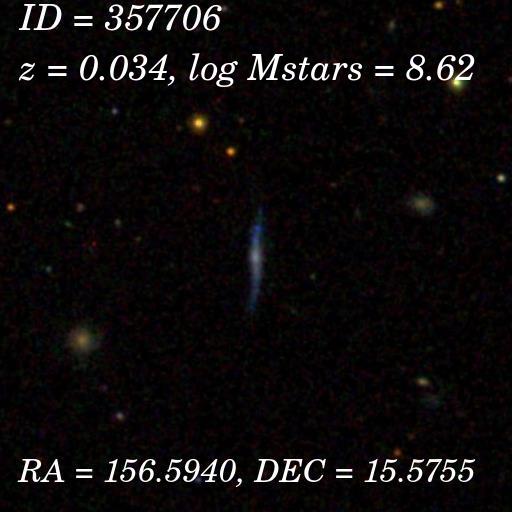} &
 \includegraphics[width=0.13\hsize]{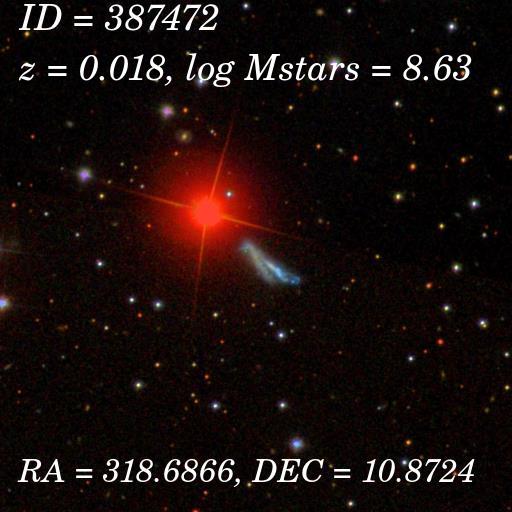} &
 \includegraphics[width=0.13\hsize]{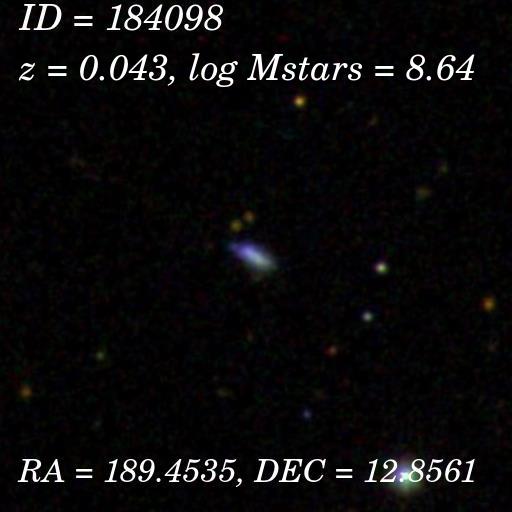} &
 \includegraphics[width=0.13\hsize]{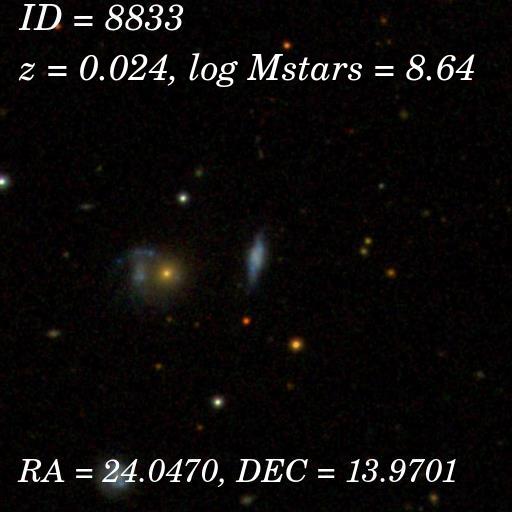} &
 \includegraphics[width=0.13\hsize]{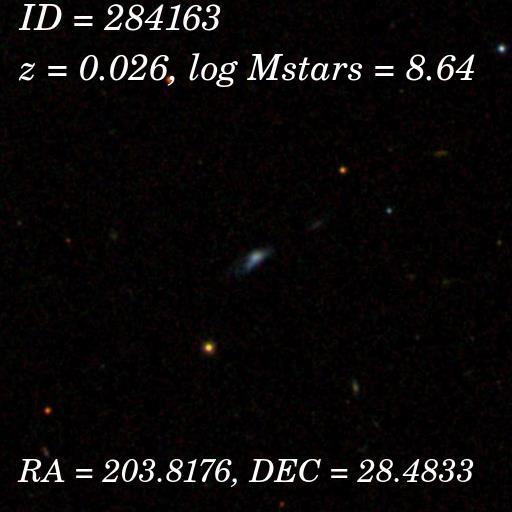} &
 \includegraphics[width=0.13\hsize]{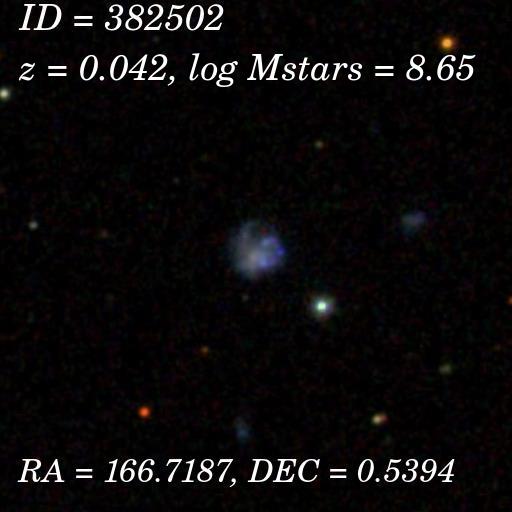} &
 \includegraphics[width=0.13\hsize]{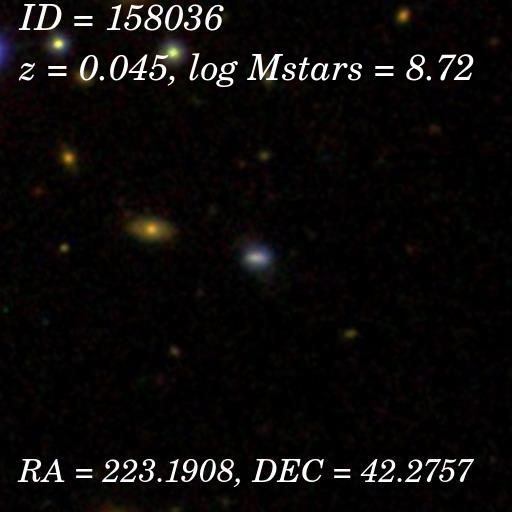} \\
 
 \includegraphics[width=0.13\hsize]{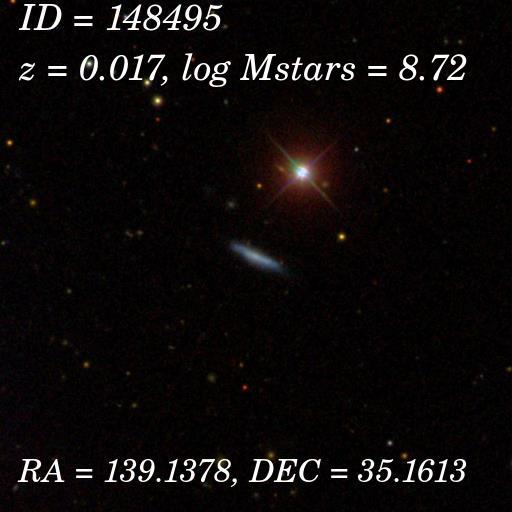} &
 \includegraphics[width=0.13\hsize]{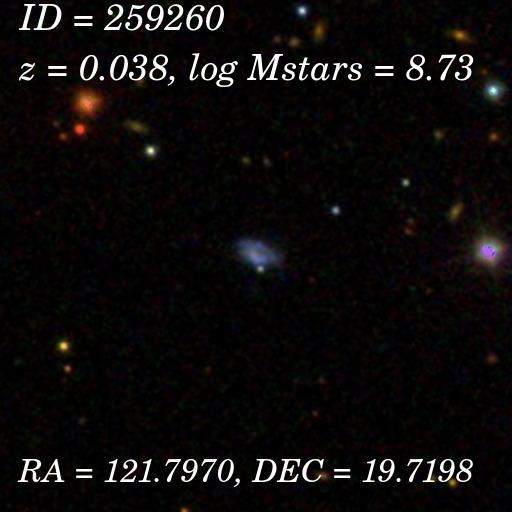} &
 \includegraphics[width=0.13\hsize]{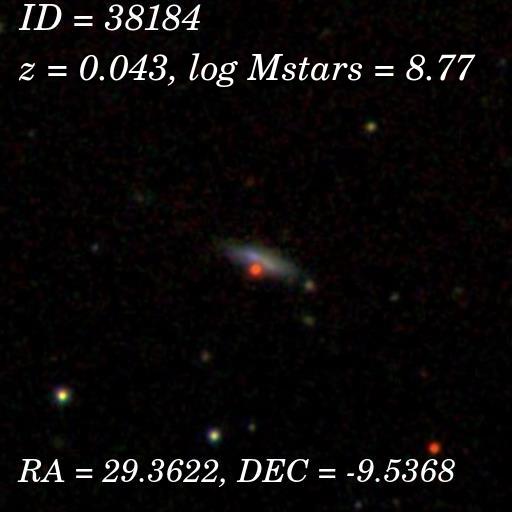} &
 \includegraphics[width=0.13\hsize]{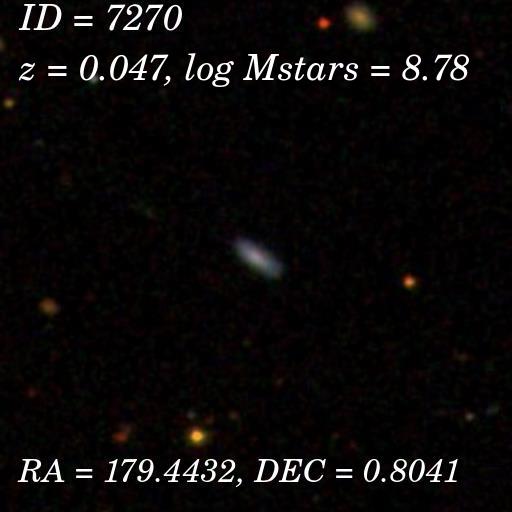} &
 \includegraphics[width=0.13\hsize]{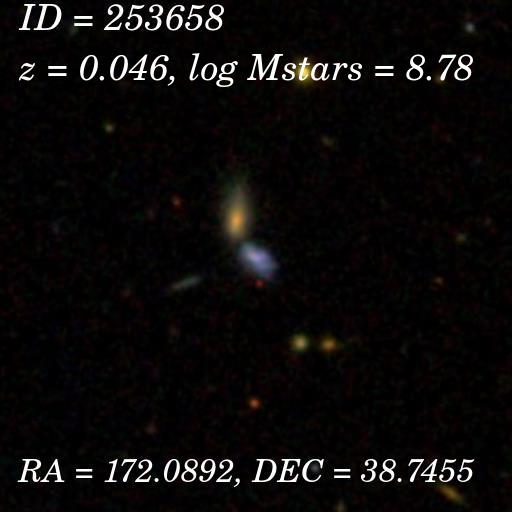} &
 \includegraphics[width=0.13\hsize]{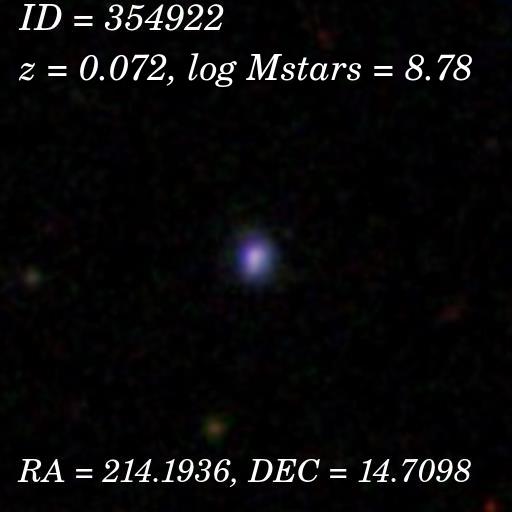} &
 \includegraphics[width=0.13\hsize]{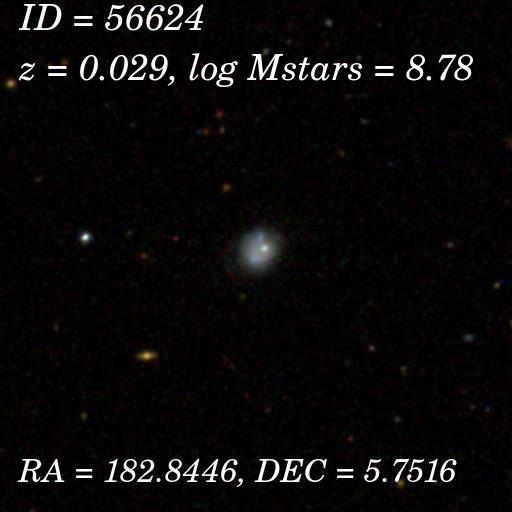} \\
 
 \includegraphics[width=0.13\hsize]{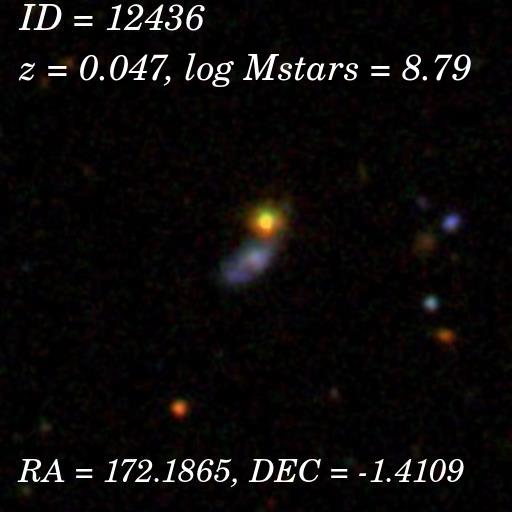} &
 \includegraphics[width=0.13\hsize]{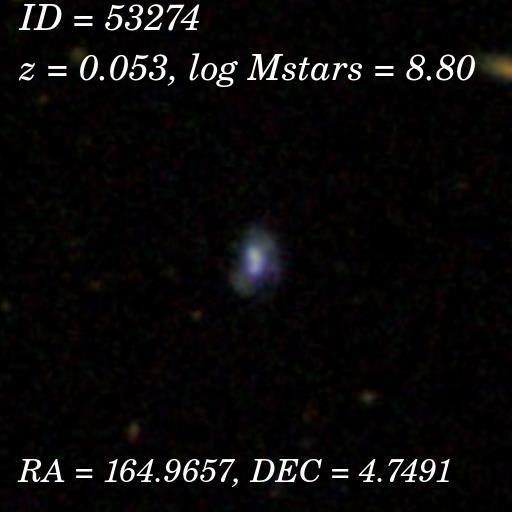} &
 \includegraphics[width=0.13\hsize]{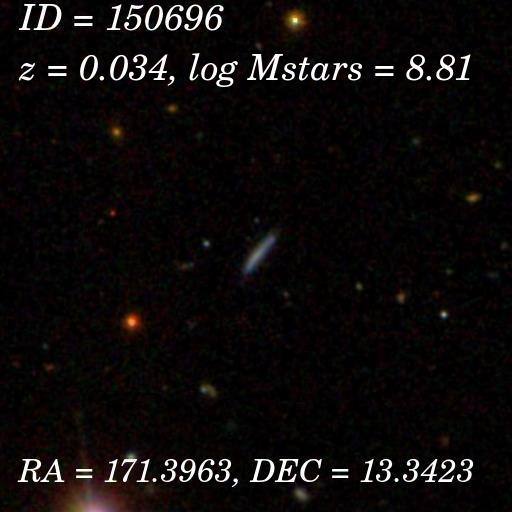} &
 \includegraphics[width=0.13\hsize]{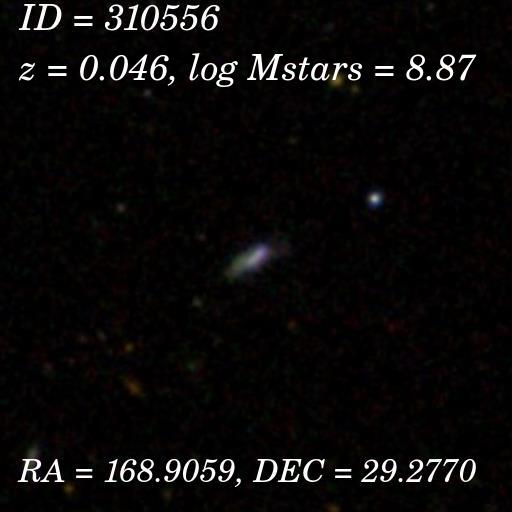} &
 \includegraphics[width=0.13\hsize]{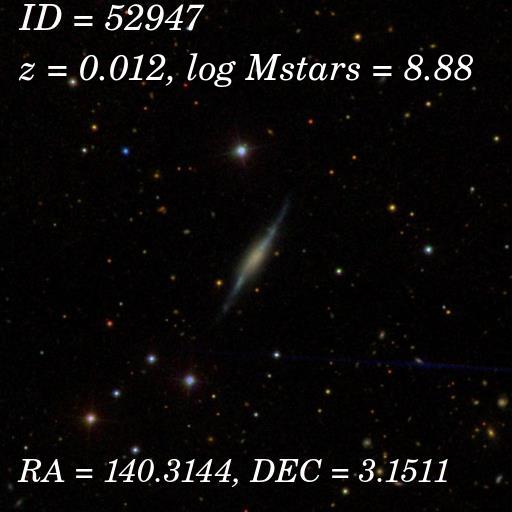} &
 \includegraphics[width=0.13\hsize]{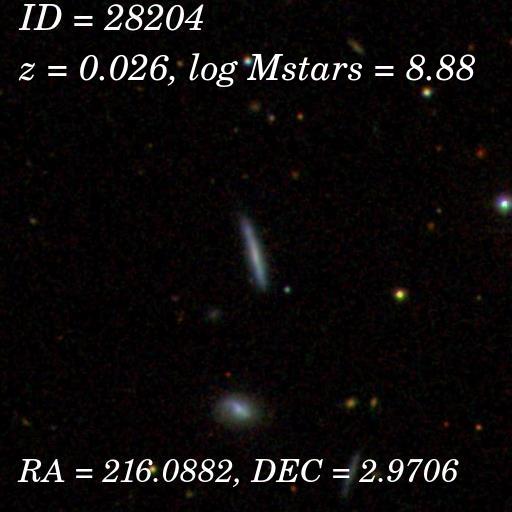} &
 \includegraphics[width=0.13\hsize]{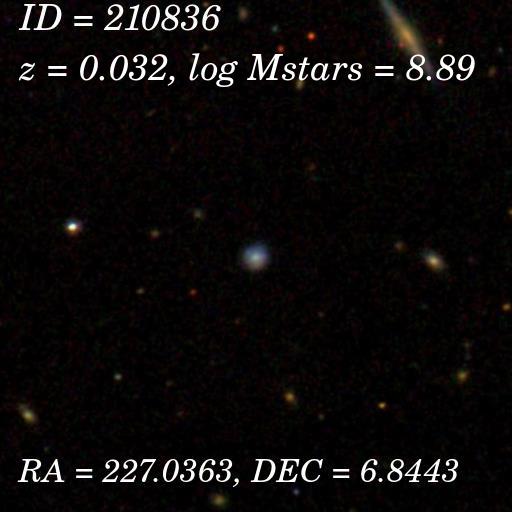} \\
 
 \includegraphics[width=0.13\hsize]{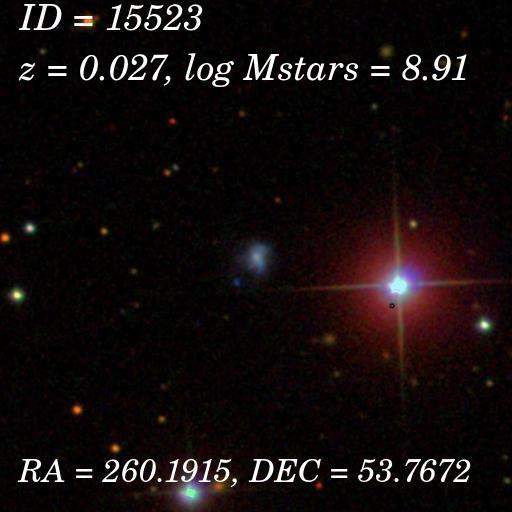} &
 \includegraphics[width=0.13\hsize]{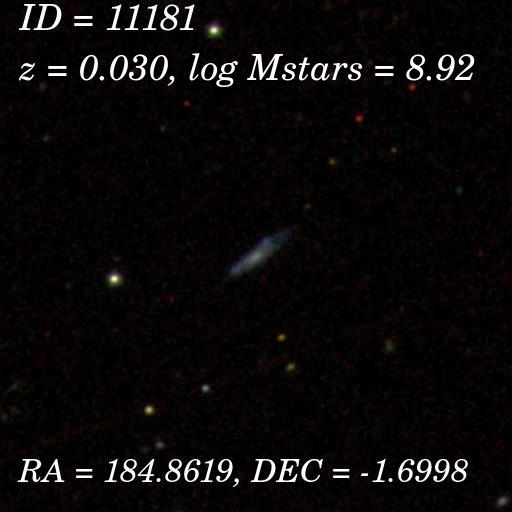} &
 \includegraphics[width=0.13\hsize]{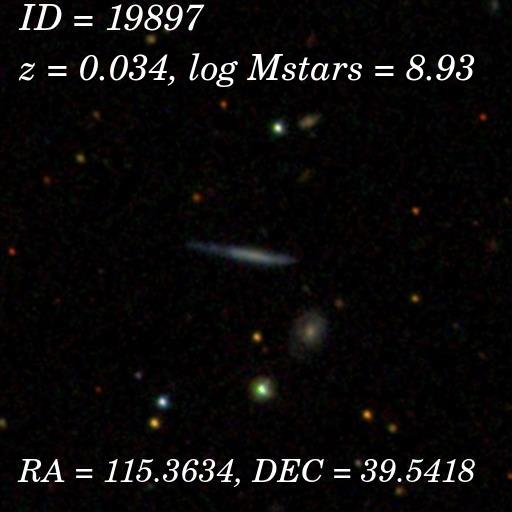} &
 \includegraphics[width=0.13\hsize]{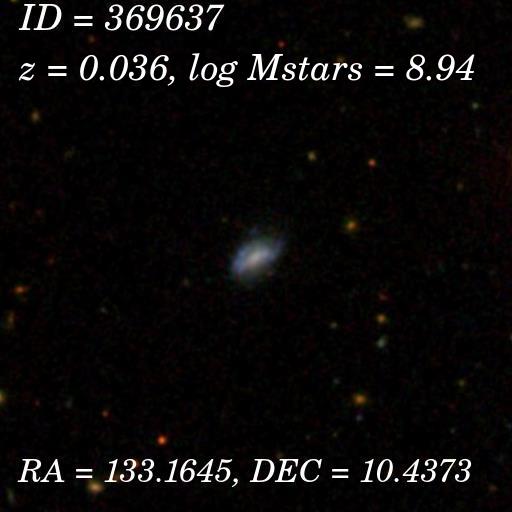} &
 \includegraphics[width=0.13\hsize]{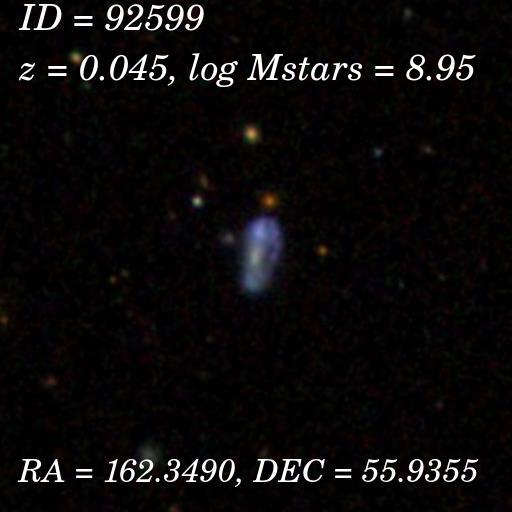} &
 \includegraphics[width=0.13\hsize]{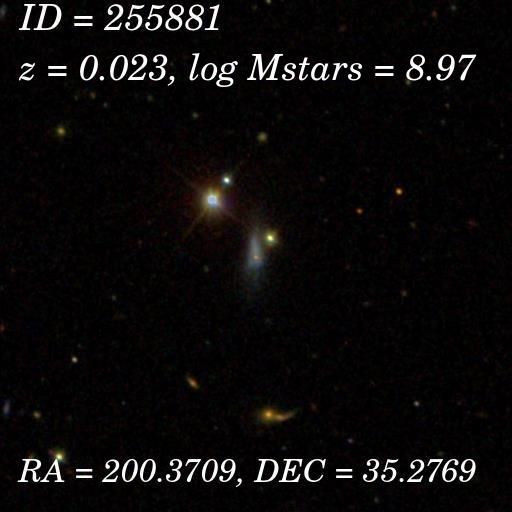} &
 \includegraphics[width=0.13\hsize]{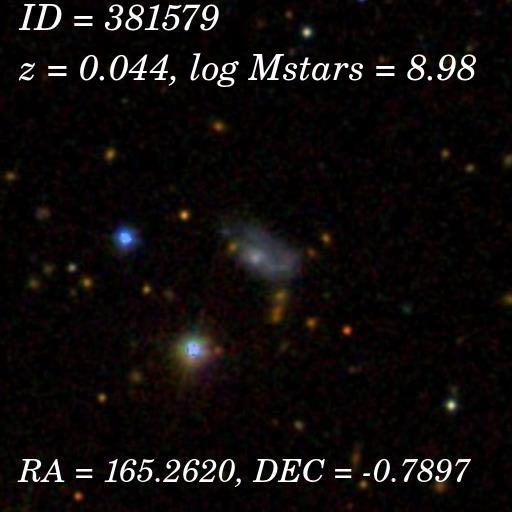} \\
 
 \includegraphics[width=0.13\hsize]{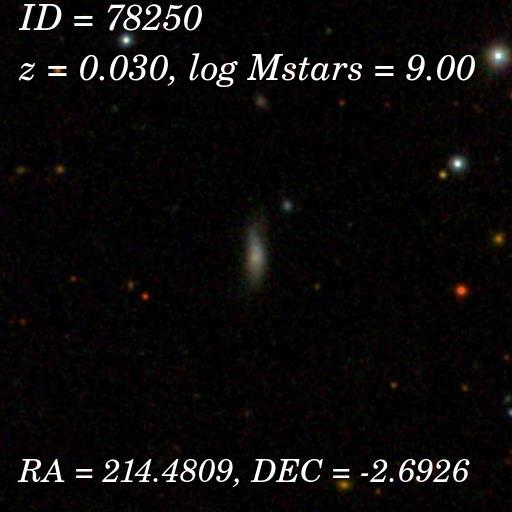} &
 \includegraphics[width=0.13\hsize]{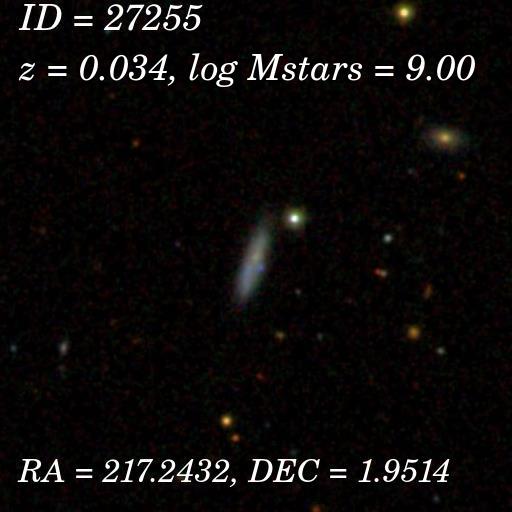} &
 \includegraphics[width=0.13\hsize]{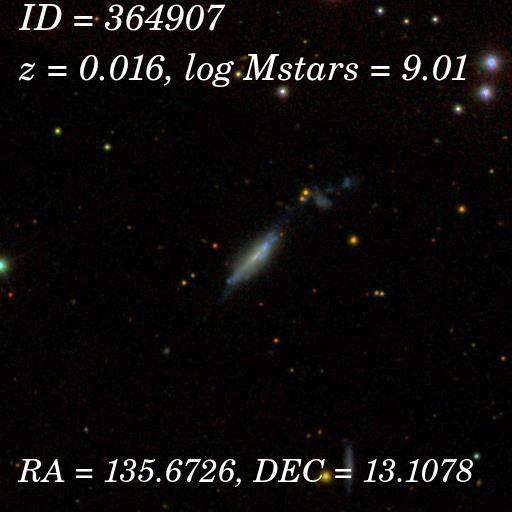} &
 \includegraphics[width=0.13\hsize]{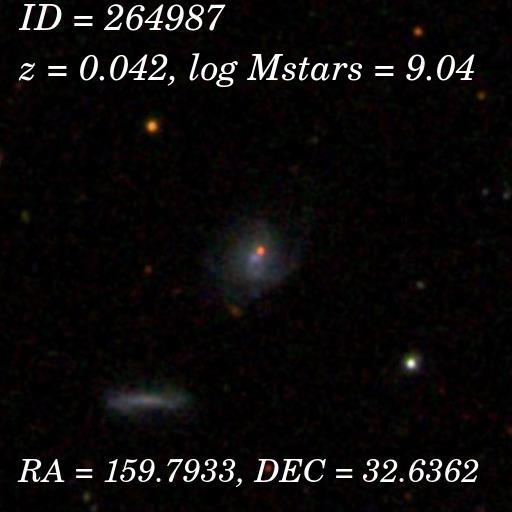} &
 \includegraphics[width=0.13\hsize]{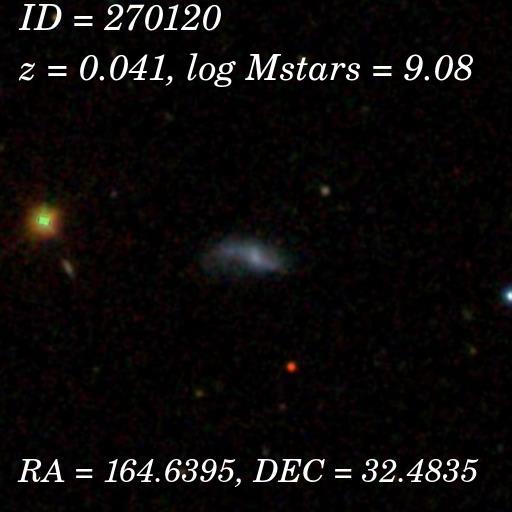} &
 \includegraphics[width=0.13\hsize]{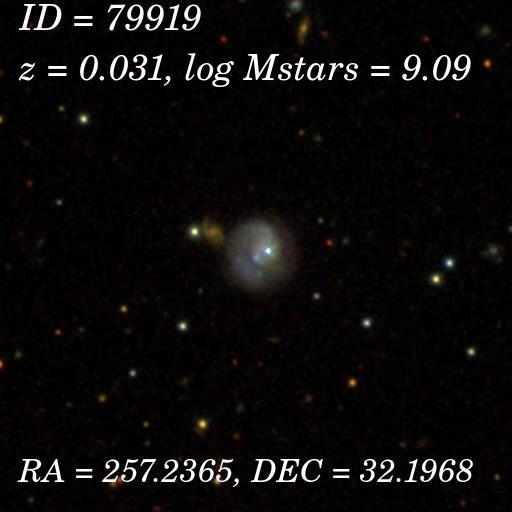} &
 \includegraphics[width=0.13\hsize]{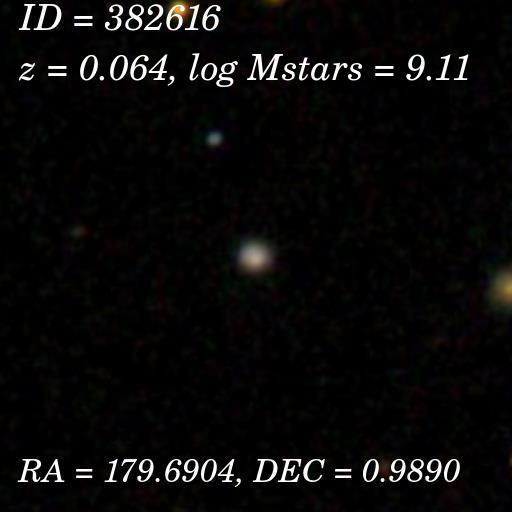} \\
 
 \includegraphics[width=0.13\hsize]{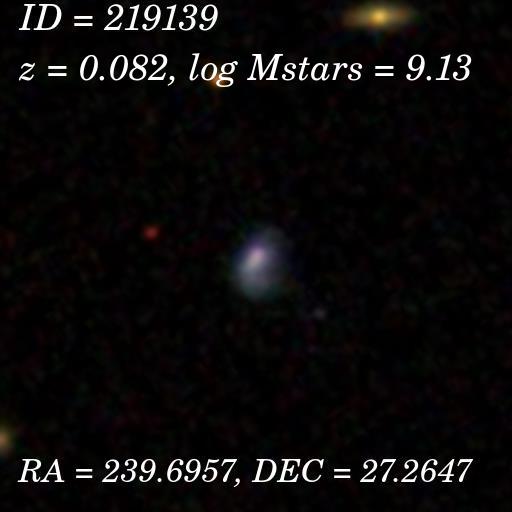} &
 \includegraphics[width=0.13\hsize]{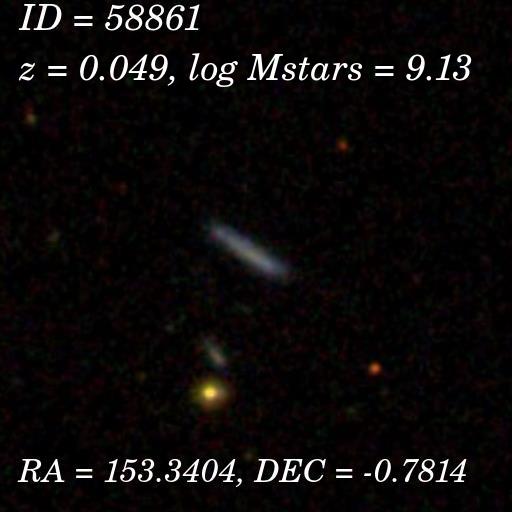} &
 \includegraphics[width=0.13\hsize]{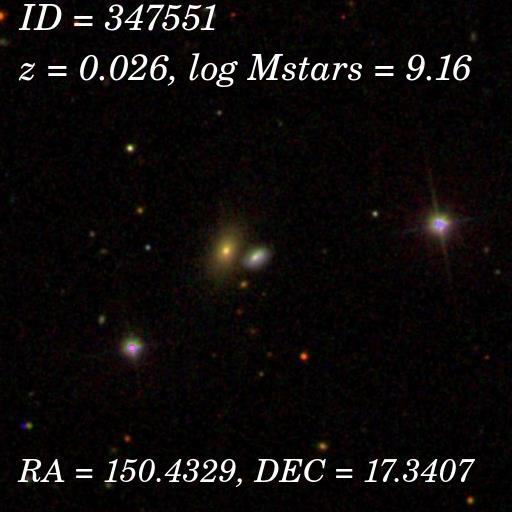} &
 \includegraphics[width=0.13\hsize]{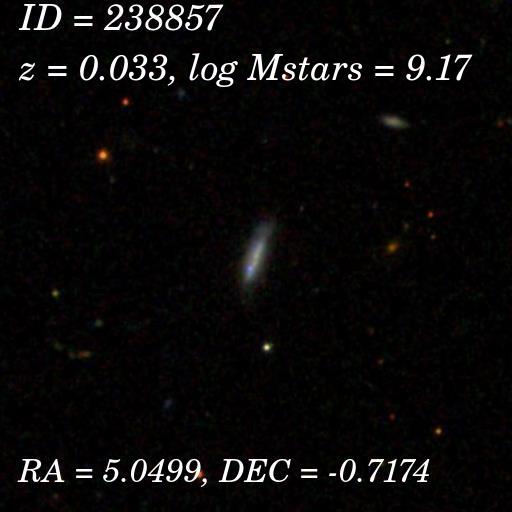} &
 \includegraphics[width=0.13\hsize]{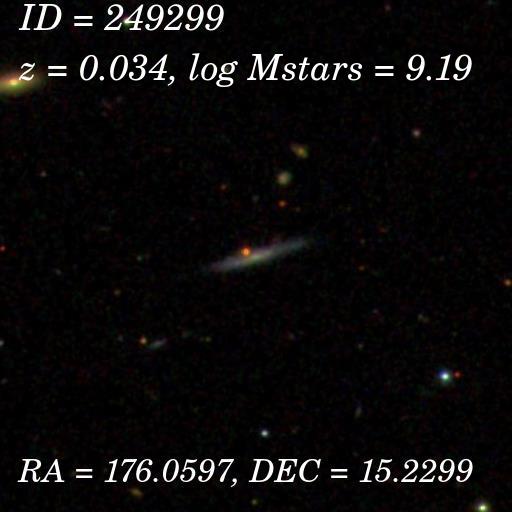} &
 \includegraphics[width=0.13\hsize]{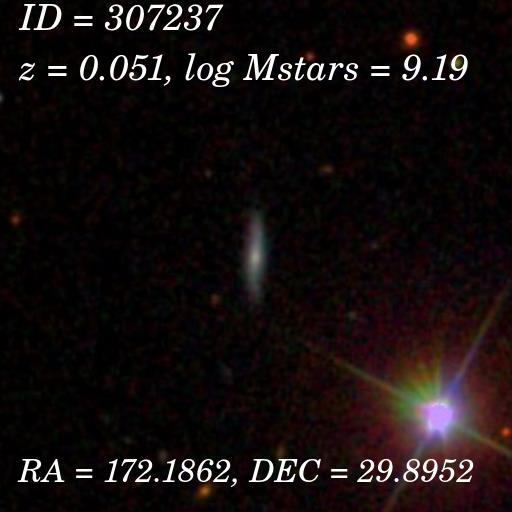} &
 \includegraphics[width=0.13\hsize]{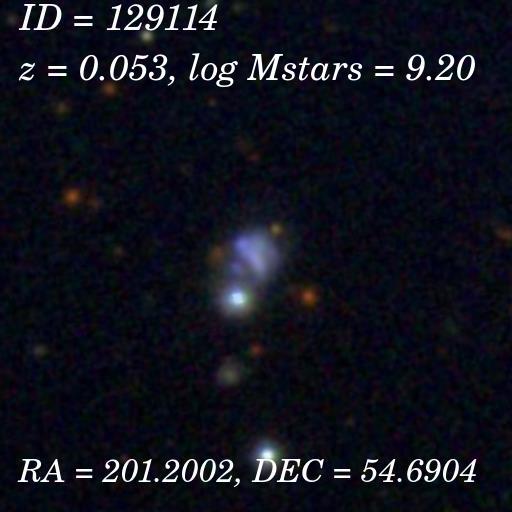} \\
\end{tabular}
\caption{SDSS images of a sample of 207 VYGs. All images are 100 x 100 kpc.}
\label{fig:vyg_images1}
\end{figure*}

\begin{figure*}
\setlength{\tabcolsep}{1pt}
\begin{tabular}{ccccccc}
 \includegraphics[width=0.13\hsize]{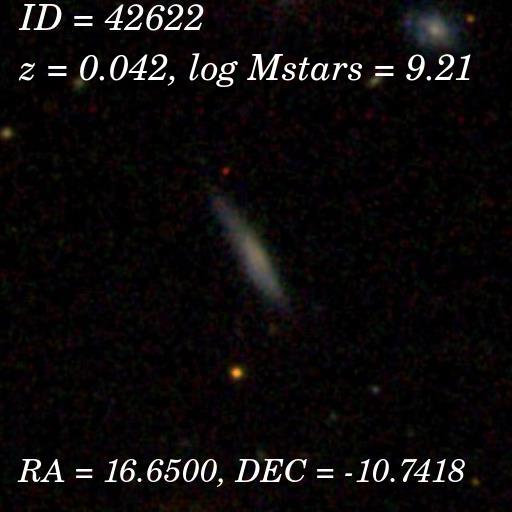} &
 \includegraphics[width=0.13\hsize]{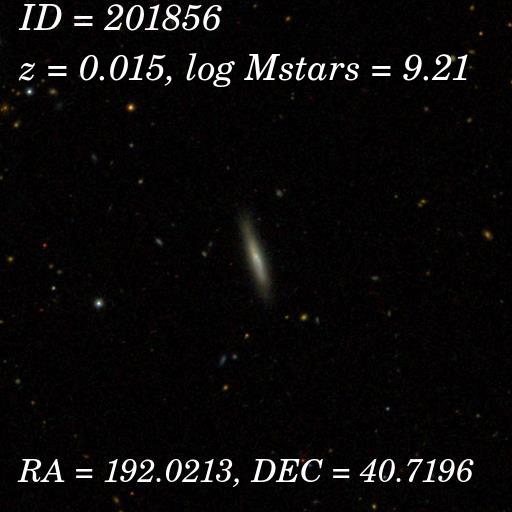} &
 \includegraphics[width=0.13\hsize]{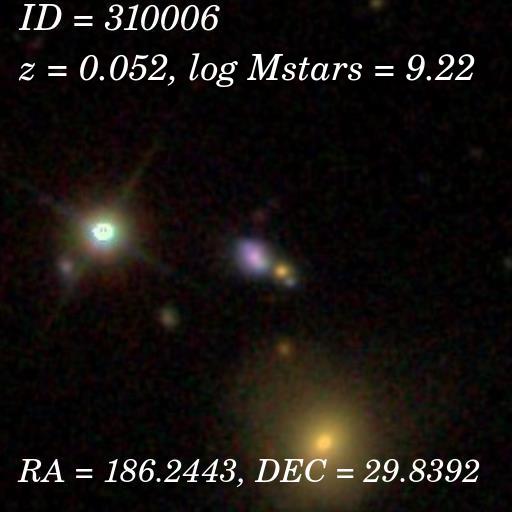} &
 \includegraphics[width=0.13\hsize]{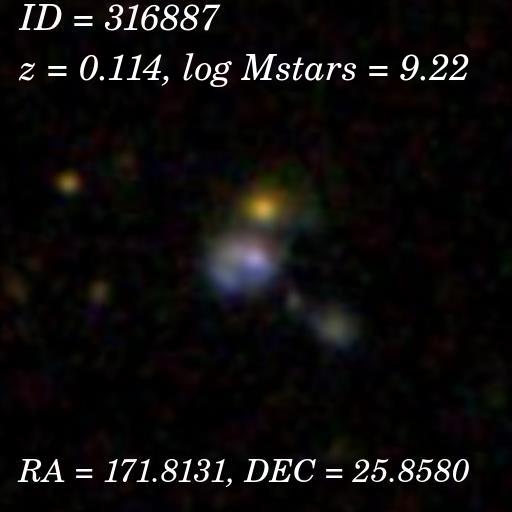} &
 \includegraphics[width=0.13\hsize]{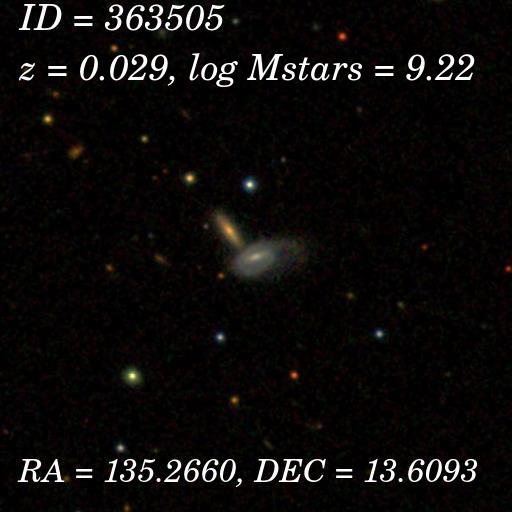} &
 \includegraphics[width=0.13\hsize]{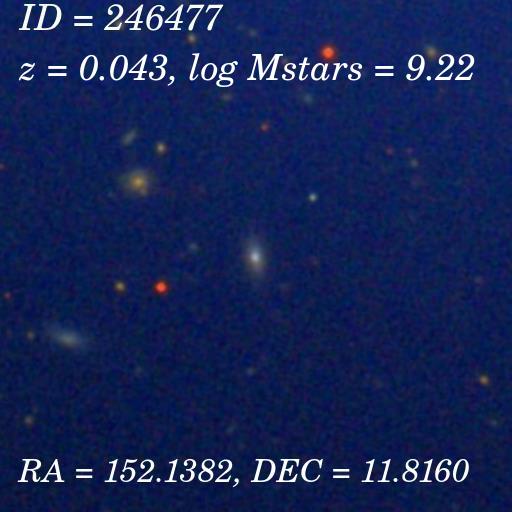} &
 \includegraphics[width=0.13\hsize]{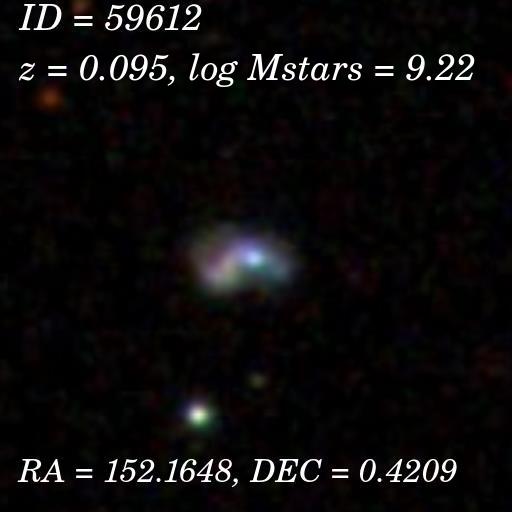} \\
 
 \includegraphics[width=0.13\hsize]{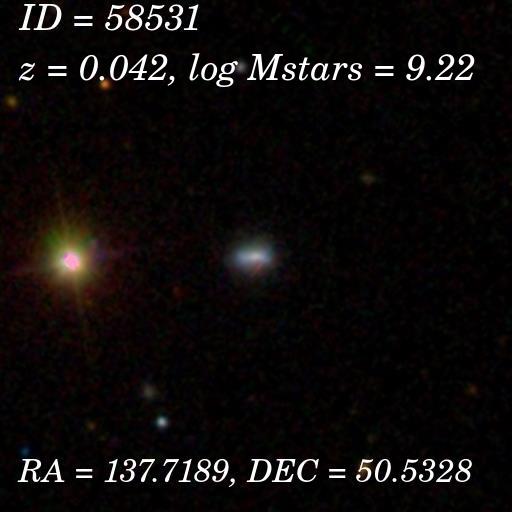} &
 \includegraphics[width=0.13\hsize]{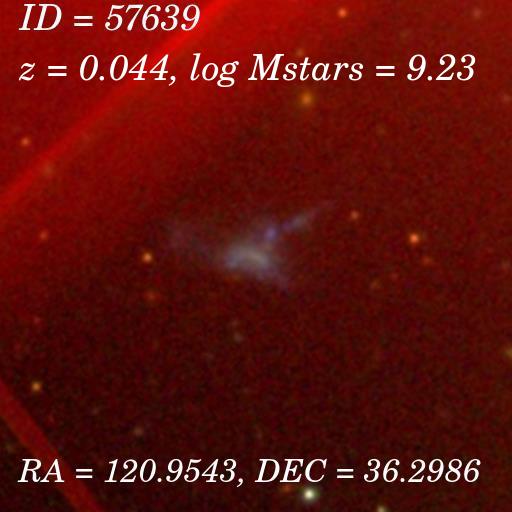} &
 \includegraphics[width=0.13\hsize]{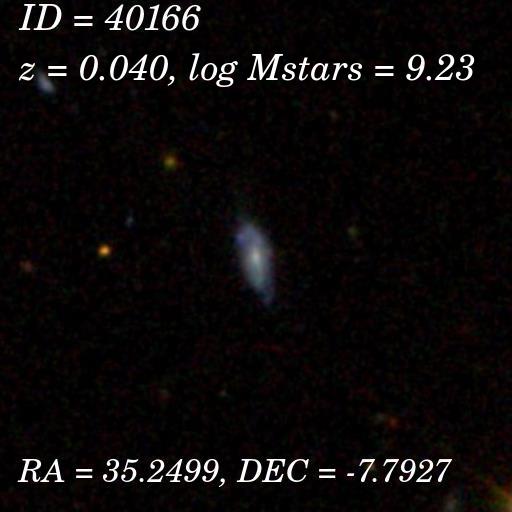} &
 \includegraphics[width=0.13\hsize]{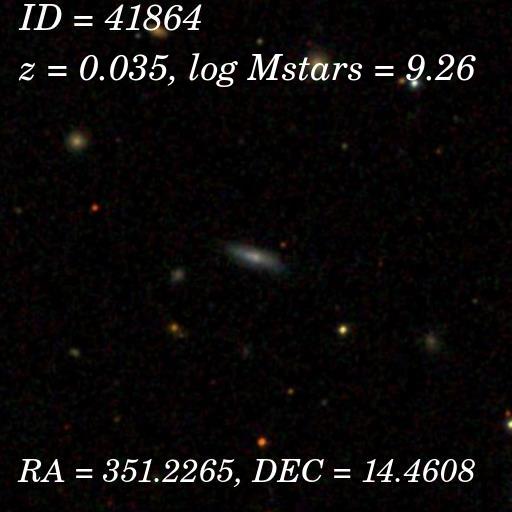} &
 \includegraphics[width=0.13\hsize]{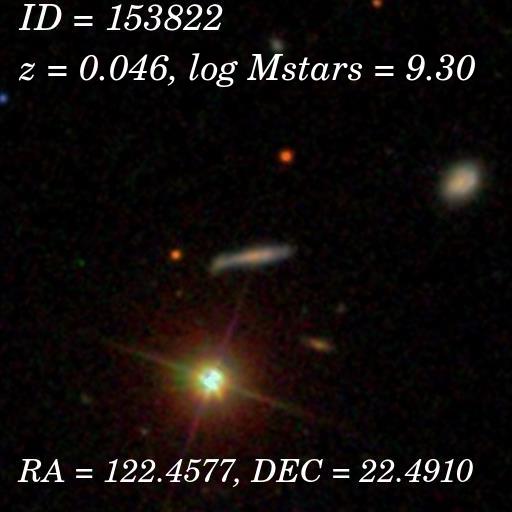} &
 \includegraphics[width=0.13\hsize]{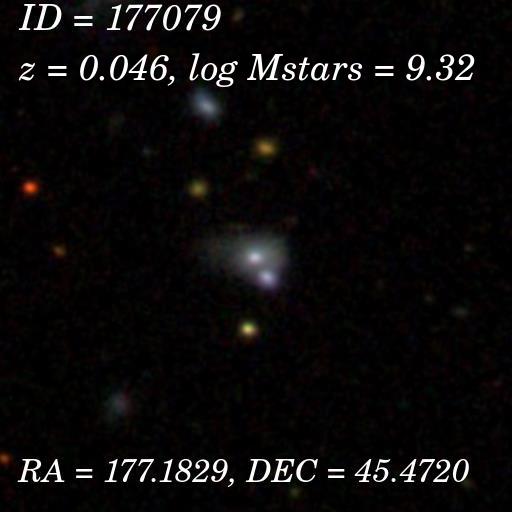} &
 \includegraphics[width=0.13\hsize]{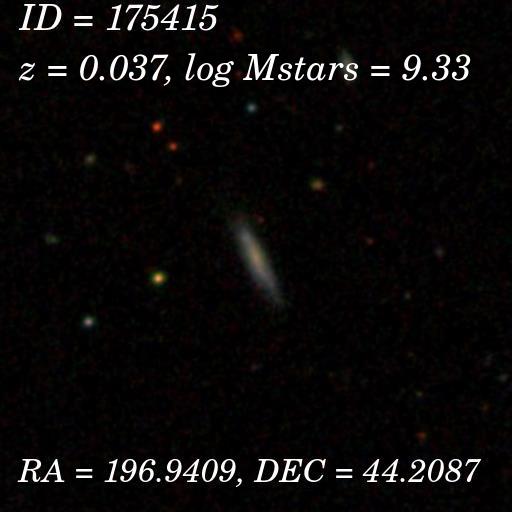} \\
 
 \includegraphics[width=0.13\hsize]{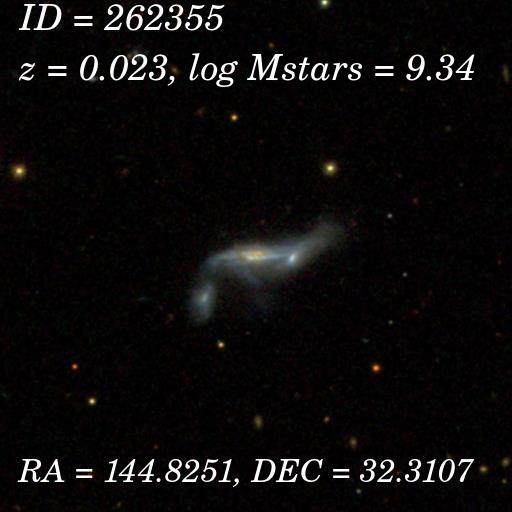} &
 \includegraphics[width=0.13\hsize]{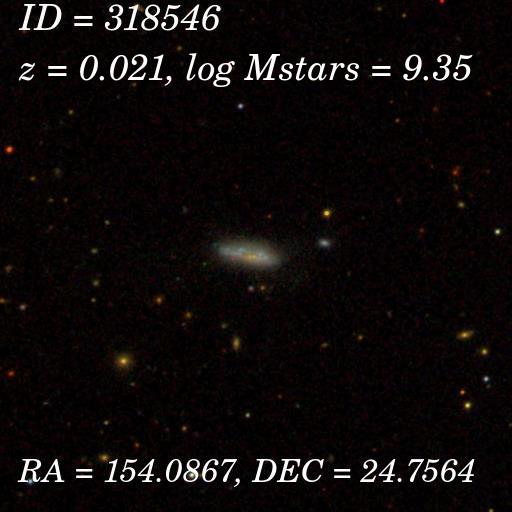} &
 \includegraphics[width=0.13\hsize]{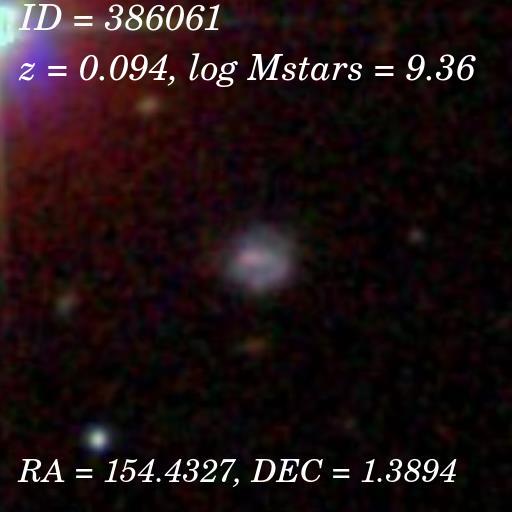} &
 \includegraphics[width=0.13\hsize]{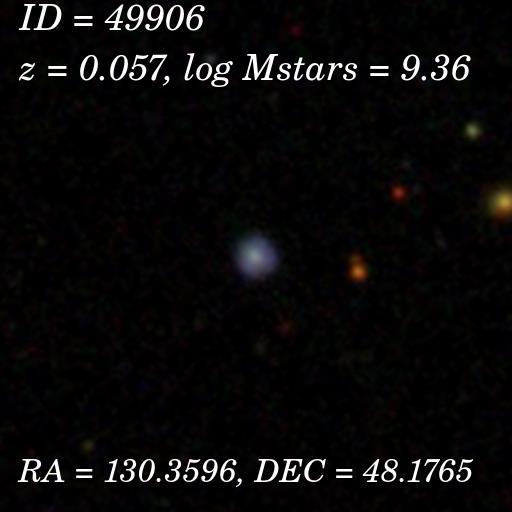} &
 \includegraphics[width=0.13\hsize]{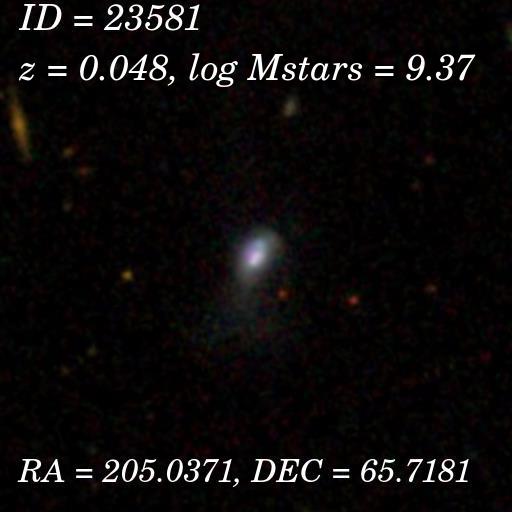} &
 \includegraphics[width=0.13\hsize]{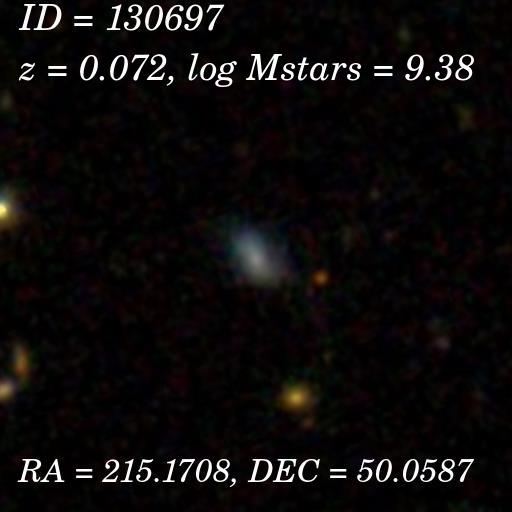} &
 \includegraphics[width=0.13\hsize]{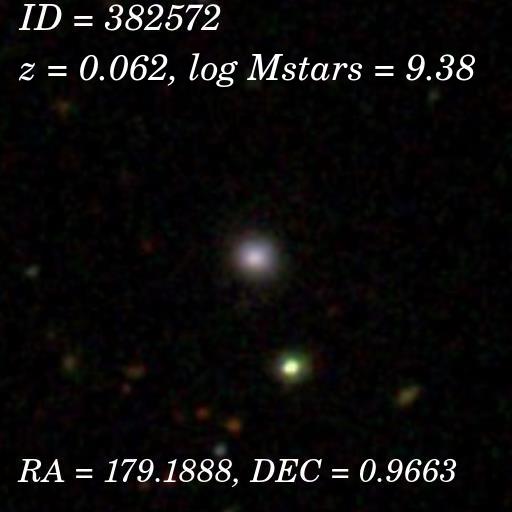} \\
 
 \includegraphics[width=0.13\hsize]{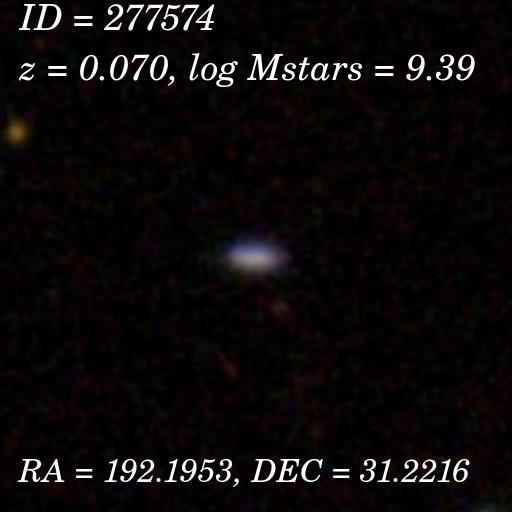} &
 \includegraphics[width=0.13\hsize]{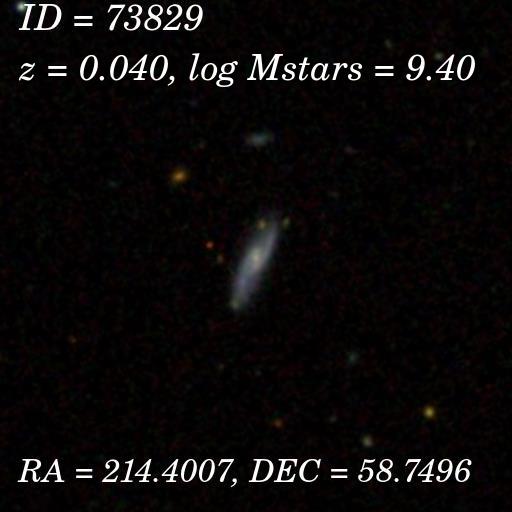} &
 \includegraphics[width=0.13\hsize]{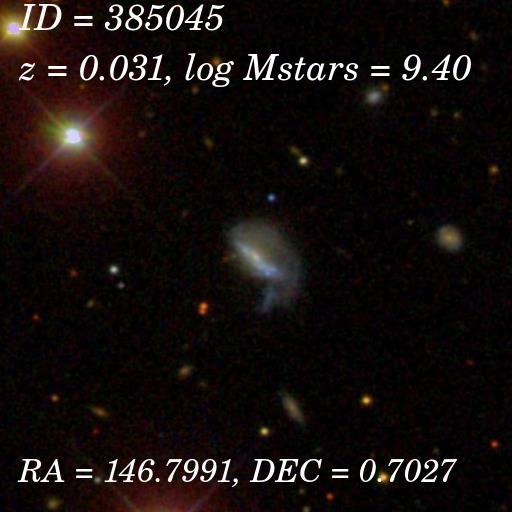} &
 \includegraphics[width=0.13\hsize]{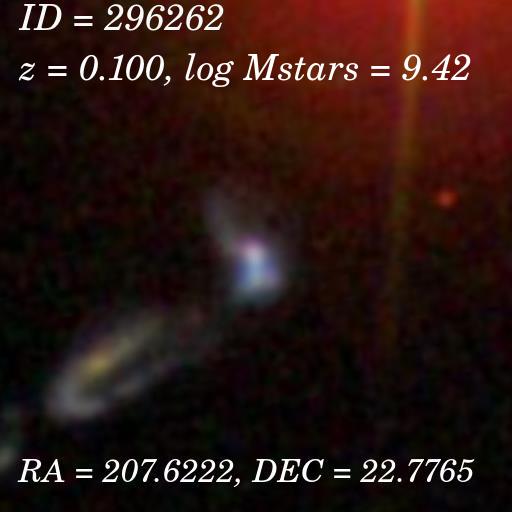} &
 \includegraphics[width=0.13\hsize]{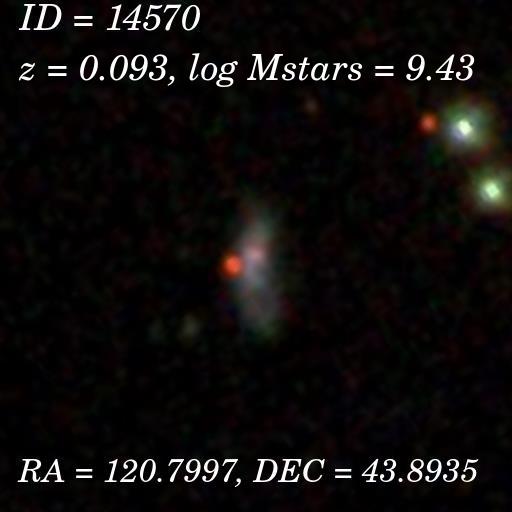} &
 \includegraphics[width=0.13\hsize]{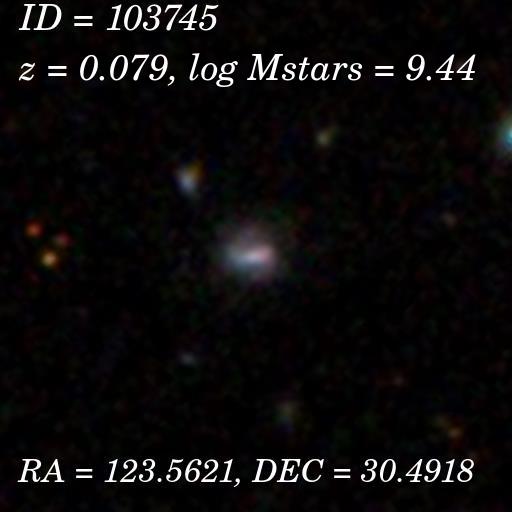} &
 \includegraphics[width=0.13\hsize]{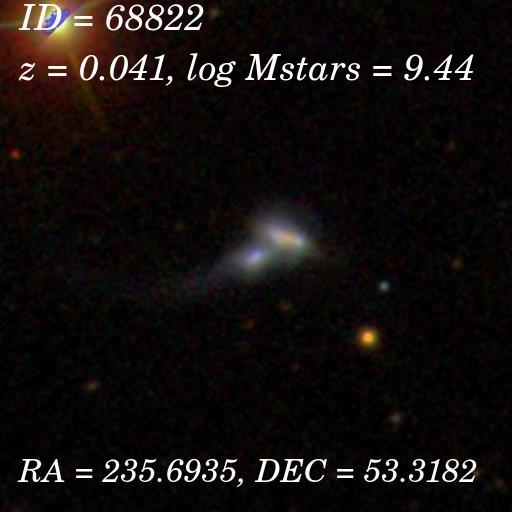} \\
 
 \includegraphics[width=0.13\hsize]{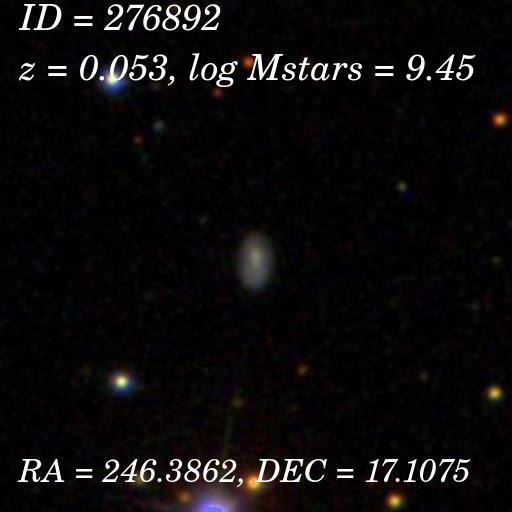} &
 \includegraphics[width=0.13\hsize]{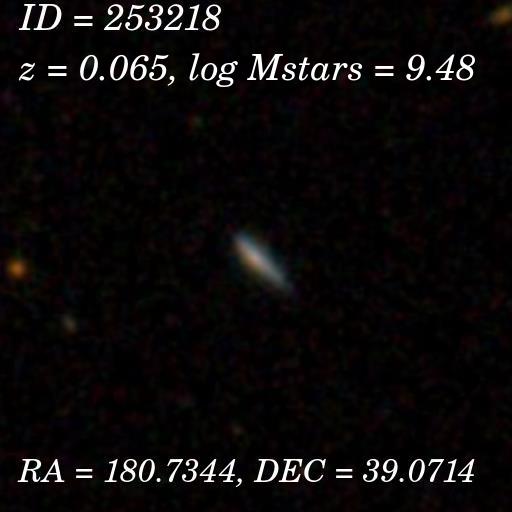} &
 \includegraphics[width=0.13\hsize]{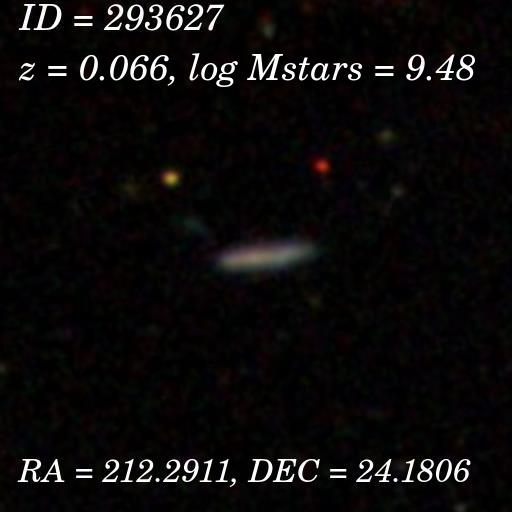} &
 \includegraphics[width=0.13\hsize]{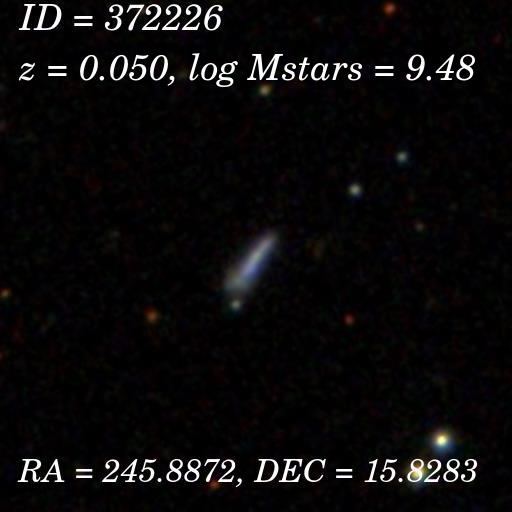} &
 \includegraphics[width=0.13\hsize]{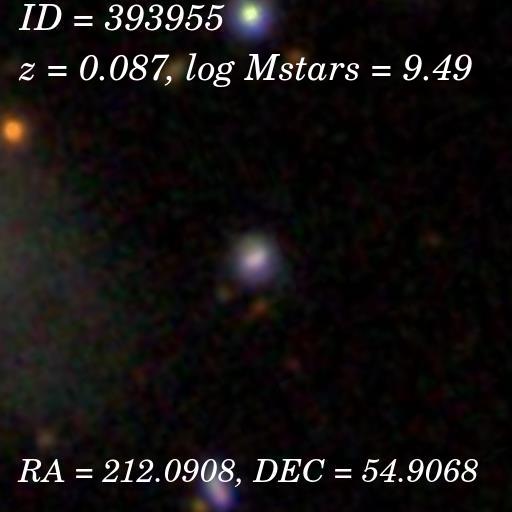} &
 \includegraphics[width=0.13\hsize]{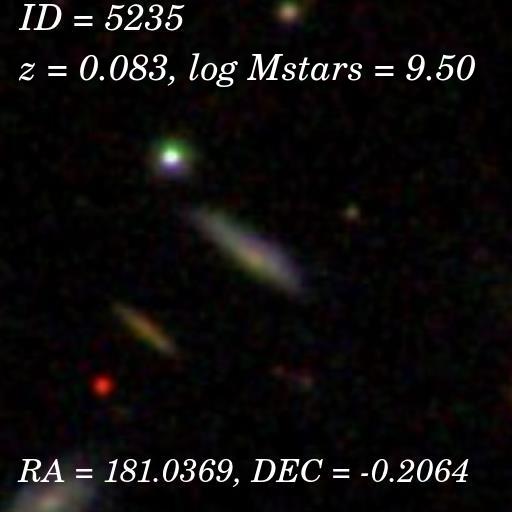} &
 \includegraphics[width=0.13\hsize]{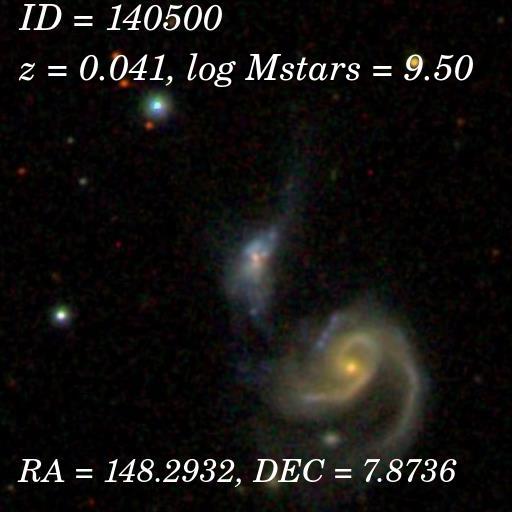} \\
 
 \includegraphics[width=0.13\hsize]{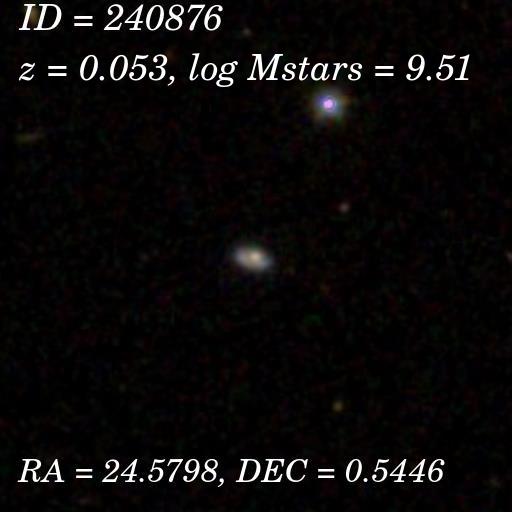} &
 \includegraphics[width=0.13\hsize]{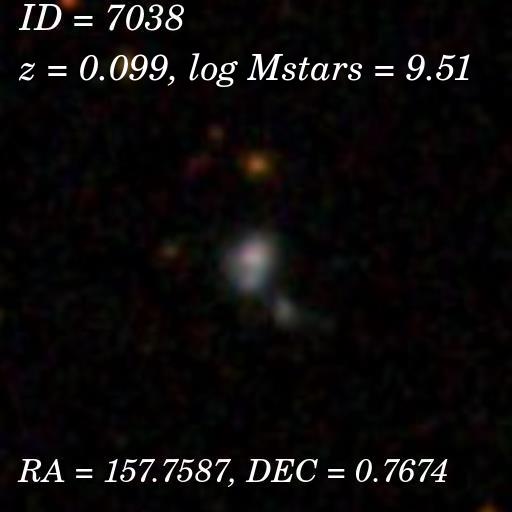} &
 \includegraphics[width=0.13\hsize]{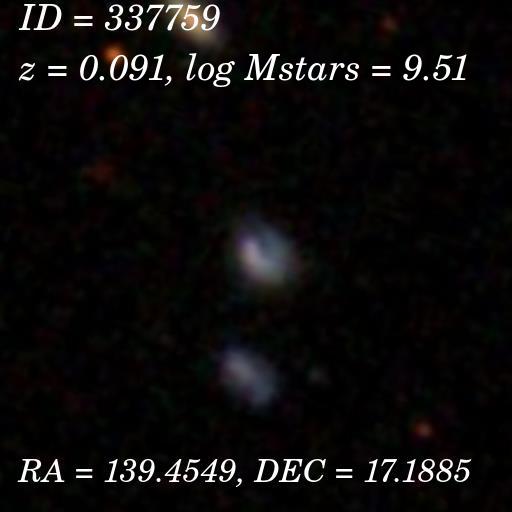} &
 \includegraphics[width=0.13\hsize]{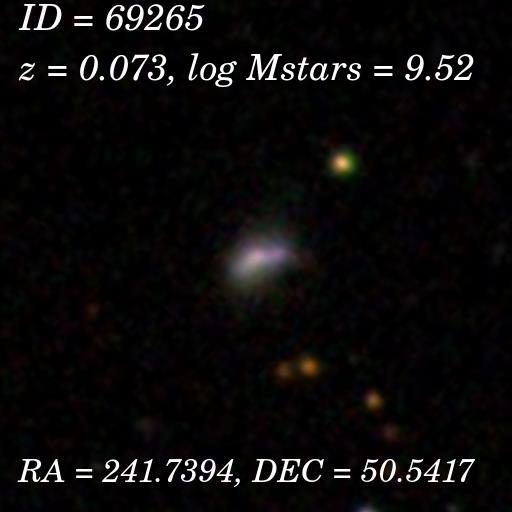} &
 \includegraphics[width=0.13\hsize]{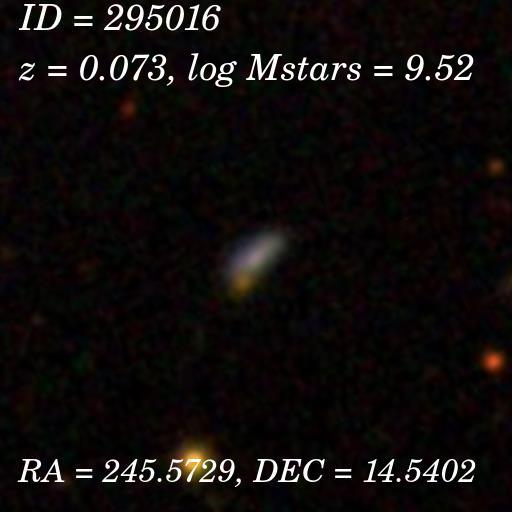} &
 \includegraphics[width=0.13\hsize]{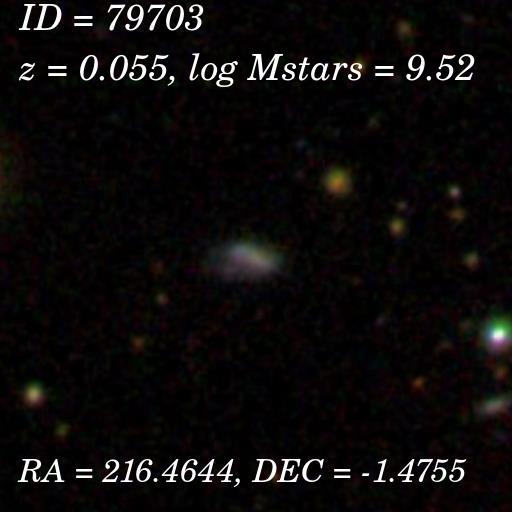} &
 \includegraphics[width=0.13\hsize]{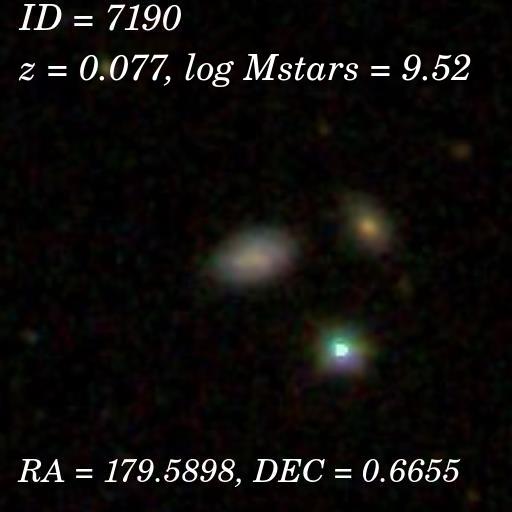} \\
 
 \includegraphics[width=0.13\hsize]{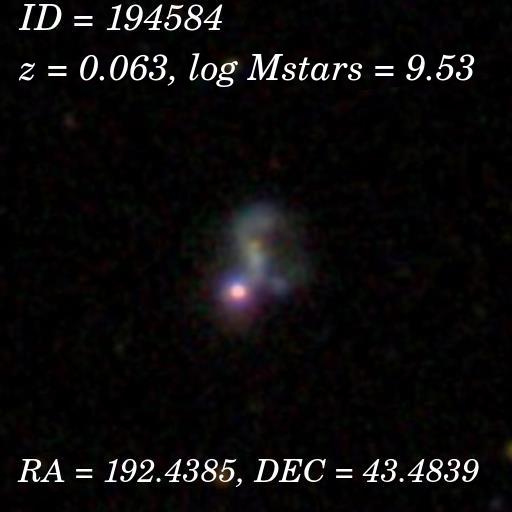} &
 \includegraphics[width=0.13\hsize]{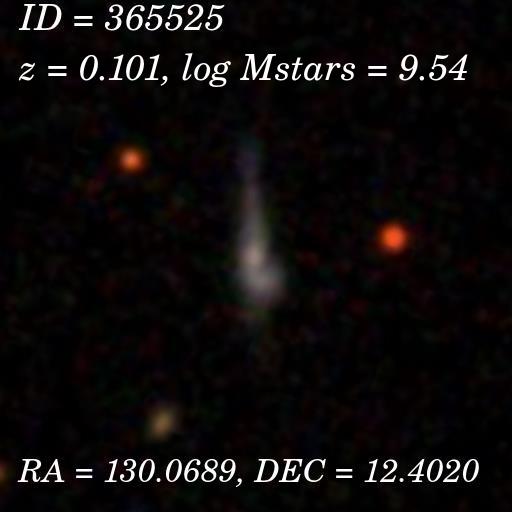} &
 \includegraphics[width=0.13\hsize]{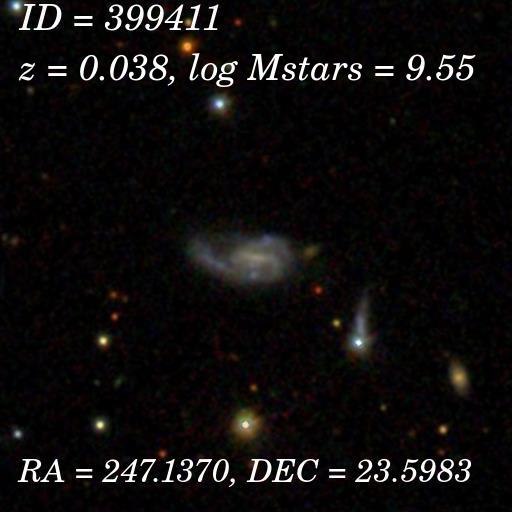} &
 \includegraphics[width=0.13\hsize]{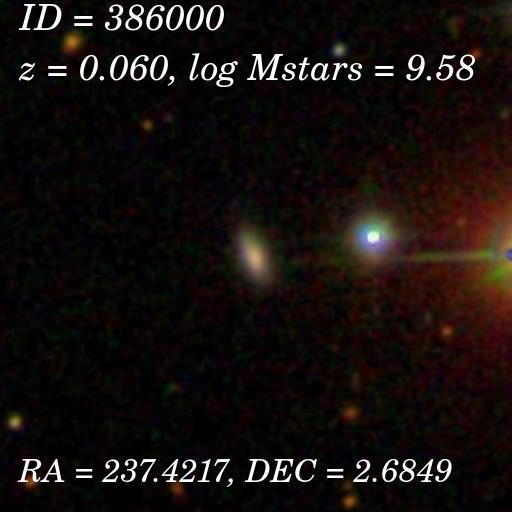} &
 \includegraphics[width=0.13\hsize]{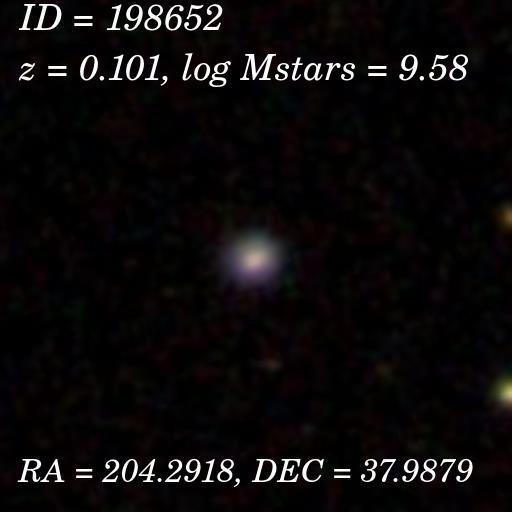} &
 \includegraphics[width=0.13\hsize]{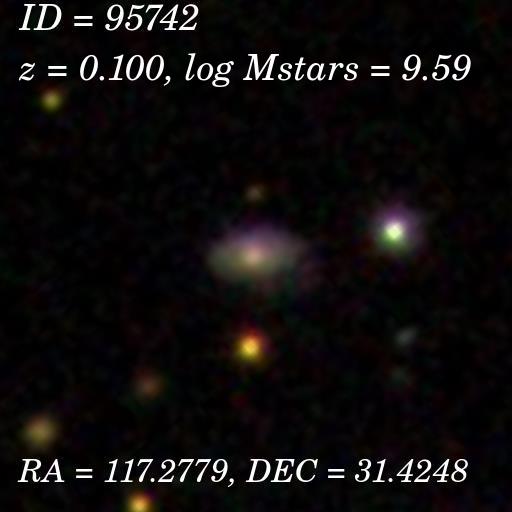} &
 \includegraphics[width=0.13\hsize]{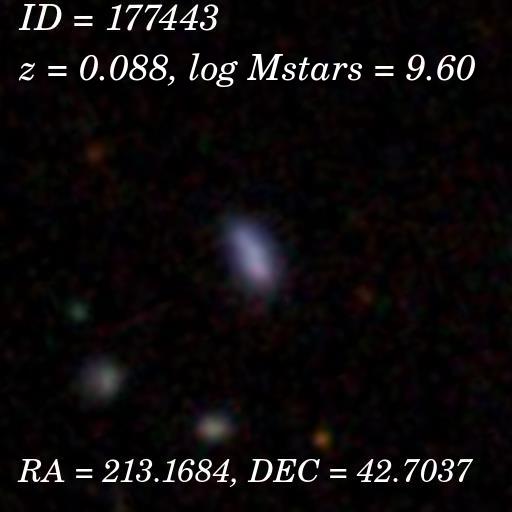} \\
 
 \includegraphics[width=0.13\hsize]{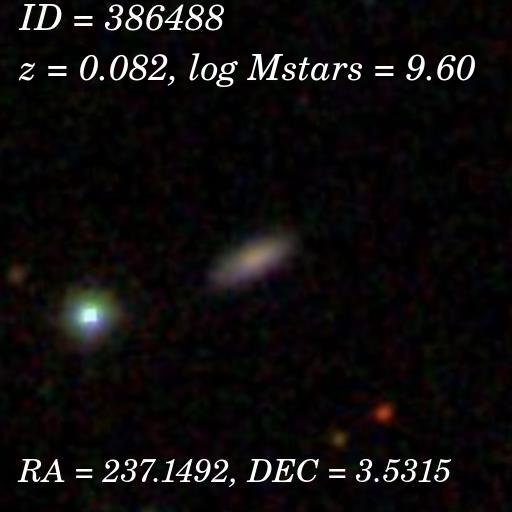} &
 \includegraphics[width=0.13\hsize]{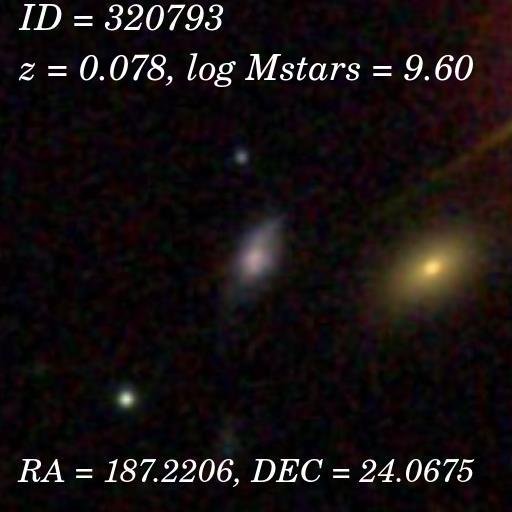} &
 \includegraphics[width=0.13\hsize]{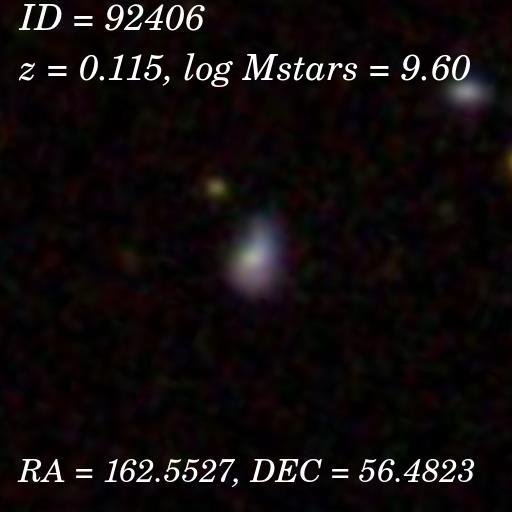} &
 \includegraphics[width=0.13\hsize]{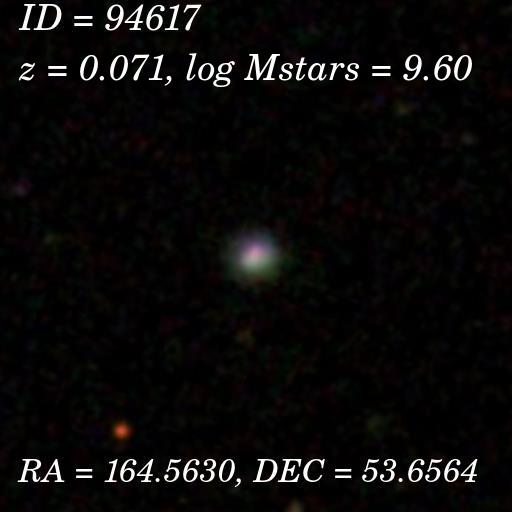} &
 \includegraphics[width=0.13\hsize]{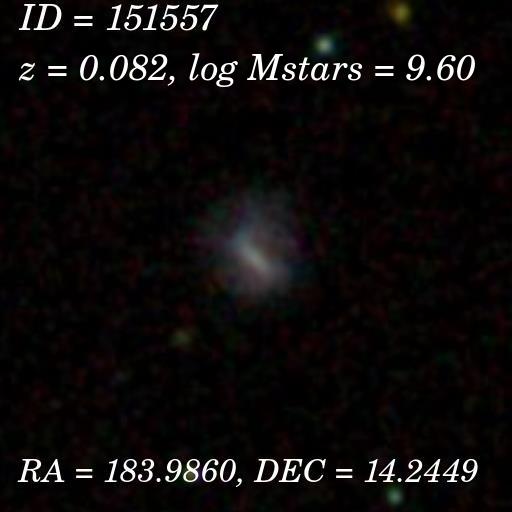} &
 \includegraphics[width=0.13\hsize]{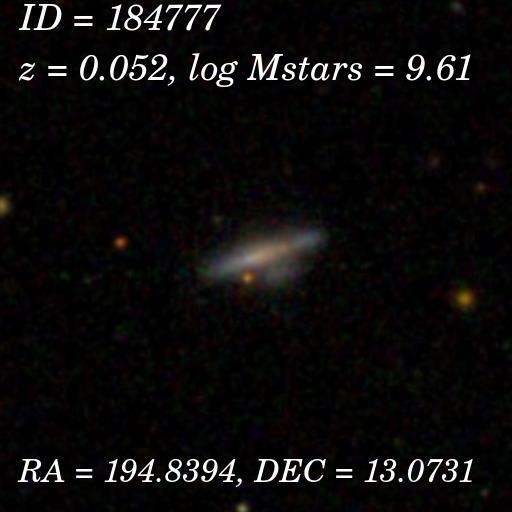} &
 \includegraphics[width=0.13\hsize]{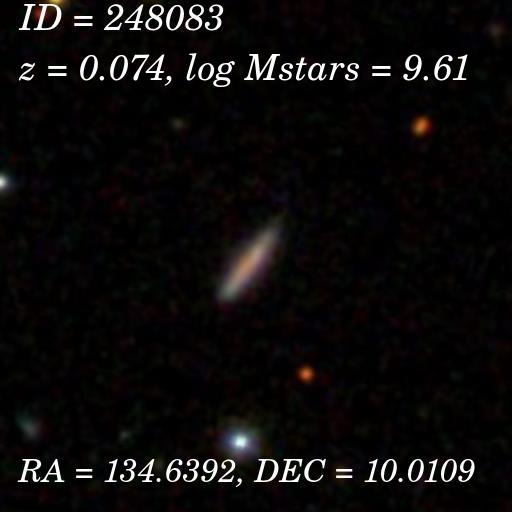} \\
 
 \includegraphics[width=0.13\hsize]{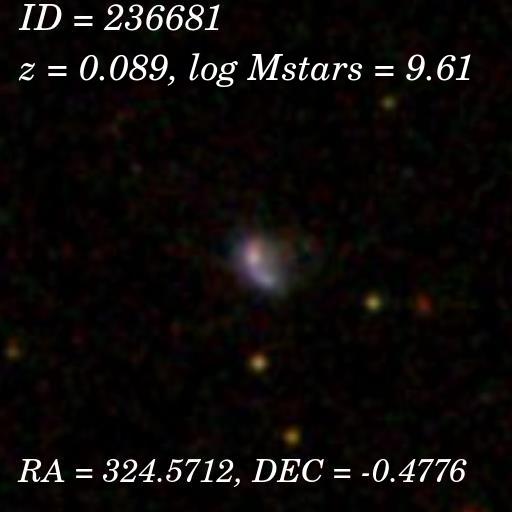} &
 \includegraphics[width=0.13\hsize]{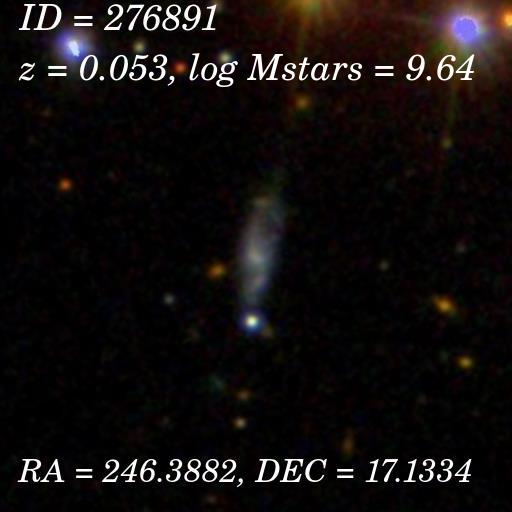} &
 \includegraphics[width=0.13\hsize]{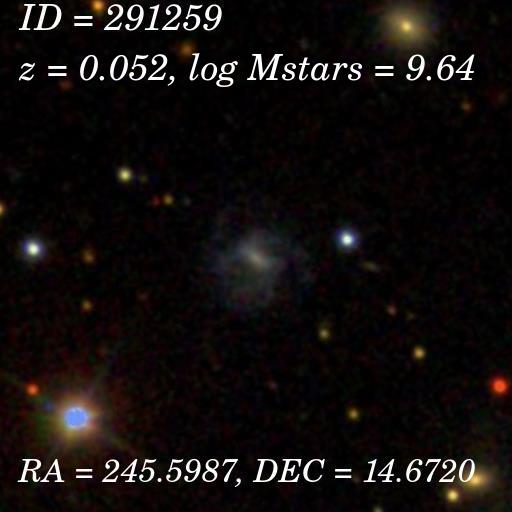} &
 \includegraphics[width=0.13\hsize]{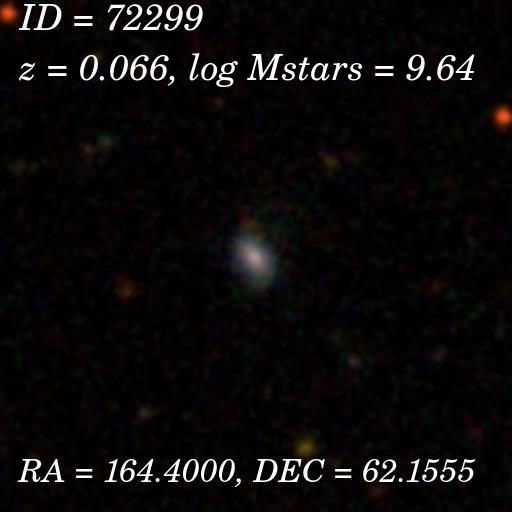} &
 \includegraphics[width=0.13\hsize]{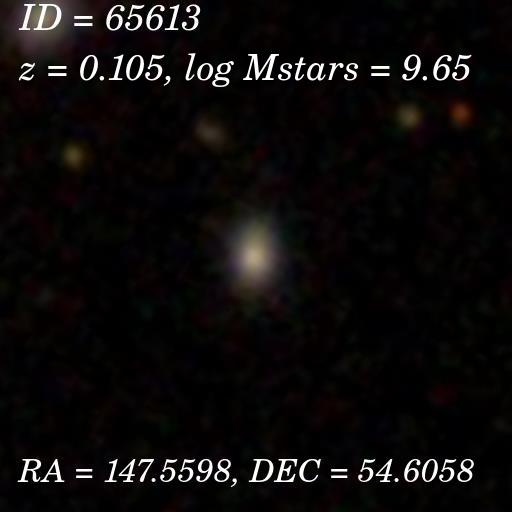} &
 \includegraphics[width=0.13\hsize]{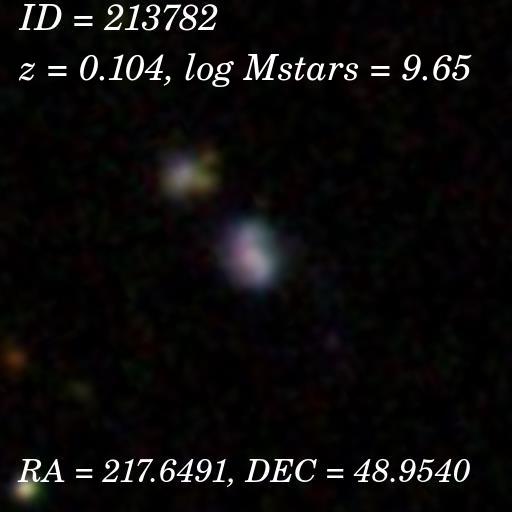} &
 \includegraphics[width=0.13\hsize]{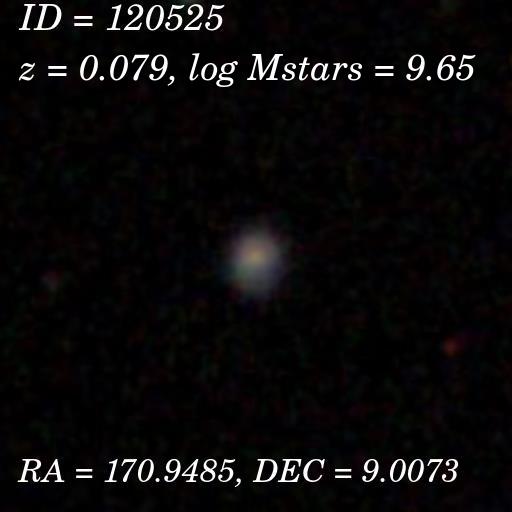} \\
 
 \end{tabular}
 \caption{SDSS images of a sample of 207 VYGs. All images are 100 x 100 kpc (\emph{cont.}).}
\label{fig:vyg_images2}
\end{figure*}

\begin{figure*}
\setlength{\tabcolsep}{1pt}
\begin{tabular}{ccccccc}
 \includegraphics[width=0.13\hsize]{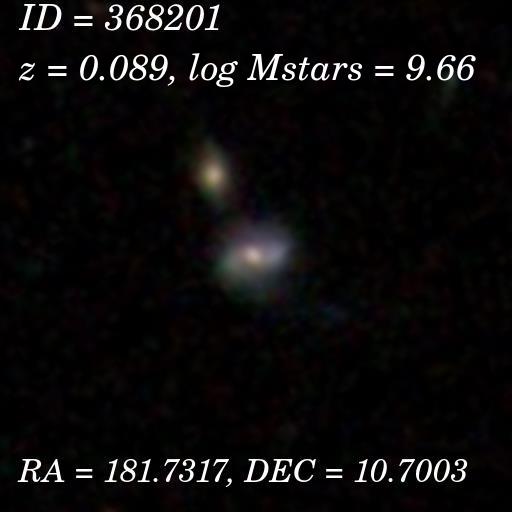} &
 \includegraphics[width=0.13\hsize]{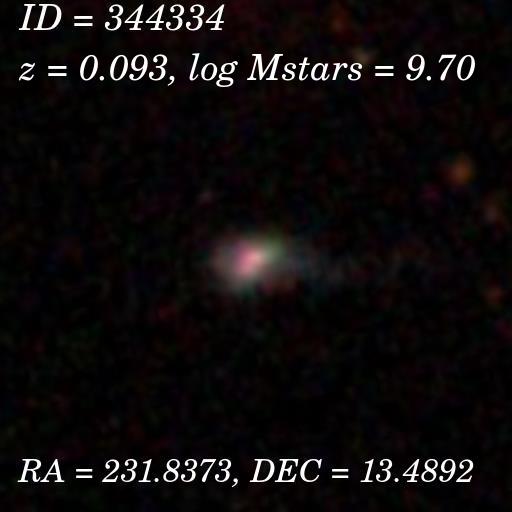} &
 \includegraphics[width=0.13\hsize]{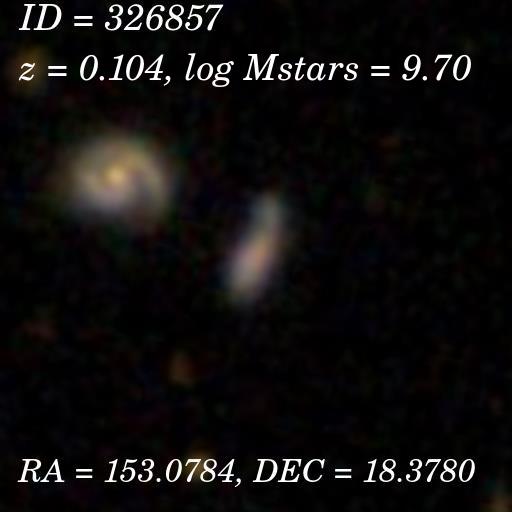} &
 \includegraphics[width=0.13\hsize]{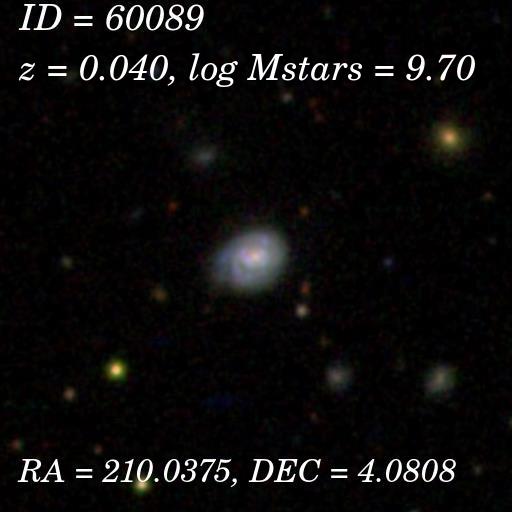} &
 \includegraphics[width=0.13\hsize]{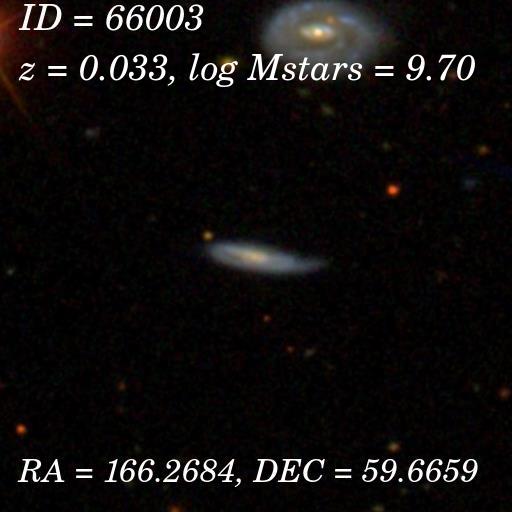} &
 \includegraphics[width=0.13\hsize]{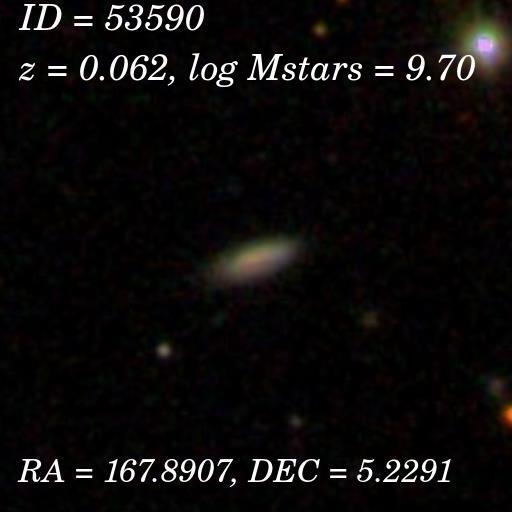} &
 \includegraphics[width=0.13\hsize]{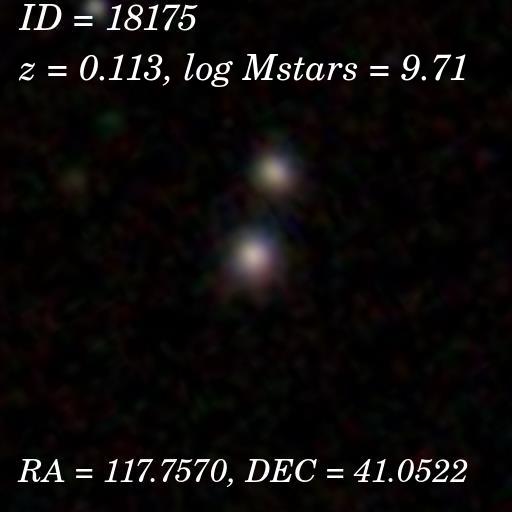} \\
 
 \includegraphics[width=0.13\hsize]{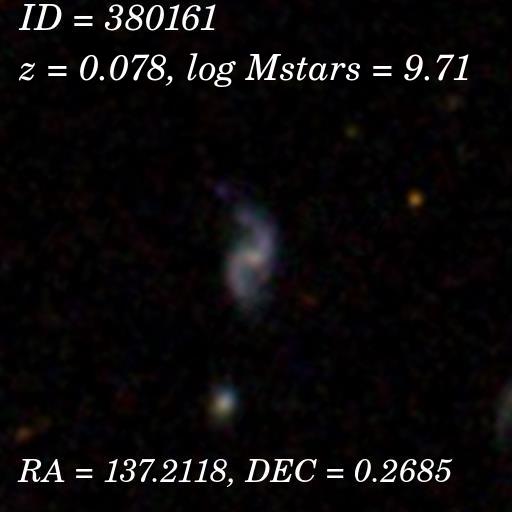} &
 \includegraphics[width=0.13\hsize]{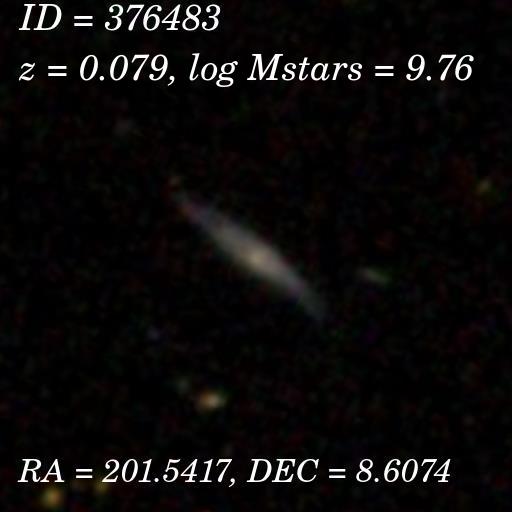} &
 \includegraphics[width=0.13\hsize]{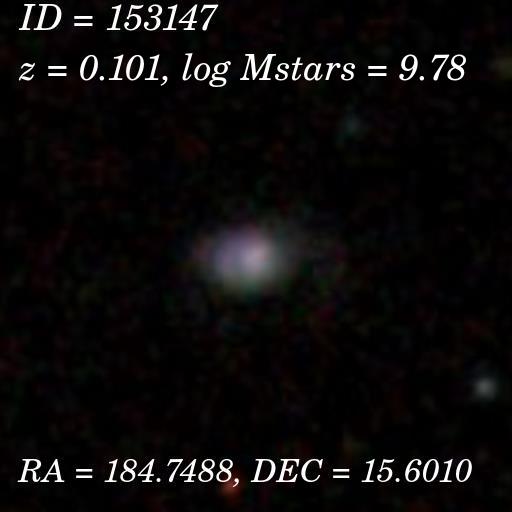} &
 \includegraphics[width=0.13\hsize]{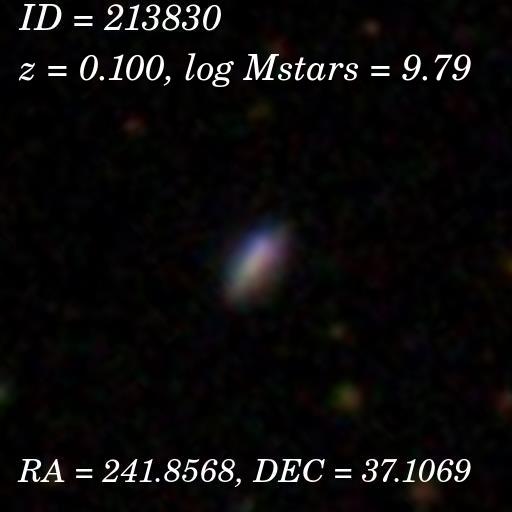} &
 \includegraphics[width=0.13\hsize]{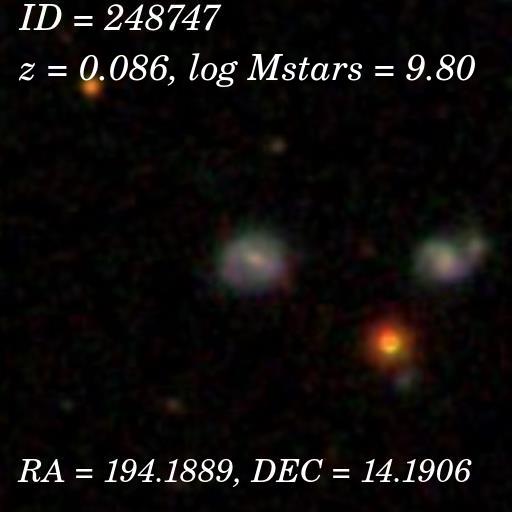} &
 \includegraphics[width=0.13\hsize]{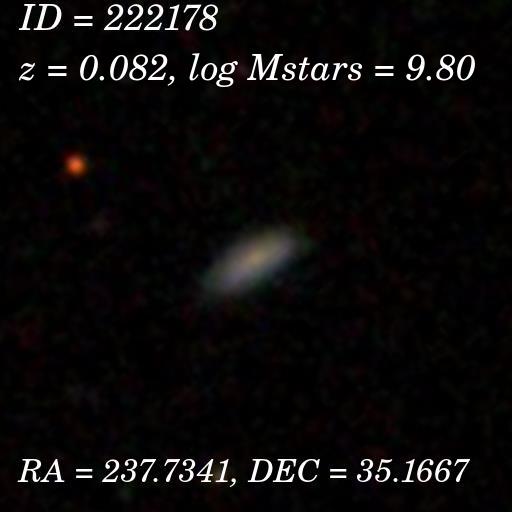} &
 \includegraphics[width=0.13\hsize]{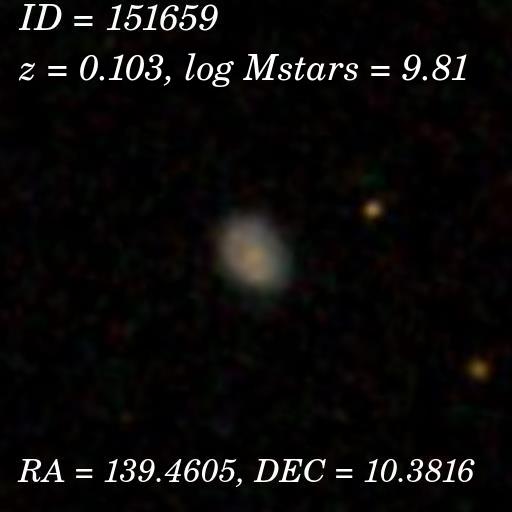} \\
 
 \includegraphics[width=0.13\hsize]{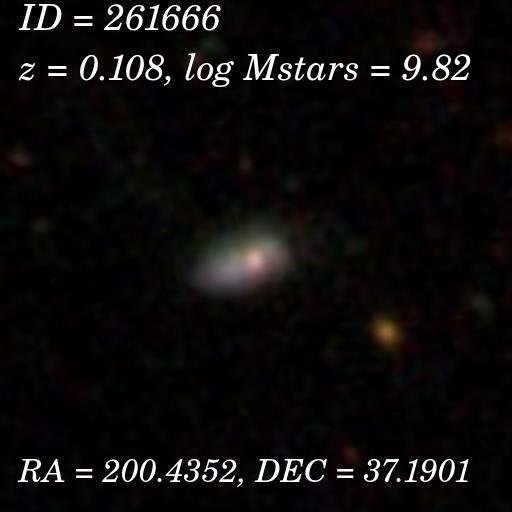} &
 \includegraphics[width=0.13\hsize]{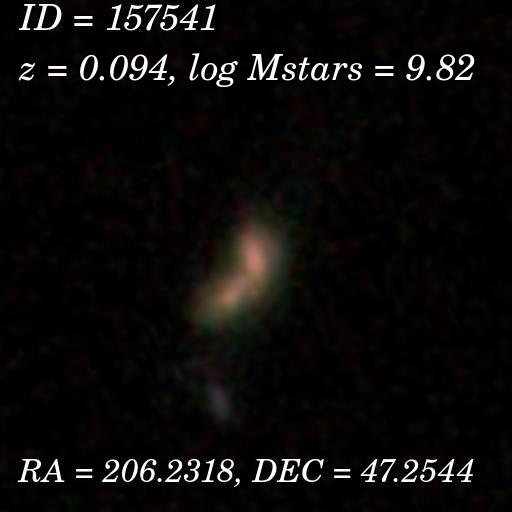} &
 \includegraphics[width=0.13\hsize]{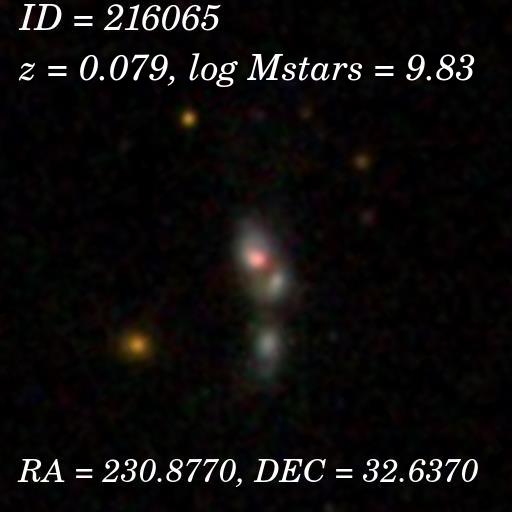} &
 \includegraphics[width=0.13\hsize]{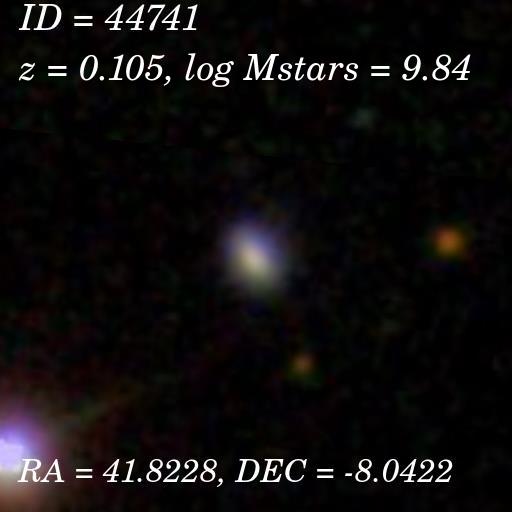} &
 \includegraphics[width=0.13\hsize]{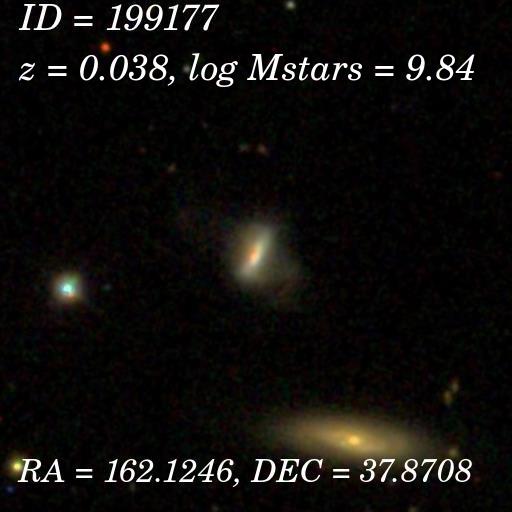} &
 \includegraphics[width=0.13\hsize]{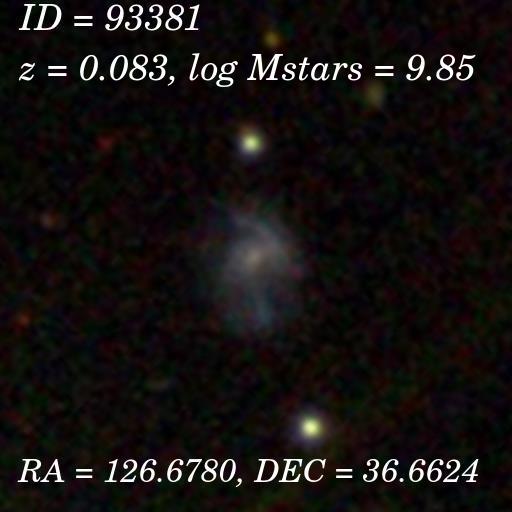} &
 \includegraphics[width=0.13\hsize]{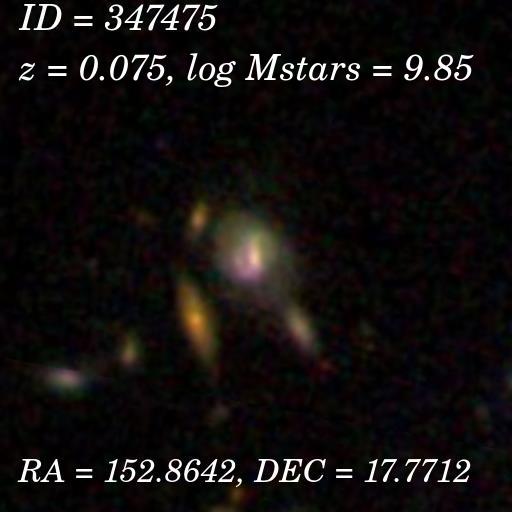} \\
 
 \includegraphics[width=0.13\hsize]{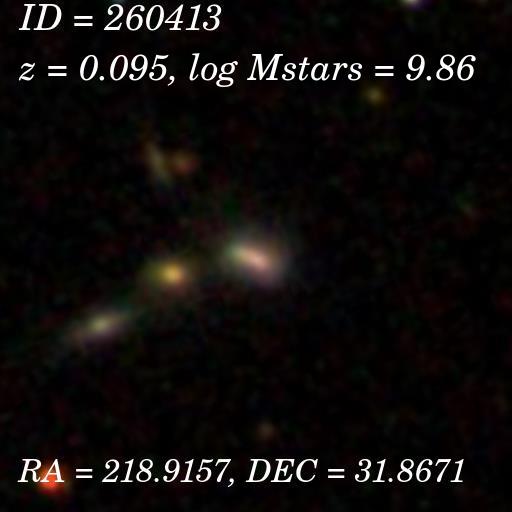} &
 \includegraphics[width=0.13\hsize]{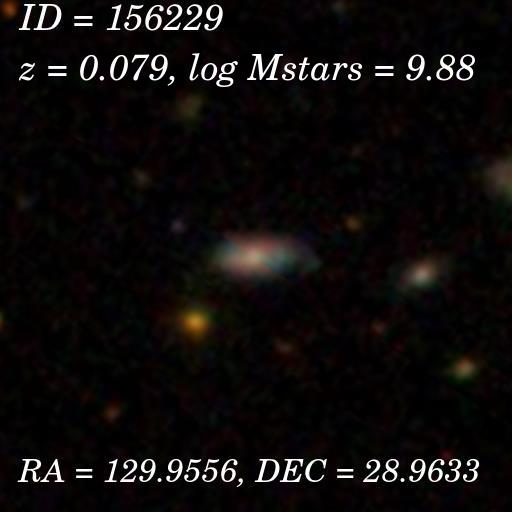} &
 \includegraphics[width=0.13\hsize]{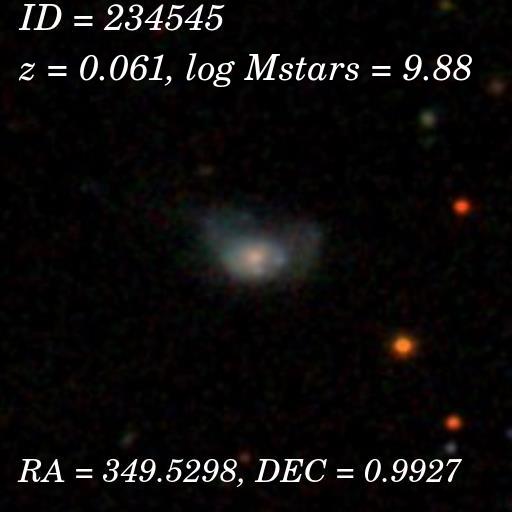} &
 \includegraphics[width=0.13\hsize]{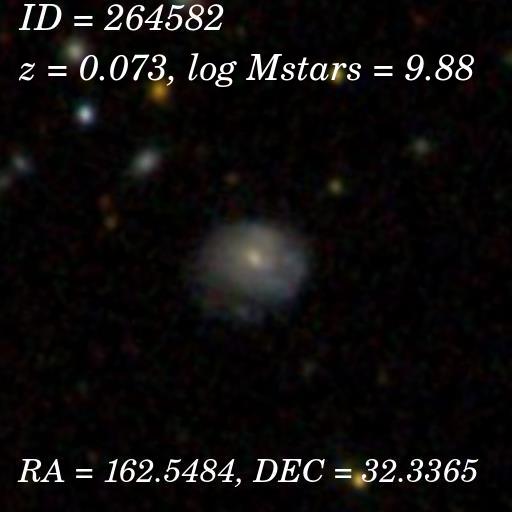} &
 \includegraphics[width=0.13\hsize]{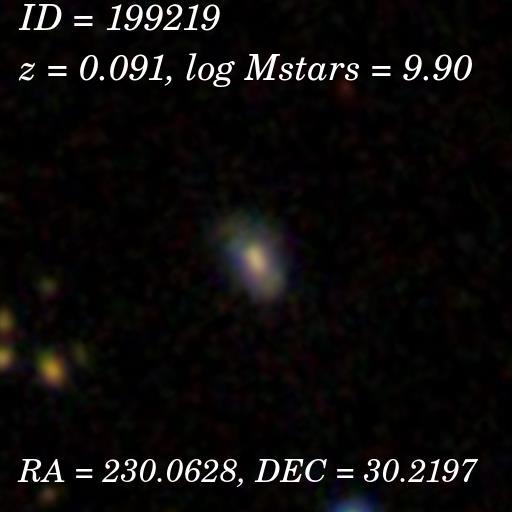} &
 \includegraphics[width=0.13\hsize]{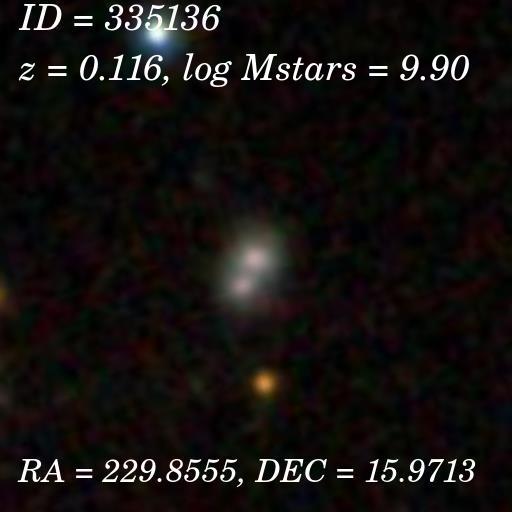} &
 \includegraphics[width=0.13\hsize]{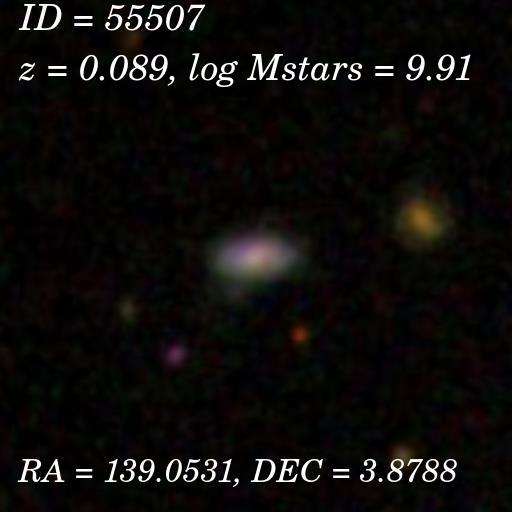} \\
 
 \includegraphics[width=0.13\hsize]{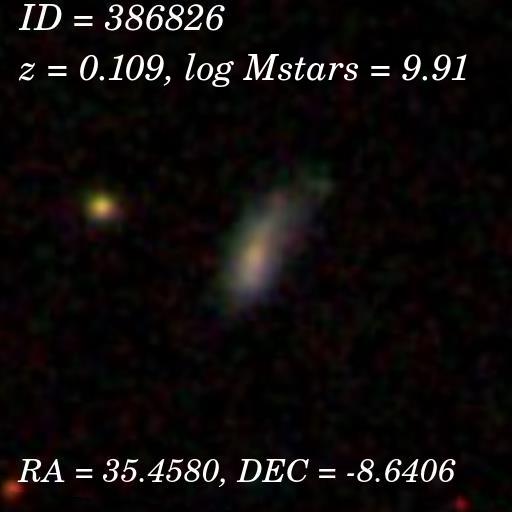} &
 \includegraphics[width=0.13\hsize]{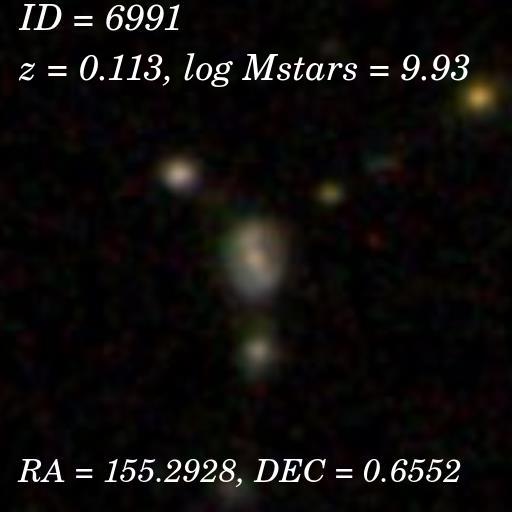} &
 \includegraphics[width=0.13\hsize]{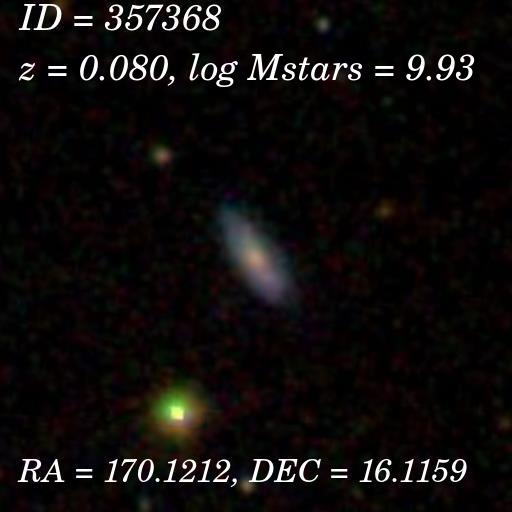} &
 \includegraphics[width=0.13\hsize]{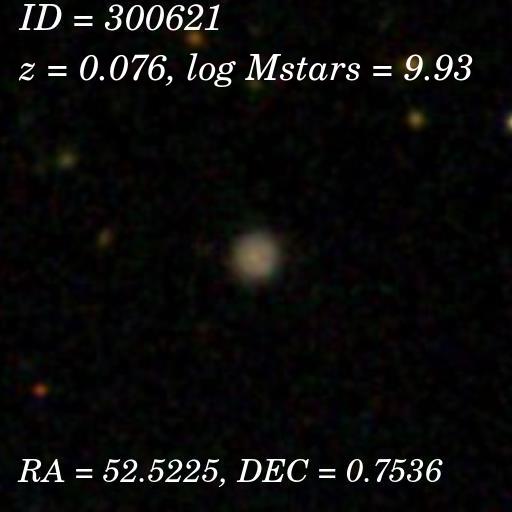} &
 \includegraphics[width=0.13\hsize]{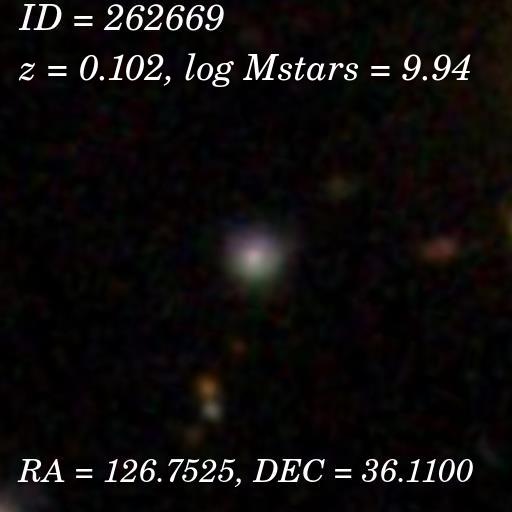} &
 \includegraphics[width=0.13\hsize]{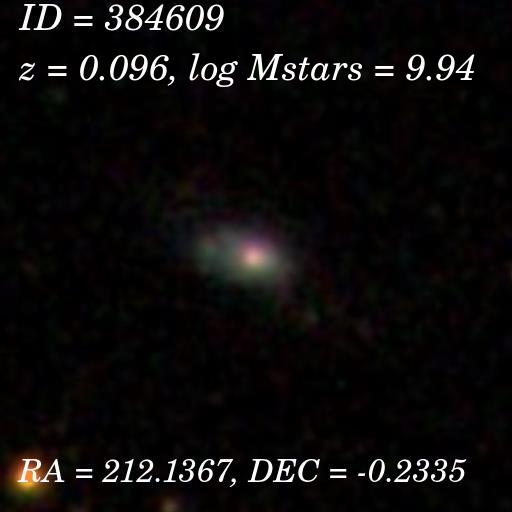} &
 \includegraphics[width=0.13\hsize]{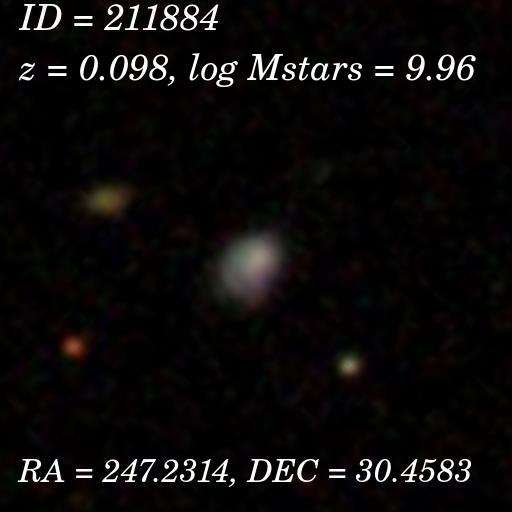} \\
 
 \includegraphics[width=0.13\hsize]{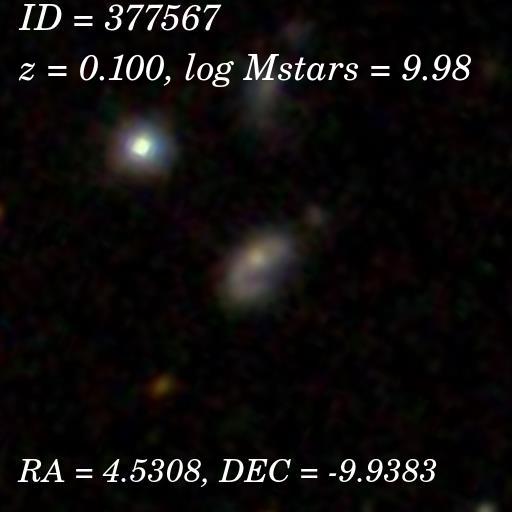} &
 \includegraphics[width=0.13\hsize]{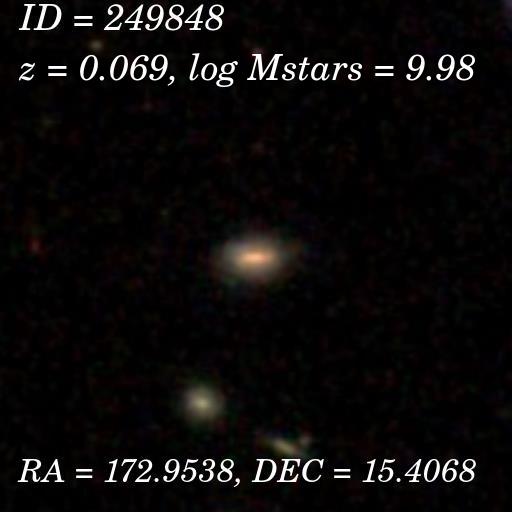} &
 \includegraphics[width=0.13\hsize]{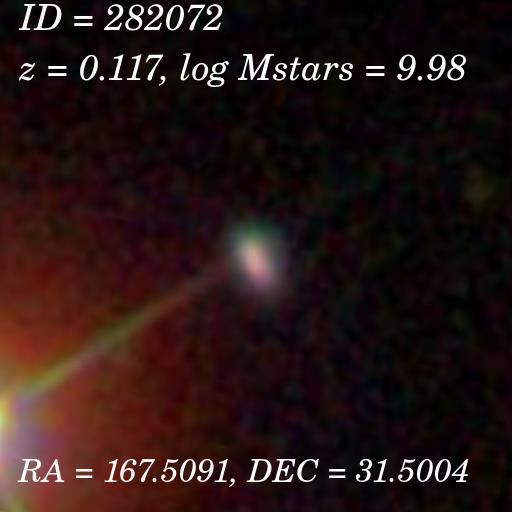} &
 \includegraphics[width=0.13\hsize]{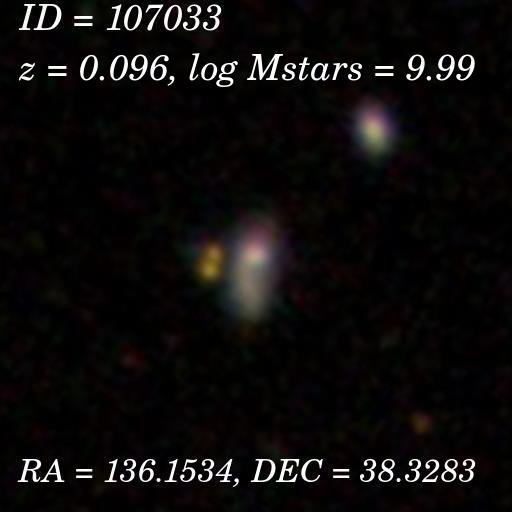} &
 \includegraphics[width=0.13\hsize]{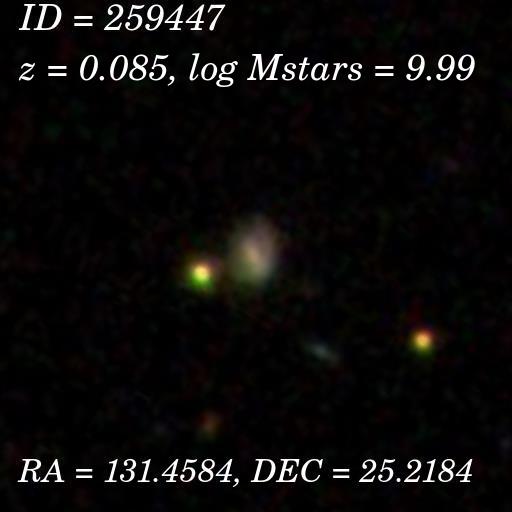} &
 \includegraphics[width=0.13\hsize]{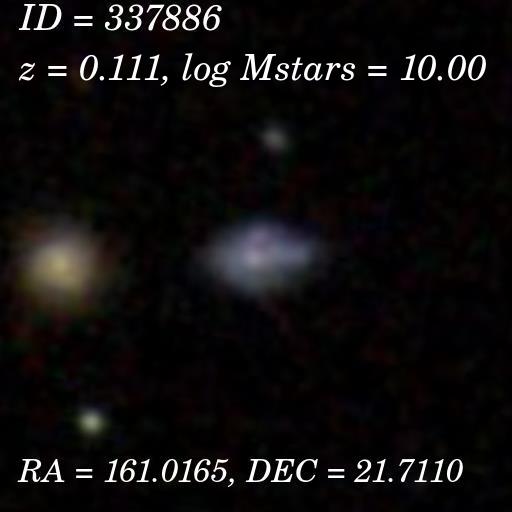} &
 \includegraphics[width=0.13\hsize]{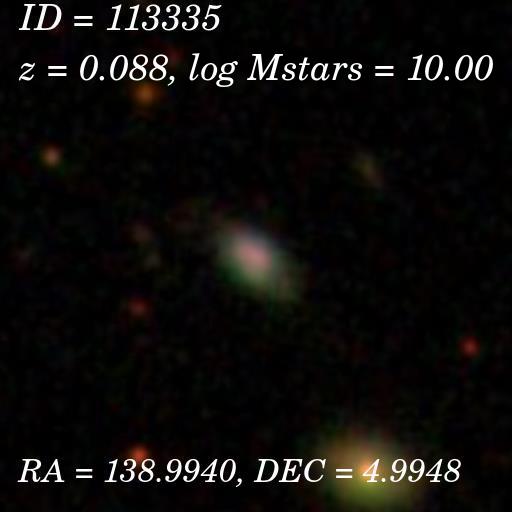} \\
 
 \includegraphics[width=0.13\hsize]{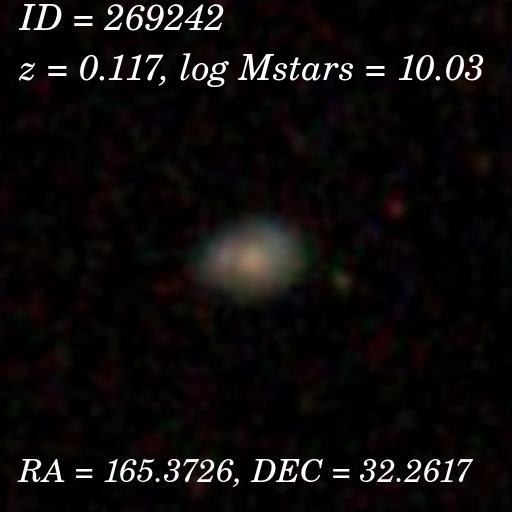} &
 \includegraphics[width=0.13\hsize]{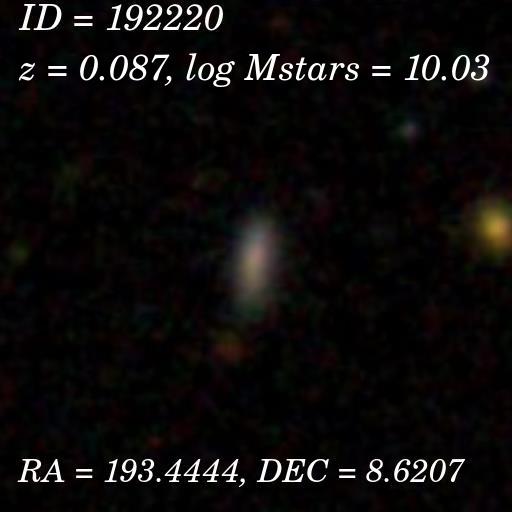} &
 \includegraphics[width=0.13\hsize]{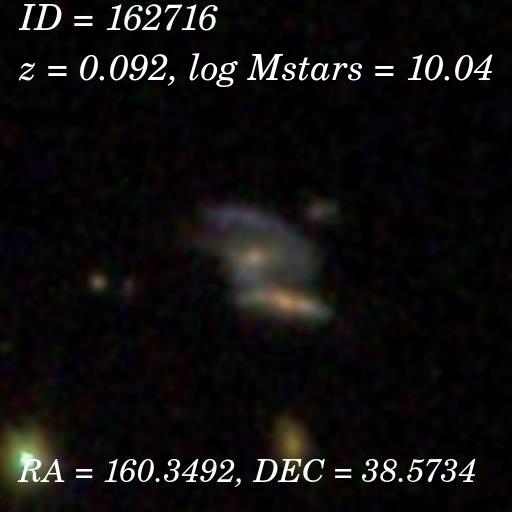} &
 \includegraphics[width=0.13\hsize]{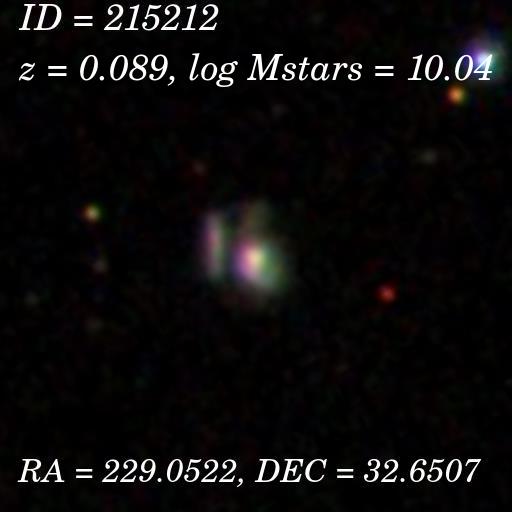} &
 \includegraphics[width=0.13\hsize]{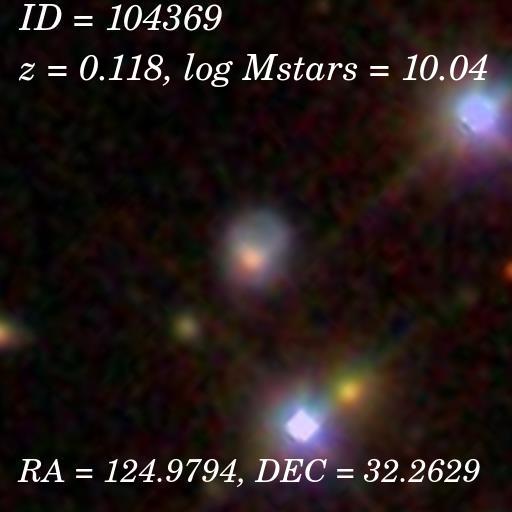} &
 \includegraphics[width=0.13\hsize]{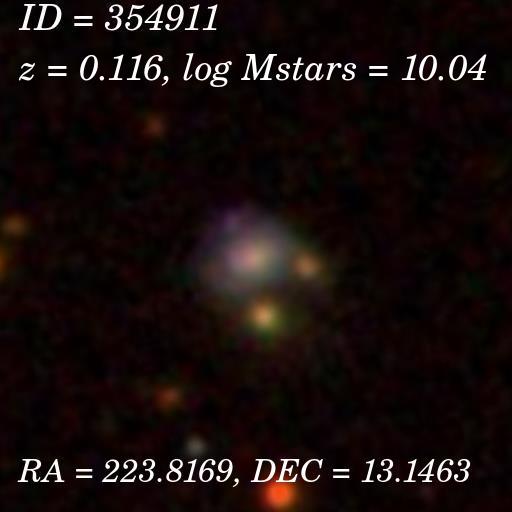} &
 \includegraphics[width=0.13\hsize]{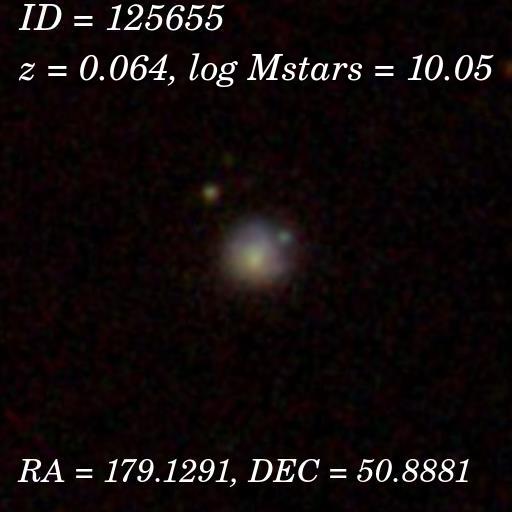} \\
 
 \includegraphics[width=0.13\hsize]{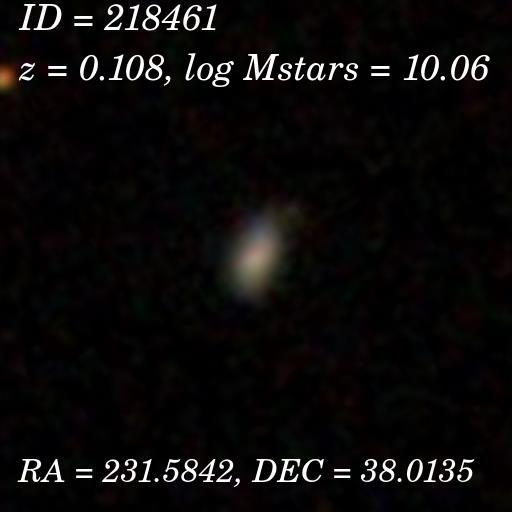} &
 \includegraphics[width=0.13\hsize]{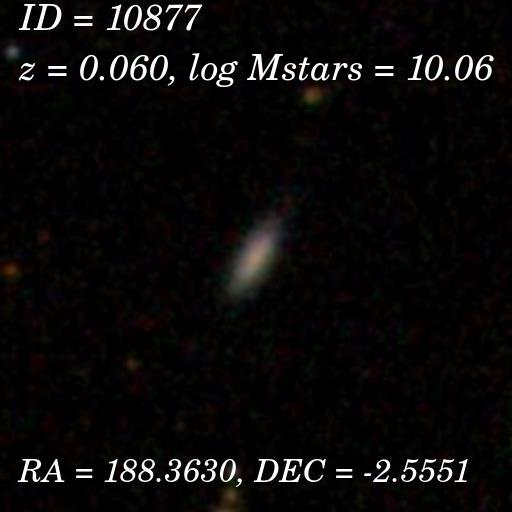} &
 \includegraphics[width=0.13\hsize]{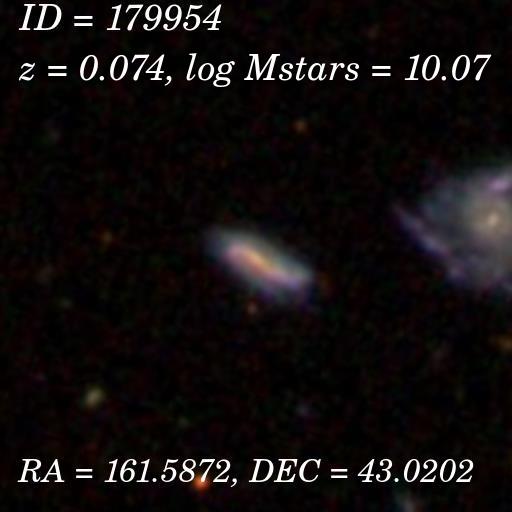} &
 \includegraphics[width=0.13\hsize]{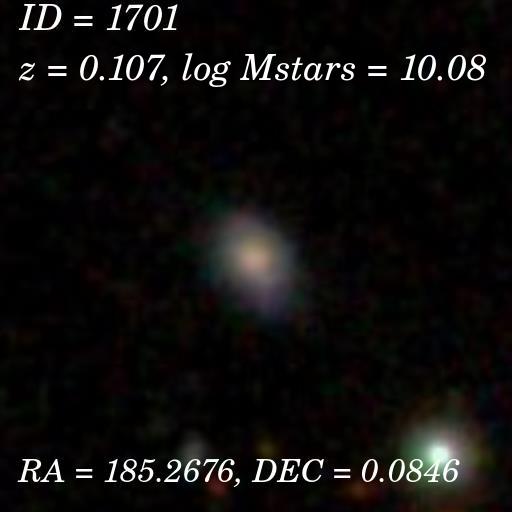} &
 \includegraphics[width=0.13\hsize]{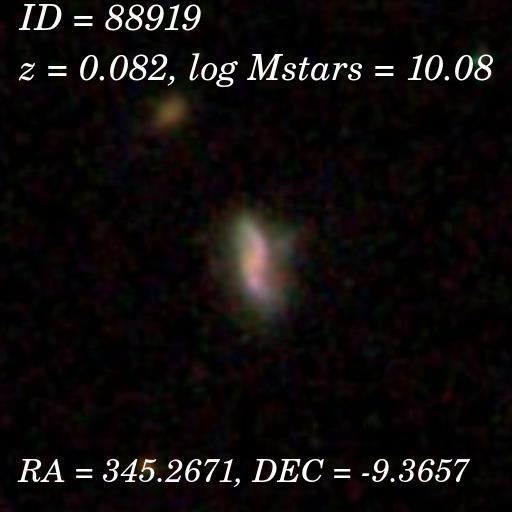} &
 \includegraphics[width=0.13\hsize]{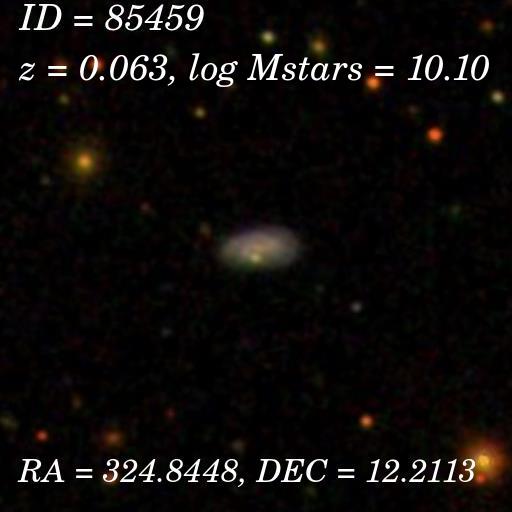} &
 \includegraphics[width=0.13\hsize]{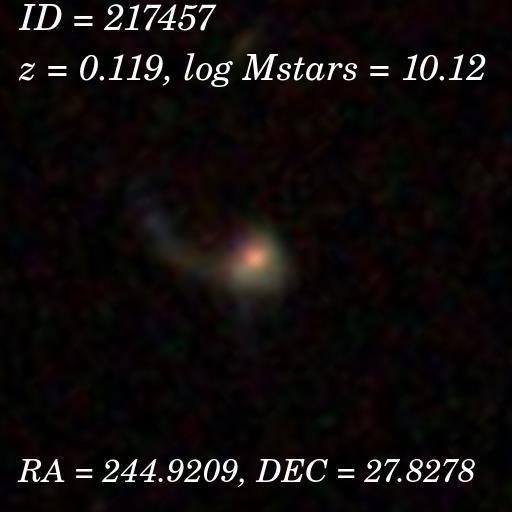} \\
 
 \includegraphics[width=0.13\hsize]{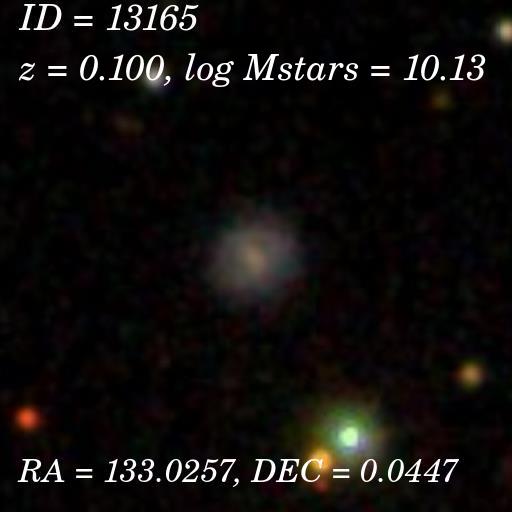} &
 \includegraphics[width=0.13\hsize]{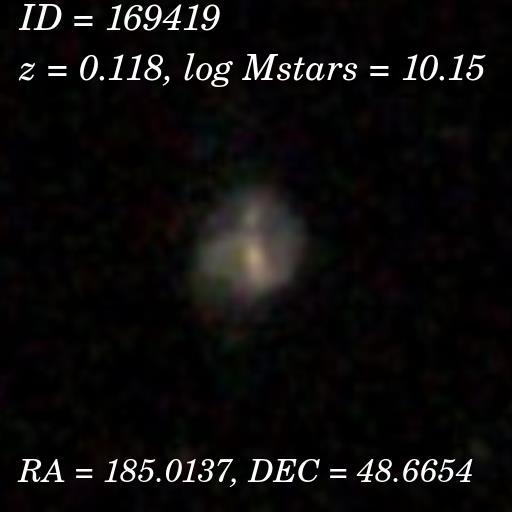} &
 \includegraphics[width=0.13\hsize]{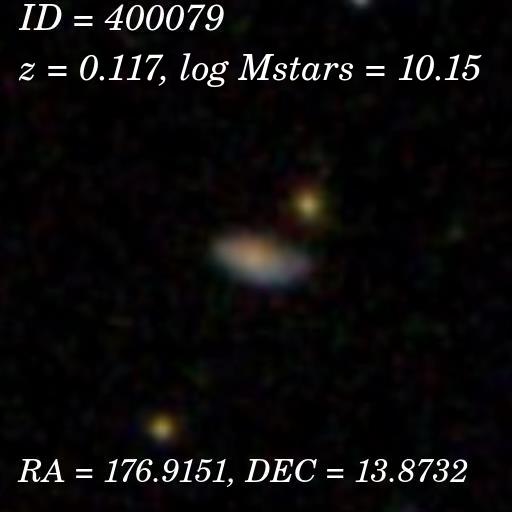} &
 \includegraphics[width=0.13\hsize]{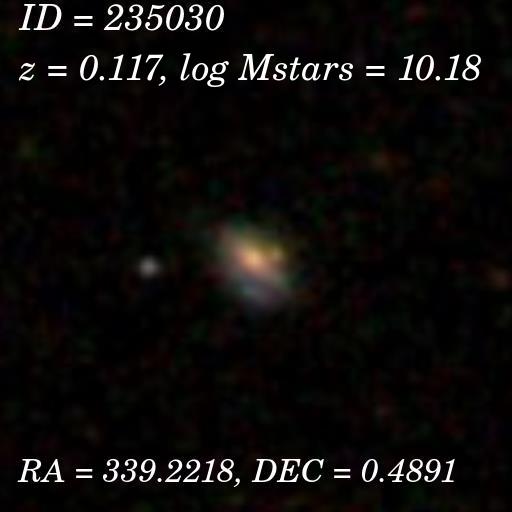} &
 \includegraphics[width=0.13\hsize]{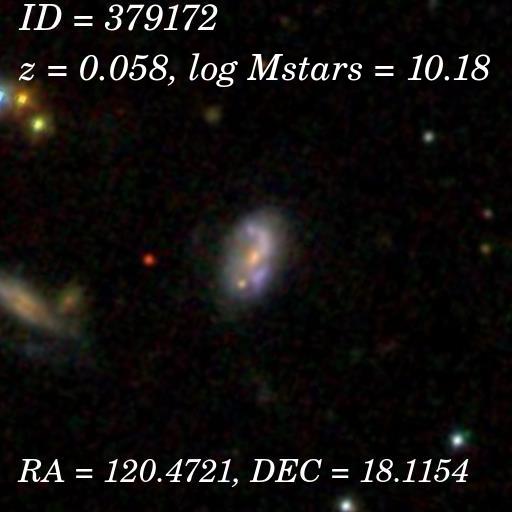} &
 \includegraphics[width=0.13\hsize]{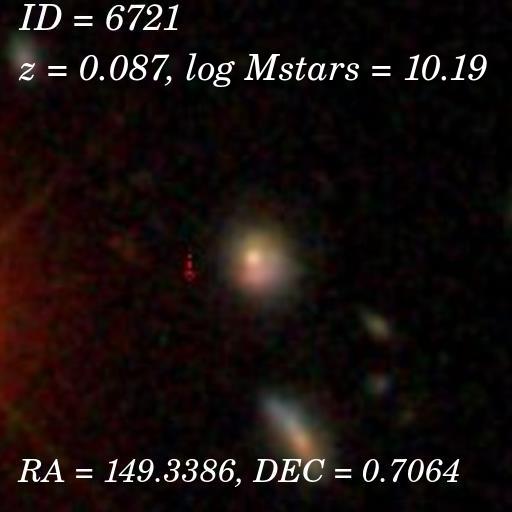} &
 \includegraphics[width=0.13\hsize]{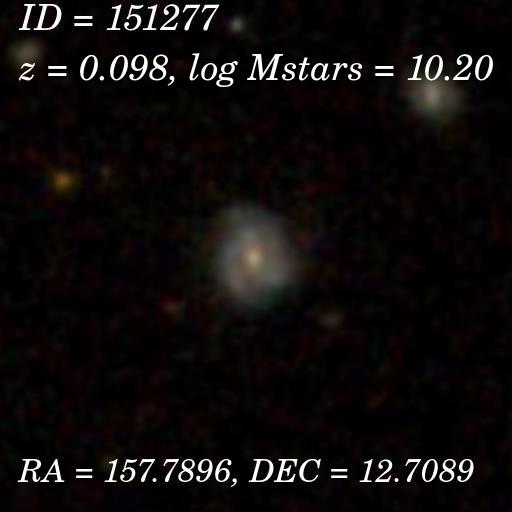} \\
 
 \end{tabular}
 \caption{SDSS images of a sample of 207 VYGs. All images are 100 x 100 kpc (\emph{cont.}).}
\label{fig:vyg_images3}
\end{figure*}

\begin{figure*}
\setlength{\tabcolsep}{1pt}
\begin{tabular}{ccccccc}
 \includegraphics[width=0.13\hsize]{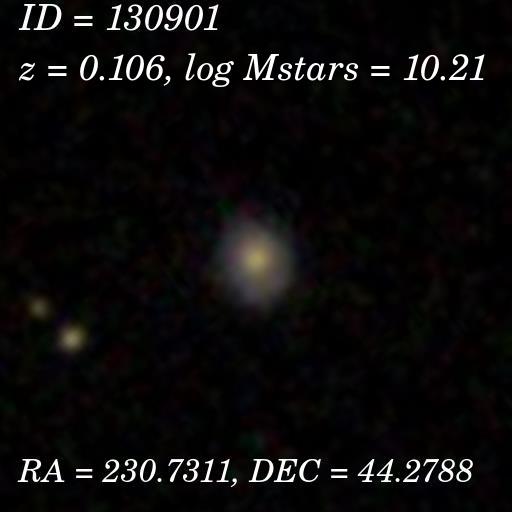} &
 \includegraphics[width=0.13\hsize]{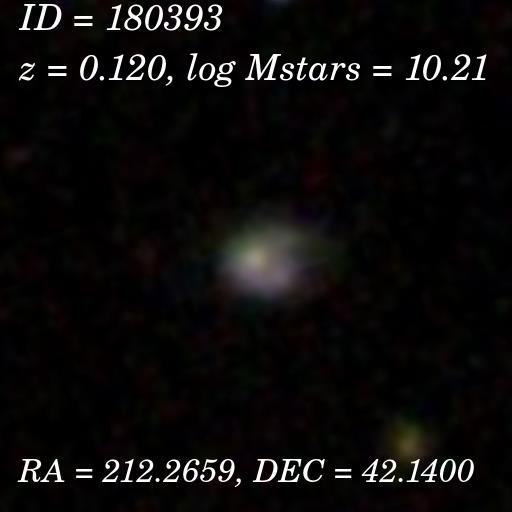} &
 \includegraphics[width=0.13\hsize]{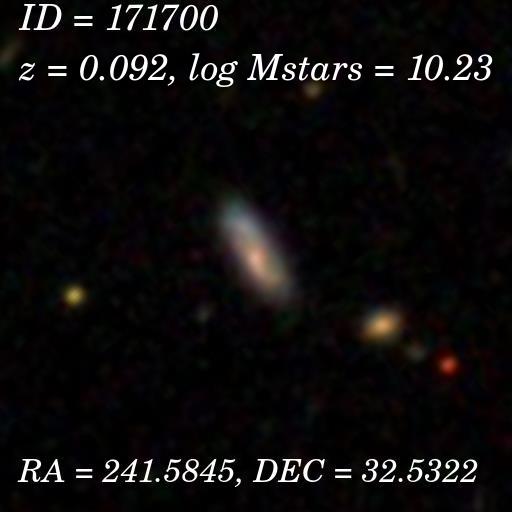} &
 \includegraphics[width=0.13\hsize]{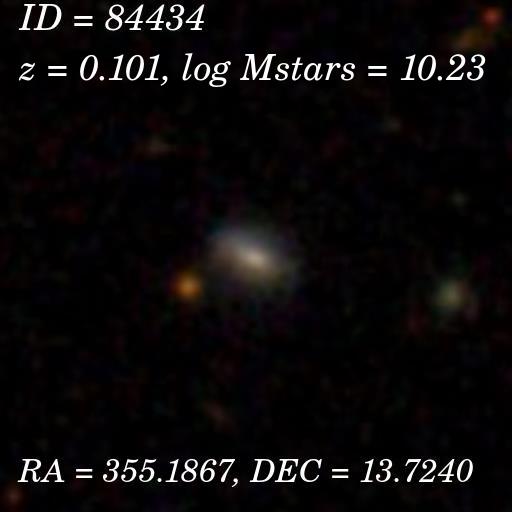} &
 \includegraphics[width=0.13\hsize]{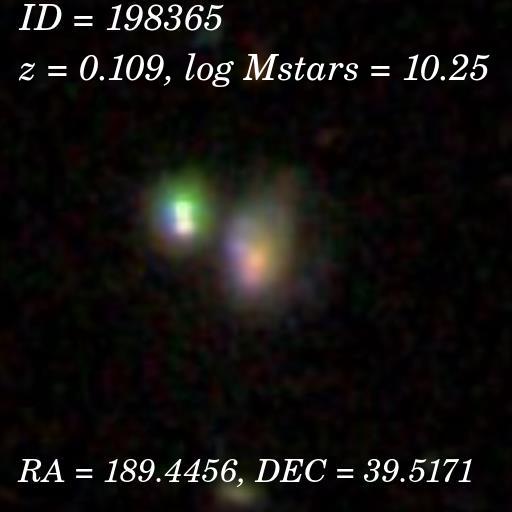} &
 \includegraphics[width=0.13\hsize]{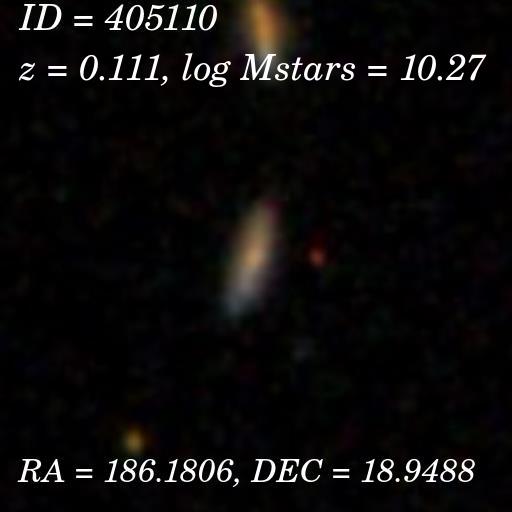} &
 \includegraphics[width=0.13\hsize]{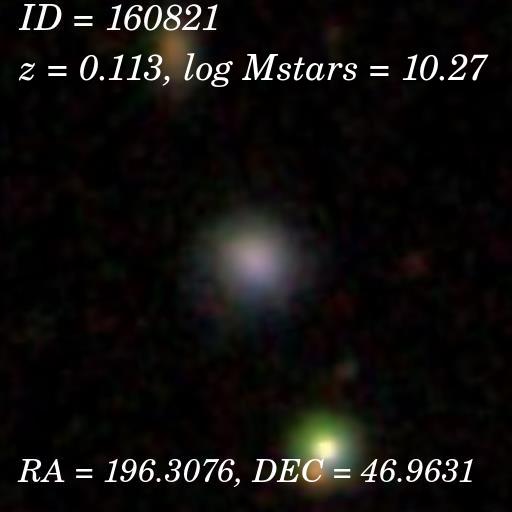} \\
 
 \includegraphics[width=0.13\hsize]{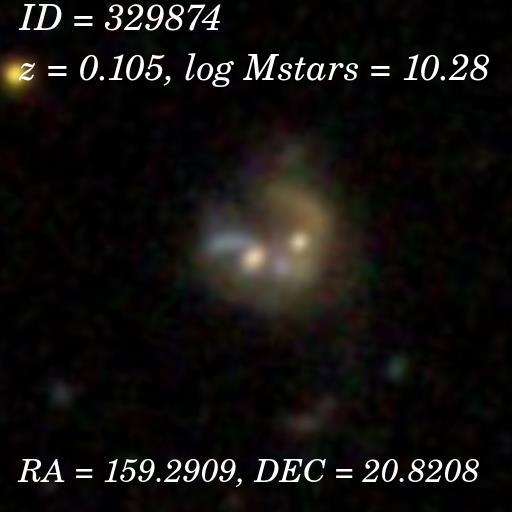} &
 \includegraphics[width=0.13\hsize]{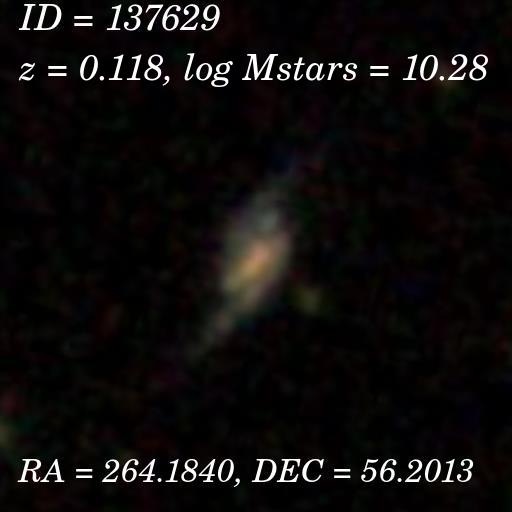} &
 \includegraphics[width=0.13\hsize]{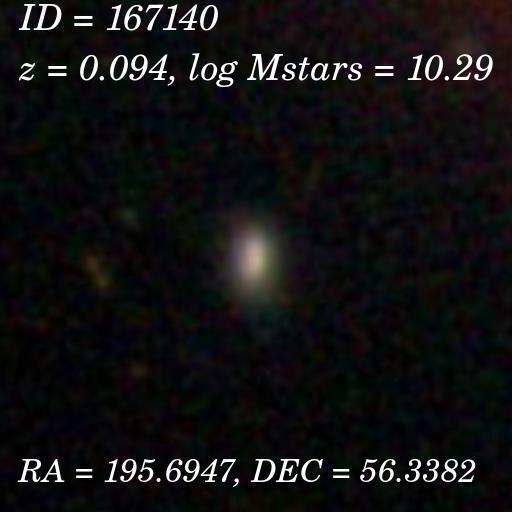} &
 \includegraphics[width=0.13\hsize]{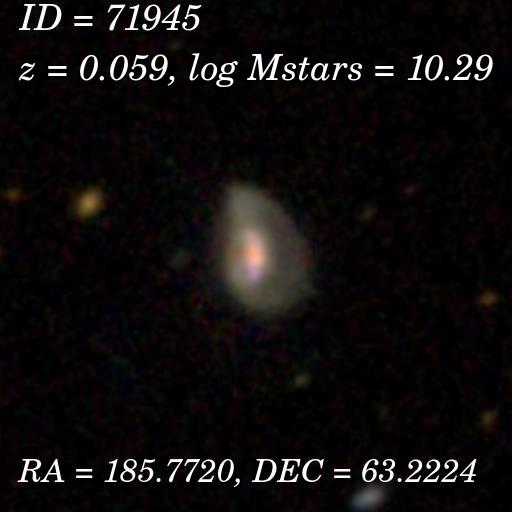} &
 \includegraphics[width=0.13\hsize]{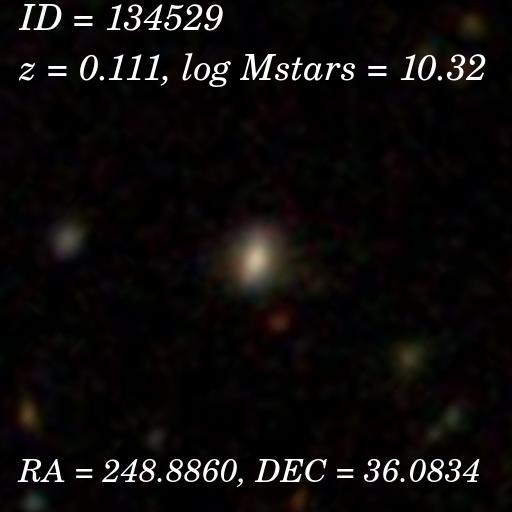} &
 \includegraphics[width=0.13\hsize]{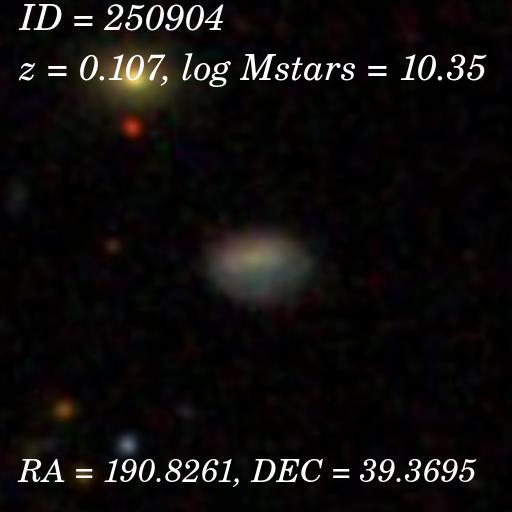} &
 \includegraphics[width=0.13\hsize]{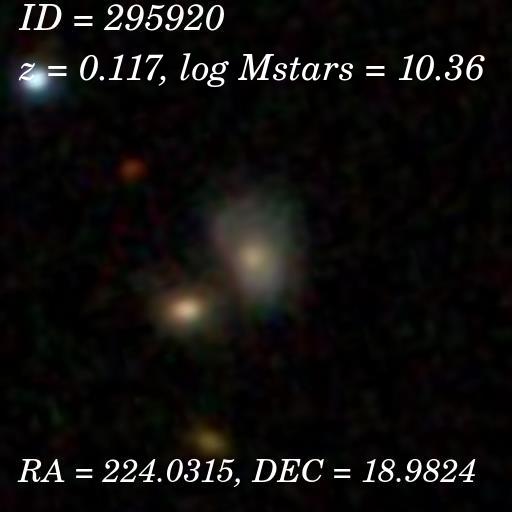} \\
 
 \includegraphics[width=0.13\hsize]{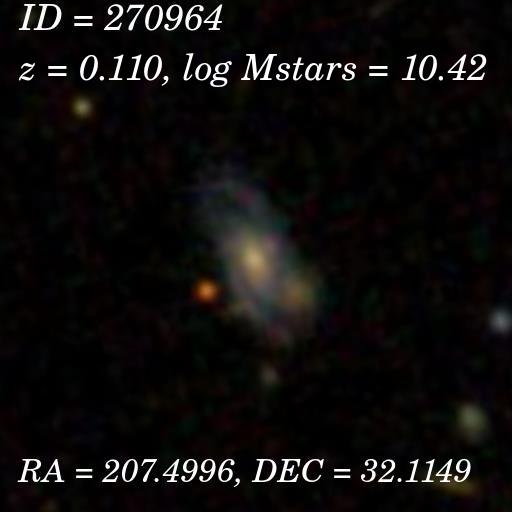} &
 \includegraphics[width=0.13\hsize]{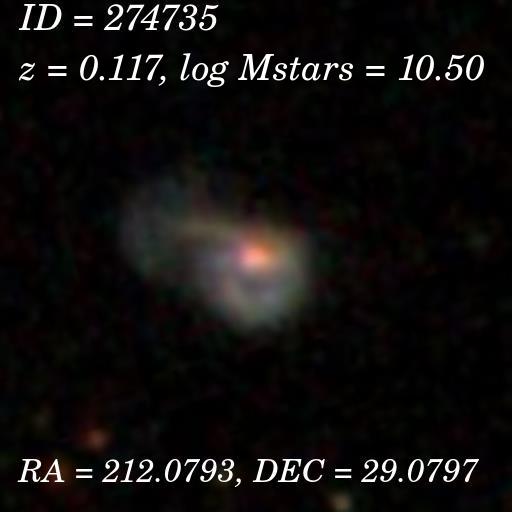} &
 \includegraphics[width=0.13\hsize]{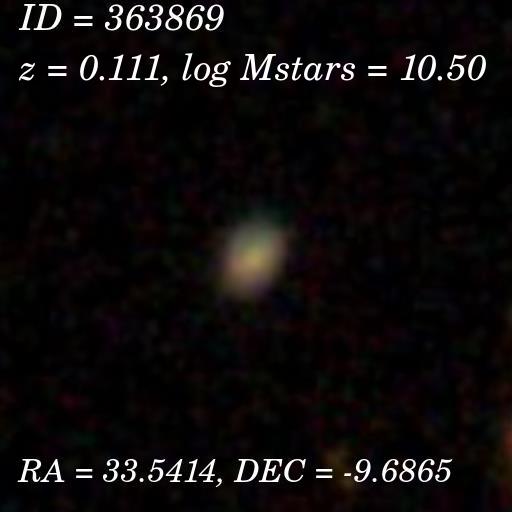} &
 \includegraphics[width=0.13\hsize]{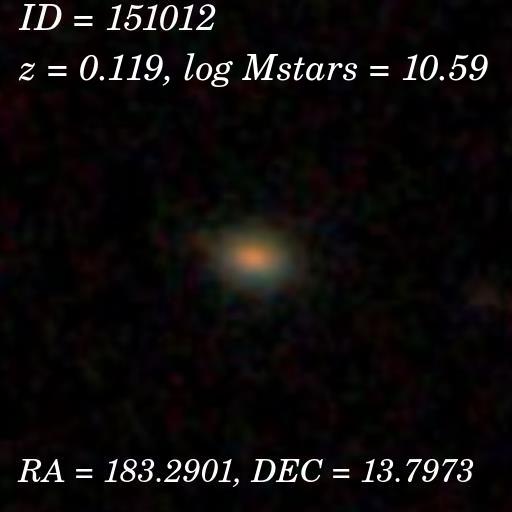} & & & \\
 
 \end{tabular}
 \caption{SDSS images of a sample of 207 VYGs. All images are 100 x 100 kpc (\emph{cont.}).}
\label{fig:vyg_images4}
\end{figure*}

Figs.~\ref{fig:vyg_images1} to \ref{fig:vyg_images4}
show the SDSS images of the 207 VYGs analysed in this study. Each
image displays the galaxy coordinates, redshift, and stellar masses derived with STARLIGHT using the \citet{Vazdekis+15} spectral models.

We provide the observed and derived properties of our VYGs and CSGs as supplementary data available online. The description of the table columns are given below.

\begin{description}
  \item {(1)} {\tt ObjID}: SDSS objID
  \item {(2)} {\tt specObjID}: SDSS specObjID
  \item {(3)} {\tt plate}: SDSS spectroscopic plate
  \item {(4)} {\tt mjd}: SDSS modified Julian date of spectroscopic observation
  \item {(5)} {\tt fiberID}: SDSS spectroscopic fibre
  \item {(6)} {\tt RA}: right ascension (J2000)
  \item {(7)} {\tt DEC}: declination (J2000)
  \item {(8)} {\tt z}: galaxy spectroscopic redshift
  \item {(9)} {\tt logMs\_Vaz15}: Present-day galaxy stellar mass obtained using the STARLIGHT code with \citet{Vazdekis+15} model
  \item {(10)} {\tt C}: galaxy concentration, defined as $\log (\theta_{90} / \theta_{50})$, where $\theta_{90}$ and $\theta_{50}$ are the radii containing 90\% and 50\% of the Petrosian flux in the $r$ band (SDSS parameters {\tt petroR90\_r} and {\tt petroR50\_r})
  \item {(11)} {\tt mu\_50}: galaxy surface brightness, computed using Eq.~\ref{eq:mu50}
  \item {(12)} {\tt logA1}: decimal logarithm  of the asymmetry parameter, computed using the code \textsc{Morfometryka}
  \item {(13)} {\tt logS1}: decimal logarithm  of the smoothness parameter, computed using the code \textsc{Morfometryka} 
  \item {(14)} {\tt logOH}: gas oxygen abundance $\log ({\rm O/H}) + 12$
  \item {(15)} {\tt f\_gas}: gas mass fraction
  \item {(16)} {\tt Av\_Balmer}: internal extinction in the $V$ band estimated from the Balmer decrement
  \item {(17)} {\tt Av\_cont}: internal extinction in the $V$ band estimated from the SPS analysis using STARLIGHT with \citet{Vazdekis+15} models
  \item {(18)} {\tt d\_1st\_kpc}: distance, in kpc, to the closest neighbour galaxy brighter than $m_r +1$, where $m_r$ is the $r$-band Petrosian magnitude of our galaxy
  \item {(19)} {\tt d\_1st\_norm}: distance, in units of $\theta_{50}$, to the closest neighbour galaxy brighter than $m_r +1$, where $m_r$ is the $r$-band Petrosian magnitude of our galaxy
\end{description}

\bsp	
\label{lastpage}

\end{document}